\documentclass[aps,prd,superscriptaddress,nofootinbib,amsfonts,amssymb,amsmath,notitlepage,twocolumn,10pt,floatfix]{revtex4-1}

\usepackage[dvipdfmx]{graphicx}
\usepackage{adjustbox}
\usepackage{amsmath}
\usepackage{amssymb} 
\usepackage{color}
\usepackage{xcolor}
\usepackage[colorlinks=true, allcolors=blue!75!black]{hyperref}
\usepackage{makecell}
\usepackage{multirow}
\usepackage{slashed}
\usepackage[caption=false]{subfig}
\usepackage{wrapfig}
\usepackage{placeins}
\usepackage{bm}
\usepackage[nameinlink]{cleveref}
\usepackage[caption=false]{subfig}

\crefname{section}{Sec.}{Secs.}
\crefname{subsection}{Sec.}{Secs.}
\crefname{equation}{Eq.}{Eqs.}
\crefname{figure}{Fig.}{Figs.}
\crefname{table}{Tab.}{Tabs.}

\Crefname{section}{Sec.}{Secs.}
\Crefname{subsection}{Sec.}{Secs.}
\Crefname{equation}{Eq.}{Eqs.}
\Crefname{figure}{Fig.}{Figs.}
\Crefname{table}{Tab.}{Tabs.}

\newcommand{\Slash}[1]{{\ooalign{\hfil/\hfil\crcr$#1$}}} 
\newcommand{\bvec}[1]{\mbox{\boldmath $#1$}}

\newcommand{\df}{\text{d}}

\newcommand{\pmat}[1]{\begin{pmatrix}#1\end{pmatrix}}

\graphicspath{{./figs/}}
\newbox{\ORCIDicon}
\sbox{\ORCIDicon}{\large
                  \includegraphics[width=0.8em]{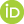}}

\allowdisplaybreaks[3]

\begin{document}

% \title{
% Chiral phase diagram with mesonic fluctuation: a symmetry-improved linear sigma model study
% }
\title{
Thermodynamics in symmetry-improved Cornwall-Jackiw-Tomboulis formalism: 
application to the low-energy effective theory of QCD 
}

\author{Yuepeng \surname{Guan}\,\href{https://orcid.org/0009-0007-8571-0931}{\usebox{\ORCIDicon}}}
\thanks{{\tt guanyp22@mails.jlu.edu.cn}} 
\affiliation{Center for Theoretical Physics and College of Physics, Jilin University, Changchun, 130012, China}

\author{Mamiya Kawaguchi\,\href{https://orcid.org/}{\usebox{\ORCIDicon}}}
\thanks{{\tt mamiya@aust.edu.cn}} 
\affiliation{ 
Center for Fundamental Physics, School of Mechanics and Physics,
Anhui University of Science and Technology, Huainan, Anhui 232001, People’s Republic of China
} 

\author{Shinya Matsuzaki\,\href{https://orcid.org/}{\usebox{\ORCIDicon}}}
\thanks{{\tt synya@jlu.edu.cn}}
\affiliation{Center for Theoretical Physics and College of Physics, Jilin University, Changchun, 130012, China}%

\author{Akio Tomiya\,\href{https://orcid.org/}{\usebox{\ORCIDicon}}}
\thanks{{\tt akio@yukawa.kyoto-u.ac.jp}} 
\affiliation{Department of Information and Mathematical Sciences, Tokyo Woman’s Christian  University, Tokyo 167-8585, Japan} 
\affiliation{RIKEN Center for Computational Science, Kobe 650-0047, Japan}

\begin{abstract}
We study the thermodynamics of the symmetry-improved Cornwall-Jackiw-Tomboulis (SICJT) formalism and apply it to a low-energy effective theory of QCD. 
In the symmetry-improved formulation, Ward-Takahashi identities are restored by auxiliary sources whose values are fixed self-consistently by the equilibrium state. 
While this construction improves the symmetry properties of the loop-wise truncated two-particle-irreducible (2PI) theory, it also makes the thermodynamic interpretation of the pressure nontrivial.
We formulate several pressure prescriptions, including the conventional vacuum-subtracted pressure, a source-matched subtraction, and a pulled-back pressure in which the explicit source-induced energy shift is removed.
Using the three-flavor linear sigma model with quarks, we analyze the equation of state, isentropic trajectories, adiabatic sound velocity, and trace anomaly across the chiral transition.
We find that the global thermodynamic structure is stable under the different prescriptions, while quantitative differences are concentrated near the crossover and first-order transition region. 
These results establish a practical framework for constructing thermodynamically consistent observables in symmetry-improved 2PI approaches.
\end{abstract}
\maketitle

\tableofcontents

% \clearpage
%%%%%%%%%%%%%%%%%%%%%%%%%%%%%%%%%%%%%%%%%%%%%%%%%%%%%%%%%%%%%%%%%%%%%%
\section{Introduction}
\label{sec:Introduction}

Understanding the thermodynamics of strongly interacting matter remains one of the central issues of quantum chromodynamics (QCD).
The QCD phase structure and equation of state are relevant to the description of the early Universe~\cite{Boyanovsky:2006bf, Addazi:2022whi}, compact stars~\cite{Fraga:2015xha, Fukushima:2025ujk}, and heavy-ion collisions~\cite{Foka:2016vta, Fukushima:2020yzx}, and they are closely tied to the restoration of chiral symmetry and the possible existence of critical phenomena at finite baryon density~\cite{Stephanov:1998dy, Stephanov:2008qz, Stephanov:2024xkn}.
At vanishing and small baryon chemical potential, lattice QCD has established that the thermal transition at the physical point is a crossover and has provided quantitatively precise information on the equation of state and related thermodynamic observables~\cite{Borsanyi:2012cr, Borsanyi:2013bia, HotQCD:2014kol, HotQCD:2018pds}.
At larger baryon density, however, first-principles calculations remain severely constrained by the sign problem, despite important progress based on Taylor expansion, analytic continuation, and related approaches~\cite{Nagata:2021ugx}.
This makes it necessary to complement lattice results with continuum and effective-theory methods when exploring the thermodynamics and phase structure of QCD away from $\mu_B=0$~\cite{Fukushima:2010bq}.

Low-energy effective theories provide one such framework.
% Models such as the Nambu-Jona-Lasinio model~\cite{Nambu:1961tp, Nambu:1961fr, Klevansky:1992qe, Vogl:1991qt}, the linear quark-meson model~\cite{Tetradis:2003qa, Schaefer:2008hk, Schaefer:2009st, Tripolt:2017zgc}, and the chiral perturbation theory~\cite{Gasser:1983yg, Scherer:2005ri} encode the chiral symmetry structure of QCD and have been widely used to study the phase diagram, critical behavior, and thermodynamic observables in regions where direct lattice simulations are difficult.
% Their attraction is twofold.
% First, they allow us to analyze the interplay between condensates, fluctuations, and the phase structure in a technically controllable setting.
% Second, they offer a useful laboratory in which one can test nonperturbative approximation schemes before embedding them into more microscopic approaches.
% At the same time, these theories are not intended to reproduce the full ultraviolet (UV) completion of QCD, and 
In such theories, thermodynamic predictions actually 
depend sensitively on how quantum and thermal fluctuations are incorporated~\cite{Braun:2018svj}. 
% The latter point is 
This point is important because mean-field and naive one-loop approximations are often insufficient when thermodynamic precision is required.
Near a crossover or phase transition, fluctuation effects may substantially modify the location and even the order of the transition~\cite{Zhang:2017icm, Osman:2024xkm, Drews:2013hha}.
Moreover, derived thermodynamic observables such as the sound velocity or the trace anomaly can be particularly sensitive to the approximation scheme and model setup~\cite{Fu:2018qsk, Hansen:2019lnf}.

In general, finite-temperature field theory often requires selective resummation beyond strict loopwise perturbation theory to obtain physically meaningful results~\cite{Andersen:2004fp, Curtin:2016urg, Niemi:2021qvp}.
This issue is familiar not only in QCD-motivated effective theories, but also in other contexts, such as the electroweak phase transition, where higher-order thermal resummation and self-consistent treatments can lead to sizable quantitative corrections relative to one-loop analyses~\cite{Niemi:2021qvp}.
%These considerations motivate the use of self-consistent resummation schemes in the study of QCD %thermodynamics as well. 

%Working with the low-energy effective theories, a candidate framework incorporating the non-perturbative dynamics in the fluctuations is provided by the 
This resummation is built in the Cornwall-Jackiw-Tomboulis (CJT) formalism, which is 
%
%effective action, or equivalently 
the two-particle-irreducible (2PI) formalism~\cite{Cornwall:1974vz}.
As a {\it $\Phi$-derivable approximation scheme} in the sense of Baym~\cite{Baym:1962sx}, the CJT formalism resums selected classes of diagrams self-consistently, and has been extensively used in finite-temperature field theory and many-body physics.
In the context of chiral effective theories, it allows one to incorporate mesonic fluctuations beyond the mean field in a manner that remains thermodynamically meaningful at the level of the underlying approximation.
A well-known drawback, however, is that a loopwise truncation of the 2PI effective action generally violates Ward-Takahashi identities (WTIs) and may therefore lead to spurious violations of global symmetries, most notably the appearance of massive would-be Goldstone modes in the spontaneously broken phase~\cite{Petropoulos:1998gt, Lenaghan:1999si, Lenaghan:2000ey, Roder:2003uz, Pilaftsis:2013xna}.

The symmetry-improved CJT (SICJT) formalism proposed in Ref.~\cite{Pilaftsis:2013xna} addresses this problem by supplementing the truncated 2PI equations with additional constraints derived from the relevant Ward identities.
This construction restores important symmetry properties and has been shown to cure the Goldstone-theorem problem in simple scalar theories, including the $O(4)$ linear sigma model (LSM)~\cite{Pilaftsis:2013xna, Mao:2013gva}.
It has also been extended to the Standard Model in the gaugeless limit, where the symmetry-improved 2PI approach provides an infrared (IR)-safe treatment of the Goldstone sector beyond conventional perturbation theory~\cite{Pilaftsis:2015bbs}.
From the perspective of QCD-inspired effective theories, the SICJT formalism is therefore particularly attractive: it preserves the nonperturbative advantages of the 2PI resummation while maintaining the symmetry constraints that are essential for a physically meaningful chiral description.

Nevertheless, once one turns to thermodynamics, the symmetry-improved formulation involves an additional conceptual issue.
In SICJT, the symmetry constraints are enforced along with the addition of auxiliary source terms, whose values are determined self-consistently together with the equilibrium state~\cite{Garbrecht:2015cla}.
Consequently, the thermodynamic interpretation of the pressure is no longer unique.
In particular, different subtraction prescriptions and different treatments of variation with respect to thermodynamic quantities, which we shall, in short, call thermodynamic derivatives, may lead to inequivalent results once the sources depend implicitly on the temperature and baryon chemical potential.
This ambiguity would be significant even at the level of the stationary equations themselves, but it becomes nontrivial and more crucial for the equation of state and for derived observables that probe thermodynamic derivatives, such as isentropic trajectories, the adiabatic sound velocity, and the trace anomaly.
Clarifying this ambiguity is necessary if the SICJT formalism is to be
used as a thermodynamically controlled framework for QCD-motivated
effective theories. 
In particular, it is important to distinguish the
intrinsic uncertainty associated with the symmetry-improving sources
from the model dependence associated with the microscopic degrees of
freedom and the truncation scheme.

In this work, we study this intrinsic issue of the SICJT formalism systematically in the three-flavor LSM with quark degrees of freedom, often dubbed the linear quark-meson model.
We formulate and compare several prescriptions for the pressure in the SICJT framework, including the conventional vacuum-subtracted pressure, a source-matched vacuum subtraction, and a pulled-back pressure defined from a ``pulled-back'' effective action in which the explicit source-induced linear tilt is removed.
%We further distinguish between total derivatives taken along the symmetry-improved manifold and %thermodynamic derivatives at fixed external source, and state that the latter provide the physically %meaningful definition of the equation of state in the present setup.
Using these prescriptions, we analyze the pressure as well as the equations of state, isentropic trajectories, adiabatic sound velocity, and trace anomaly, and compare them with the corresponding results in the conventional CJT formalism.
Our results show that the thermodynamic structure is robust overall in the thermal evolution, whichever pressure prescriptions are applied, whereas quantitative differences can be seen near the crossover and first-order transition region, where the treatment of the added sources is most consequential.

This paper is organized as follows.
In \Cref{sec:LinearSigmaModelWithQuarks}, we introduce the three-flavor LSM with quarks used as the low-energy effective theory of QCD.
In \Cref{sec:2PIFormalismAndSymmetryImprovement}, we summarize the CJT formalism and its symmetry-improved extension. %, and also other possible approximation schemes below these levels.
\Cref{sec:Thermodynamics} is devoted to the construction of the self-consistent SICJT effective action, and related thermodynamics, with particular remarks on the pressure prescriptions and the definition of thermodynamic derivatives.
In \Cref{sec:Results}, we present the numerical analysis of the chiral phase structure and thermodynamics, including the pressure, isentropic trajectories, adiabatic sound velocity, and trace anomaly.
Finally, \Cref{sec:Conclusion} contains our conclusions and outlook.

%%%%%%%%%%%%%%%%%%%%%%%%%%%%%%%%%%%%%%%%%%%%%%%%%%%%%%%%%%%%%%%%%%%%%%
\section{Three-flavor linear sigma model with quarks}
\label{sec:LinearSigmaModelWithQuarks}

In this section, we introduce the low-energy effective description of QCD that will serve as a reference model for our general analysis of symmetry-improved thermodynamics.
To this end, we consider the three-flavor LSM and discuss its thermodynamic properties within the 2PI framework based on the symmetry-improved CJT formalism.

The Lagrangian of this model in Euclidean spacetime is given by
\begin{align}
    \mathcal{L}_{qm} = \mathcal{L}_{q} + \mathcal{L}_{m},
\end{align}
where the quark and meson parts are given by
\begin{gather}
    \mathcal{L}_{q} = \bar q_i \Big[ \Slash{\partial} \delta_{ij} + \mu_{ij} \gamma_4 + g 
    %\big( T^a \big)_{ij} (\sigma_a + i \gamma_5 \pi_a) \Big] 
    \Phi_{ij} P_R + {\rm h.c.} \Big]
    q_j, \nonumber\\
    \mathcal{L}_{m} = \operatorname{tr} \Big[ \partial_\mu \Phi^\dagger \partial_\mu \Phi \Big] + V_m\big(\Phi,\Phi^\dagger\big) \ ,
    \label{eq:lagrangianQandM}
\end{gather}
where $\Slash{\partial} = \gamma_\mu \partial_\mu$ is the Dirac operator in Euclidean spacetime, $P_{L/R}$ is the left-/right-hand projection matrix in the Dirac space, and $i,j = 1,2,3$ stands for the flavor indices.
In \Cref{eq:lagrangianQandM}, the quark field content contains three flavors of quarks $q_i = (u^{\mathsf T}, d^{\mathsf T}, s^{\mathsf T})^{\mathsf T}$, and the mesonic field content is given by $\Phi = \phi_a T^a = ( \sigma_a + i \pi_a ) T^a$, in which $\sigma_a$ and $\pi_a$ denote the scalar and pseudo-scalar fields, respectively. 
%In the context, 
Hereafter, we shall use $\varphi_a$ to represent either the scalar $\sigma_a$ or the pseudo-scalar $\pi_a$ for convenience. 
The matrices $T^a = \lambda^a/2$ are the generators of $U(3)$ in the fundamental representation, in which $\lambda^{a=1,\cdots,8}$ are the Gell-Mann matrices, and $\lambda^0 = \sqrt{2/3} \, \textbf{1}_{3 \times 3}$.
Those generators satisfy the following algebraic relations
\begin{gather}
    \operatorname{tr}[T^a T^b] = \frac{1}{2} \delta^{ab}, \nonumber\\
    [T^a, T^b] = i f^{ab}_{\quad c} T^c, \nonumber\\
    \{ T^a, T^b \} = d^{ab}_{\quad c} T^c,
\end{gather}
where $f^{ab}_{\quad c}$ and $d^{ab}_{\quad c}$ denote the antisymmetric and symmetric structure constants for the $SU(3)$ group (for $a,b,c = 1, \cdots, 8$), while $f^{ab}_{\quad 0} = 0$ and $d^{ab}_{\quad 0} = \sqrt{2/3} \delta^{ab}$ are satisfied for the zeroth component.
The quark chemical potential matrix $\mu_{ij}$ is chosen as
\begin{align}
    \mu_{ij} = \operatorname{diag} \{ \mu_B/3, \mu_B/3, \mu_B/3 \} \ ,
\end{align}
where $\mu_B$ is the baryon chemical potential.

The mesonic potential $V\big(\Phi,\Phi^\dagger\big)$ in \Cref{eq:lagrangianQandM} is given by
\begin{align}
    V_m\big(\Phi,\Phi^\dagger\big) = V_0 + V_{\rm SB} + V_{\rm anom}. 
    \label{eq:LSMTreePotential}
\end{align}
%
%where individually, 
The $SU(3)_L \times SU(3)_R$ symmetric part $V_0$ is given by
\begin{align}
    V_0 = \mu^2 \operatorname{tr} \big[ \Phi^\dagger\Phi \big] + \lambda_1 \operatorname{tr} \big[ \big( \Phi^\dagger\Phi \big)^2 \big] + \lambda_2 \big( \operatorname{tr} \big[ \Phi^\dagger\Phi \big] \big)^2 \, ,
\end{align}
which encodes the information of the underlying QCD dynamics within this truncated mesonic effective description.
The explicit chiral breaking part $V_{\rm SB}$ is: 
\begin{align}
    V_{\rm SB} = -c \operatorname{tr} \big[ \mathcal{M} \Phi^\dagger + \mathcal{M}^\dagger \Phi \big],
    \label{eq:TadpoleTerm}
\end{align}
where $\mathcal{M} \equiv \operatorname{diag} \{ m_u, m_d, m_s \}$ to incorporate the proper explicit-chiral breaking into mesons arising from the current quark masses present in QCD. 
The $U(1)_A$ anomaly part is: 
\begin{align}
    V_{\rm anom} = - B \big( \det\big[ \Phi \big] + \det\big[ \Phi^\dagger \big] \big),
\end{align}
%
% which is the so-called Kobayashi-Maskawa-’t\,Hooft (KMT) term.
which represents the anomaly determinant term dubbed the
Kobayashi--Maskawa--'t\,Hooft (KMT) interaction~\cite{Kobayashi:1970ji,Kobayashi:1971qz,tHooft:1976rip,tHooft:1976snw}.

When the 't\,Hooft fermion determinant vertex~\cite{tHooft:1976snw} is present in the underlying quark theory, the KMT term can be understood as the leading contribution obtained after
bosonization, e.g., in the stationary-phase approximation.
For more details on the bosonization of the 't\,Hooft fermion determinant interaction and the induced mesonic effective interactions, we refer readers
to Refs.~\cite{Osipov:2002wj, Osipov:2003xu, Osipov:2006xa}.

Given the LSM with quarks, the dynamical chiral symmetry breaking (D$\chi$SB) is characterized by the expectation value of the order parameters defined by the flavor-chiral transformations 
\begin{gather}
    \Phi \rightarrow g_L \cdot \Phi \cdot g_R^\dagger, \nonumber\\
    q_{L/R} \rightarrow g_{L/R} q_{L/R},
    \label{eq:chiralTransformation}
\end{gather}
where $g_{L/R} \in SU(3)_{L/R}$ is the element of the fundamental representation, with $\theta_L = - \theta_R \equiv \theta_A$. 
The order parameters are then parameterized as
\begin{align}
    \langle \Phi \rangle \equiv \bar \Phi = \bar \sigma_a T^a,
    \label{eq:MeanFieldBackground}
\end{align}
without loss of generality. 
In the present model, the isospin symmetry is adopted in the current work via the quark mass difference, i.e., $m_u = m_d \equiv m_l \neq m_s$.  
Consequently, the vacuum expectation values (VEVs) of the $\sigma$ fields in Eq.(\ref{eq:MeanFieldBackground}) read 
\begin{align}
    \bar \sigma_a T^a = \bar \sigma_0 T^0 + \bar \sigma_8 T^8 = \operatorname{diag} \{ \bar \Phi_1, \bar \Phi_1, \bar \Phi_3 \},
\end{align}
with the following relations
\begin{align}
    \bar \Phi_1 &= \frac{1}{\sqrt{6}} \bar \sigma_0 + \frac{1}{2\sqrt{3}} \bar \sigma_8, \nonumber\\
    \bar \Phi_3 &= \frac{1}{\sqrt{6}} \bar \sigma_0 - \frac{1}{\sqrt{3}} \bar \sigma_8.
    \label{eq:FieldBasisTrans}
\end{align}
The chiral order parameters of the underlying theory are the light- and strange-quark condensates.
In the present low-energy description, they are represented by the background fields $\bar \Phi_1$ and $\bar \Phi_3$ as
\begin{align}
    &\langle \bar \ell \ell \rangle = \left\langle \frac{\partial \mathcal L_{qm}}{\partial m_l} \right\rangle = -2 c \bar\Phi_1 \, , 
    \nonumber\\
    &\langle \bar s s \rangle = \left\langle \frac{\partial \mathcal L_{qm}}{\partial m_s} \right\rangle = -2 c \bar\Phi_3 \, .
\label{eq:QuarkCondensates}
\end{align}

Around the background field $\bar \Phi$, the tree-level equations of motion of mesons are obtained as
\begin{align}
    \frac{\delta S_{m}}{\delta \Phi_1} \Biggl|_{\Phi = \bar\Phi} = 0, \qquad \frac{\delta S_{m}}{\delta \Phi_3} \Biggl|_{\Phi = \bar\Phi} = 0.
    \label{eq:StationaryCondAtTree}
\end{align}
where $S_{m/q} = \int_x \mathcal L_{m/q}$ is the mesonic/quarkonic part of the Euclidean action.
Hereafter, we use the shorthand notation for the integral in the Euclidean spacetime, which reads
\begin{align}
    \int_x \equiv \int_0^\beta \df \tau \int \df^3 x \, , \quad \int_{x,y} \equiv \int_x \int_y \, ,
\end{align}
where $\beta \equiv T^{-1}$, and $T$ denotes the temperature.
The explicit form of the tree-level mesonic potential $V_m$ expanded around the $(\bar\sigma_0, \bar\sigma_8)$-background and the stationary conditions~\eqref{eq:StationaryCondAtTree} can be found in \Cref{app:StationaryConditionsTreeLevel}. 

The mesonic curvature masses are given by
\begin{align}
    \Big[ m_S^2(\bar\sigma) \Big]^{ab} &= \frac{\partial^2 V_{m}}{\partial \sigma_a \partial \sigma_b} \Biggl|_{\sigma_a = \bar\sigma_a}, \nonumber\\
    \Big[ m_P^2(\bar\sigma) \Big]^{ab} &= \frac{\partial^2 V_{m}}{\partial \pi_a \partial \pi_b} \Biggl|_{\sigma_a = \bar\sigma_a}.
    \label{eq:mesonMassSpectraTree}
\end{align}
To obtain the physical meson mass spectra, the $(0,8)$ components of mass matrices in \Cref{eq:mesonMassSpectraTree} are diagonalized by the mixing angles $\theta_S^0$ and $\theta_P^0$.
The explicit form of the curvature masses~\eqref{eq:mesonMassSpectraTree}, the expression of the mixing angles, and the correspondence between the mass matrix elements and the physical meson masses are provided in  \Cref{app:CurvatureMassesAtVacuum}.

The tree-level two-point function of quarks in momentum space reads
\begin{align}
    &\big( \Delta^q_0(p,q) \big)^{-1}_{ij} \nonumber\\
    &\qquad \equiv \frac{\delta^2 S_q}{\delta \bar q_i(p) \delta q_j(q)} \nonumber\\
    &\qquad = \delta_{p,q} \big[ i \Slash{p} \delta_{ij} + \mu_{ij} \gamma_4 + g \,  (\sigma_a + i \gamma_5 \pi_a) \big( T^a \big)_{ij} \big] \, ,
    \label{eq:quarkPropagatorInverse}
\end{align}
where $\delta_{p,q}$ is the delta function for energy-momentum conservation in $3+1$-dimensional spacetime, and $\Slash{p} = p_\mu \gamma_\mu$.

%%%%%%%%%%%%%%%%%%%%%%%%%%%%%%%%%%%%%%%%%%%%%%%%%%%%%%%%%%%%%%%%%%%%%%
\section{CJT formalism and symmetry improvement}
\label{sec:2PIFormalismAndSymmetryImprovement}

\subsection{Construction of the mesonic CJT effective action}

\subsubsection{Mesonic 2PI effective action}

In the current work, we utilize the CJT formalism with the 2PI effective action and discuss its symmetry-improvement to study the thermodynamic properties of the model formulated in \Cref{sec:LinearSigmaModelWithQuarks}.
To this end, we introduce the local sources
\begin{align}
    j = \pmat{J(x), &J^\dagger(x), &\eta(x), &\bar \eta(x)}
    \label{eq:generalSourceContent}
\end{align}
coupled with the field contents 
\begin{align}
\mathcal X_m = \pmat{\Phi^\dagger(x), &\Phi(x), &\bar q(x), &q(x)},
\end{align}
where $m$ denotes the general field indices. 
We also consider the following bilocal source coupled with the mesonic field: 
\begin{align}
    \Phi^\dagger \cdot K \cdot \Phi = \int_{x,y} \operatorname{tr} \Big[ \Phi^\dagger(x) K(x,y) \Phi(y) \Big],
\end{align}
where the dot between functions denotes the contraction in the spacetime.
The 2PI effective action is defined as the double-Legendre transformation
\begin{align}
    \Gamma[\mathcal X_{\rm cl},\Delta] = \sup_{j,K} &\biggl[ j^m \cdot \mathcal X_{{\rm cl},m} + \frac{1}{2} K^{mn} \cdot \Big( \Delta_{mn} + \mathcal X_{{\rm cl},m} \mathcal X_{{\rm cl},n} \Big) \nonumber\\
    &\hspace{5ex} - W[j,K] \biggl] \ ,
    \label{eq:define2PIEffectiveAction}
\end{align}
where $\mathcal X_{{\rm cl}} \equiv \langle \mathcal X \rangle$ denotes the external field background, and $\Delta$ the undetermined meson propagators. 
We shall drop the subscripts of the background fields and simply use $\chi$ or $\phi$ in addressing the effective actions, just for readability. 
In \Cref{eq:define2PIEffectiveAction}, we have used the same convention as in Ref.~\cite{Blaizot:2021ikl}, so that 
\begin{align}
    W[j,K] = \log Z[j,K] \ ,
\end{align}
where $Z[j,K]$ is the generating functional.

Since the action $S_{qm}$ is quadratic in the quark fields, we shall integrate them before introducing the 2PI effective potential.
At the level of the path integral, this reads
\begin{align}
    Z &= \int \mathcal D \Phi \, \mathcal D \Phi^\dagger  \, e^{-S_m[\Phi, \Phi^\dagger]} \int \mathcal D q \, \mathcal D\bar q \, e^{-S_q[q, \bar q, \Phi, \Phi^\dagger]}
    \nonumber\\
    &= \int \mathcal D \Phi \, \mathcal D \Phi^\dagger  \, e^{-S_m[\Phi, \Phi^\dagger]- \Delta \Gamma_{\bar q q}[\Phi, \Phi^\dagger]} 
    \nonumber\\
    &\equiv \int \mathcal D \Phi \, \mathcal D \Phi^\dagger  \, e^{-S_m^{\rm eff}[\Phi, \Phi^\dagger]} \, ,
    \label{eq:generatingFUnctionalAfterIntegrateQuarks}
\end{align}
where
\begin{align}
    \Delta \Gamma_{\bar q q}[\Phi, \Phi^\dagger] = - \log \operatorname{Det} \big[ ( \Delta_0^q[\Phi,\Phi^\dagger])^{-1} \big]\,. 
\end{align}
%
%is the logarithm of the Dirac determinant, 
%which contributes to the UV action before integrating out the mesonic fluctuations.
%Concerning the above path integral, which only consists of the mesonic fluctuation explicitly, we shall replace
With this prescription, the field content is now only the mesons, so that in the resulting mesonic 2PI formulation, we make the following replacement: 
\begin{align}
    \Gamma[\mathcal X,\Delta] \to \Gamma[\phi,\Delta]
\,. 
\end{align}
We define the short-hand notations for the mesonic momentum integral
\begin{align}
    \int_k f(k) \equiv T \sum_n\int \frac{\df^3 \bvec{k}}{(2\pi)^3} f(\omega_n,\bvec{k}) \, ,
    \label{eq:mesonicMomentumIntegral}
\end{align}
and for the fermionic momentum integral
\begin{align}
   \int_p f(p) \equiv T \sum_n\int \frac{\df^3 \bvec{p}}{(2\pi)^3} f(\tilde \nu_n,\bvec{p}) \, ,
   \label{eq:fermioinicMomIntegral}
\end{align}
where $\omega_n = 2n\pi T$ and $\tilde \nu_n = (2n+1)\pi T - i\mu_f$ are the bosonic and fermionic Matsubara frequencies, respectively. 

%%%%%%%%%%%%%%
%%%%%%%%%%%%%%
\begin{figure}[t!]
    \centering
    \includegraphics[width=0.7\linewidth]{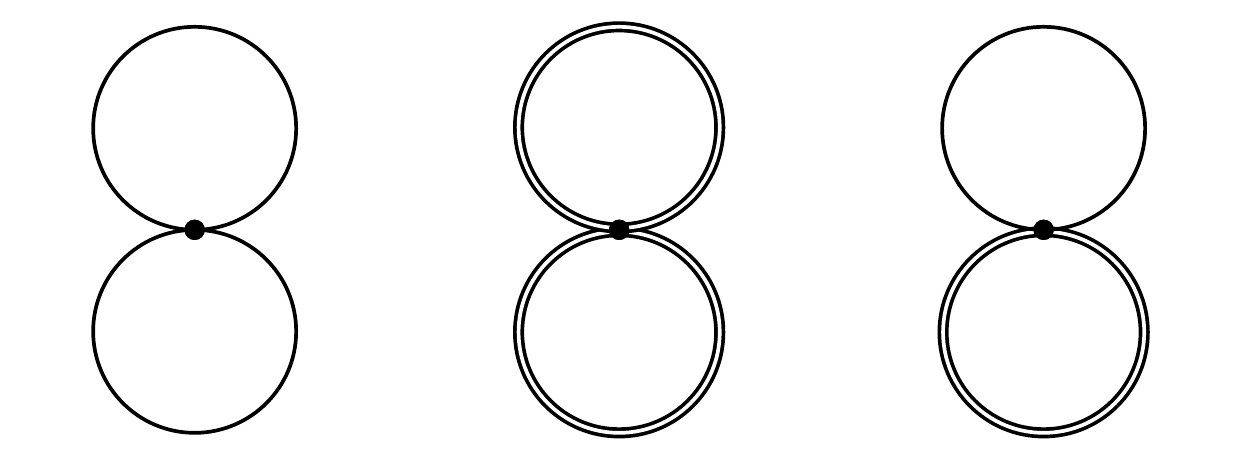}
    \caption{
    Feynman diagrams contributing to $V_2$ in \Cref{eq:V2UnderHartree}.
    The solid lines represent the scalar propagators $[G_S]_{ab}$, and the double-solid lines the pseudo-scalar propagators $[G_P]_{ab}$.
    The black dots denote the bare vertices, which are expressed as $\mathcal{F}^{abcd}$ and $\mathcal{H}^{abcd}$ in the text, respectively. 
    }
    \label{fig:HartreeApprox}
\end{figure}
%%%%%%%%%%%%%%
%%%%%%%%%%%%%%

Using \Cref{eq:fermioinicMomIntegral}, and 
taking the leading order of the derivative expansion for the one-particle-irreducible (1PI) effective potential, the quark contribution reads~\cite{Kapusta:2006pm, Schaefer:2008hk}
\begin{align}
    \Delta \Gamma_{\bar q q} &= \int_x  \Omega_{\bar q q}(\Phi,\Phi^\dagger) + \cdots 
    \, , \notag \\ 
%\end{align}
%the quark one-loop integral of the effective potential reads
%
%\begin{align}
{\rm with} \qquad 
    \Omega_{\bar q q} 
&= - \int_p \operatorname{tr} \log \big( \Delta^q_0(p) \big)^{-1} \nonumber\\
    &= \nu_c \sum_{f = u,d,s} \int_0^\infty \frac{\df^3 \bvec{p}}{(2\pi)^3} \biggl\{ T \log \Big[ 1 - n_{q,f}(T,\mu_f) \Big] \nonumber\\
    &\qquad + T \log \Big[ 1 - n_{\bar q,f}(T,\mu_f) \Big] + E_{q,f} \biggl\}, 
    \label{eq:Omegaqq}
\end{align}
where $\nu_c = 2 N_c = 6$ is the internal quark degrees of freedom, and 
\begin{align}
    n_{q,f}(T,\mu_f) = \Big[ 1 + \exp\{ (E_{q,f} - \mu_f)/T \} \Big]^{-1}
\end{align}
is the Fermi-Dirac distribution function, and $n_{\bar q,f}(T,\mu_f) = n_{q,f}(T,-\mu_f)$. 
$E_{q,f}$ is the single-particle dispersion relation for the fermions given by
\begin{align}
    E_{q,f} = \sqrt{|\bvec{p}|^2 + m_f^2},
\end{align}
where the flavor-dependent mass $m_f^2$ is evaluated from the eigenvalues of the meson field-induced mass matrix  
\begin{align}
    M^2_{ij} &= g^2 \big( T^a \big)_{ik} (\sigma_a + i \gamma_5 \pi_a)\big( T^b \big)_{kj} (\sigma_a - i \gamma_5 \pi_a) \nonumber\\
    &= g^2 \, P_R \, \big[ \Phi^\dagger \Phi \big]_{ij} + g^2 \, P_L \, \big[ \Phi \Phi^\dagger \big]_{ij},
\end{align}
%
%where 
with $P_{L/R} = (1 \mp \gamma_5)/2$ being 
%stands for 
the left/right-hand projectors in the Dirac spinor space. 
The constituent quark masses induced from the mesonic mean-field background read
\begin{align}
    M_l = g \, \bar \Phi_1, \quad M_s = g \, \bar \Phi_3 \ ,
\end{align}
where the subscripts $l$ and $s$ stand for the light and strange quarks, respectively.

Regarding $S_m^{\rm eff}$ in \Cref{eq:generatingFUnctionalAfterIntegrateQuarks} as the ultraviolet (UV) action, the mesonic CJT potential with quark loop contribution, in the presence of the homogeneous mean-field background in Eq.~\eqref{eq:MeanFieldBackground}, is read as the loop expansion:  
\begin{align}
    &V_{\rm CJT}[\sigma, G_S, G_P] \nonumber\\
    =& V(\sigma) + \frac{1}{2} \int_k \operatorname{tr} \Big[ \log G_S^{-1}(k) + \log G_P^{-1}(k) \Big] \nonumber\\
    &+ \frac{1}{2} \int_k \operatorname{tr} \Big[ \bar G_S^{-1}(k;\sigma) G_S(k) + \bar G_P^{-1}(k;\sigma) G_P(k) - 2 \cdot \textbf{1}_{\rm adj} \Big] \nonumber\\
    &+ V_2[\sigma,G_S, G_P] + \Omega_{\bar q q}(\sigma),
    \label{eq:GeneralCJTPotnetialMainText}
\end{align}
where $\sigma$ denotes the homogeneous mean-field background, and $\textbf{1}_{\rm adj}$ is the unit matrix in the adjoint representation of $U(3)$ having the size of $N_f^2 \times N_f^2$, and the scalar (pseudo-scalar) propagator is represented as $G_S$ ($G_P$). These $G_S$ and $G_P$ are treated as independent variables in the current formalism. 
%In \Cref{eq:GeneralCJTPotnetialMainText}, 
$\bar G_S^{-1}(k;\sigma)$ and $\bar G_P^{-1}(k;\sigma)$ are the mesonic inverse-propagators constructed from the tree-level meson terms and the quark one-loop contributions, which read  
\begin{align}
    \Big[ \bar G_S^{-1}(k;\sigma) \Big] ^{ab}&= k^2 \delta^{ab} + \Big[ m_S^2(\sigma) \Big]^{ab} + \frac{\partial^2 \Omega_{\bar q q}}{\partial \sigma_a \partial \sigma_b} \Biggl|_{\Phi = \bar\Phi}, \nonumber\\
    \Big[ \bar G_P^{-1}(k;\sigma) \Big]^{ab} &= k^2 \delta^{ab} + \Big[ m_P^2(\sigma) \Big]^{ab} + \frac{\partial^2 \Omega_{\bar q q}}{\partial \pi_a \partial \pi_b} \Biggl|_{\Phi = \bar\Phi}.
    \label{eq:TreeLevelMesonTwoPoint}
\end{align}
The explicit form of the derivatives of $\Omega_{\bar q q}$ can be found in \Cref{app:quarkloopCOntributions}.

For the 2PI diagram contributions, $V_2$ present in \Cref{eq:GeneralCJTPotnetialMainText}, we adopt the Hartree approximation, which is truncated at the two-loop level as follows: 
\begin{align}
    V_2[G_S, G_P] &= \mathcal{F}^{abcd} \int_k [G_S(k)]_{ab} \int_q [G_S(q)]_{cd}
    \nonumber\\
    &+ \mathcal{F}^{abcd} \int_k [G_P(k)]_{ab} \int_q [G_P(q)]_{cd}
    \nonumber\\
    &+ 2 \mathcal{H}^{abcd} \int_k [G_S(k)]_{ab} \int_q [G_P(q)]_{cd},
    \label{eq:V2UnderHartreeMainText}
\end{align}
where $\mathcal{F}^{abcd}$ and $\mathcal{H}^{abcd}$ consist of the four-meson couplings $\lambda_1$ and $\lambda_2$, which are defined in \Cref{eq:definitionOfHandF}.
The corresponding Feynman diagrams are listed in \Cref{fig:HartreeApprox}.

%With the setup given as above, the daisy-type diagrams are resummed systematically by %solving the gap equation of the meson propagators, i.e., 
%\Cref{eq:gapEquationsInCJTMainText}.

\subsubsection{Self-consistency of the quark sector}

Here, we further comment on the approximation scheme adopted in the present formulation.
%A possible source of confusion is that 
%One might think somewhat self-inconsistent that 
First of all, note that the quark propagator $\Delta_0^q$ in Eq.(\ref{eq:Omegaqq}) keeps the 
background-field dependence of $\sigma$, which is evaluated at tree-level. 
%form in the mesonic CJT effective action. 
This is understood as a consequence of the two-step construction of the generating
functional: the quark fields are first integrated out for a fixed mesonic configuration,
and the resulting mesonic theory is then treated within the CJT formalism. 

To see this more explicitly, let us temporarily reintroduce the Grassmann-valued sources $\eta$ and $\bar\eta$ coupled to the quark fields. 
For a fixed mesonic configuration, the quark part is Gaussian, and the quark propagator is given by the inverse of the corresponding Dirac operator. 
After integrating out the quarks, the full quark two-point function can still be reconstructed as an insertion in the remaining mesonic path integral:
\begin{align}
    &\big[\Delta^q(x,y)\big]_{ij}
    \nonumber\\
    &=
    \frac{1}{Z}
    \frac{\overrightarrow{\delta}}{\delta \bar\eta_i(x)}
    Z[\eta,\bar\eta]
    \frac{\overleftarrow{\delta}}{\delta \eta_j(y)}
    \bigg|_{\eta,\bar\eta=0}
    \nonumber\\
    &=
    \frac{1}{Z}
    \int \mathcal D\Phi\,\mathcal D\Phi^\dagger\,
    \det\!\left[\Delta_0^{q,-1}[\Phi,\Phi^\dagger]\right]\,
    e^{-S_m[\Phi,\Phi^\dagger]}
    \nonumber\\
    &\qquad\times\big[\Delta_0^q[\Phi,\Phi^\dagger](x,y)\big]_{ij}
    \nonumber\\
    &\equiv
    \Big\langle
    \big[\Delta_0^q[\Phi,\Phi^\dagger](x,y)\big]_{ij}
    \Big\rangle_{\phi}.
    \label{eq:quantumDressedQuarkPropagator}
\end{align}
Here $\langle \cdots \rangle_\phi$ denotes the expectation value with respect to the effective mesonic weight after the quarks have been integrated out. 
This expression makes clear that the full quark propagator is not simply the inverse Dirac operator evaluated at the mean meson field. 
Rather, it is the mesonic-fluctuation average at the fixed-background $\phi$.
To connect this expression with the usual Dyson-Schwinger form, we decompose the mesonic field into its expectation value and fluctuation,
\begin{align}
    \Delta_0^{q,-1} = i \slashed{\tilde p} + g \bar \phi + g \tilde \phi 
    \equiv \Delta_0^{q,-1}[\bar \phi] + g \tilde \phi \, ,
    \label{eq:fluctuationFieldSeperation}
\end{align}
where $\bar \phi = \langle \phi \rangle_\phi$ is the mesonic background field, and $\langle \tilde \phi \rangle_\phi = 0$.
In \Cref{eq:fluctuationFieldSeperation}, we have omitted the group structure of the mesonic fields for the sake of clarity.
Noting the following identity
\begin{align}
    \left( \Delta_0^{q,-1}[\bar \phi] + g \tilde \phi \right) \Delta_0^{q} = \textbf{1} \, ,
\end{align}
and taking the average among the mesonic fluctuations, one obtains
\begin{align}
    \Delta_0^{q,-1}[\bar \phi] \Delta^{q} + g  \big \langle \tilde \phi \, \Delta^{q}_0 \big \rangle_\phi = \textbf{1} \, .
    \label{eq:earlyDSE}
\end{align}
The second term on the left-hand side of \Cref{eq:earlyDSE} is read as the response of the quark propagator to the source of the fluctuating meson field.
Thus, it is the connected three-point function in Euclidean spacetime, 
\begin{align}
    \big \langle \tilde \phi \, \Delta^{q}_0 \big \rangle_\phi = - \Delta^\phi \, \Delta^q \, \Gamma^{(3)}_{\bar q \phi q} \, \Delta^q \, ,
    \label{eq:threePointFunctionKernel}
\end{align}
where $\Delta^\phi$ is the full mesonic propagator, and $\Gamma^{(3)}_{\bar q \phi q}$ is the 1PI Yukawa vertex.
Substituting \Cref{eq:threePointFunctionKernel} into \Cref{eq:earlyDSE}, and multiplying by right by $\Delta^{q,-1}$ on both sides, one obtains the Dyson-Schwinger equation of the quark two-point function in the corresponding Yukawa theory
\begin{align}
    \Delta^{q,-1} = \Delta_0^{q,-1}[\bar \phi] - g \Delta^{q} \Delta^{\phi} \Gamma^{(3)}_{\bar q \phi q} \, ,
\end{align}
which consists of the mesonic rainbow diagrams. 
Thus, the mesonic fluctuation corrections to the quark propagator are present in the exact theory. 
In the present approximation, however, these corrections are not fed back into the mesonic CJT functional, so that the quark determinant and the quark-loop contribution to the mesonic inverse propagator are evaluated merely by $\Delta_0^q[\bar\phi]$. 
This is in the sense in which the quark sector is treated at the background-field level.

A more complete treatment would require introducing a bi-local source for the quark fields and deriving a quark-meson 2PI effective action. 
In such an extended formulation, both the mesonic and the quark propagators are variational variables, and their stationarity conditions lead to coupled gap equations. 
Such a treatment would self-consistently feed the mesonic fluctuation corrections to the quark propagator back into the mesonic sector. 
By contrast, the present mesonic CJT construction keeps only the background-field quark propagator in the fermion determinant and in the quark-loop contribution to the mesonic two-point function.

To further clarify the relation between those two approximation schemes, it is useful to %refer to an analogy 
borrow terms used in the functional renormalization group (fRG). 
In the fRG approach, one introduces the scale-dependent 1PI effective average action $\Gamma_k$, which interpolates between the microscopic action at the UV scale ($S$) and the full quantum effective action in the IR ($\Gamma$), like 
\begin{align}
    \Gamma_{k\to \Lambda} \simeq S,
    \qquad
    \Gamma_{k\to 0} = \Gamma \, .
\end{align}
Its scale dependence is determined by the Wetterich equation~\cite{Wetterich:1992yh}
\begin{align}
    \partial_k \Gamma_k
    =
    \frac{1}{2}
    \mathrm{STr}
    \left[
        \left(
            \Gamma_k^{(2)}+R_k
        \right)^{-1}
        \partial_k R_k
    \right] \, ,
    \label{eq:WetterichEquation}
\end{align}
where $R_k$ is the IR regulator and $\mathrm{STr}$ denotes the supertrace over all internal, spacetime or momentum, and field-space indices. 
Although \Cref{eq:WetterichEquation} is formally exact, its practical use requires specifying an ansatz for $\Gamma_k$, and the chosen ansatz determines which correlation functions evolve self-consistently.  
For example, in a local potential approximation for the quark-meson model, one may parameterize the quark two-point function as
\begin{align}
    \Gamma_{k,\bar q q}^{(2)}(p;\bar\phi)
    =
    i\slashed p + g_k\bar\phi \, .
\end{align}
At this truncation level, the quark mass is identified as the Yukawa-induced local background mass. 
Mesonic and quark fluctuations may still contribute to the flow of the effective potential and other retained couplings, but the mesonic fluctuation corrections to the quark propagator itself are truncated out in the evaluation of the quark propagator. %not solved 
%as an independent part of the truncation. 
This is directly analogous to the approximation adopted in the present mesonic CJT construction, where the quark determinant and the quark-loop contribution to the mesonic two-point function are evaluated at the background-field quark propagator level. 

A more refined fRG truncation instead promotes the quark two-point function to a running parameter. 
For instance, one may write
\begin{align}
    \Gamma_{k,\bar q q}^{(2)}(p)
    =
    i\slashed p
    +
    m_{q,k} \, .
\end{align}
In such a formulation, the background value $g_\Lambda\bar\phi$ is treated as the initial condition for the flow equation, i.e., the classical level, of the scalar part of the inverse quark propagator, while mesonic fluctuations generate additional corrections onto the flow of $\Gamma_{k,\bar q q}^{(2)}$. 
This is analogous in spirit to a quark-meson 2PI formulation in which the quark propagator is also treated as a variational quantity and is solved self-consistently together with the mesonic propagators.

\subsection{Gap equations of meson propagators}
 \label{sec:GapEquationsOfMesonPropagators}
 
We now come back to the effective action $V_{\rm CJT}$ in Eq.~(\ref{eq:GeneralCJTPotnetialMainText}). 
The solutions of the meson correlators, $\mathcal G_S$ and $ \mathcal G_P$, are determined by solving the quantum equations of motion with vanishing external sources
\begin{align}
    \frac{\delta  V_{\rm CJT}[\sigma, G_S, G_P]}{\delta [G_S]_{ab}}\Biggl|_{\sigma = \bar \sigma, G_S = {\mathcal G_S}, G_P = {\mathcal G_P}} &= 0, \nonumber\\
    \frac{\delta  V_{\rm CJT}[\sigma, G_S, G_P]}{\delta [G_P]_{ab}}\Biggl|_{\sigma = \bar \sigma, G_S = {\mathcal G_S}, G_P = {\mathcal G_P}} &= 0.
    \label{eq:CJTGapEquations}
\end{align}
Within the Hartree approximation, 
the meson propagators do not get wavefunction renormalization. 
Thus, we parameterize the meson correlators with the momentum-independent dynamical mass terms $M_S$ and $M_P$, which read
\begin{align}
    \Big[ {\mathcal G_S}^{-1}(k) \Big]^{ab} &= k^2 \delta^{ab} + \Big[ M_S^2 \Big]^{ab}, \nonumber\\
    \Big[ {\mathcal G_P}^{-1}(k) \Big]^{ab} &= k^2 \delta^{ab} + \Big[ M_P^2 \Big]^{ab}.
    \label{eq:parameterizationOfFullPropagatorsMainText}
\end{align}
Inserting \Cref{eq:TreeLevelMesonTwoPoint,eq:parameterizationOfFullPropagatorsMainText} into \Cref{eq:CJTGapEquations}, the gap equations of $M_S$ and $M_P$ are evaluated as 
\begin{align}
    \Big[ M_S^2 \Big]^{ab} =& \Big[ m_S^2(\bar\sigma) \Big]^{ab} + \frac{\partial^2 \Omega_{\bar q q}}{\partial \sigma_a \partial \sigma_b} \Biggl|_{\Phi = \bar\Phi} \nonumber\\
    &+ 4 \mathcal{F}^{abcd} \int_k [{\mathcal G_S}(k)]_{cd} + 4 \mathcal{H}^{abcd} \int_k [{\mathcal G_P}(k)]_{cd}, \nonumber\\
    \Big[ M_P^2 \Big]^{ab} =& \Big[ m_P^2(\bar\sigma) \Big]^{ab} + \frac{\partial^2 \Omega_{\bar q q}}{\partial \pi_a \partial \pi_b} \Biggl|_{\Phi = \bar\Phi} \nonumber\\
    &+ 4 \mathcal{F}^{abcd} \int_k [{\mathcal G_P}(k)]_{cd} + 4 \mathcal{H}^{abcd} \int_k [{\mathcal G_S}(k)]_{cd}.
    \label{eq:gapEquationsInCJTMainText}
\end{align}
Note that by solving these gap equations iteratively, the daisy-type diagrams are also resummed in a self-consistent way with the thermal loop contributions from quarks.

As in the vacuum case, we transform the meson mass matrix from the original basis with indices $(a,b, \cdots)$ to the diagonal basis with indices $(i,j, \cdots)$ through the orthogonal  transformation
\begin{align}
    \Big[ \tilde M_{S/P}^2 \Big]_{(i)} \delta^{ij} = \big( O_{S/P}^{\prime,-1} \big)^{i}_{\,\, a} \Big[ M_{S/P}^2 \Big]^{ab} \big( O_{S/P}^\prime \big)_{b}^{\,\, j}.
\end{align}
We refer readers to \Cref{app:CJTgapEquations} for more details.

\subsection{Stationary conditions and symmetry improvement}
\label{sec:StationaryConditionsAndSymmetryImprovement}

In this subsection, we discuss various schemes for determining the VEVs of the mesonic fields.

We begin by considering a one-loop improved version of the 1PI formalism.
In this approach, the stationary conditions of $\bar\sigma$ are extended by replacing the tree-level meson propagators with the resummed propagators discussed in \Cref{sec:GapEquationsOfMesonPropagators}.
This yields the following equations:
\begin{align}
    0 &= \frac{\partial V_m(\bar\sigma)}{\partial \bar\sigma_0}  + \frac{\partial \Omega_{\bar q q}}{\partial \sigma_0} \Biggl|_{\Phi = \bar\Phi}
    \nonumber\\
    &\quad - 3 \mathcal A^{0bc} \biggl[ \int_k [{\mathcal G_S}(k)]_{cb} - \int_k [{\mathcal G_P}(k)]_{cb} \biggl] \nonumber\\
    &\quad + 4 \mathcal F^{0bcd} \bar\sigma_b \int_k [{\mathcal G_S}(k)]_{dc} + 4 \mathcal H^{0bcd} \bar\sigma_b \int_k [{\mathcal G_P}(k)]_{dc} \, , 
    \nonumber\\
    0 &= \frac{\partial V_m(\bar\sigma)}{\partial \bar\sigma_8}  + \frac{\partial \Omega_{\bar q q}}{\partial \sigma_8} \Biggl|_{\Phi = \bar\Phi}
    \nonumber\\
    &\quad - 3 \mathcal A^{8bc} \biggl[ \int_k [{\mathcal G_S}(k)]_{cb} - \int_k [{\mathcal G_P}(k)]_{cb} \biggl] \nonumber\\
    &\quad + 4 \mathcal F^{8bcd} \bar\sigma_b \int_k [{\mathcal G_S}(k)]_{dc} + 4 \mathcal H^{8bcd} \bar\sigma_b \int_k [{\mathcal G_P}(k)]_{dc} \, , 
    \label{eq:stationaryConditions1PIScheme}
\end{align}
where the definition of the KMT-term-induced vertex $\mathcal A^{abc}$ can be found in \Cref{eq:definitionOfG}.
In each line of \Cref{eq:stationaryConditions1PIScheme}, the first term arises from the tree-level mesonic potential, the second to fourth terms represent the mesonic one-loop contributions, and the last term comes from the quark one-loop potential.

%%%%%%%%%%%%%%
%%%%%%%%%%%%%%
\begin{figure}[t!]
    \centering
    \includegraphics[width=0.5\linewidth]{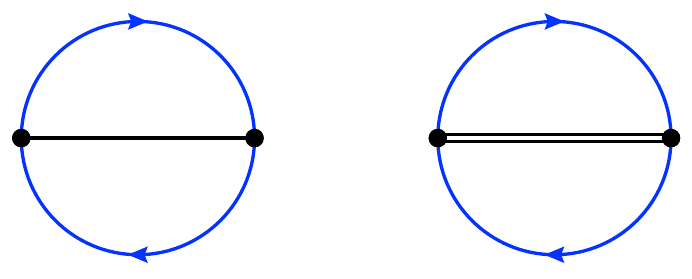}
    \caption{
    Two-loop diagrams of the effective quark contributions in Eq.~\labelcref{eq:quarkLoopContricutionToCJT} to the 2PI effective action in the current work.
    The blue lines with an arrow denote the quark propagators, the solid lines and the double-solid lines represent the scalar propagators $[G_S]_{ab}$ and the pseudo-scalar propagators $[G_P]_{ab}$, respectively, in the same manner as in \Cref{fig:HartreeApprox}. 
    The black dots represent the bare vertices arising from the Yukawa interaction term. 
    }
    \label{fig:QuarkMesonTwoLoop}
\end{figure}
%%%%%%%%%%%%%%
%%%%%%%%%%%%%%

In the CJT formalism, the VEVs of the field $\bar\sigma$ are determined by solving the stationary condition of the CJT potential, i.e.,
\begin{align}
    \frac{\partial V_{\rm CJT}[\sigma, G_S, G_P]}{\partial \sigma_a} \Biggl|_{\sigma = \bar \sigma, G_S = {\mathcal G_S}, G_P = {\mathcal G_P}} &= 0,
    \label{eq:CJTstationaryConditions}
\end{align}
By inserting the solution of the coupled meson gap equation in Eq.~\eqref{eq:gapEquationsInCJTMainText}, the stationary conditions of the background fields $(\bar \sigma_0, \bar \sigma_8)$ take the form 
\begin{widetext}
\begin{align}
    0 &= \frac{\partial V_m(\bar\sigma)}{\partial \bar\sigma_0} - 3 \mathcal A^{0bc} \biggl[ \int_k [{\mathcal G_S}(k)]_{cb} - \int_k [{\mathcal G_P}(k)]_{cb} \biggl] + \frac{1}{2} \frac{\partial^3 \Omega_{\bar q q}}{\partial \sigma_0 \partial \sigma_a \partial \sigma_b} \Biggl|_{\Phi = \bar\Phi} \int_k [{\mathcal G_S}(k)]_{ba} + \frac{1}{2} \frac{\partial^3 \Omega_{\bar q q}}{\partial \sigma_0 \partial \pi_a \partial \pi_b} \Biggl|_{\Phi = \bar\Phi} \int_k [{\mathcal G_P}(k)]_{ba} \nonumber\\
    &\qquad + 4 \mathcal F^{0bcd} \bar\sigma_b \int_k [{\mathcal G_S}(k)]_{dc} + 4 \mathcal H^{0bcd} \bar\sigma_b \int_k [{\mathcal G_P}(k)]_{dc} + \frac{\partial \Omega_{\bar q q}}{\partial \sigma_0} \Biggl|_{\Phi = \bar\Phi}, \nonumber\\
    0 &= \frac{\partial V_m(\bar\sigma)}{\partial \bar\sigma_8} - 3 \mathcal A^{8bc} \biggl[ \int_k [{\mathcal G_S}(k)]_{cb} - \int_k [{\mathcal G_P}(k)]_{cb} \biggl] + \frac{1}{2} \frac{\partial^3 \Omega_{\bar q q}}{\partial \sigma_8 \partial \sigma_a \partial \sigma_b} \Biggl|_{\Phi = \bar\Phi} \int_k [{\mathcal G_S}(k)]_{ba} + \frac{1}{2} \frac{\partial^3 \Omega_{\bar q q}}{\partial \sigma_8 \partial \pi_a \partial \pi_b} \Biggl|_{\Phi = \bar\Phi} \int_k [{\mathcal G_P}(k)]_{ba} \nonumber\\
    &\qquad + 4 \mathcal F^{8bcd} \bar\sigma_b \int_k [{\mathcal G_S}(k)]_{dc} + 4 \mathcal H^{8bcd} \bar\sigma_b \int_k [{\mathcal G_P}(k)]_{dc} + \frac{\partial \Omega_{\bar q q}}{\partial \sigma_8} \Biggl|_{\Phi = \bar\Phi}.
    \label{eq:stationaryCOnditionsInCJTMainText}
\end{align}
\end{widetext}
Compared to the stationary conditions with one-loop contributions in Eq.~\eqref{eq:stationaryConditions1PIScheme}, in \Cref{eq:stationaryCOnditionsInCJTMainText} there are additional 3-point vertex terms induced from quark loops: $\partial^3 \Omega_{\bar q q}/\partial \sigma_i \partial \varphi_a \partial \varphi_b$. 
At the level of the 2PI effective potential, this additional term comes from a two-loop diagram contribution, which is encoded in the following mesonic Gaussian contraction term
\begin{align}
    &\frac{1}{2} \int_k \operatorname{tr} \Big[ \bar G_S^{-1}(k;\sigma) G_S(k) + \bar G_P^{-1}(k;\sigma) G_P(k) \Big]
    \nonumber\\
    &\ni \frac{1}{2} \int_k \operatorname{tr} \Big[ \frac{\partial^2 \Omega_{\bar q q}}{\partial \sigma_a \partial \sigma_b} \Biggl|_{\Phi = \bar\Phi} G_S(k) + \frac{\partial^2 \Omega_{\bar q q}}{\partial \pi_a \partial \pi_b} \Biggl|_{\Phi = \bar\Phi} G_P(k) \Big] \, . 
    \label{eq:quarkLoopContricutionToCJT}
\end{align}
This contribution is diagrammatically depicted in \Cref{fig:QuarkMesonTwoLoop}. 
In the current work, the external-momentum-dependent quantum corrections to the mesonic fields are dropped within the Hartree-Fock approximation, and the quark-loop contributions are reduced to the potential level independent of momenta.

In the SICJT formalism~\cite{Pilaftsis:2013xna, Mao:2013gva, Guan:2025tmf}, instead of using the stationary conditions in \Cref{eq:stationaryCOnditionsInCJTMainText}, the background field values $(\bar \sigma_0, \bar \sigma_8)$ are determined by the following relations, which are equivalent to the Gell-Mann-Oakes-Renner (GMOR) relations in the context of the LSM:
\begin{align}
    M_\pi^2 \bar\Phi_1 &= c m_l, \nonumber\\
    M_K^2 \Big( \bar \Phi_1 + \bar \Phi_3 \Big) &= c m_l + cm_s.
    \label{eq:GMORrelations1PIMainText}
\end{align} 
These relations are derived from the modified WTIs associated with the chiral rotations, whose derivations are summarized in \Cref{app:AWTIsandGMORs}.
A similar GMOR in the three-flavor LSM can also be found in Ref.~\cite{Guan:2025tmf}. 

\begin{figure}[t!]
    \centering
    \includegraphics[width=0.7\linewidth]{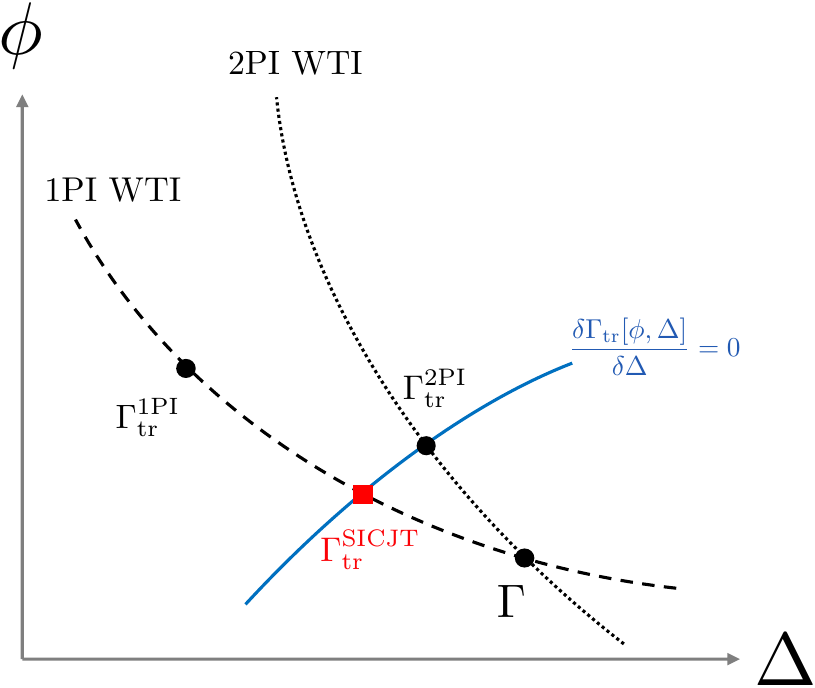}
    \caption{
    A schematic illustration of the SICJT formalism at the level of the equations of motion, as given in Ref.~\cite{Guan:2025tmf}.
    The figure is shown in the $(\phi,\Delta)$ plane, where $\phi$ denotes the sourced field VEV and $\Delta$ the sourced propagator, which together parameterize the effective actions.
    The black blobs (from right to left) indicate the solutions obtained from the full quantum effective action $\Gamma$, the truncated 1PI effective action $\Gamma_{\rm tr}^{\rm 1PI}$, and the truncated 2PI effective action $\Gamma_{\rm tr}^{\rm 2PI}$.
    The black dashed and dotted curves represent the constrained trajectories associated with the 1PI and 2PI formalisms, respectively, while the blue solid curve denotes the propagator gap equation for arbitrary $\phi$.
    The red square marks the solution realized within the SICJT formalism.
    }
    \label{fig:ScenarioOfSICJT}
\end{figure}

\Cref{eq:GMORrelations1PIMainText} reflects the so-called \textit{threshold property}~\cite{Pilaftsis:2013xna} of the Nambu-Goldstone (NG) bosons, which ensures the lightness of the NG boson in the chirally broken phase, or before the threshold of the chiral restoration occurs. 
By imposing \Cref{eq:GMORrelations1PIMainText} instead of the stationary conditions in \Cref{eq:stationaryCOnditionsInCJTMainText},  the problem of violation of the NG theorem caused by the introduction of the loopwise truncation in the 2PI formalism is thus resolved~\cite{Pilaftsis:2013xna}. 
We show a schematic sketch of the symmetry improvement in \Cref{fig:ScenarioOfSICJT}.  A self-consistent formulation at the level of the effective action based on the external sources is presented in \Cref{sec:SourceBasedFormulationOfSICJT}.

%%%%%%%%%%%%%%
%%%%%%%%%%%%%%
%%%%%%%%%%%%%%
%%%%%%%%%%%%%%

The three approaches in \Cref{fig:ScenarioOfSICJT} 
extend the analysis beyond the purely quark one-loop effective potential by incorporating mesonic fluctuations. In contrast, if only the quark one-loop contribution is retained, the stationary conditions for the mesonic background fields $(\bar \sigma_0, \bar \sigma_8)$ reduce to
\begin{align}
    &0 = \frac{\partial V_m(\bar\sigma)}{\partial \bar\sigma_0} + \frac{\partial \Omega_{\bar q q}}{\partial \sigma_0} \Biggl|_{\Phi = \bar\Phi}, \nonumber\\
    &0 = \frac{\partial V_m(\bar\sigma)}{\partial \bar\sigma_8} + \frac{\partial \Omega_{\bar q q}}{\partial \sigma_8} \Biggl|_{\Phi = \bar\Phi}.
    \label{eq:stationaryConditionsFermiLoop}
\end{align}
In the later section, we compare the resulting chiral phase diagrams obtained from the different approaches and examine how mesonic fluctuations influence the nature of the chiral phase transition.
For completeness, the necessary formulations of the CJT formalism, together with the explicit expressions for the quark-loop potential $\Omega_{\bar q q}$ and its derivatives with respect to the mesonic fields %$\varphi_a$ 
are summarized in \Cref{app:CJTFormalismAtFinitTemperature}.

\subsection{Renormalization}
 \label{sec:Renormalization}

%When formulating the schemes 
In the CJT formalism, UV divergences arise in the loop integrals.
A renormalization scheme must therefore be introduced to remove these divergences and extract physically meaningful results.
In this work, we employ the same renormalization scheme as in Refs.~\cite{Scavenius:2000qd,Lenaghan:2000ey,Schaefer:2008hk, Kawaguchi:2020qvg, Guan:2025tmf}, where the renormalized vacuum part of the loop corrections is absorbed into the tree-level parameters and subtracted accordingly, such that only the minimal medium effects are retained.
Although this scheme neglects part of the field-background dependence of the effective action, it remains adequate for a qualitative analysis.

Fermions contribute to the effective action only through the quark one-loop potential $\Omega_{\bar q q}$ and its derivatives, evaluated in a homogeneous mesonic background.
We therefore discard the vacuum contribution,
\begin{align}
    \Omega_{\bar q q}\Big|_{\rm vac} = \nu_c \sum_{f = u,d,s} \int_0^\infty \frac{\df^3 \bvec{p}}{(2\pi)^3}  E_{q,f},
\end{align}
and retain only the thermal part of the quark one-loop potential,
\begin{align}
    \Omega^R_{\bar q q} &= \Omega_{\bar q q}\Big|_{T}
    \nonumber\\
    &= \nu_c T \sum_{f = u,d,s} \int_0^\infty \frac{\df^3 \bvec{p}}{(2\pi)^3} \biggl\{ \log \Big[ 1 - n_{q,f}(T,\mu_f) \Big] \nonumber\\
    &\qquad + \log \Big[ 1 - n_{\bar q,f}(T,\mu_f) \Big] \biggl\}.
\end{align}

In the mesonic sector, two types of one-loop terms appear in the 2PI effective action:
the trace-log terms,
\begin{align}
    \int_k \operatorname{tr} \log [\tilde{\mathcal G}_S^{-1}(k)]_{(i)}, 
    \quad 
    \int_k \operatorname{tr} \log [\tilde{\mathcal G}_P^{-1}(k)]_{(i)} ,
\end{align}
and the tadpole integrals,
\begin{align}
    \int_k [\tilde{\mathcal G}_S(k)]_{(i)} 
    \quad 
    \int_k [\tilde{\mathcal G}_P(k)]_{(i)} \, .
\end{align}
In the Hartree approximation, the meson propagators are parameterized as in \Cref{eq:parameterizationOfFullPropagatorsMainText}, so these loop integrals take simple forms.
Taking the scalar propagator $[\tilde{\mathcal G}_S]_{(i)} $ as an example, one obtains
\begin{align}
    &\int_k \operatorname{tr} \log [\tilde{\mathcal G}_S^{-1}(k)]_{(i)} 
    \nonumber\\
    &\qquad= 
    T \sum_{n = -\infty}^{\infty} \int \frac{\df^3 \bvec{k}}{(2 \pi)^3} \, \log \left[  \omega_n^2 + \Big(  \epsilon^{S,(i)}_{\bvec{k}} \Big)^2 \right]
    \nonumber\\
    &\qquad= 
    \int \frac{\df^3 \bvec{k}}{(2 \pi)^3} \, \left\{  \epsilon^{S,(i)}_{\bvec{k}} - 2 T \, \log \left[ 1 + n_B\left(\epsilon^{S,(i)}_{\bvec{k}}\right) \right] \right\}, 
    \\
    &\int_k [\tilde{\mathcal G}_S(k)]_{(i)} 
    \nonumber\\
    &\qquad= 
    T \sum_{n = -\infty}^{\infty} \int \frac{\df^3 \bvec{k}}{(2 \pi)^3} \, \left[  \omega_n^2 + \Big(  \epsilon^{S,(i)}_{\bvec{k}} \Big)^2 \right]^{-1}
    \nonumber\\
    &\qquad= 
    \int \frac{\df^3 \bvec{k}}{(2\pi)^3} \frac{1}{2\epsilon^{S,(i)}_{\bvec{k}}} \left[ 1 + 2 n_B\left(\epsilon^{S,(i)}_{\bvec{k}}\right) \right],
\end{align}
where $n_B$ denotes the Bose-Einstein distribution function,
\begin{align}
     n_B\left(\epsilon^{S/P,(i)}_{\bvec{k}}\right) = \left( \exp \left[ \epsilon^{S/P,(i)}_{\bvec{k}}/T \right] -1 \right)^{-1},
\end{align}
and the mesonic dispersion relation is given by
\begin{align}
    \epsilon^{S/P,(i)}_{\bvec{k}} \equiv \sqrt{|{\bvec{k}}|^2 + \Big[ \tilde M_{S/P}^2 \Big]_{(i)} }.
\end{align}

As in the fermionic sector, we discard the vacuum contributions and retain only the thermal parts in the mesonic sector,
\begin{align}
    &\int_k \operatorname{tr} \log [\tilde{\mathcal G}_S^{-1}(k)]_{(i)} 
    \nonumber\\
    &\qquad= 
    -2T\int \frac{\df^3 \bvec{k}}{(2 \pi)^3} \, \, \log \left[ 1 + n_B\left(\epsilon^{S,(i)}_{\bvec{k}}\right) \right] , 
    \nonumber\\
    &\int_k [\tilde{\mathcal G}_S(k)]_{(i)} 
    = 
    \int \frac{\df^3 \bvec{k}}{(2\pi)^3} \frac{n_B\left(\epsilon^{S,(i)}_{\bvec{k}}\right)}{\epsilon^{S,(i)}_{\bvec{k}}} .
\end{align}
These subtracted loop integrals can be used directly in the stationary conditions, \Cref{eq:stationaryCOnditionsInCJTMainText}, and in the gap equations, \Cref{eq:gapEquationsInCJTMainText}. 
We have verified the consistency between the subtracted 2PI effective action and the corresponding equations of motion by comparing the solutions of these equations with the numerical minima of the CJT effective potential.
This completes the present formulation of both the CJT and SICJT formalisms.

%%%%%%%%%%%%%%%%%%%%%%%%%%%%%%%%%%%%%%%%%%%%%%%%%%%%%%%%%%%%%%%%%%%%%%
\section{Symmetry-improved 2PI action and thermodynamics}
\label{sec:Thermodynamics}

In this section, we discuss the implementation of symmetry improvement at the level of the self-consistent 2PI action and formulate the corresponding thermodynamics.
As addressed in \Cref{sec:StationaryConditionsAndSymmetryImprovement}, the stationary condition of the 2PI effective potential~\labelcref{eq:stationaryCOnditionsInCJTMainText} is not imposed directly.
Instead, the ground state of the system is determined by the GMOR relations~\labelcref{eq:GMORrelations1PIMainText} induced by the WTIs.
In Ref.~\cite{Pilaftsis:2013xna}, the symmetry-improved formalism is realized by means of a Lagrange multiplier.
However, if one extends the formulation with nonzero quark masses, beyond the chiral limit, the gap equations would be modified by the constraint term.
This is because, at a finite current quark mass, the singular behavior of the multiplier no longer arises, and the right-hand side of the gap equation~\labelcref{eq:CJTGapEquations} remains finite.
Therefore, a reformulation of the symmetry-improved 2PI effective action is necessary, as it provides the basis for a consistent thermodynamic formulation.

%%%%%%%%%%%%%%%%%%%%%%%%%%%%%%%%%%%%%%%%%%%%%%%%%%%%%%%%%%%%%%%%%%%%%%
\subsection{Source-based formulation of SICJT beyond the chiral limit}
\label{sec:SourceBasedFormulationOfSICJT}

In this work, we adopt an alternative but equivalent interpretation of the symmetry improvement at the level of the 2PI effective action, which can be directly applied to the cases with a finite current quark mass term.
To this end, we note again that the stationary conditions~\labelcref{eq:stationaryCOnditionsInCJTMainText} are not imposed; rather, they can be expressed by introducing the local external sources $j^m$ to the thermodynamic system, such that the equations of motion for the background fields become
\begin{align}
    \frac{\delta \Gamma[\phi, \Delta]}{\delta \phi_m} = j^m\,. 
    \label{eq:2PIEOMwithSource}
\end{align}
While the bi-local source $K^{mn}$ is kept vanishing, the gap equation remains unchanged: 
\begin{align}
    \frac{\delta \Gamma[\phi, \Delta]}{\delta \Delta_{mn}} = 0.
    \label{eq:2PIEOMkeepRvanish}
\end{align}
The source $j^m$ is not fixed at this point but is determined only after additional IR constraints are imposed.
For the three-flavor LSM, we introduce two independent external sources, $J^1$ and $J^3$, coupled to the mesonic background fields $\Phi_1$ and $\Phi_3$.
In the following context, we use $j$ to stand for the source coupled to the general field contents, and $J$ for the source coupled to the mesonic fields only, which follows the convention in \Cref{eq:generalSourceContent}.

In the symmetry-improvement procedure, the external sources are then fixed by the GMOR relations~\labelcref{eq:GMORrelations1PIMainText}.
In principle, this is achieved by considering the 2PI version of the WTIs,
\begin{align}
    &\frac{\delta^2 \Gamma[\phi, \Delta]}{\delta \phi_i \delta \phi_m} \cdot d_m^{\,\, n} \phi_n 
    + 
    \frac{\delta}{\delta \phi_i}\biggl[
    \biggl\langle \frac{\delta S}{\delta \mathcal{M}} \biggl\rangle_J \cdot \delta_\epsilon \mathcal{M} 
    + 
    {\rm h.c.}
    \biggl] 
    \nonumber\\
    &=  
    -\frac{\delta \Gamma[\phi, \Delta]}{\delta \phi_m} \cdot d_m^{\,\, i} 
    - 
    \frac{\delta^2 \Gamma[\phi, \Delta]}{\delta \phi_i \delta \Delta_{mn}} \cdot \Big( d_n^{\,\, k} \Delta_{mk} + d_m^{\,\, k} \Delta_{kn} \Big) \ ,
    \label{eq:2PIWTIderivative}
\end{align}
where we have taken an additional functional derivative with respect to the field $\phi_i$ and suppressed the spacetime labels.
In \Cref{eq:2PIWTIderivative}, $d_m^{\,\, n}$ denotes the matrix representation of the infinitesimal symmetry shift in the field space, i.e., $ \delta_\epsilon \phi_m = d_m^{\,\, n} \phi_n$, $S$ is the classical action, and $\mathcal M$ is the symmetry-breaking spurion field, which is the current quark mass matrix defined in \Cref{eq:TadpoleTerm}.
For the derivation, we refer readers to \Cref{app:AWTIsandGMORs}.

One notices that the left-hand side of \Cref{eq:2PIWTIderivative} yields the 1PI version of the WTIs.
To ensure the 1PI WTIs, the sources are locally fixed such that the terms in the second line of \Cref{eq:2PIWTIderivative} are forced to cancel each other once the equations of motion \labelcref{eq:2PIEOMwithSource} are applied.
We write the set of sources that satisfy this constraint as
\begin{align}
    \{ J^1_{\rm SI}, J^2_{\rm SI}, \cdots  \} \ ,
\end{align}
where “SI” stands for %``symmetry-improved'' or 
``symmetry-improvment''. 
Once the sources are fixed in this way, the 1PI WTIs are restored.
The stationary conditions are then replaced by functional relations between the truncated 2PI effective action and the external sources, while the system is stabilized by the 1PI WTIs together with the unmodified gap equations.
This completes the source-based formulation of symmetry improvement at the level of the 2PI effective action. 
The equilibrium conditions are then specified by the unmodified gap equations together with the symmetry-improvement constraints. 
%, although multiple solution branches may still occur in the first-order region.

After deriving the present source-based formulation, we became aware of the external-sources approach of Garbrecht et al.~\cite{Garbrecht:2015cla}, which realizes symmetry improvement through non-vanishing external sources. 
The two constructions are closely related in spirit. 
In the present work, we take one further step by focusing on the thermodynamic formulation and its ambiguities beyond the chiral limit, which has not been discussed in the literature.

%%%%%%%%%%%%%%%%%%%%%%%%%%%%%%%%%%%%%%%%%%%%%%%%%%%%%%%%%%%%%%%%%%%%%%
\subsection{Pressure and equation of state in the sourced system}

In the presence of the external sources, all thermodynamic observables are modified by the background of these sources. 
This is understood by enabling the external forces or external fields to couple to the system.
Consider the definition of the grand canonical potential
\begin{align}
    \Omega(\mathcal V_3, T, \mu_B, j) = - \beta^{-1} \log Z[j] \ ,
\end{align}
where $Z[j]$ is the corresponding partition function, $\mathcal V_3 = \int \df^3 x$ denotes the three-dimensional space volume. 
% and $\beta = T^{-1}$.
According to \Cref{eq:define2PIEffectiveAction}, when the fields and propagators are evaluated at the solutions for the stationary conditions~\labelcref{eq:2PIEOMwithSource} and the gap equations~\labelcref{eq:2PIEOMkeepRvanish} in the presence of the external sources, the resulting effective action is related to the grand canonical potential $\Omega(\mathcal V_3, T, \mu_B, j)$ by
\begin{align}
    \Gamma[\bar \phi, \bar \Delta] = j^m \cdot \bar \phi_m + \beta \Omega(\mathcal V_3, T, \mu_B, j) \ .
    \label{eq:constructingGrandPotential}
\end{align}
This allows us to construct the thermodynamic observables from the 2PI effective action with non-vanishing $j^m$.

\subsubsection{Pressure and thermodynamic ambiguity}
\label{sec:PressureAndThermodynamicAmbiguity}

In this work, we adopt the truncated 2PI effective action $\Gamma_{\rm tr}$ in Euclidean spacetime along with the homogeneous field background $\bar \sigma(x) = \bar \sigma$. The CJT effective potential for the three-flavor LSM is thus directly read off from $\Gamma_{\rm tr}$ as 
\begin{align}
    \Gamma_{\rm tr}[\bar \sigma, \mathcal G_S, \mathcal G_P] = \beta \ \mathcal V_3 \ V_{\rm CJT}[\bar \sigma, \mathcal G_S, \mathcal G_P] \ .
    \label{eq:2PIEAandEP}
\end{align}
In the grand canonical ensemble, the raw pressure is obtained by the scaling law between the grand canonical potential $\Omega$ and the volume of the space $\mathcal V_3$ as 
\begin{align}
    P_{\rm raw} = - \frac{\Omega}{\mathcal V_3} \, . 
    \label{eq:definingThePressure} 
\end{align}
From \Cref{eq:constructingGrandPotential,eq:2PIEAandEP,eq:definingThePressure}, the raw pressure thus reads 
\begin{align}
    P_{\rm raw} = - V_{\rm CJT}(\bar \sigma, \mathcal G_S, \mathcal G_P) + J^1_{\rm SI} \bar \Phi_1 + J^3_{\rm SI} \bar \Phi_3 \ ,
    \label{eq:PressureWithSourcesLSM}
\end{align}
where $\bar \Phi_1$ and $\bar \Phi_3$ are related to $\bar \sigma_0$ and $\bar \sigma_8$ through \Cref{eq:FieldBasisTrans}, and the sources are read off from the field derivatives of the CJT potential at the symmetry-improved ground states as 
\begin{gather}
    J^1_{\rm SI} = \frac{\partial V_{\rm CJT}}{\partial \Phi_1} \Biggl|_{\sigma = \bar \sigma, G_S = {\mathcal G_S}, G_P = {\mathcal G_P}} \ , \nonumber\\
    J^3_{\rm SI} = \frac{\partial V_{\rm CJT}}{\partial \Phi_3} \Biggl|_{\sigma = \bar \sigma, G_S = {\mathcal G_S}, G_P = {\mathcal G_P}} \ .
\end{gather}

To demonstrate the equivalence between the current formulation and the one using Lagrange's multiplier in the chiral limit, we perform an explicit comparison of the raw pressure with the expression obtained in Ref.~\cite{Pilaftsis:2013xna}. 
In the literature, the 2PI effective action for a general theory is formulated in Minkowskian spacetime as 
\begin{align}
    \Gamma_{\rm tr} = - \int_x \ V_{\rm eff} + \cdots \, , 
\end{align}
where the spacetime volume is given by
\begin{align}
    \int_x \equiv \int \df t \int \df^3 x = -i \beta \ \mathcal V_3 \, ,
\end{align}
and $V_{\rm eff}$ denotes the effective potential of he theory.
The ``symmetry-improved'' 2PI effective action $\tilde \Gamma$, grand-canonical potential $\tilde \Omega$, and the pressure $P$ are:  
\begin{gather}
    \tilde \Gamma = \Gamma_{\rm tr} - \ell_y \ \phi_x^H \ \Delta_{xy}^{-1,GG} \, , \nonumber\\
    \tilde \Omega = - P \ \mathcal V_3 = - i \beta^{-1} \ \tilde \Gamma \, .
\end{gather}
The multiplication of the local Lagrange's multiplier $\ell_x$ by the inverse NG-boson propagator $\Delta_{xy}^{-1,GG}$ at the solutions converges into the regular values~\cite{Pilaftsis:2013xna}
\begin{align}
    \int_y \ell_y \ \Delta_{xy}^{-1, GG} = l_0 \ m^2 \ , 
\end{align}
where $m^2$ is a dimensionful parameter introduced when defining the regularized multiplier $l_0$.
The VEV of $\phi^H$ is obtained by stabilizing the symmetry-improved 2PI effective action
\begin{align}
    \frac{\delta \tilde \Gamma}{\delta \phi^H_x}\Biggl|_{\phi^H = v} = 0.
\end{align}
Noting that the external source in the original effective action $\Gamma_{\rm tr}$ is expressed as 
\begin{align}
    J = - \frac{\delta \Gamma_{\rm tr}}{\delta \phi_x^H} \Biggl|_{\phi^H = v} = - l_0 \ m^2 \ , 
\end{align}
we obtain that the raw pressure in this formulation satisfies
\begin{align}
    P_{\rm raw} &= i (\beta \mathcal V_3)^{-1} \ \tilde \Gamma = - V_{\rm eff} - l_0 m^2 v \nonumber\\
    &= -V_{\rm eff} + J v \ ,
\end{align}
which matches \Cref{eq:PressureWithSourcesLSM}.

To define the physically meaningful thermodynamics, normalizing the raw pressure by subtraction is necessary.
This subtraction is justified by the fact that a constant shift of the effective action does not affect the dynamics or the equations of motion when the external variables are fixed.
By contrast, comparisons among different equilibrium states, such as those involved in the evaluation of the sound velocity and the trace anomaly, require a sensible reference point for the pressure.
This in turn leads to a thermodynamic ambiguity in systems where the sources are not prescribed as external fields, but are instead determined intrinsically by the dynamics of the system.

%%%%%%%%%%%%%%
%%%%%%%%%%%%%%
\begin{figure}[t]
    \centering
    \includegraphics[width=0.65\linewidth]{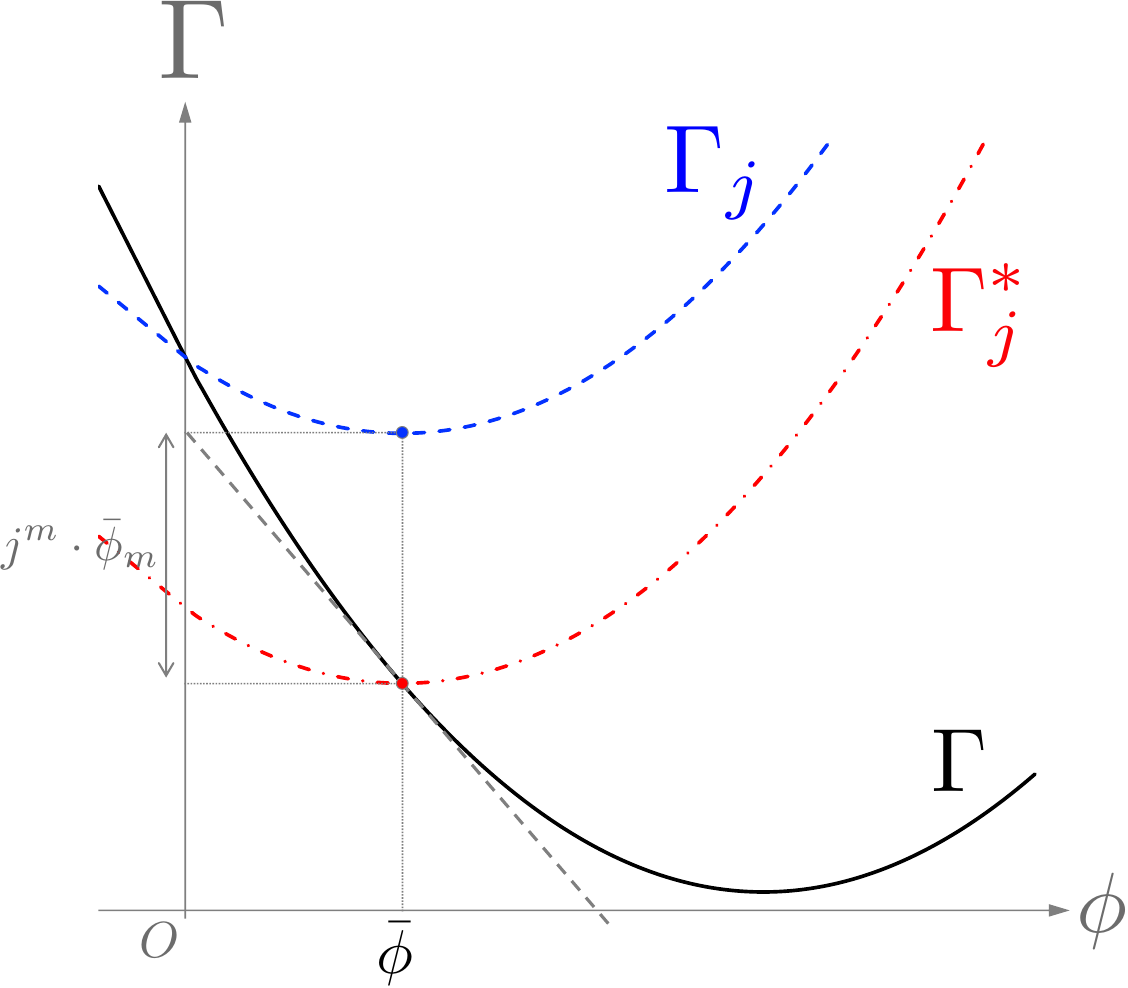}
    \caption{
    Schematic illustration of the original effective action $\Gamma$, the source-shifted effective action $\Gamma_j$, and the pulled-back effective action $\Gamma_j^*$ along the field direction $\phi$. 
    The source-shifted effective action $\Gamma_j$ is obtained by adding the linear tilt induced by the external source, such that the self-consistent solution $\bar\phi$ becomes a stationary point. 
    The pulled-back effective action $\Gamma_j^*$ differs from $\Gamma_j$ by a constant shift $j^m\!\cdot\!\bar\phi_m$, which removes the explicit source-induced energy offset while preserving the same stationary point and local fluctuation structure. 
    The vertical and horizontal dashed lines indicate the on-shell value and the location of $\bar\phi$, respectively.
}
    \label{fig:PulledBackGamma}
\end{figure}
%%%%%%%%%%%%%%
%%%%%%%%%%%%%%

In the present work, we consider three different prescriptions for the normalization of the pressure.
The first is obtained by subtracting the vacuum value of the raw pressure, $P_{\rm raw}(0, 0)$, from the raw pressure at finite $T$ and $\mu_B$.
We refer to the pressure normalized this way as the vacuum-subtracted pressure, dubbed $P_{\rm vac}$: 
\begin{align}
    P_{\rm vac} &\equiv P_{\rm raw}(T, \mu_B) - P_{\rm raw}(0, 0) 
    \nonumber\\
    &= -\frac{1}{\mathcal V_3} \Big[ \Omega(\mathcal V_3, T, \mu_B, j) - \Omega(\mathcal V_3, 0, 0, 0) \Big].
\label{P-vac}
\end{align}
This prescription is the standard one in sourceless systems and is widely adopted in thermodynamic analysis
In the current formulation, the loop corrections at the vacuum are turned off, leaving the external sources to vanish.
We rewrite $P_{\rm vac}$ in Eq.(\ref{P-vac}) in a more practical form: 
\begin{align}
    P_{\rm vac} &= V_{m}(\bar \Phi^{(0)}) - V_{\rm CJT}(\bar \sigma, \mathcal G_S, \mathcal G_P) 
    \nonumber\\
    &\qquad+ J^1_{\rm SI} \bar \Phi_1 + J^3_{\rm SI} \bar \Phi_3,
    \label{eq:vacuumSubtractedPressure}
\end{align}
where $V_{m}(\bar \Phi^{(0)}) $ denotes the tree-level mesonic potential evaluated at the background fields $\bar \Phi^{(0)}$ and $T=\mu_B=0$.

The second prescription is to subtract the vacuum pressure evaluated at the same source configuration $j$. 
We refer to this pressure as the source-matched vacuum-subtracted pressure, denoted as $P_{\rm vac}^{(j)}$:
\begin{align}
    P_{\rm vac}^{(j)} &\equiv P_{\rm raw}(T,\mu_B)-P_{\rm raw}(0,0;j)
    \nonumber\\
    &= -\frac{1}{\mathcal V_3}\Big[\Omega(\mathcal V_3,T,\mu_B,j)-\Omega(\mathcal V_3,0,0,j)\Big]. \label{P-vac-j}
\end{align}
This prescription removes the vacuum offset contributions, while keeping the source configuration fixed, and therefore allows comparing the finite-temperature system with the vacuum system under the same source supply. 
$P_{\rm vac}^{(j)}$ in Eq.~(\ref{P-vac-j}) is further evaluated and takes more explicit form 
\begin{align}
    P_{\rm vac}^{(j)} &= V_{m}(\bar \Phi^{(0)}) - V_{\rm CJT}(\bar \sigma, \mathcal G_S, \mathcal G_P) 
    \nonumber\\
    &\qquad + J^1_{\rm SI} \Big(\bar \Phi_1 - \bar \Phi_1^{(0)} \Big) + J^3_{\rm SI} \Big(\bar \Phi_3 - \bar \Phi_3^{(0)} \Big)
    \label{eq:vacuumSubtractedPressureWithSource}
\end{align}

In light of the third prescription, we first consider the effective action in the presence of external sources.
For nonvanishing $j^m$, one may define
\begin{align}
    \Gamma_{j}[\phi, \Delta] \equiv \Gamma[\phi, \Delta] - j^m \cdot \phi_m,
\end{align}
such that the equation of motion for the expectation value of $\phi_m$ takes the sourceless form
\begin{align}
    \frac{\delta \Gamma_{j}[\phi, \Delta]}{\delta \phi_m}\Biggl|_{\phi = \bar \phi} = 0.
\end{align}
The raw pressure $P_{\rm raw}$ as well as the symmetry-improved 2PI action introduced in Ref.~\cite{Pilaftsis:2013xna} are defined by the on-shell value of the source-shifted effective action $\Gamma_j$.
This shift also modifies the on-shell value of the effective action by the field-dependent amount $j^m \cdot \bar \phi_m$.
Such a field-dependent shift is usually introduced when $j^m$ is regarded as a genuine external source, otherwise not. 
%but becomes less well motivated when the source is not externally prescribed and 
However, this shift can be determined self-consistently by intrinsic dynamics, as in the SICJT formalism.  
%For this reason, 
Thus, we further introduce a ``pulled-back'' effective action, denoted as $ \Gamma_{j}^*$:  
\begin{align}
    \Gamma_{j}^*[\phi, \Delta] &\equiv \Gamma_{j}[\phi, \Delta] + j^m \cdot \bar \phi_m,
\end{align}
or equivalently,
\begin{align}
    \Gamma_{j}^*[\phi, \Delta] = \Gamma[\phi, \Delta] - j^m \cdot ( \phi_m - \bar \phi_m).
    \label{eq:starGammaDefinition}
\end{align}
This ``pulled-back" removes the source-induced linear tilt at the self-consistent solution as well as the additional energy shift, while preserving the stationary condition and the local fluctuation structure. 
This procedure is illustrated in \Cref{fig:PulledBackGamma}.
Then, we define the ``pulled-back'' pressure as the thermodynamic quantity associated with the on-shell value of the pulled-back effective action $ \Gamma_{j}^*$, namely, 
\begin{align}
    P^* \equiv -\frac{1}{\mathcal V_3} \Big[ \Omega^*(\mathcal V_3,T,\mu_B) - \Omega^*(\mathcal V_3,0,0) \Big],
    \label{eq:starPressureDefinition}
\end{align}
where the ``pulled-back'' thermodynamic potential is defined as 
\begin{align}
     \Omega^*(T,\mu_B) \equiv \frac{1}{\beta} \Gamma^*_j[\bar \phi, \bar \Delta].
     \label{eq:starOmegaDefinition}
\end{align}
With this $P^*$ given, the artificial energy shift induced by the explicit source $j^m$ is removed, while its implicit influence through the self-consistent solution remains. 
This feature can be seen more explicitly by further rewriting $P^*$ as 
\begin{align}
    P^* = V_{m}(\bar \Phi^{(0)}) - V_{\rm CJT}(\bar \sigma, \mathcal G_S, \mathcal G_P),
    \label{eq:starPressureInLSM}
\end{align}
which follows from \Cref{eq:2PIEAandEP,eq:starGammaDefinition,eq:starPressureDefinition,eq:starOmegaDefinition} and coincides with the pressure in the sourceless system.

Among these three prescriptions, comparison of pressure at different values of $T$ and $\mu_B$ becomes uncertain, due to the explicit $(T,\mu_B)$-dependence of the external sources
\begin{align}
    J^1_{\rm SI} = J^1_{\rm SI}(T, \mu_B), \quad J^3_{\rm SI} = J^3_{\rm SI}(T, \mu_B).
    \label{eq:TmuDpendenceOfJs}
\end{align}
This ambiguity is directly reflected in the spinodal region associated with the first-order chiral phase transition.
A more detailed discussion is given in \Cref{sec:ChiOrderParamAndPhaseDiagram}.

\subsubsection{Thermodynamic observables on the SI manifold}
\label{sec:ThermodynamicObservablesOnTheSIManifold}

Once the pressure of the SI system is defined, the remaining thermodynamic observables are obtained as follows: 
\begin{gather}
    s = \left( \frac{\partial P}{\partial T} \right)_{\mu_B,\, j},
    \qquad
    n_B = \left( \frac{\partial P}{\partial \mu_B} \right)_{T,\, j},
    \nonumber\\
    \varepsilon = -P + Ts + \mu_B n_B \, ,
\end{gather}
where $s$, $n_B$, and $\varepsilon$ denote the entropy density, baryon number density, and energy density, respectively. 
The quantities held fixed in the differentiation are indicated by the subscripts.
However, an additional ambiguity arises when the derivatives with respect to the thermodynamic variables are evaluated: as discussed in the last subsection, the external sources are introduced to cancel the extra terms that appear in the 2PI WTIs but are absent in their 1PI counterparts.
Consequently, at the SI thermal equilibrium, the sources become functions of  $T$ and $\mu_B$, as in \Cref{eq:TmuDpendenceOfJs}.

\begin{figure}[t]
    \centering
    \includegraphics[width=0.7\linewidth]{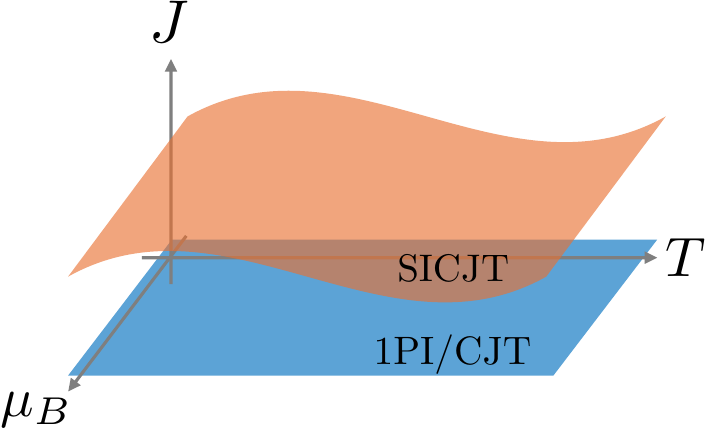}
    \caption{
    The sketch of the equilibria for different formulations.
    The axes denote the thermodynamic parameters such as the baryon chemical potential $\mu_B$, the temperature $T$, and the external sources $J$.
    The blue plane corresponds to the thermal equilibrium defined in the 1PI formalism as well as in the standard 2PI formalism, while the SICJT formalism gives the orange surface.
    }
    \label{fig:SICJTSketchpdf}
\end{figure}

The medium dependence of the sources affects thermodynamics if one adopts the conventional total-derivative prescription, 
\begin{align}
     \biggl( \frac{\partial P}{\partial X_i} \biggl)_{\rm SI} \equiv \frac{\df }{\df X_i} \biggl[
        P\bigl(T,\mu_B; j = J_{\mathrm{SI}}(T,\mu_B) \bigr)
    \biggr],
     \label{eq:EOStotalDer}
\end{align}
where the medium-dependent $J_{\mathrm{SI}}(T,\mu_B) $ contributes in part to the physical system, and the pressure $P$ is differentiated along the SI ``manifold" as illustrated for the effective action later in Fig.~\ref{fig:SICJTSketchpdf}. 
In \Cref{eq:EOStotalDer}, $X_i = T, \mu_B, \cdots$ denotes the all possible thermodynamic parameters.
One may retain the original thermodynamic definition and perform the differentiation at a fixed external source, imposing the SI condition only after the differentiation.
This way corresponds to subtracting the contribution induced by the implicit variation of the SI sources: 
\begin{align}
    \left( \frac{\partial P}{\partial X_i} \right)_{\mathrm{SI}}
    &\equiv
    \frac{\df}{\df X_i}
    \biggl[
        P\bigl(T,\mu_B; j = J_{\mathrm{SI}}(T,\mu_B) \bigr)
    \biggr]
    \nonumber\\
    &\qquad -
    \left.
    \frac{\partial P}{\partial j^m}
    \right|_{j = J_{\mathrm{SI}}}
    \frac{\df J_{\mathrm{SI}}^m(T,\mu_B)
    )}{\df X_i} \, .
    \label{eq:EOSpartialDer}
\end{align}
This subtraction ambiguity reflects two different interpretations of the source background: whether the SI source should be regarded as an intrinsic part of the thermodynamic system, or as an auxiliary construction introduced to enforce the symmetry constraints.
Our numerical analysis supports the latter interpretation in the present formulation. 
With the total-derivative prescription in \Cref{eq:EOStotalDer}, unphysical behavior may arise, such as negative entropy density or negative baryon number density in regions with positive $\mu_B$. 
By contrast, the fixed-source prescription in \Cref{eq:EOSpartialDer} removes these pathologies and yields physically reasonable isentropic trajectories. 
Therefore, in the following, we adopt \Cref{eq:EOSpartialDer} as the definition of the thermodynamic derivatives used to construct the equation of state.

Depending on the different normalization prescriptions for $P$, 
introduced in the previous subsection, the adopted derivative definition  \Cref{eq:EOSpartialDer} gives  
\begin{align}
    \left( \frac{\partial P_{\rm vac}}{\partial X_i} \right)_{\mathrm{SI}}
    &\equiv
    \frac{\df}{\df X_i}
    \biggl[
        P_{\rm vac}\bigl(T,\mu_B; j = J_{\mathrm{SI}} \bigr)
    \biggr]
    \nonumber\\
    &\quad -
    \bar \Phi_1 
    \frac{\df J_{\mathrm{SI}}^1}{\df X_i}
    -
    \bar \Phi_3 
    \frac{\df J_{\mathrm{SI}}^3}{\df X_i}\, ,
   \notag \\
    \left( \frac{\partial P_{\rm vac}^{(j)}}{\partial X_i} \right)_{\mathrm{SI}}
    &\equiv
    \frac{\df}{\df X_i}
    \biggl[
        P_{\rm vac}^{(j)}\bigl(T,\mu_B; j = J_{\mathrm{SI}} \bigr)
    \biggr]
    \nonumber\\
    &\quad -
    \Big(\bar \Phi_1 - \bar \Phi_1^{(0)} \Big)
    \frac{\df J_{\mathrm{SI}}^1}{\df X_i}
    -
    \Big(\bar \Phi_3 - \bar \Phi_3^{(0)} \Big)
    \frac{\df J_{\mathrm{SI}}^3}{\df X_i}\, ,
    \notag \\
    \left( \frac{\partial P^*}{\partial X_i} \right)_{\mathrm{SI}}
    &\equiv
    \frac{\df}{\df X_i}
    \biggl[
        P^*\bigl(T,\mu_B; j = J_{\mathrm{SI}} \bigr)
    \biggr]\, , \label{deriv-P}
\end{align}
for \Cref{eq:vacuumSubtractedPressure,eq:vacuumSubtractedPressureWithSource,eq:starPressureInLSM}, respectively. 
Substituting \Cref{eq:vacuumSubtractedPressure,eq:vacuumSubtractedPressureWithSource,eq:starPressureInLSM} into Eq.(\ref{deriv-P}), we get 
\begin{align}
    \left( \frac{\partial P_{\rm vac}}{\partial X_i} \right)_{\mathrm{SI}}
    &=
    \left( \frac{\partial P_{\rm vac}^{(j)}}{\partial X_i} \right)_{\mathrm{SI}}
    \nonumber\\
    &= \frac{\df}{\df X_i} \Big[ V_{m}(\bar \Phi^{(0)}) - V_{\rm CJT}(\bar \sigma, \mathcal G_S, \mathcal G_P) \Big]
    \nonumber\\
    &\qquad+ J^1_{\rm SI} \frac{\df \bar \Phi_1}{\df X_i} + J^3_{\rm SI} \frac{\df \bar \Phi_3}{\df X_i} \, ,
    \notag \\
    \left( \frac{\partial P^*}{\partial X_i} \right)_{\mathrm{SI}}
    &=
    \frac{\df}{\df X_i} \Big[ V_{m}(\bar \Phi^{(0)}) - V_{\rm CJT}(\bar \sigma, \mathcal G_S, \mathcal G_P) \Big] \,.
\end{align}
%
%The fixed-source derivatives of $P_{\rm vac}$ and $P_{\rm vac}^{(j)}$ are identical.
%This is because the difference associated with the explicit source dependence has been %removed by construction, while the remaining dependence on the condensates differs %only by a constant vacuum subtraction.

%%%%%%%%%%%%%%
%%%%%%%%%%%%%%
%%%%%%%%%%%%%%
%%%%%%%%%%%%%%

For higher-order derivatives, such as susceptibilities, \Cref{eq:EOSpartialDer} can be generalized to an arbitrary thermodynamic quantity $\varsigma$ as
\begin{align}
    \left( \frac{\partial \varsigma}{\partial X_i} \right)_{\mathrm{SI}}
    &\equiv
    \frac{\df}{\df X_i}
    \biggl[
        \varsigma\bigl(T,\mu_B; j = J_{\mathrm{SI}} \bigr)
    \biggr]
    \nonumber\\
    &\qquad -
    \left.
    \frac{\delta \varsigma}{\delta j^m}
    \right|_{j = J_{\mathrm{SI}}}
    \cdot
    \frac{\df J_{\mathrm{SI}}^m}{\df X_i} \, ,
    \label{eq:EOSHigherpartialDer}
\end{align}
where $\delta/\delta j^m$ denotes the functional derivative, and the spacetime integration has been abbreviated into the symbol ``$\cdot$''.
Accordingly, the second derivatives of the different pressure prescriptions take the forms 
\begin{align}
    \left( \frac{\partial^2 P_{\rm vac}}{\partial X_i \partial X_j} \right)_{\mathrm{SI}}
    &=
    \left( \frac{\partial^2 P_{\rm vac}^{(j)}}{\partial X_i \partial X_j} \right)_{\mathrm{SI}}
    \nonumber\\
    &= \frac{\df^2}{\df X_i \df X_j} \Big[ V_{m}(\bar \Phi^{(0)}) - V_{\rm CJT}(\bar \sigma, \mathcal G_S, \mathcal G_P) \Big]
    \nonumber\\
    &\qquad+ J^1_{\rm SI} \frac{\df^2 \bar \Phi_1}{\df X_i \df X_j} + J^3_{\rm SI} \frac{\df^2 \bar \Phi_3}{\df X_i \df X_j} \, ,
    \\
    \left( \frac{\partial^2 P^*}{\partial X_i \partial X_j} \right)_{\mathrm{SI}}
    &=
    \frac{\df^2}{\df X_i \df X_j} \Big[ V_{m}(\bar \Phi^{(0)}) - V_{\rm CJT}(\bar \sigma, \mathcal G_S, \mathcal G_P) \Big] \,.
\end{align}
With this prescription, the Maxwell relation
\begin{align}
    \left( \frac{\partial s}{\partial \mu_B} \right)_{\mathrm{SI}} = \left( \frac{\partial n_B}{\partial T} \right)_{\mathrm{SI}}
\end{align}
is satisfied in all three normalization criteria.

%This completes the current discussion on thermodynamic observables within the symmetry-%improved formalism. 
In the following section, we analyze the thermodynamics of the SI system 
based on the three different descriptions for pressure and related equations of state. 
The schematic relations among the three different cases are depicted in \Cref{fig:SICJTSketchpdf}.
In the 1PI formalism, the GMOR relation is satisfied irrespective of the external sources, whereas in the 2PI formalism, it is fulfilled only for particular source values.
Accordingly, the equilibrium states are defined differently with or without the presence of external sources.

%%%%%%%%%%%%%%
%%%%%%%%%%%%%%
\begin{figure*}
    \centering
    \includegraphics[width=0.45\linewidth]{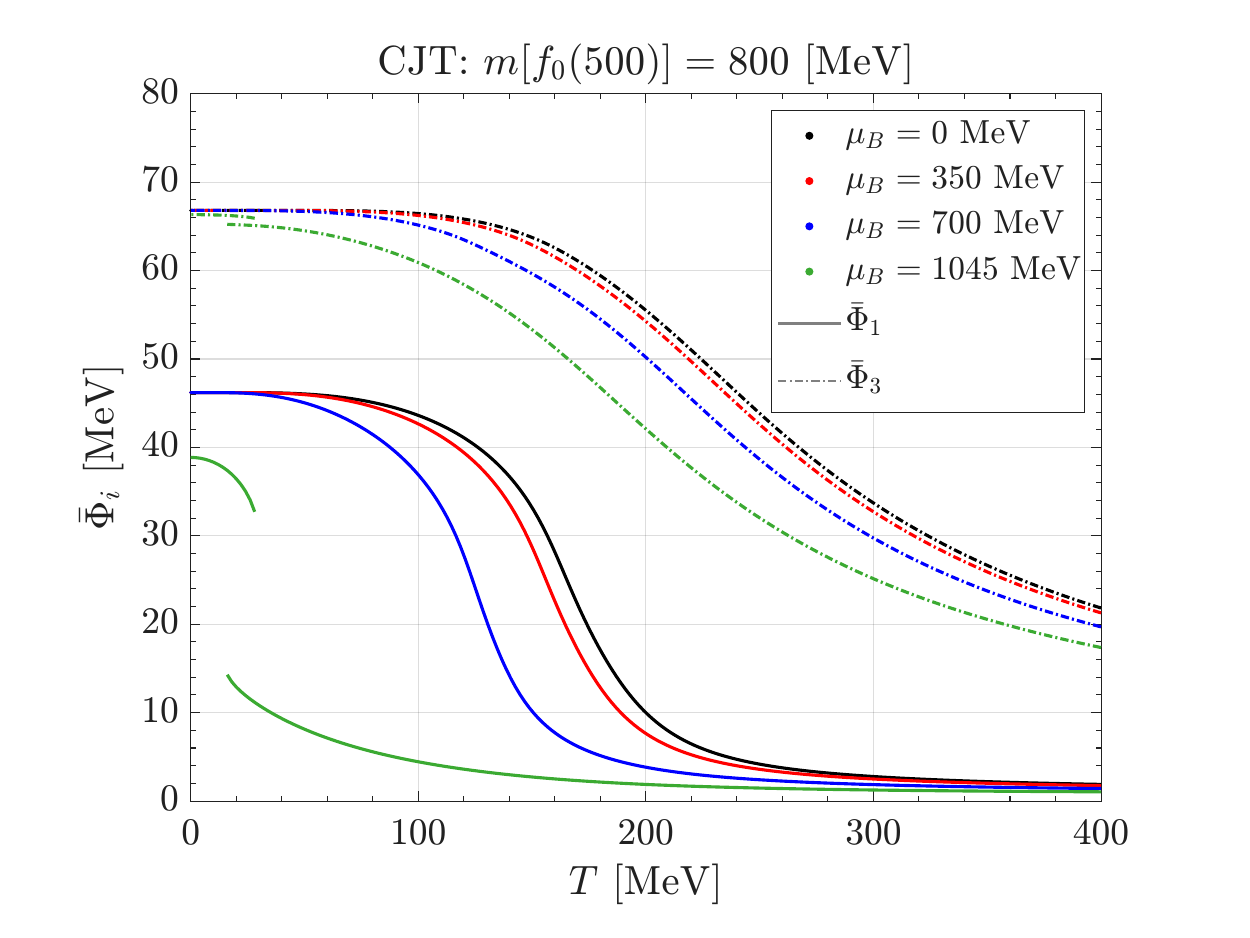}
    \includegraphics[width=0.45\linewidth]{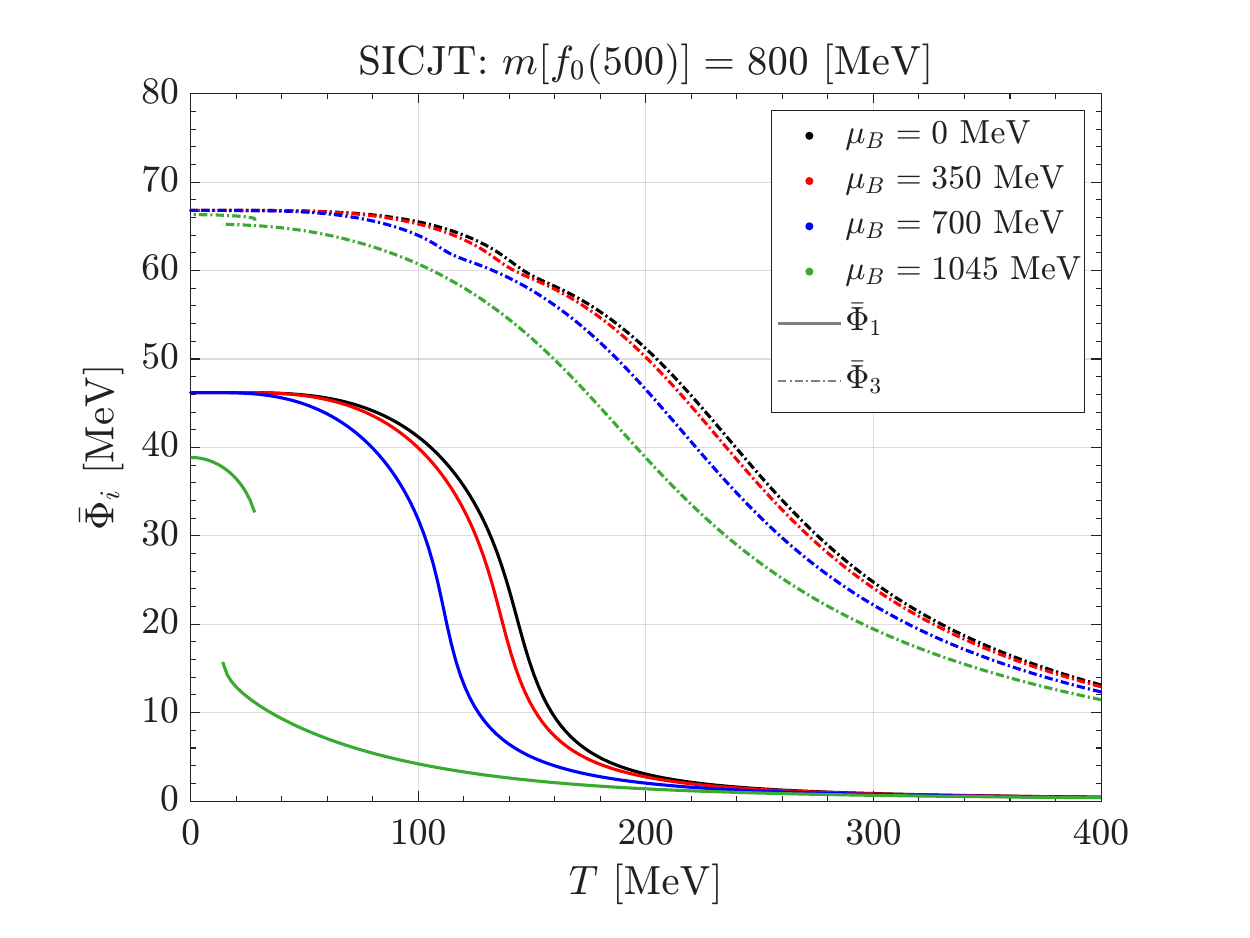}
    \caption{
    Temperature dependence of the chiral order parameters $\bar \Phi_1$ (solid lines) and $\bar \Phi_3$ (dashed lines) in the CJT approach (left panel) and the SICJT approach (right panel), for several values of the baryon chemical potential $\mu_B$, indicated by different colors.
    }
    \label{fig:Condensates}
\end{figure*}
%%%%%%%%%%%%%%
%%%%%%%%%%%%%%

%%%%%%%%%%%%%%
%%%%%%%%%%%%%%
\begin{figure*}
    \centering
    \includegraphics[width=0.32\linewidth]{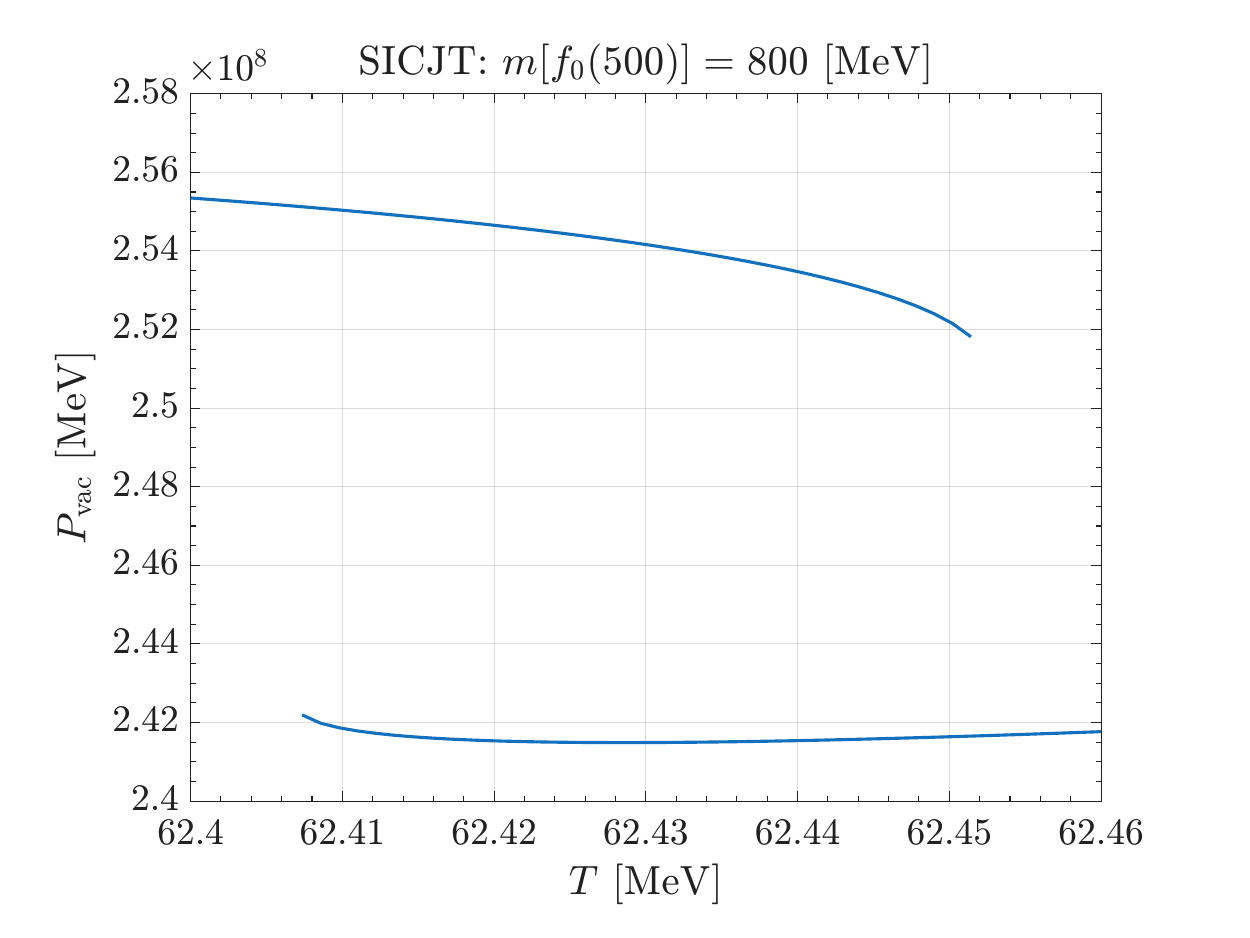}
    \includegraphics[width=0.32\linewidth]{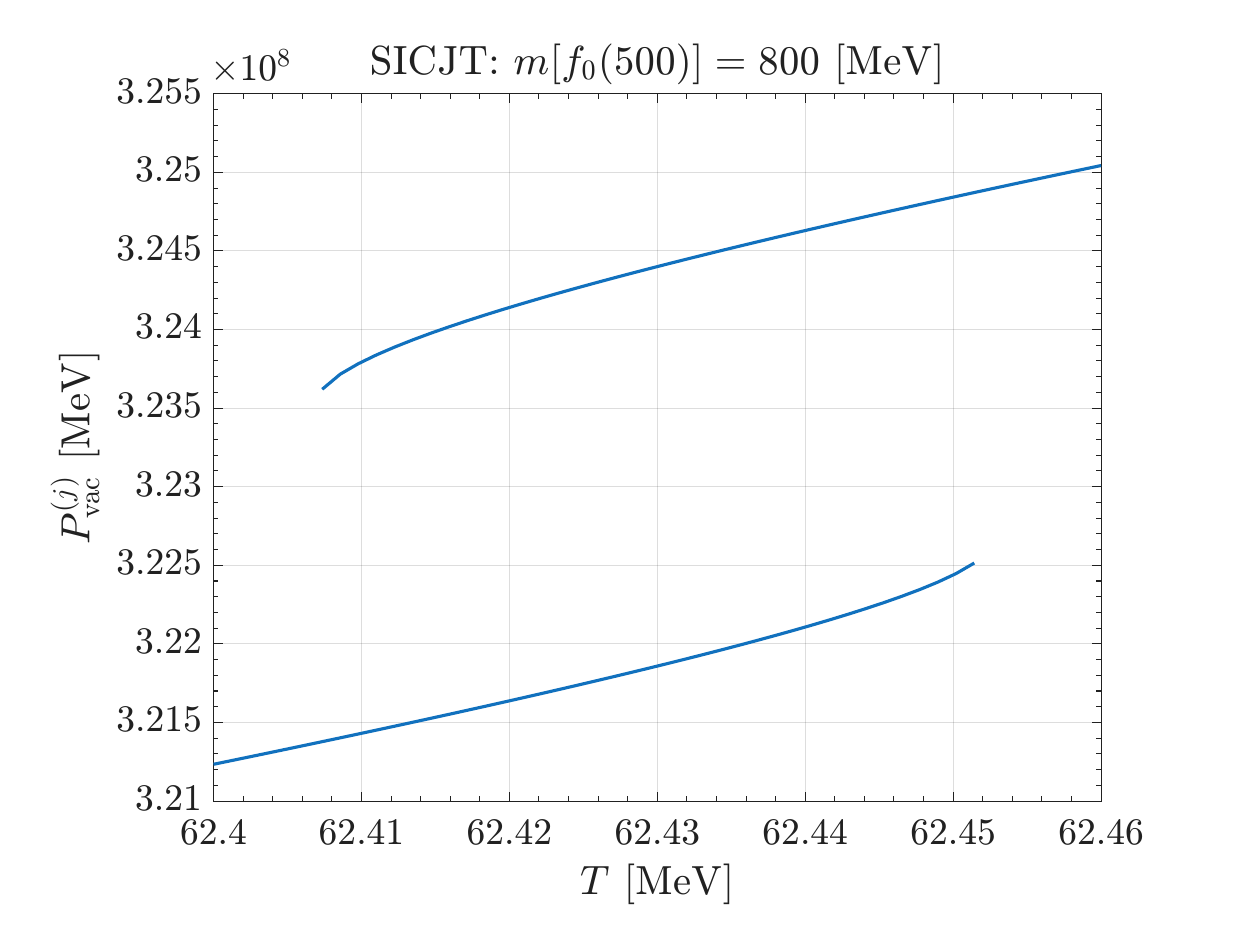}
    \includegraphics[width=0.32\linewidth]{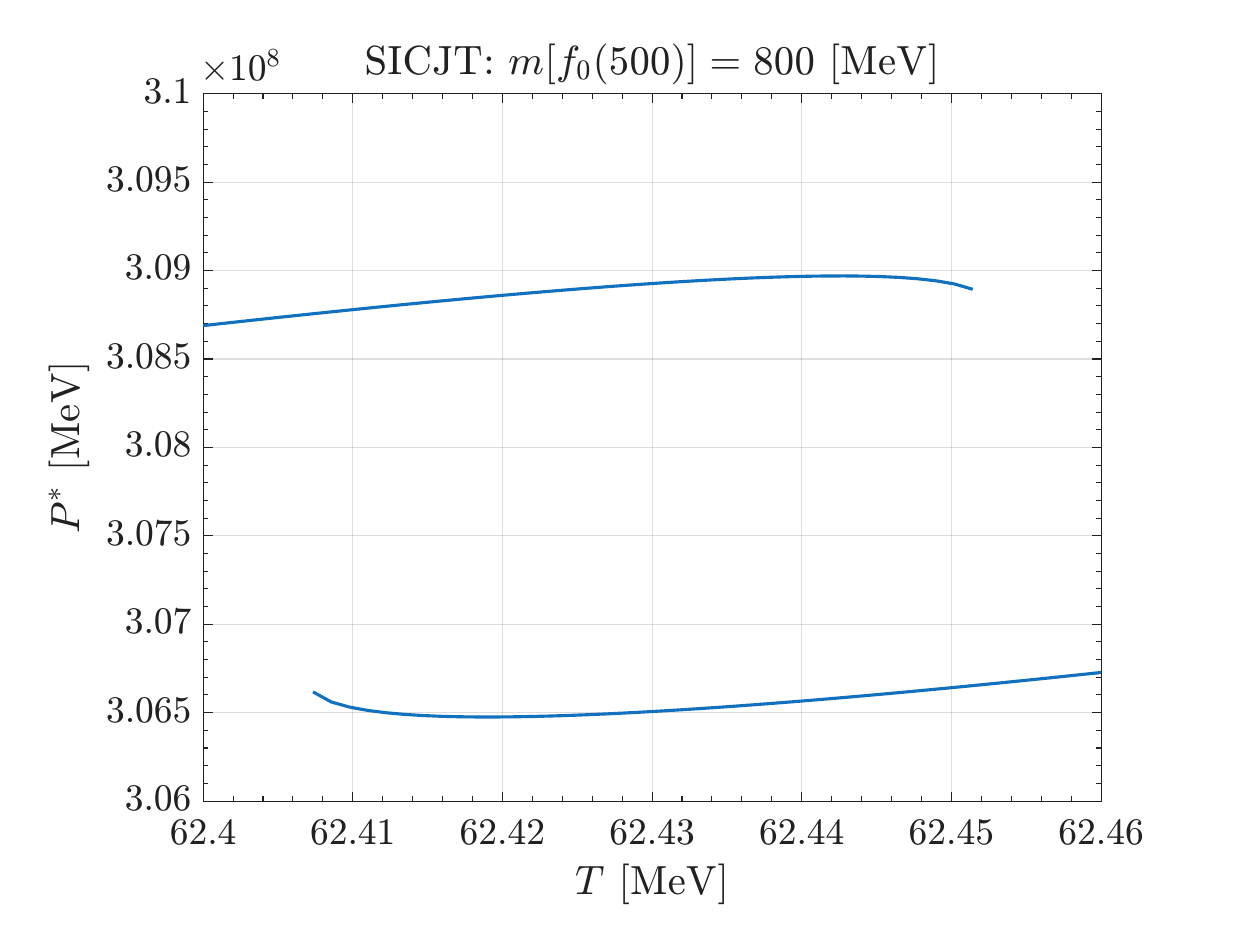}
    \caption{
    Temperature dependence of the pressures under three different prescriptions at $\mu_B = 970 \ {\rm MeV}$. 
    }
    \label{fig:SpinodalPressures}
\end{figure*}
%%%%%%%%%%%%%%
%%%%%%%%%%%%%%

%%%%%%%%%%%%%%
%%%%%%%%%%%%%%
\begin{figure*}
    \centering
    \includegraphics[width=0.32\linewidth]{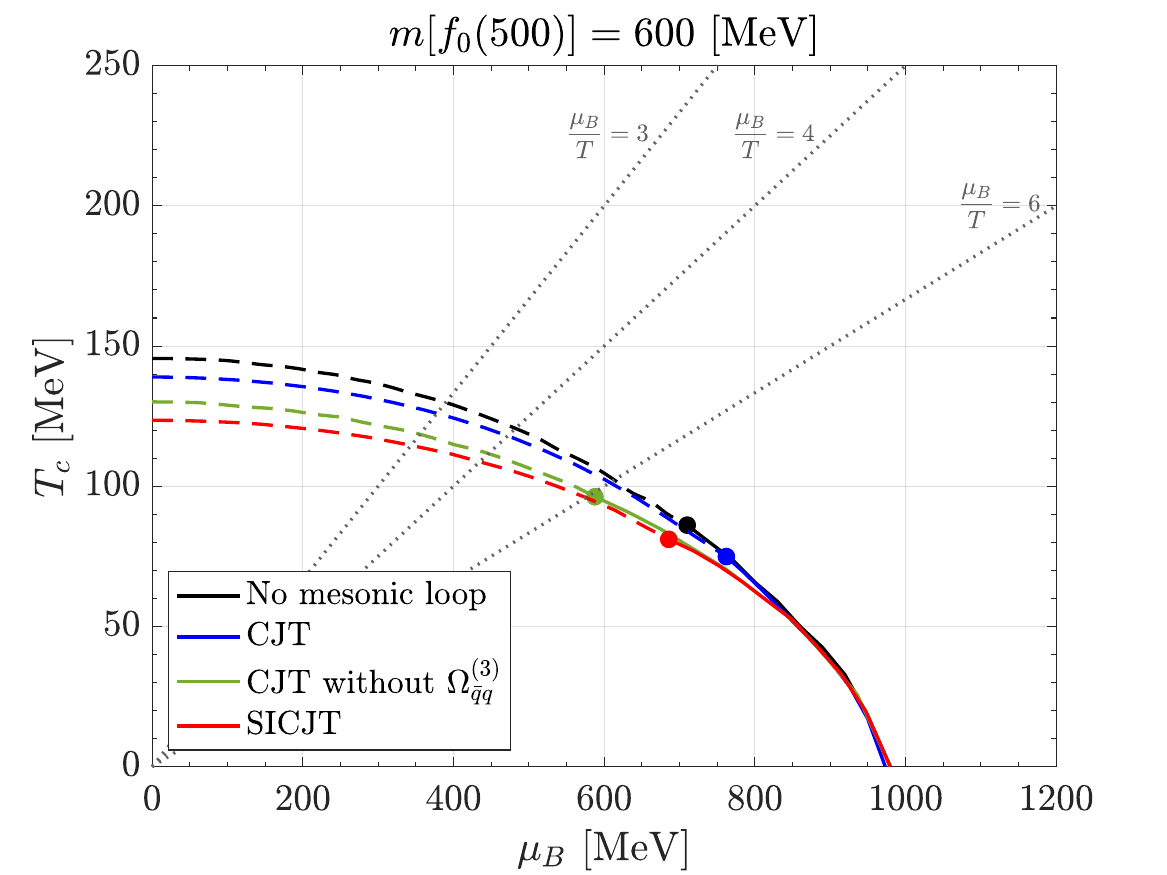}
    \includegraphics[width=0.32\linewidth]{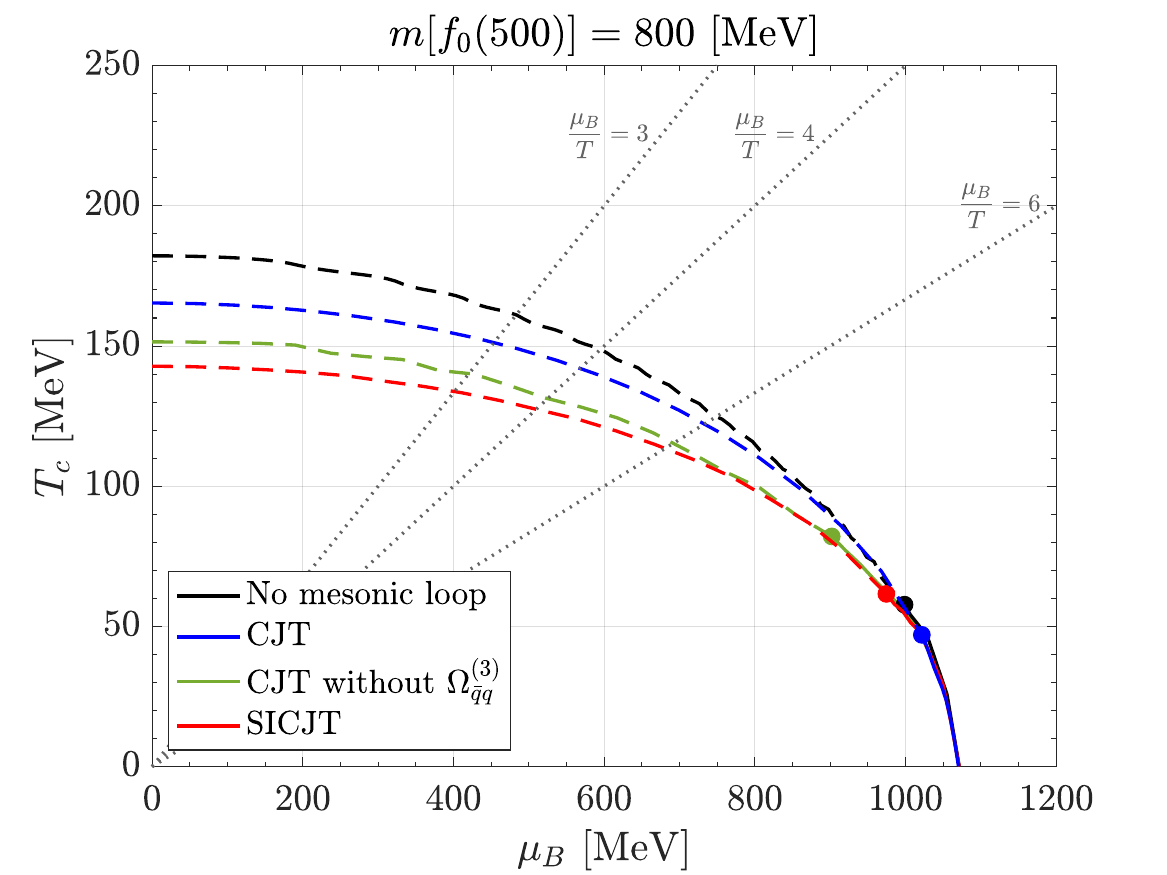}
    \includegraphics[width=0.32\linewidth]{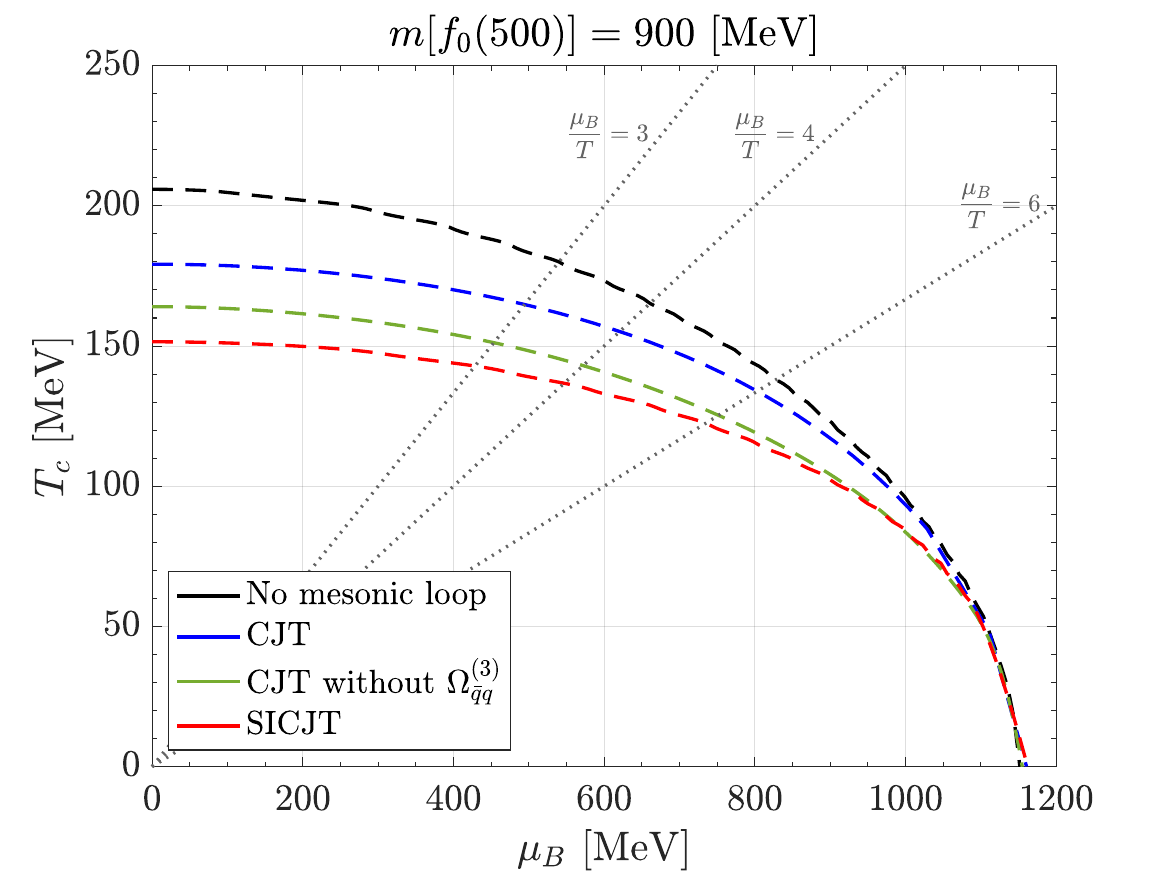}
    \caption{
    Phase diagrams of the D$\chi$SB in the $(\mu_B, T)$ plane from different approaches introduced in \Cref{sec:StationaryConditionsAndSymmetryImprovement}.
    From left to right panels, the $f_0(500)$ masses are taken to be $m[f_0(500)] = 600 \ {\rm MeV}$, $m[f_0(500)] = 800 \ {\rm MeV}$, and $m[f_0(500)] = 900 \ {\rm MeV}$.
    For each panel, the solid lines denote the first-order phase boundaries, the solid blobs denote the position of the critical end points, and the dashed lines the pseudo-critical temperature lines of the chiral crossover. 
    }
    \label{fig:PhaseDiagram}
\end{figure*}
%%%%%%%%%%%%%%
%%%%%%%%%%%%%%

%%%%%%%%%%%%%%
%%%%%%%%%%%%%%
\begin{figure*}
    \centering
    \includegraphics[width=0.45\linewidth]{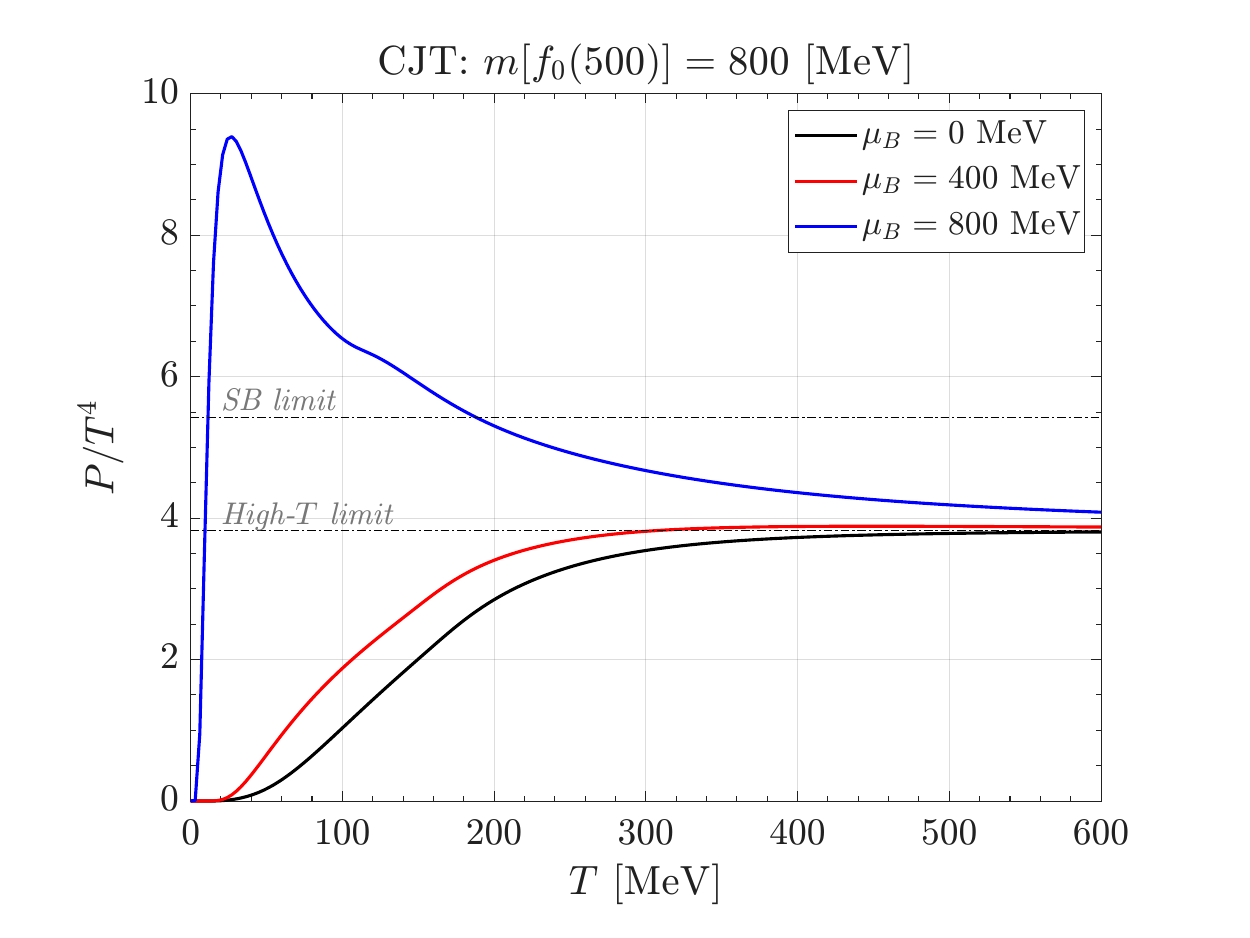}
    \includegraphics[width=0.45\linewidth]{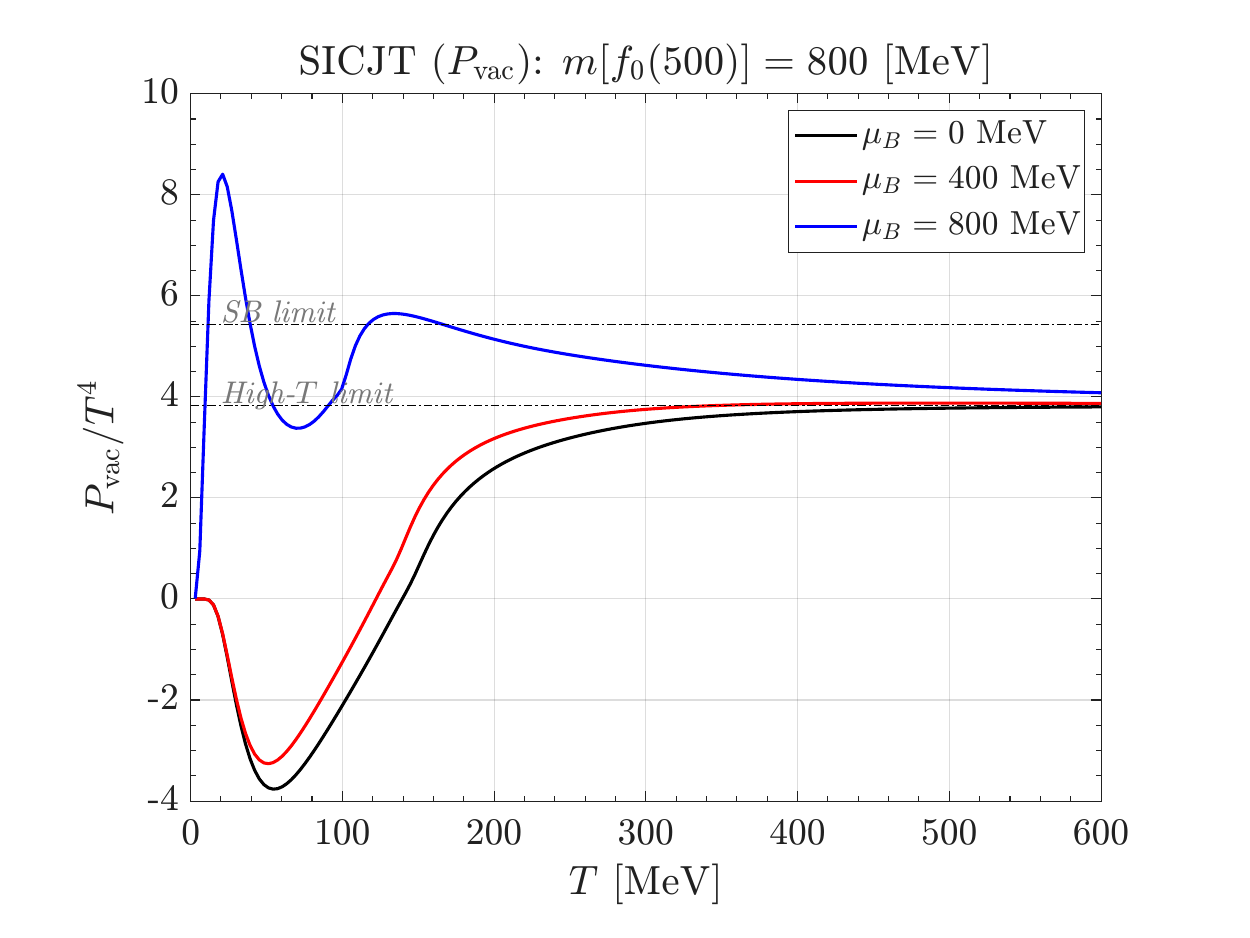}
    \includegraphics[width=0.45\linewidth]{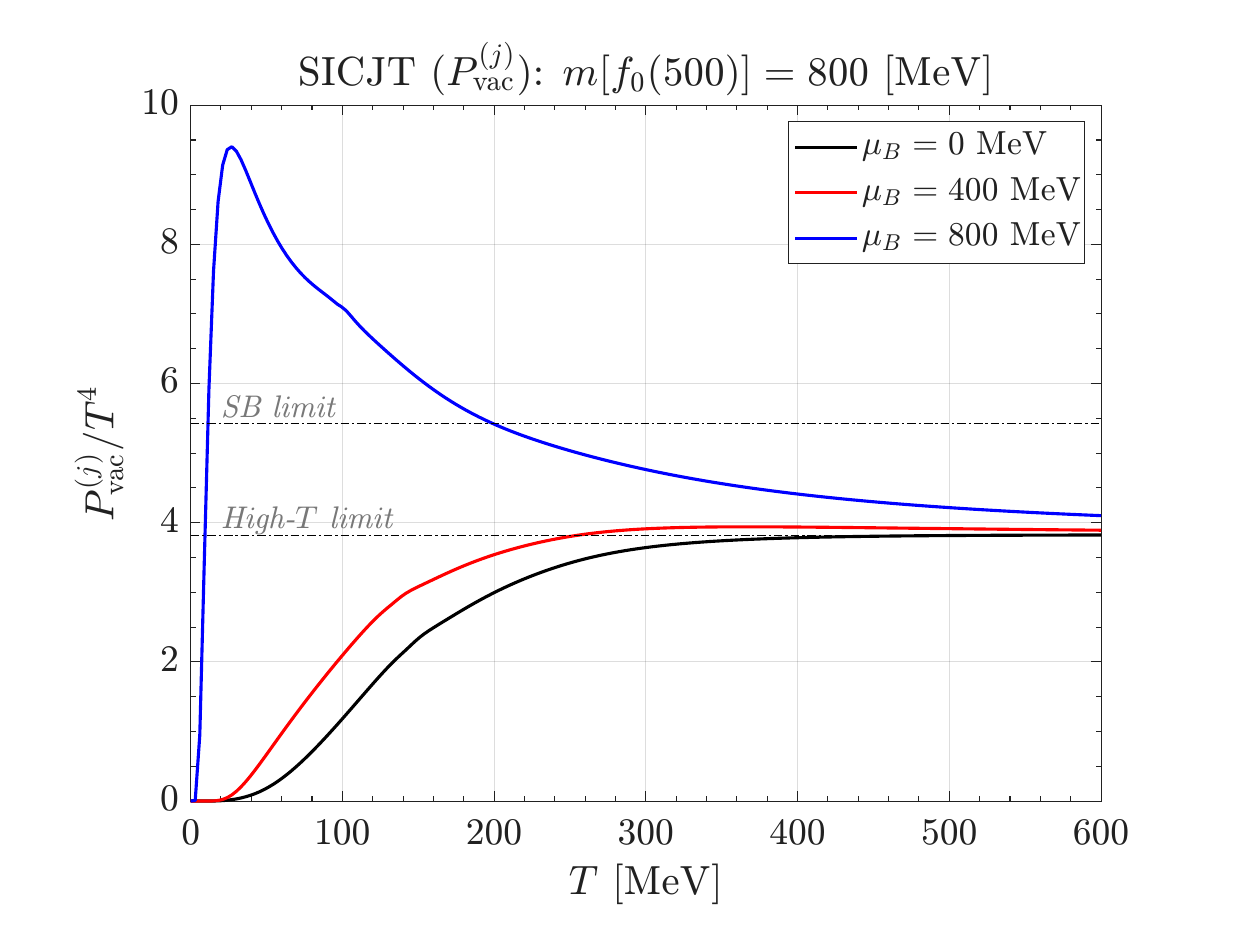}
    \includegraphics[width=0.45\linewidth]{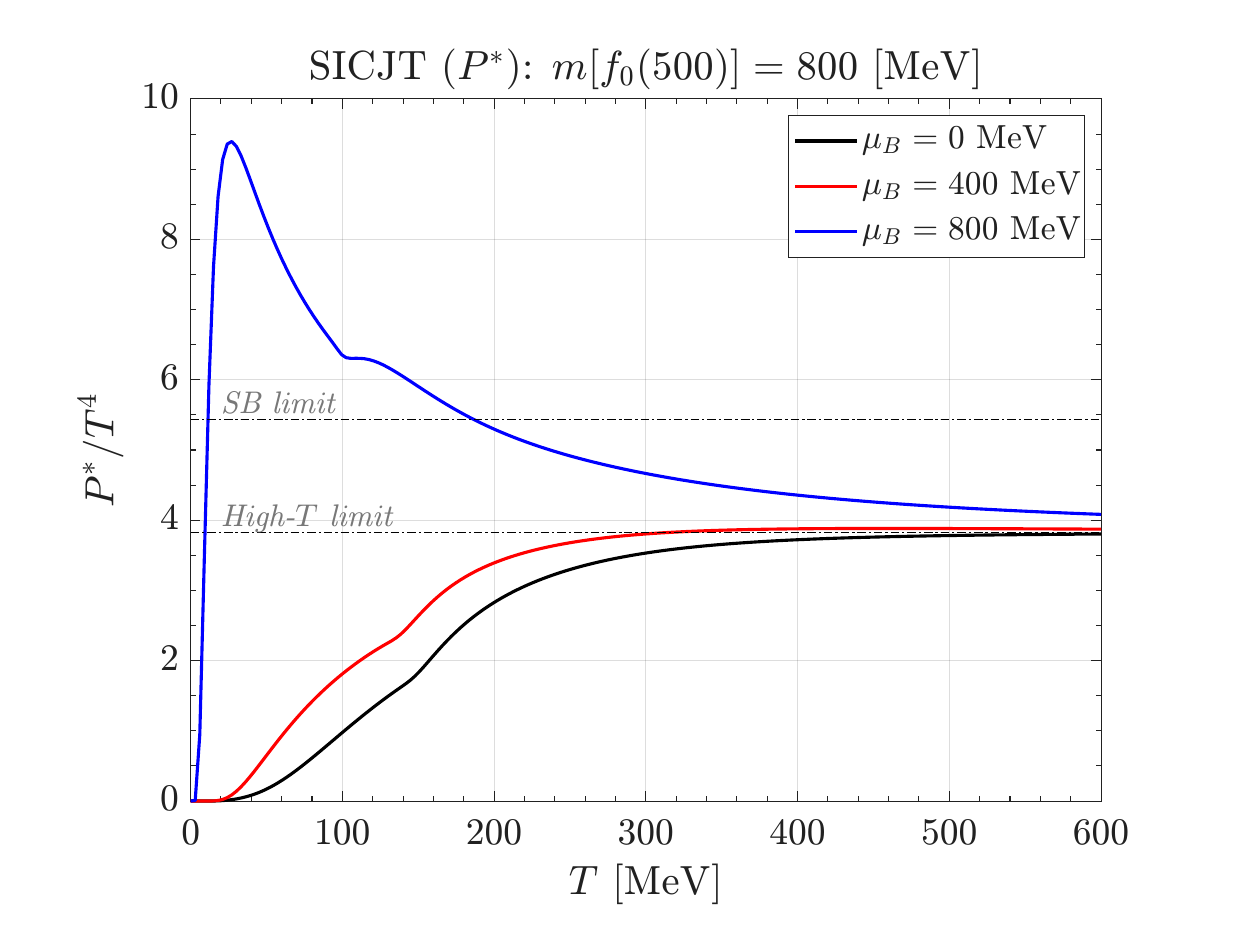}
    \caption{
    The pressures, normalized in three different ways as in the text, are plotted as a function of temperature for several fixed values of the baryon chemical potential $\mu_B$ (solid lines). 
    The different panels display the results obtained in different formalisms (CJT or SICJT) and with different pressure prescriptions ($P_{\rm vac}$, or $P_{\rm vac}^{(j)}$, $P^*$). 
    The horizontal dashed lines denote the model-specific high-temperature limit and the Stefan-Boltzmann (SB) limit. More detailed discussions are given in the main text. 
    }
    \label{fig:HighTPressure}
\end{figure*}
%%%%%%%%%%%%%%
%%%%%%%%%%%%%%

%%%%%%%%%%%%%%
%%%%%%%%%%%%%%
\begin{figure*}
    \centering
    \includegraphics[width=0.32\linewidth]{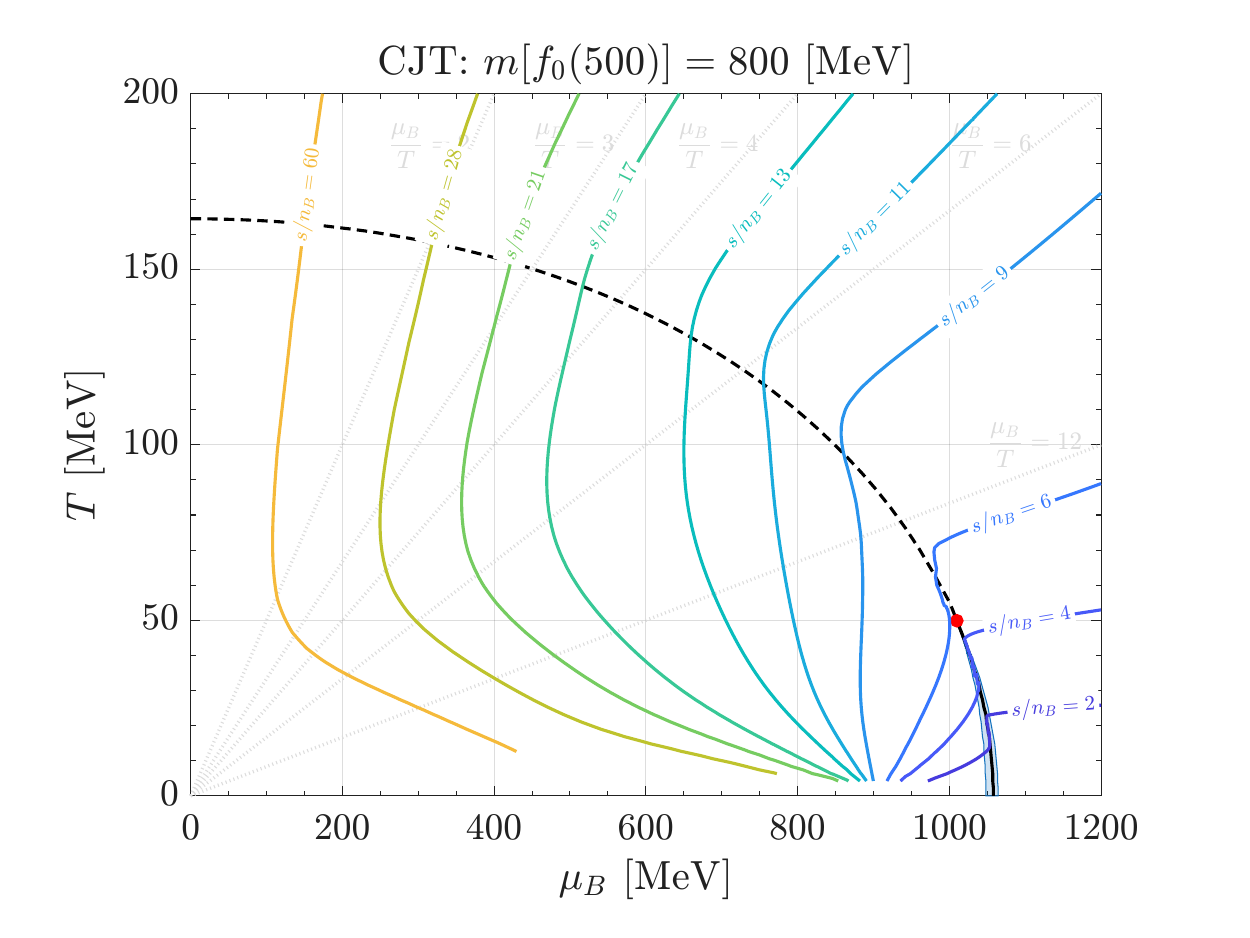}
    \includegraphics[width=0.32\linewidth]{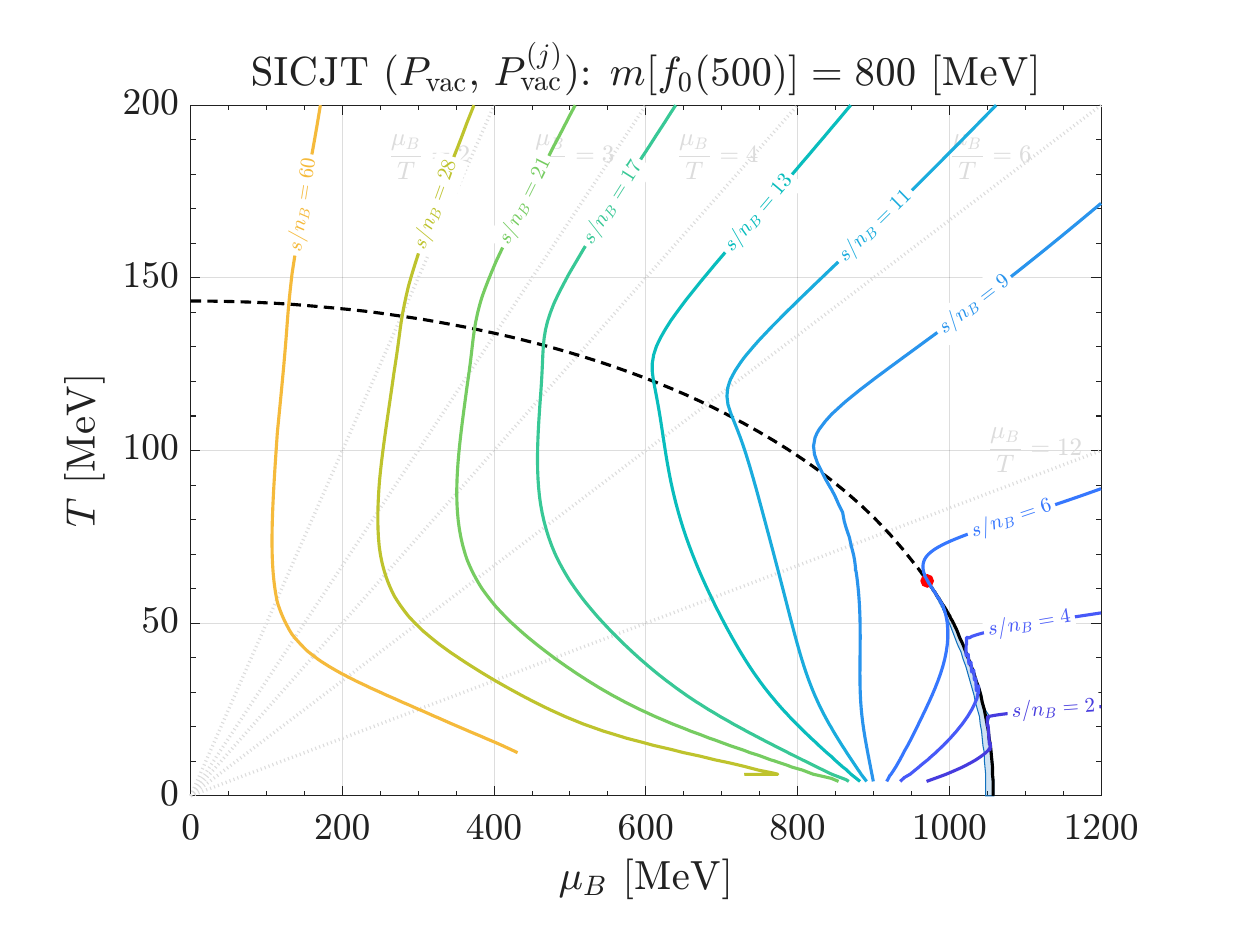}
    \includegraphics[width=0.32\linewidth]{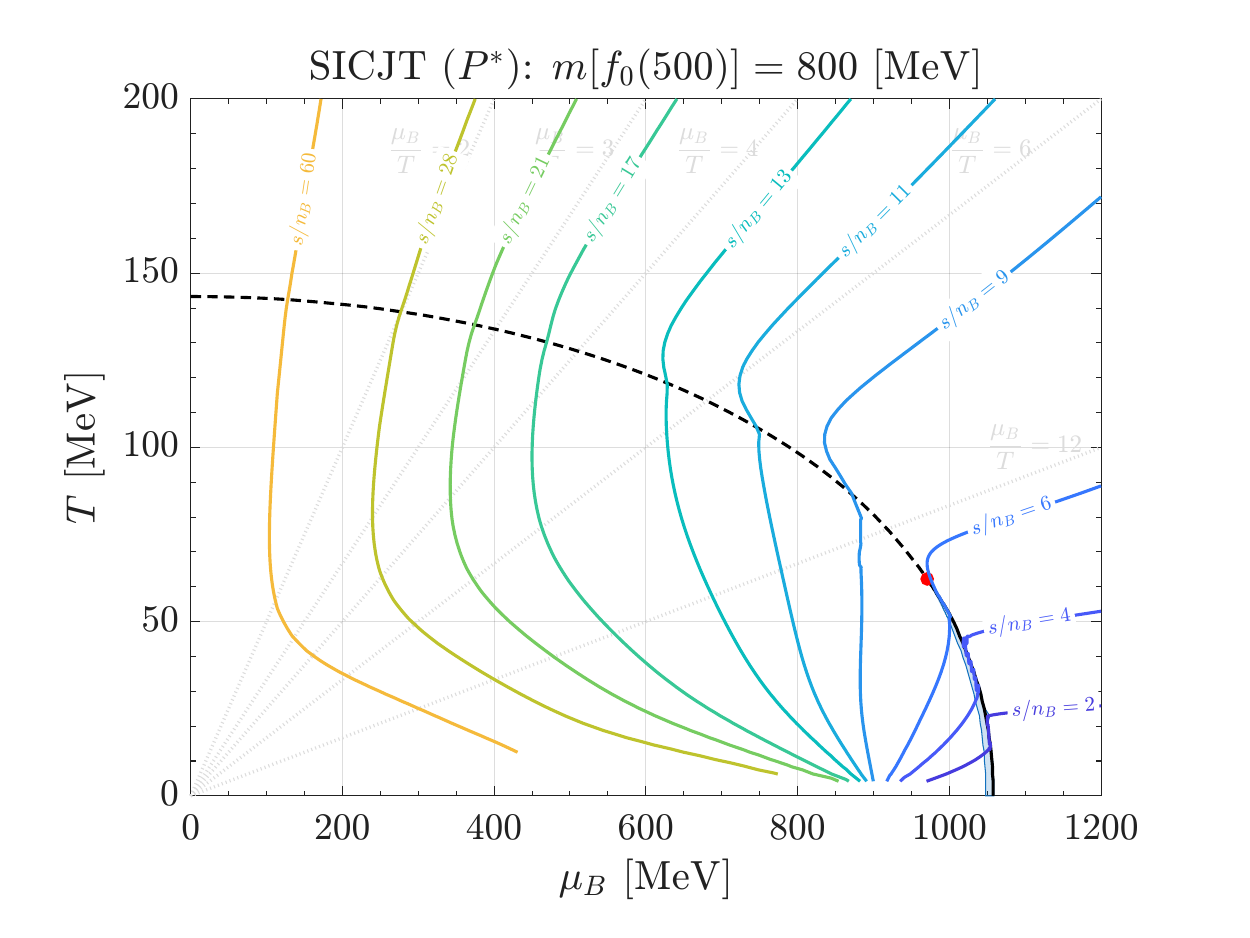}
    \caption{
    Isentropic trajectories (solid colored lines) in the phase diagram for $s/n_B = [60, 28, 21, 17, 13, 11, 9, 6, 4, 2]$. 
    The different panels show the results obtained in different formalisms and with different pressure prescriptions. 
    The solid and dashed lines denote the first-order phase and crossover boundaries, respectively, while the critical endpoint is marked by the red dot. 
    The blue shaded region indicates the spinodal region.
    }
    \label{fig:isentropicTrajectory}
\end{figure*}
%%%%%%%%%%%%%%
%%%%%%%%%%%%%%

%%%%%%%%%%%%%%%%%%%%%%%%%%%%%%%%%%%%%%%%%%%%%%%%%%%%%%%%%%%%%%%%%%%%%%
\section{Numerical analysis}
\label{sec:Results}

In this section, we present the results on the numerical evaluation of the chiral phase diagram and related thermodynamics obtained from the LSM using the approaches introduced in \Cref{sec:2PIFormalismAndSymmetryImprovement}. 
For numerical implementations, we refer readers to Ref.~\cite{Guan:2025tmf}.

\subsection{Parameter setup}
\label{sec:ParamSetups}

In this work, the model is specified by seven independent parameters,
\begin{align}
    g, \quad \mu^2, \quad \lambda_1, \quad \lambda_2, \quad c m_l, \quad c m_s, \quad B .
\end{align}
The parameters entering the mesonic potential $V_m$ are fixed at tree level by reproducing vacuum mesonic observables, with the vacuum quantum fluctuation contribution subtracted. 
Following Ref.~\cite{Schaefer:2008hk}, we consider three parameter sets. 
They share the same decay-constant inputs,
\begin{align}
    f_\pi = 92.4~{\rm MeV} \, , 
    \qquad 
    f_K = 113.02~{\rm MeV} \, ,
\end{align}
which are related to the mesonic field VEVs through 
\begin{align}
    f_\pi = 2 \bar \Phi_1 \, ,
    \qquad
    f_K = \bar \Phi_1 + \bar \Phi_3 \, .
\end{align}
The common mesonic parameters are
\begin{gather}
    c m_l = (95.82)^3~{\rm MeV^3} \, ,
    \quad
    c m_s = (299.71)^3~{\rm MeV^3} \, , 
    \nonumber\\
    \lambda_1 = 46.48 \, , 
    \quad
    B = 4807.84~{\rm MeV} \, .
\end{gather}
The three parameter sets mainly differ in the choice of the light scalar-isoscalar mass $m[f_0(500)]$. 
This choice fixes the remaining set-dependent parameters $\mu^2$ and $\lambda_2$, and also affects the heavier scalar-isoscalar mass and the scalar mixing angle. 
The resulting values are summarized in \Cref{tab:ParameterSets}.

{
\renewcommand{\arraystretch}{1.25}
\begin{table}[t]
\caption{
Parameter sets adopted from Ref.~\cite{Schaefer:2008hk}. 
The three sets differ primarily by the input value of $m[f_0(500)]$.
}
\label{tab:ParameterSets}
\begin{ruledtabular}
\begin{tabular}{c c c c c c}
Set 
& \begin{tabular}{c}
    $\mu^2$ \\
    $[{\rm MeV}^2]$
  \end{tabular}
& $\lambda_2$
& \begin{tabular}{c}
    $m[f_0(500)]$ \\
    $[{\rm MeV}]$
  \end{tabular}
& \begin{tabular}{c}
    $m[f_0(980)]$ \\
    $[{\rm MeV}]$
  \end{tabular}
& $\theta_S^0$
\\
\hline
I   
& $(342.52)^2$    
& $1.40$  
& $600$ 
& $1221.1$ 
& $19.9^\circ$ 
\\
II  
& $-(306.26)^2$  
& $13.49$ 
& $800$ 
& $1278.0$ 
& $30.7^\circ$ 
\\
III 
& $-(520.80)^2$  
& $23.65$ 
& $900$ 
& $1348.0$ 
& $40.9^\circ$ 
\end{tabular}
\end{ruledtabular}
\end{table}
}

For all three sets, the remaining vacuum meson spectrum is kept unchanged,
\begin{gather}
    m_{a_0} = 1028.7~{\rm MeV} \, , 
    \quad
    m_\kappa = 1124.3~{\rm MeV} \, , 
    \nonumber\\
    m_{\eta^\prime} = 963.0~{\rm MeV} \, , 
    \quad
    m_\eta = 539.0~{\rm MeV} \, , 
    \nonumber\\
    m_\pi = 138~{\rm MeV} \, , 
    \quad
    m_K = 496~{\rm MeV} \, ,
\end{gather}
and the pseudoscalar mixing angle is
\begin{align}
    \theta_P^0 = -5.0^\circ .
\end{align}
The Yukawa coupling is fixed around $g \sim 6.5$ so that the light constituent quark mass yields $M_l = 300~{\rm MeV}$.
Thus, the three parameter sets provide a controlled way to examine the sensitivity of the thermodynamics to the scalar-isoscalar sector, especially to the poorly constrained vacuum mass of $f_0(500)$. 
In the following analysis, we use parameter set II, corresponding to $m[f_0(500)] = 800~{\rm MeV}$, as our primary reference.

\subsection{Chiral order parameters and phase diagram}
\label{sec:ChiOrderParamAndPhaseDiagram}

%We first briefly examine the results of the chiral phase transition and its phase diagram properties.

In \Cref{fig:Condensates}, we show the temperature dependence of the chiral order parameters \labelcref{eq:MeanFieldBackground} from the CJT and the SICJT formalisms for different values of baryon chemical potential $\mu_B$. 
In both formalisms, the temperature dependence of the light $\Phi_1$ and strange $\Phi_3$ condensates exhibits the expected hierarchical behavior, as also seen in other approaches. 
The crossover pseudo-critical temperature lines $T_{\rm pc}(\mu_B)$ (dashed lines) are identified as  the peak place for the thermal susceptibility of the light-quark condensate
\begin{align}
    \chi_{T,\ell} \equiv - \frac{\partial \langle \bar \ell \ell \rangle}{\partial T}.
    \label{chiT}
\end{align}
As $\mu_B$ increases, the drop in condensate becomes progressively sharper, and the corresponding transition shifts to lower temperatures, signaling that the theory approaches the first-order region where the different phase branches develop. 
Comparing the CJT and SICJT results at the same $\mu_B$, one observes that the SI construction modifies both the location and the sharpness of the condensate melting. This indicates that the restoration pattern is quantitatively sensitive to the manner in which the chiral symmetry constraints are imposed. 
%With the chiral order parameters determined over the $(\mu_B, T)$ plane, the phase structure associated %with chiral symmetry restoration can be inferred. 

For orientation with respect to other effective and first-principles approaches, the zero-density condensate behavior and the corresponding pseudocritical temperatures are compared with several representative results from other approaches in \Cref{app:chiralCondensatesCompare}. 
Since the definitions and normalizations of the condensates are not identical in all approaches, the comparison is used to characterize the crossover location and shape, rather than as a point-by-point identification of the order parameters.

For the first-order phase boundary (FOPB), we compare the pressures of coexisting phase branches 
at fixed $T$ and $\mu_B$. 
As noted at the end of \Cref{sec:PressureAndThermodynamicAmbiguity}, the various pressure prescriptions in the SICJT approach involve a common ambiguity: in the spinodal region, the pressures of different branches do not necessarily cross each other.
This situation can be seen in \Cref{fig:SpinodalPressures} for different prescriptions of the pressure in the SICJT at $\mu_B = 970$ MeV. 
This issue originates from the fact that the SICJT procedure displaces the solutions associated with the saddle-point structure of the original truncated CJT effective action $\Gamma_{\rm tr}$.
Consequently, a branch-by-branch comparison of the pressure, or of the effective potential, becomes uncertain in the spinodal region, since different branches generally correspond to different intrinsic source backgrounds $J_{\rm SI}$. 
In the present work, the FOPB in the SICJT formalism is defined by making the comparison based on $P_{\rm vac}$ for demonstration.

In \Cref{fig:PhaseDiagram}, we present the chiral phase diagrams obtained from the parameter sets I, II, and III specified in~\Cref{tab:ParameterSets}.
Since these parameter sets differ mainly in the input value of the light scalar mass $m[f_0(500)]$, the resulting changes in the phase boundaries directly reflect the sensitivity of the chiral phase structure to the scalar meson sector. 
The dashed pseudo-critical lines are extracted from the maxima of the thermal susceptibility of the light-quark condensate in Eq.(\ref{chiT}), while the solid lines denote the first-order phase boundaries and the solid dot the corresponding critical end point (CEP). 
As $m[f_0(500)]$ increases, the pseudo-critical temperature at $\mu_B=0$ increases, whereas the CEP moves toward larger $\mu_B$ and lower $T$, and eventually disappears from the phase diagram. 

The differences among the phase diagrams obtained from the various approaches can be traced back to the distinct prescriptions used to determine the equilibrium background.
In the pure fermionic one-loop case, the phase boundaries reproduce the benchmark result in Ref.~\cite{Schaefer:2008hk}.
Once mesonic fluctuations are included, the phase boundaries are modified quantitatively. 
In particular, the difference between the CJT and SICJT results reflects the fact that the SICJT formalism replaces the stationary conditions of the truncated 2PI effective potential by the GMOR-type symmetry-improvement constraints, and therefore probes a different realization of chiral symmetry at the truncated level.

As emphasized above, the SICJT first-order boundary shown in \Cref{fig:PhaseDiagram} should be understood as the result extracted from the $P_{\rm vac}$ prescription, rather than as a prescription-independent thermodynamic phase boundary. 
In contrast, the crossover line is comparatively more robust because it is determined directly from the susceptibility of the light condensate.

\subsection{Equations of state}
\label{sec:EquationsOfState}

In this subsection, we compare the equations of state obtained from the different pressure prescriptions introduced in \Cref{sec:Thermodynamics} within the SICJT formalism. 
The equation of state and other thermodynamic quantities associated with 
energy density and pressure are sensitive to the particle contents of the theory. 
In lattice QCD, gluons govern there as a large regular background at high $T$~\cite{HotQCD:2014kol}, while the present LSM with quarks disregards gluons. 
%Lattice QCD results on the equation of state are thus dominated by the gluonic degrees of freedom in the thermal bath, so 
Thus, quantitatively comparing the result of our current work with the lattice one 
will be misleading, hence we will not do that in this section, and all the subsequent sections~\Cref{sec:IsentropicTraj,sec:SoundVel,sec:TraceAnomaly}. 
In Appendix~\Cref{app:EOSCompare} we present a qualitative comparison with lattice QCD only on the pseudo-critical regime of the chiral crossover transition.

In \Cref{fig:HighTPressure}, we show the normalized pressure, $P/T^4$, as a function of temperature at fixed baryon chemical potentials $\mu_B=0$, $400$, and $800$ MeV.
The CJT pressure ($P$), the pulled-back pressure $P^*$, and the source-matched vacuum-subtracted pressure $P_{\rm vac}^{(j)}$ exhibit very similar behavior over most of the temperature range, in particular in the intermediate and high temperature regions.
By contrast, $P_{\rm vac}$ based on the conventional vacuum-subtracted prescription shows 
a qualitatively different low-temperature behavior from the other three cases, which can drop below zero during thermal evolution. 
This indicates that the $P_{\rm vac}$ prescription is much more sensitive to the source background, whereas $P^*$ and $P_{\rm vac}^{(j)}$ are fairly insensitive to the source background and seem to capture the conventional thermal evolution while maintaining positive pressure.

In the low-temperature region at $\mu_B = 800$ MeV, there exists a peak structure for all four cases. 
We recognize this universal behavior as the pre-stage of the non-vanishing pressure induced by sufficiently high $\mu_B$ at $T=0$, where the normalized pressure $P/T^4$ tends to blow up from $T=0$. 
% in the low-temperature region.
This phenomenon reflects the so-called Silver Blaze property~\cite{Cohen:2003kd, Fu:2016tey} of the pressure, which keeps zero below the threshold value of the baryon chemical potential at $\mu_B = 3 M_l \sim 900$ MeV.

At high temperature, none of those pressure prescriptions realizes the SB limit.
Instead, all four pressures converge to the identical asymptotic value specific to the present model, $P/T^4 \sim 4$.
The reason is that, within the present quark-meson description, the mesons are treated as fundamental dynamical particles, and the corresponding meson mass-to-temperature ratio
\begin{align}
    r_T^\phi \equiv \left( \frac{\textrm{meson mass}}{T} \right)
\end{align}
does not vanish in the limit $T\to\infty$, but rather approaches a finite and numerically sizable constant. 
For the parameter set with $m[f_0(500)] = 800$ MeV, $N_c=3$, and $N_f=3$, we have 
\begin{align}
    r_\infty^\phi \simeq 3.6674 \, .
\end{align}
As a consequence, the mesonic thermal loops remain non-negligible even in the asymptotically large-$T$ regime, so that the pressure approaches a constant different from the SB value. 
Including both the mesonic loop contribution and the interaction terms, in the high-temperature limit, we have 
\begin{align}
    \lim_{T\to\infty}\frac{P_{\rm High\text{-}T}}{T^4} \simeq 3.8207 \, .
\end{align}
The detailed derivation of this asymptotic behavior is summarized in \Cref{app:HighTemperatureLimit}.

%We give more comments on the deviation from the SB limit.
%In the present work, the mesonic fields are introduced as fundamental degrees of freedom rather than %emergent resonance states.

It is also instructive to make a comparison among the different pressure prescriptions from a thermodynamic point of view. 
Since only pressure differences are physically significant, the possibility of negative $P_{\rm vac}$ at low temperature is not excluded by itself. 
However, such a case can strongly distort derived observables that are sensitive to shifts in the pressure. 
This is particularly relevant for quantities such as the trace anomaly.
In this sense, the closer agreement of $P^*$ and $P_{\rm vac}^{(j)}$ with the CJT baseline provides further support to interpret the source background as an auxiliary term rather than as an actual source applied to the physical medium. 
To be more explicit, we summarize the results on the baryon number density $n_B$, the entropy density $s$, and the energy density $\varepsilon$ in \Cref{app:ResultsOnEquationsOfState}, for CJT and SICJT formalisms with different pressure prescriptions.

The equation of state and related thermodynamic observables are sensitive to the microscopic degrees of freedom included in the theory, as has been noted in the beginning of \Cref{sec:EquationsOfState}. 
The present linear quark-meson model contains no explicit gluonic or Polyakov-loop sector and treats the mesonic fields as fundamental propagating degrees of freedom. 
In contrast to the present meson model description, in QCD-based approaches with dynamically generated hadronic modes, the bosonic sector contribution to pressure is expected to be highly suppressed in the high-temperature regime.
This trend is reflected, for example, in fRG studies with dynamical hadronization, where the mesonic wave-function renormalization decreases with increasing temperature; see, e.g., Fig. 12 in Ref.~\cite{Fu:2019hdw}. 
If the meson wavefunction renormalization tends to vanish asymptotically,
\begin{align}
    Z_\phi \to 0 \, ,
\end{align}
the renormalized meson mass-to-temperature ratio is expected to diverge: 
\begin{align}
    r_T^{\phi,R} = \frac{r_T^\phi}{\sqrt{Z_\phi}} \to \infty \, ,
\end{align}
so that the mesonic contribution becomes effectively decoupled.
It is therefore not expected to reproduce the absolute normalization or the genuine high-temperature limit of full QCD.

Nevertheless, a comparison around the crossover remains useful for
separating the effects of the thermodynamic prescription from those of
the underlying model content. In \Cref{app:EOSCompare}, we
therefore compare the zero-density CJT and SICJT results with several results including quark-meson model, Polyakov-quark-meson model, and QCD from various qualitative approaches.
Both the conventional temperature-normalized quantities and quantities
normalized by the asymptotic limit appropriate to each approach are
shown. 
These comparisons are intended as external consistency tests of
the crossover shape and thermodynamic evolution, rather than as a direct precision calibration of the individual SICJT pressure prescriptions.

\subsection{Isentropic trajectories}
\label{sec:IsentropicTraj}

%In this subsection, we discuss 
The isentropic trajectories provide a sensitive diagnostic of the equation of state. % at the level of its first thermodynamic derivatives.
Those trajectories are also phenomenologically relevant, as they approximately trace the adiabatic evolution of strongly interacting matter in the phase diagram.

In \Cref{fig:isentropicTrajectory}, we show the isentropic trajectories for
\begin{align}
    \frac{s}{n_B} = [60,\,28,\,21,\,17,\,13,\,11,\,9,\,6,\,4,\,2] \, .
\end{align}
The overall shape of the trajectories is universal and robust against the different pressure prescriptions.
In particular, trajectories with large $s/n_B$ remain in the crossover region and vary smoothly across the phase diagram, whereas those with smaller $s/n_B$ are strongly distorted in the vicinity of the CEP and the first-order boundary. 

%%%%%%%%%%%%%%
%%%%%%%%%%%%%%
\begin{figure*}[t]
    \centering
    \includegraphics[width=0.32\linewidth]{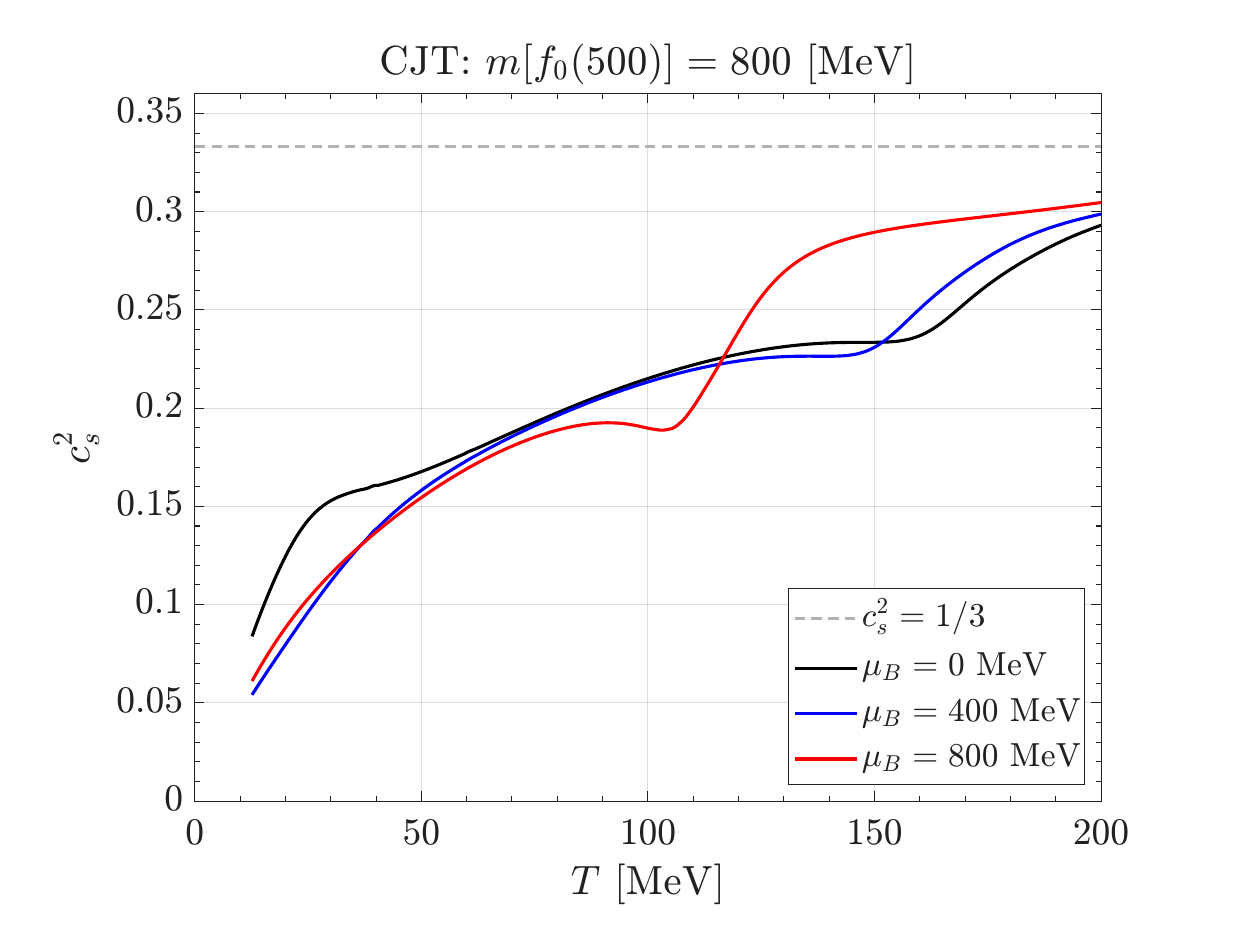}
    \includegraphics[width=0.32\linewidth]{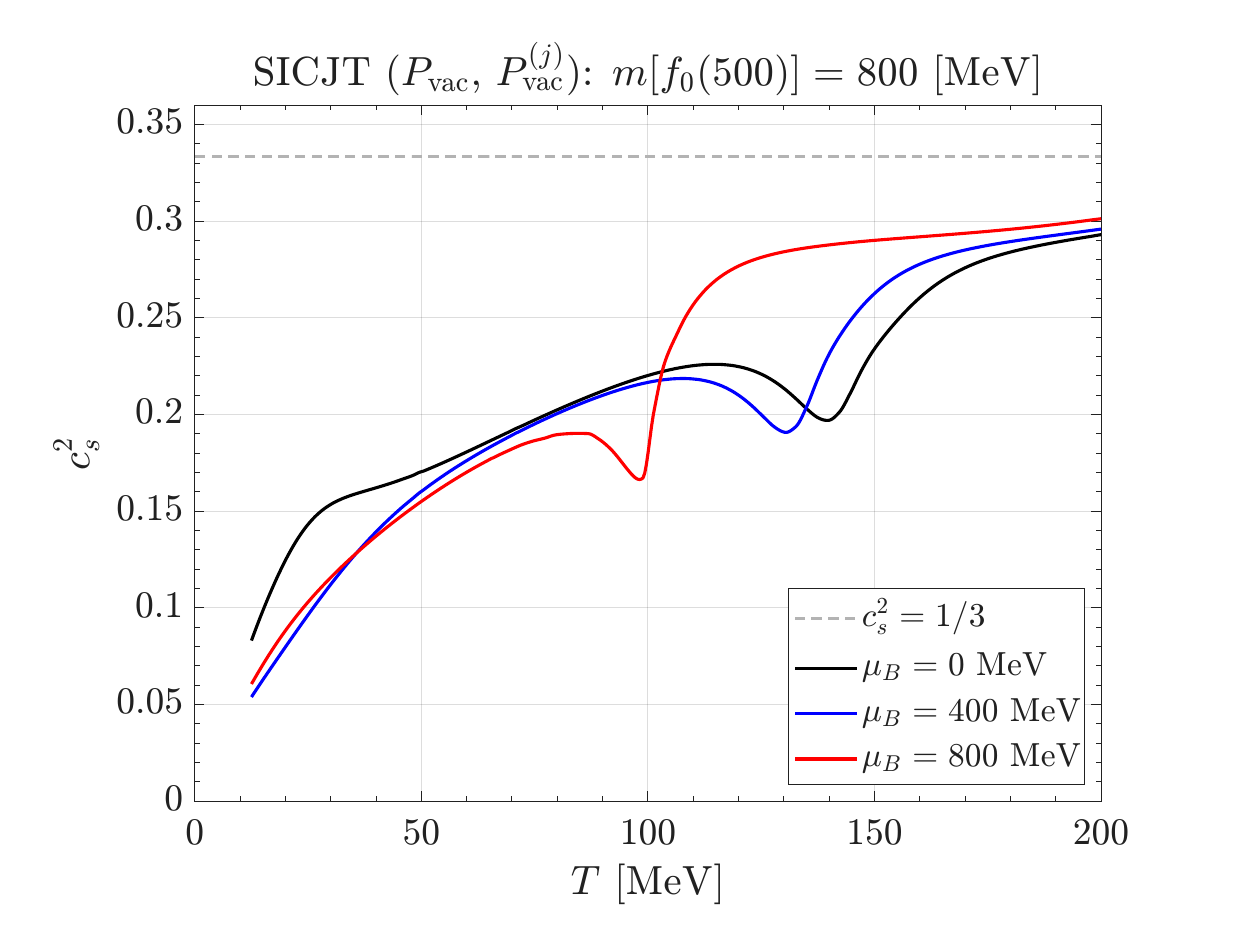}
    \includegraphics[width=0.32\linewidth]{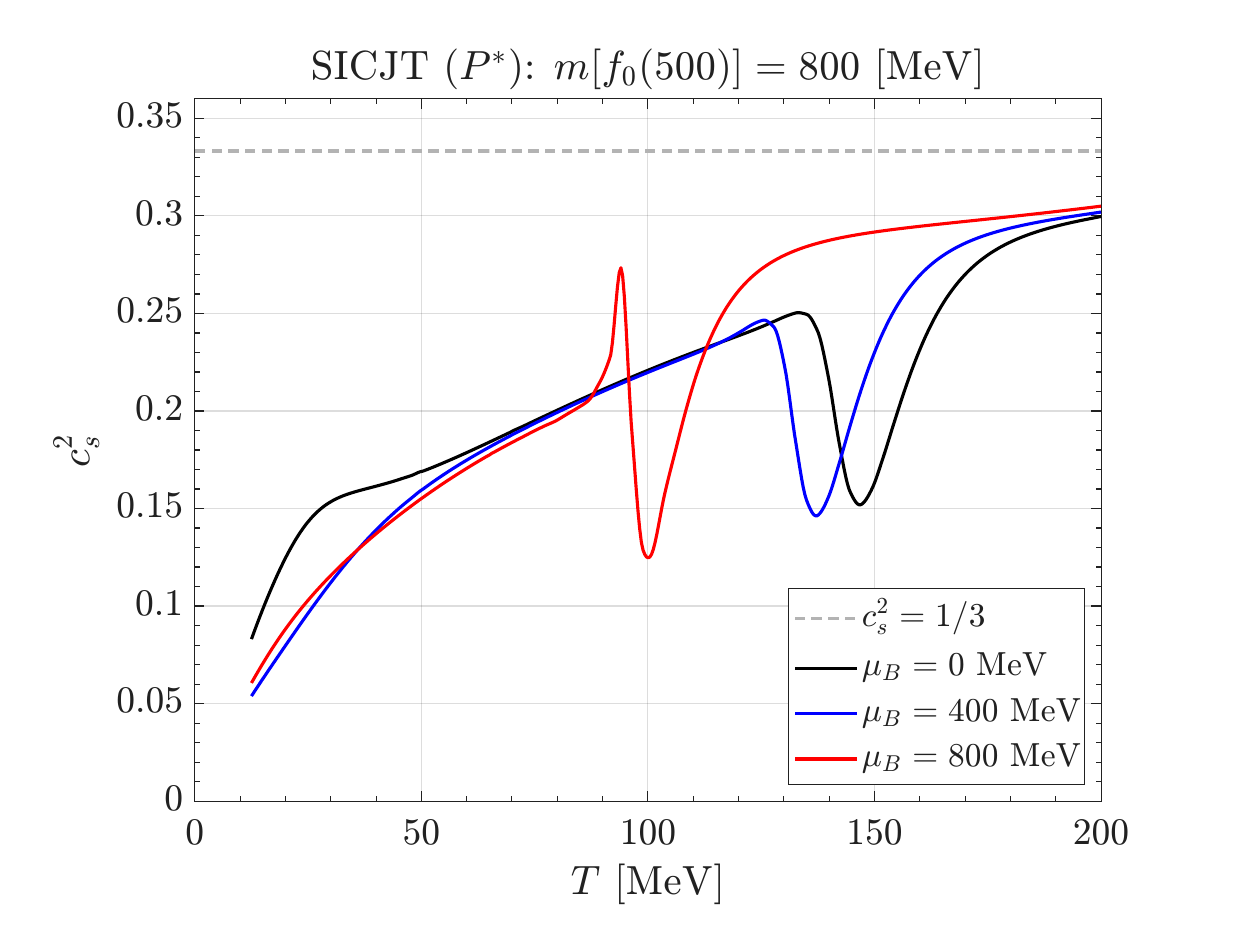}
    \caption{
    Adiabatic sound velocities (colored-solid lines) as a function of temperature for several fixed values of the baryon chemical potential
    The different panels show the results obtained in different formalisms and with different pressure prescriptions. 
    The gray-dashed line denotes the conformal and ultrarelativistic limit with $c_s^2 = 1/3$.
    }
    \label{fig:soundVelocity}
\end{figure*}
%%%%%%%%%%%%%%
%%%%%%%%%%%%%%

The quantitative sensitivity to the pressure prescription is therefore concentrated in the low-$s/n_B$, large-$\mu_B$ region.
Comparing the results of the CJT and SICJT cases, the SI formulation shifts the phase boundary and the CEP, and the low-$s/n_B$ trajectories follow this change accordingly. 
%Moreover, 
The trajectories obtained from $P_{\rm vac}$ and $P_{\rm vac}^{(j)}$ are exactly identical, which should be so by construction. 
The quantitative difference between $P^*$ and $P_{\rm vac}/P_{\rm vac}^{(j)}$ is observed mainly for $6<s/n_B \lesssim 13$ around the phase boundary.

\subsection{Sound velocity}
\label{sec:SoundVel}

The squared adiabatic sound velocity is defined by
\begin{align}
    c_s^2 \equiv \left( \frac{\partial P}{\partial \varepsilon} \right)_\varsigma
    =
    \frac{\frac{\partial(P,\varsigma)}{\partial(T,\mu_B)}}{\frac{\partial(\varepsilon,\varsigma)}{\partial(T,\mu_B)}} \, ,
\end{align}
where $\varsigma = s/n_B$ is kept fixed when taking the derivative, and $\frac{\partial(\,,\,)}{\partial(\,,\,)}$ denotes the Jacobian determinant.
%Inserting the thermodynamic quantities, one obtains 
The right-hand side is evaluated as 
\begin{align}
    c_s^2
    =
    \frac{
        n_B^2 \frac{\partial s}{\partial T}
        -
        n_B s \left(
            \frac{\partial s}{\partial \mu_B}
            +
            \frac{\partial n_B}{\partial T}
        \right)
        +
        s^2 \frac{\partial n_B}{\partial \mu_B}
    }{
        (Ts+\mu_B n_B)
        \left(
            \frac{\partial s}{\partial T}
            \frac{\partial n_B}{\partial \mu_B}
            -
            \frac{\partial s}{\partial \mu_B}
            \frac{\partial n_B}{\partial T}
        \right)
    } \, .
    \label{eq:explicitSoundVelocity}
\end{align}
%

%%%%%%%%%%%%%%
%%%%%%%%%%%%%%
\begin{figure}[b]
    \centering
    \includegraphics[width=0.9\linewidth]{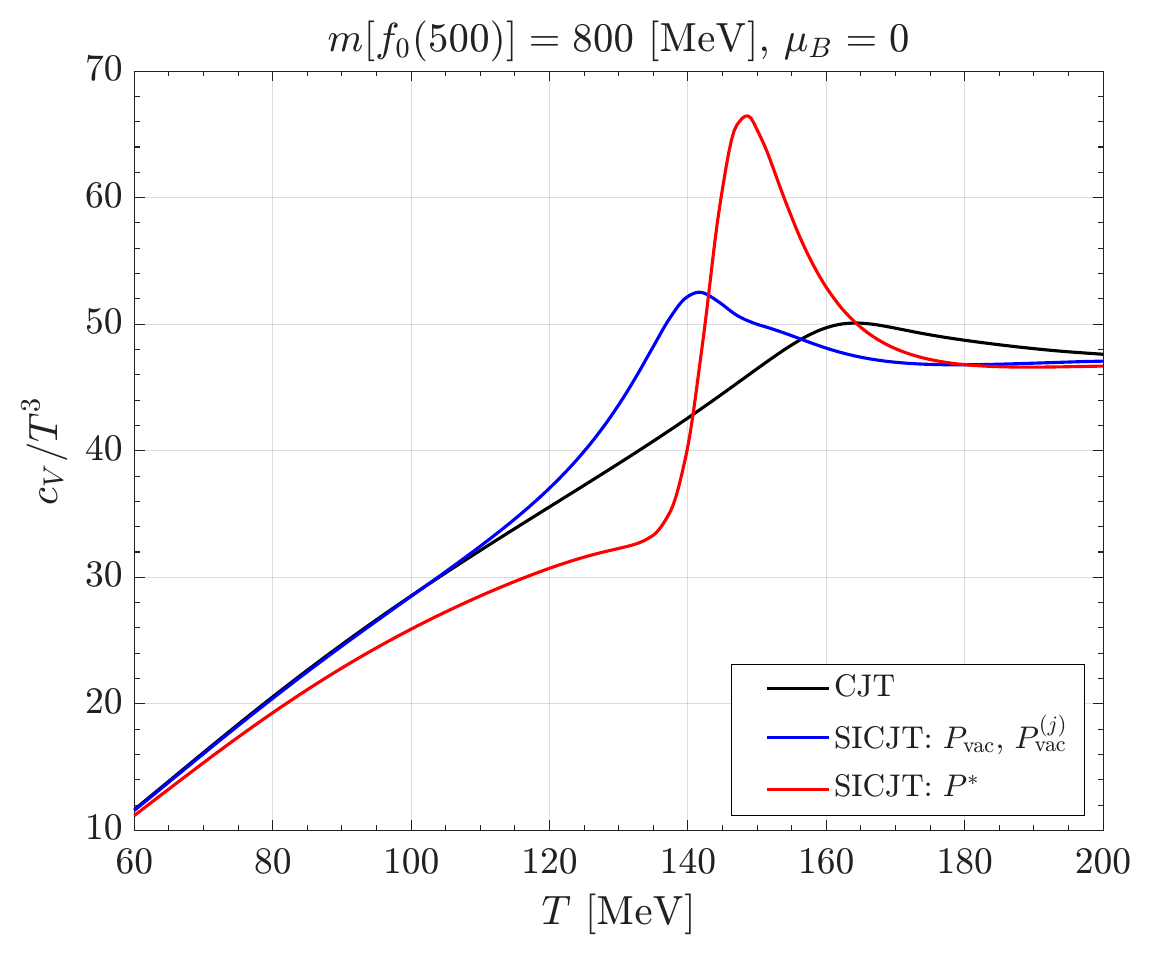}
    \caption{
    Normalized volumetric densities of the specific heat capacity $c_V$ from different formalisms as a function of temperature at $\mu_B = 0$.
    }
    \label{fig:specificHeat}
\end{figure}
%%%%%%%%%%%%%%
%%%%%%%%%%%%%%

In \Cref{fig:soundVelocity}, we show the temperature dependence of the adiabatic sound velocity squared $c_s^2$ at fixed baryon chemical potentials in the CJT formalism and for the different pressure prescriptions in the SICJT approach.
In all cases, $c_s^2$ remains positive in the temperature range shown and increases toward the high-temperature regime, while approaching the conformal value in large $T$ asymptotics to be 
\begin{align}
    c_s^2 = \frac{1}{3} \, .
\end{align}
On a qualitative level, the CJT result exhibits a relatively smooth thermal evolution.
For $\mu_B=0$ and $400$ MeV, the sound velocity increases monotonically until a mild shoulder structure shows up in the crossover region, while for $\mu_B=800$ MeV, a more visible softening develops around the phase boundary before $c_s^2$ grows again at higher temperature.
This behavior reflects the expected softening of the equation of state in the vicinity of the chiral transition, which becomes more pronounced as the system moves deeper into the baryon-rich region.

Comparing the different pressure prescriptions in the SICJT formalism, one finds that the main difference is seen in the transition region, while the low- and high-temperature limits remain rather similar. 
For the ``pulled-back'' pressure $P^*$, the nonmonotonic structure becomes significantly sharper than the one observed in the CJT case.
In particular, the $\mu_B=800$ MeV curve develops a pronounced peak-dip structure around $T\sim 100$ MeV, while the $\mu_B=400$ and $0$ MeV curves display deeper minima around $T\sim 140$-$150$ MeV. 
This indicates that the ``pulled-back'' prescription enhances the softening of the equation of state in the vicinity of the phase boundary and leads to a stronger separation for the different chemical-potential slices. 

By construction, the results associated with $P_{\rm vac}$ and $P_{\rm vac}^{(j)}$ remain identical and exhibit a smoother behavior of the sound velocity than the one with $P^*$. 
Although the same qualitative softening pattern is present, the corresponding minima are shallower and less pronounced.
This shows that the quantitative sensitivity of $c_s^2$ to the pressure prescription is reflected mainly in the crossover-to-first-order transition region. 
In contrast, outside this regime, the different prescriptions yield quite similar $T$ dependence of sound velocities.

%%%%%%%%%%%%%%
%%%%%%%%%%%%%%
\begin{figure*}[t]
    \centering
    \includegraphics[width=0.45\linewidth]{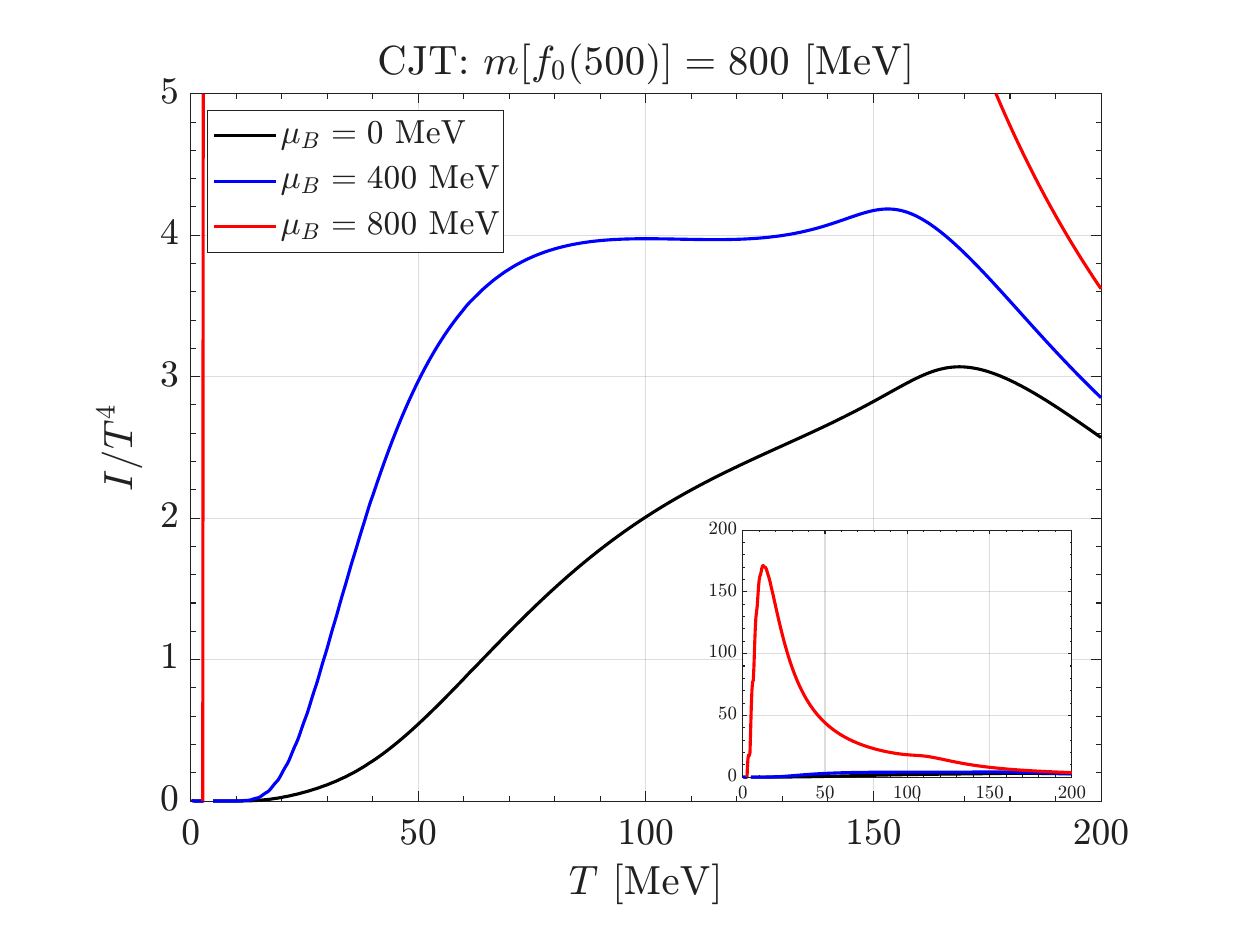}
    \includegraphics[width=0.45\linewidth]{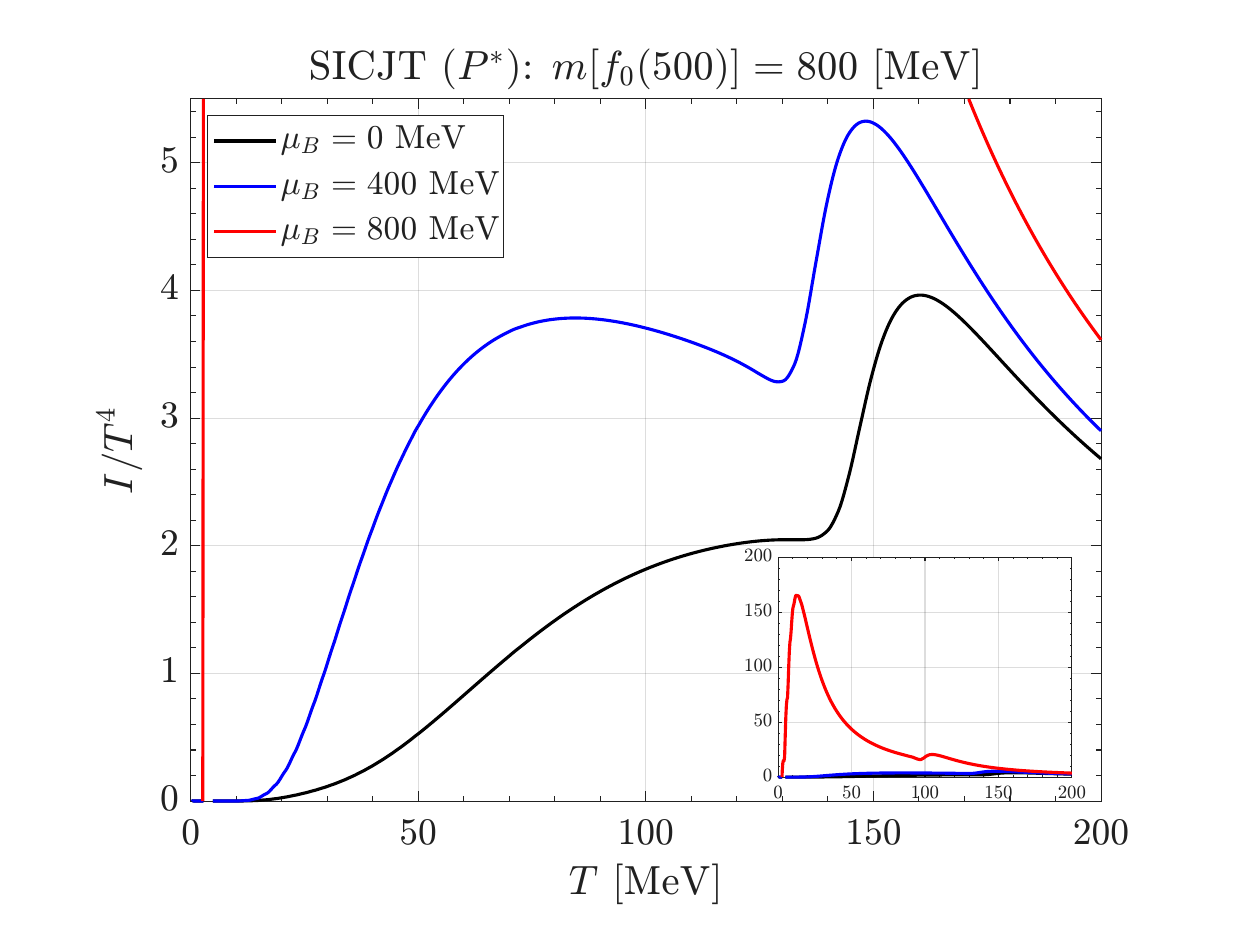}
    \includegraphics[width=0.45\linewidth]{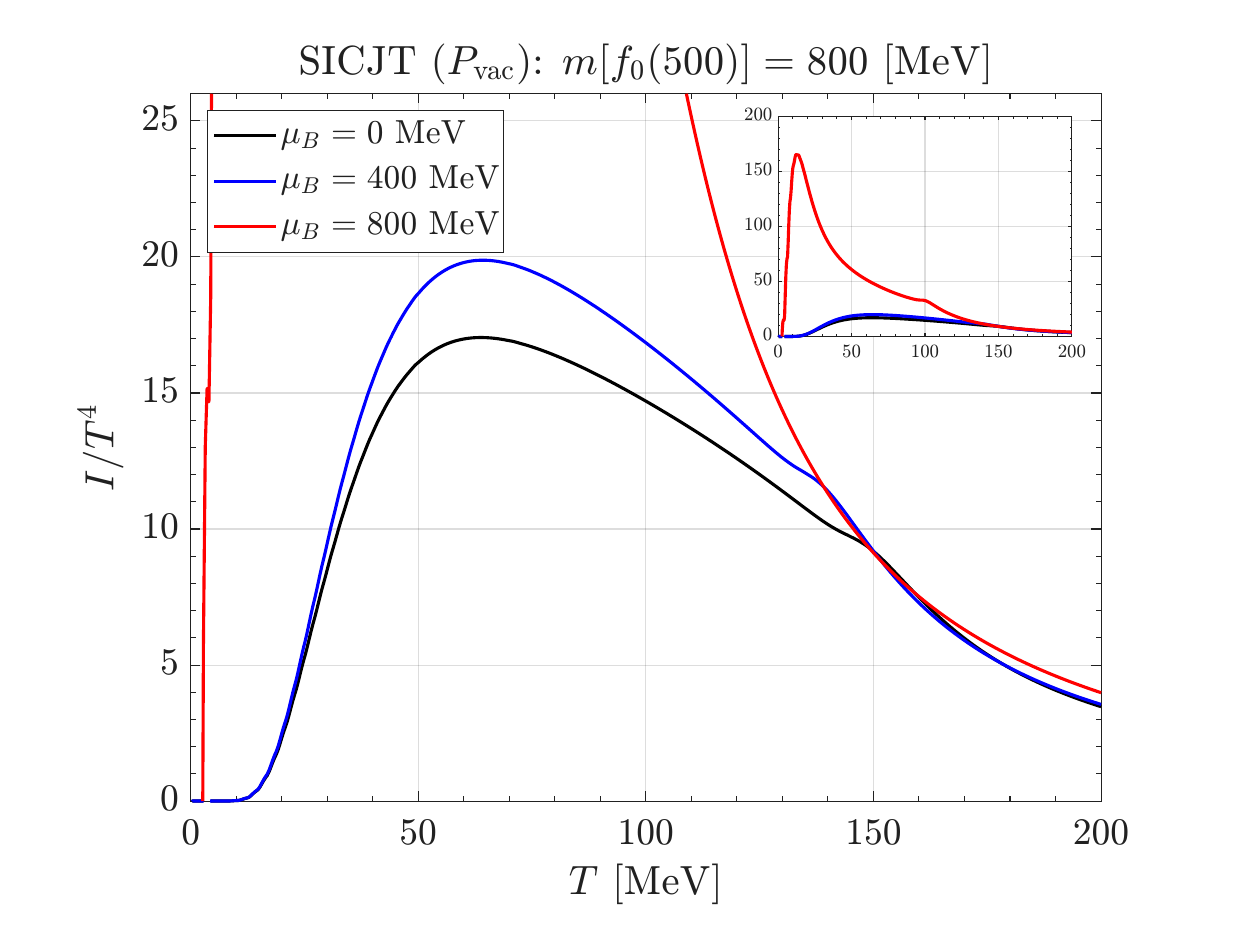}
    \includegraphics[width=0.45\linewidth]{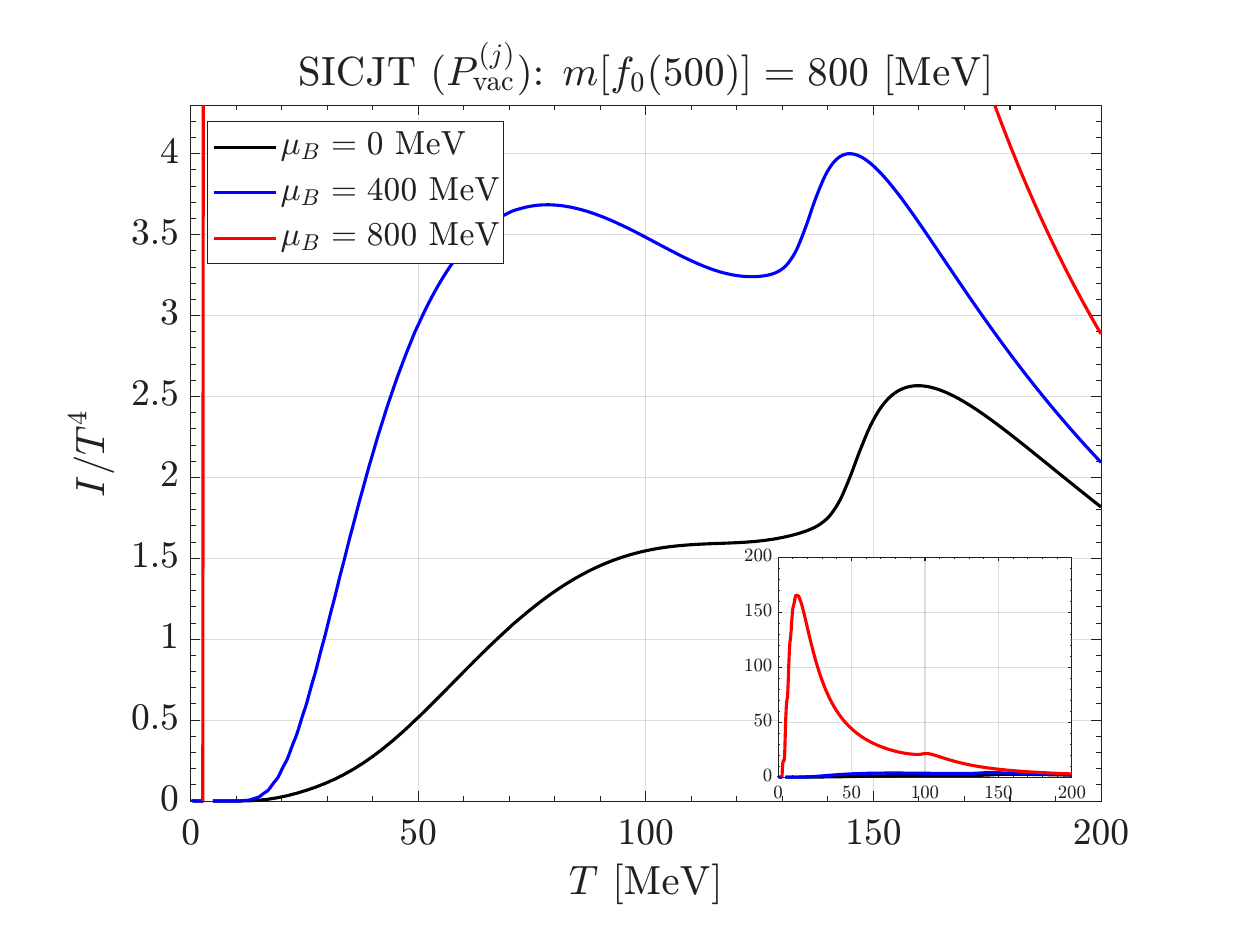}
    \caption{
    Normalized trace anomaly $I/T^4$ as a function of temperature for fixed baryon chemical potentials. 
    The different panels show the results obtained in the CJT formalism and from the different pressure prescriptions in the SICJT approach. 
    The inset in each panel enlarges the plot region on the y-axis.
    }
    \label{fig:traceAnomaly}
\end{figure*}
%%%%%%%%%%%%%%
%%%%%%%%%%%%%%

In the limit of the vanishing baryon chemical potential $\mu_B = 0$, \Cref{eq:explicitSoundVelocity} has a more intuitive form.
In this case, one notices that $\partial s / \partial \mu_B = 0$ since the entropy density forms an even function of $\mu_B$.
Then, \Cref{eq:explicitSoundVelocity} is reduced to
\begin{align}
    c_s^2 \Big|_{\mu_B = 0} = \frac{s}{T \frac{\partial s}{\partial T}} = \frac{s}{c_V} \, ,
    \label{eq:soundVelocityAndSpecificHeat}
\end{align}
where $c_V = T \, \partial s/ \partial T$ is the volumetric density of the specific heat capacity, which is equivalently the susceptibility of the energy density with respect to temperature, characterizing the energy fluctuations.
This expression shows that the softening of the sound velocity is not determined solely by the specific heat. 
Rather, the dip structure of $c_s^2$ reflects the competition between the growth of the entropy density and that of the specific heat. 
In normalized units, \Cref{eq:soundVelocityAndSpecificHeat} may be viewed as the ratio between $s/T^3$ and $c_V/T^3$. 
Around the pseudo-critical temperature $T_{pc}$, the normalized specific heat typically develops a pronounced enhancement associated with energy-density fluctuations. 
If this enhancement is stronger than the simultaneous increase in the normalized entropy density, the ratio $s/c_V$ decreases and a local minimum of $c_s^2$ appears. 
A similar qualitative softening behavior has also been observed in lattice QCD, see, e.g., Refs.~\cite {Borsanyi:2012cr, HotQCD:2014kol}.

In \Cref{fig:specificHeat}, we show the normalized specific heat capacities obtained from the different formalisms as functions of temperature at $\mu_B=0$. 
The comparison is intended to clarify the origin of the sound-velocity structure within the present quark-meson model. 
% As has been noted at the beginning of \Cref{sec:EquationsOfState}, a direct quantitative comparison with lattice QCD or other nonperturbative QCD-based studies should be made with caution, since those calculations often probe sufficiently high-temperature regimes where the relevant microscopic degrees of freedom and the approach to the QCD high-temperature limit differ from those of the present low-energy effective model. 
% For a related study in the Polyakov LSM at the quark-loop level, we refer the reader to Ref.~\cite{Mao:2009aq}.
A direct comparison with lattice QCD and representative functional calculations is also presented in \Cref{app:EOSCompare}. 
As noted previously in the beginning of \Cref{sec:EquationsOfState}, because the microscopic degrees of freedom and asymptotic normalizations differ among these approaches, we focus primarily on the location and depth of the crossover softening and on the evolution toward the corresponding high-temperature limit, rather than on an unqualified point-by-point comparison.

% Overall, the sound-velocity analysis leads to the same conclusion as the pressure and isentropic-trajectory studies.
% The global thermodynamic behavior is robust, while the quantitative differences among the prescriptions are concentrated near the phase boundary.
% In particular, the ``pulled-back'' prescription enhances the local softening of the equation of state, whereas $P_{\rm vac}$ and $P_{\rm vac}^{(j)}$ yield a smoother behavior.

\subsection{Trace anomaly}
\label{sec:TraceAnomaly}

The trace anomaly is defined by
\begin{align}
    I \equiv \varepsilon - 3P = Ts + \mu_B n_B - 4P \, .
\end{align}
In \Cref{fig:traceAnomaly}, we show the normalized interaction measure $I/T^4$ as a function of temperature at fixed baryon chemical potential for the CJT formalism and for the different pressure prescriptions in the SICJT formalism.

In the CJT case, the interaction measure exhibits a moderate and relatively broad enhancement in the crossover region.
At $\mu_B=0$, the peak is located around $T\sim 170$ MeV, which is shifted to a lower temperature with a higher peak as $\mu_B$ increases. 
%while for larger $\mu_B$ the enhancement becomes stronger and shifts toward lower temperature. 
%For the largest chemical-potential slice, a strong low-temperature rise is also visible in the inset.
This qualitative trend can also be seen in the SICJT case for all types of pressure prescription, where the peak shape around the transition region becomes sharper than in the SICJT case.
In particular, $P^*$ produces a more distinct peak-dip-peak pattern around $T\sim 130$-$160$ MeV, whereas $P_{\rm vac}^{(j)}$ produces a smoother curve, which yields $I/T^4$ numerically close to the CJT result. 
In contrast, $P_{\rm vac}$ leads to a qualitatively different behavior.
For $\mu_B=0$ and $400$ MeV, $I/T^4$ develops a large, broad maximum around $T\sim 60$-$70$ MeV, with a magnitude much larger than the one obtained in the other prescriptions. 
This reflects the fact that the trace anomaly is highly sensitive to pressure normalization in the presence of medium-dependent SI sources.
Since $I$ contains the combination $-4P$, an additive shift in pressure is directly amplified in the interaction measure.
The prominent enhancement found for $P_{\rm vac}$, therefore, appears to be largely driven by the subtraction prescription rather than by the intrinsic thermal dynamics. 

% Overall, the trace-anomaly analysis supports the same conclusion as the pressure, isentropic-trajectory, and sound-velocity studies.
% The global thermodynamic behavior is robust, but the quantitative structure near the transition region depends sensitively on how the SI source contribution is treated.
% Among the SICJT prescriptions, $P_{\rm vac}^{(j)}$ provides the closest overall behavior to the CJT baseline, while $P^*$ enhances the nonmonotonic structure in the crossover region and $P_{\rm vac}$ yields a substantially distorted interaction measure.

%%%%%%%%%%%%%%%%%%%%%%%%%%%%%%%%%%%%%%%%%%%%%%%%%%%%%%%%%%%%%%%%%%%%%%
\section{Conclusion}
\label{sec:Conclusion}

In this work, we have investigated the thermodynamics of the SICJT formalism applied to the three-flavor LSM with quarks as a low-energy effective theory of QCD.
Our main motivation was that, although the SICJT construction restores the relevant symmetry constraints through self-consistently determined auxiliary sources, the thermodynamic interpretation of the resulting equilibrium state is not unique once the pressure and its derivatives are considered.

To clarify this issue, we formulated several prescriptions for the pressure in the SICJT framework, including the conventional vacuum-subtracted pressure $P_{\rm vac}$, the source-matched vacuum-subtracted pressure $P_{\rm vac}^{(j)}$, and the pulled-back pressure $P^*$ defined from the pulled-back effective action.
We further distinguished between total derivatives taken along the symmetry-improved manifold and thermodynamic derivatives evaluated at a fixed external source.
This separation is essential in the present setup, since the symmetry-improving sources acquire a nontrivial $(T,\mu_B)$ dependence at equilibrium and therefore contribute explicitly to thermodynamic derivatives unless they are treated as auxiliary quantities.

Our numerical analysis shows that the global thermodynamic structure is robust under the different pressure prescriptions; for example, the overall qualitative behavior of the isentropic trajectories remains stable.
At the same time, quantitative differences arise in the vicinity of the crossover and first-order transition region, where the source dependence becomes most relevant.
This is especially visible in source-sensitive observables such as the adiabatic sound velocity and the trace anomaly.

The conventional vacuum-subtracted prescription $P_{\rm vac}$ appears less suitable for normalization-sensitive thermodynamic observables in the present source-based SICJT formulation. 
Although a negative pressure by itself is not inconsistent, the large source-induced offset in $P_{\rm vac}$ is strongly amplified in quantities such as the trace anomaly.
By contrast, the source-matched prescription $P_{\rm vac}^{(j)}$ and the pulled-back prescription $P^*$ yield thermodynamic behavior much closer to the CJT baseline.
These results support the interpretation that the symmetry-improving source should be regarded as an auxiliary construction rather than as an intrinsic part of the physical medium, and that the physically meaningful thermodynamic derivatives are those evaluated at a fixed external source.

We have also discussed the high-temperature asymptotics within the present quark-meson realization.
Because the mesonic fields are treated as fundamental propagating degrees of freedom, the model does not approach the conventional quark-gluon SB limit.
Instead, all pressure prescriptions converge to the same model-dependent asymptotic constant.

The present analysis provides a practical framework for constructing thermodynamically consistent observables in symmetry-improved 2PI approaches.
At the same time, several important extensions remain to be explored.
A natural next step is to go beyond the present potential approximation and include corrections to the kinetic sector, i.e., wave-function renormalization factors $Z_\phi$ for the mesonic fields.
Such an extension is expected to improve the description of the medium dependence of the propagating modes, but it also requires a proper and fully consistent renormalization of the corresponding derivative terms.
From a broader perspective, the inclusion of nontrivial $Z_\phi$ factors may also open the possibility of accessing more intricate momentum-dependent structures, such as the emergence of the composite degrees of freedom~\cite{Blaschke:2017boi}, or the moat regime~\cite{Rennecke:2021ovl, Fu:2024rto}, within the symmetry-improved framework.
It would furthermore be interesting to study how these improvements affect the thermodynamics near the critical region and whether they lead to a more realistic connection between low-energy effective descriptions and the asymptotic behavior expected from QCD.

\section*{Acknowledgments} 
%%%%%%
This work was supported in part by the National Science Foundation of China (NSFC) under Grant Nos. 11747308, 11975108, 12047569, and the Seeds Funding of Jilin University (S.M.). 
%%%%%%
The work by M.K. is supported by RFIS-NSFC under Grant No. W2433019.
%%%%%%
The work of A.T. was partially supported by JSPS  KAKENHI Grant Numbers 20K14479, 22K03539, 22H05112, and 22H05111, and MEXT as ``Program for Promoting Researches on the Supercomputer Fugaku'' (Simulation for basic science: approaching the new quantum era; Grant Number JPMXP1020230411, and Search for physics beyond the standard model using large-scale lattice QCD simulation and development of AI technology toward next-generation lattice QCD; Grant Number JPMXP1020230409).  
%%%%%%

% \clearpage
\onecolumngrid
\appendix
%%%%%%%%%%%%%%%%%%%%%%%%%%%%%%%%%%%%%%%%%%%%%%%%%%%%%%%%%%%%%%%%%%%%%%
\section{Vacuum phenomenology of linear sigma model}
\label[appendix]{app:VacuumPhenomenologyOfLinearSigmaModel}

In this section of the appendix, we summarize the related formulations of the LSM with quarks given in \Cref{eq:lagrangianQandM} at tree-level.

\subsection{Stationary conditions}
\label[appendix]{app:StationaryConditionsTreeLevel}

At tree-level, the potential \eqref{eq:LSMTreePotential} around the $(\bar\Phi_1,\bar\Phi_3)$-background reads
\begin{align}
    V_m(\bar\Phi) = \mu^2 \Big( 2 \bar \Phi_1^2 + \bar\Phi_3^2 \Big) + \lambda_1 \Big( 2 \bar \Phi_1^4 + \bar\Phi_3^4 \Big) + \lambda_2 \Big( 2 \bar \Phi_1^2 + \bar\Phi_3^2 \Big)^2 - 2B \bar \Phi_1^2 \bar \Phi_3 - 2\Big( 2 cm_l \bar\Phi_1 + cm_s \bar\Phi_3 \Big) ,
\end{align}
for which the slopes of the classical action read
\begin{align}
    \frac{\delta S_{m}}{\delta \Phi_1} \Biggl|_{\Phi = \bar\Phi} &= 4 \Big( \mu^2 \bar\Phi_1 + 2 \lambda_1 \bar\Phi_1^3 + 4 \lambda_2 \bar\Phi_1^3 + 2 \lambda_2 \bar\Phi_1 \bar\Phi_3^2 - B \bar\Phi_1 \bar\Phi_3 - cm_l \Big), \nonumber\\
    \frac{\delta S_{m}}{\delta \Phi_3} \Biggl|_{\Phi = \bar\Phi} &= 2 \Big( \mu^2 \bar\Phi_3 + 2 \lambda_1 \bar\Phi_3^3 + 4 \lambda_2 \bar\Phi_1^2 \bar\Phi_3 - B \bar\Phi_1^2 - cm_s \Big).
    \label{eq:SlopPhibasis}
\end{align}
Around the $(\bar\sigma_0,\bar\sigma_8)$-background, the tree-level potential reads
\begin{align}
    V_m(\bar\sigma) =& \frac{1}{2} \mu^2 \Big( \bar\sigma_0^2 + \bar\sigma_8^2 \Big) + \lambda_1 \biggl[ \frac{1}{12} \bar\sigma_0^4 + \frac{1}{2} \bar\sigma_0^2 \bar\sigma_8^2 - \frac{1}{3\sqrt{2}} \bar\sigma_0 \bar\sigma_8^3 + \frac{1}{8} \bar\sigma_8^4 \biggl] + \frac{\lambda_2}{4} \Big( \bar\sigma_0^2 + \bar\sigma_8^2 \Big)^2 \nonumber\\
    &- B \biggl[ \frac{1}{3\sqrt{6}}\bar\sigma_0^3 - \frac{1}{2\sqrt{6}}\bar\sigma_0 \bar\sigma_8^2 - \frac{1}{6\sqrt{3}}\bar\sigma_8^3 \biggl] - \sqrt{\frac{2}{3}} (2cm_l + cm_s)\bar\sigma_0 - \frac{2}{\sqrt{3}} (cm_l - cm_s)\bar\sigma_8,
    \label{eq:treePotentialAroundMeanField}
\end{align}
for which the slopes of the classical action read
\begin{align}
    \frac{\delta S_{m}}{\delta \sigma_0} \Biggl|_{\sigma_a = \bar\sigma_a} =& \mu^2 \bar\sigma_0 + \lambda_1 \biggl[ \frac{1}{3} \bar\sigma_0^3 + \bar\sigma_0 \bar\sigma_8^2 - \frac{1}{3\sqrt{2}}\bar\sigma_8^3  \biggl] + \lambda_2 \Big( \bar\sigma_0^2 + \bar\sigma_8^2 \Big) \bar\sigma_0 - B \biggl[ \frac{1}{\sqrt{6}}\bar\sigma_0^2 - \frac{1}{2\sqrt{6}} \bar\sigma_8^2 \biggl] \nonumber\\
    &- \sqrt{\frac{2}{3}} (2cm_l + cm_s), \nonumber\\
    \frac{\delta S_{m}}{\delta \sigma_8} \Biggl|_{\sigma_a = \bar\sigma_a} =& \mu^2 \bar\sigma_8 + \lambda_1 \biggl[ \bar\sigma_0^2 \bar\sigma_8 - \frac{1}{\sqrt{2}} \bar\sigma_0 \bar\sigma_8^2 + \frac{1}{2} \bar\sigma_8^3 \biggl] + \lambda_2 \Big( \bar\sigma_0^2 + \bar\sigma_8^2 \Big) \bar\sigma_8 + B \biggl[ \frac{1}{\sqrt{6}}\bar\sigma_0 \bar\sigma_8 + \frac{1}{2\sqrt{3}}\bar\sigma_8^2 \biggl] \nonumber\\
    &- \frac{2}{\sqrt{3}} (cm_l - cm_s).
\end{align}
These two sets of slopes are correlated with each other through the linear transformation arising from the chain rule
\begin{align}
    \pmat{\frac{\delta S_{m}}{\delta \Phi_1}  \\ \frac{\delta S_{m}}{\delta \Phi_3}}
    =
    \pmat{2\sqrt{\frac{2}{3}} & \frac{2}{\sqrt{3}} \\ \sqrt{\frac{2}{3}} & -\frac{2}{\sqrt{3}} }
    \pmat{\frac{\delta S_{m}}{\delta \sigma_0} \\ \frac{\delta S_{m}}{\delta \sigma_8}}.
    \label{eq:linearTransOfSlops}
\end{align}

At the classical level, the equilibrium criteria, or the stationary conditions, read
\begin{align}
    \frac{\delta S_{m}}{\delta \Phi_1} \Biggl|_{\Phi = \bar\Phi} = 0, \qquad \frac{\delta S_{m}}{\delta \Phi_3} \Biggl|_{\Phi = \bar\Phi} = 0.
\end{align}
which is equivalent to
\begin{align}
    \frac{\delta S_{m}}{\delta \sigma_0} \Biggl|_{\sigma_a = \bar\sigma_a} = 0, \qquad \frac{\delta S_{m}}{\delta \sigma_8} \Biggl|_{\sigma_a = \bar\sigma_a} = 0
\end{align}
through \Cref{eq:linearTransOfSlops}.

\subsection{Curvature masses}
\label[appendix]{app:CurvatureMassesAtVacuum}

The Hessian matrix is found by taking the second functional derivative with respect to the fluctuating modes around the mean-field background.
Throughout this work, we take the scalars $\sigma_a$ and pseudo-scalars $\pi_a$ as dynamical degrees of freedom, thus
\begin{align}
    \Big[ \bar G_{S,{\rm tree}}^{-1}(k, k^\prime; \bar\sigma) \Big]^{ab} &= \frac{\delta^2 S_{m}}{\delta \sigma_a(-k^\prime) \delta \sigma_b(k)} = (2\pi)^4 \delta^{(4)}(k - k^\prime)\biggl( k^2 \delta^{ab} + \Big[ m_S^2(\bar\sigma) \Big]^{ab} \biggl) \nonumber\\
    &\equiv (2\pi)^4 \delta^{(4)}(k - k^\prime) \Big[ \bar G_{S,{\rm tree}}^{-1}(k; \bar\sigma) \Big]^{ab}, \nonumber\\
    \Big[ \bar G_{P,{\rm tree}}^{-1}(k, k^\prime; \bar\sigma) \Big]^{ab} &= \frac{\delta^2 S_{m}}{\delta \pi_a(-k^\prime) \delta \pi_b(k)} = (2\pi)^4 \delta^{(4)}(k - k^\prime)\biggl( k^2 \delta^{ab} + \Big[ m_P^2(\bar\sigma) \Big]^{ab} \biggl) \nonumber\\
    &\equiv (2\pi)^4 \delta^{(4)}(k - k^\prime) \Big[ \bar G_{P,{\rm tree}}^{-1}(k; \bar\sigma) \Big]^{ab},
    \label{eq:TreePropagators}
\end{align}
where the tree-level curvature masses are defined by
\begin{align}
    \Big[ m_S^2(\bar\sigma) \Big]^{ab} = \frac{\partial^2 V_{m}}{\partial \sigma_a \partial \sigma_b} \Biggl|_{\sigma_a = \bar\sigma_a} , \qquad \Big[ m_P^2(\bar\sigma) \Big]^{ab} = \frac{\partial^2 V_{m}}{\partial \pi_a \partial \pi_b} \Biggl|_{\sigma_a = \bar\sigma_a}.
\end{align}
Notice that the mixing matrix elements such as $ \frac{\delta^2 S_{m}}{\delta \pi_a(-k^\prime) \delta \sigma_b(k)}$ vanish since the theory preserves the parity-even nature.
The scalar masses read
\begin{align}
    \Big[ m_S^2(\bar\sigma) \Big]^{00} &= \mu^2 + \lambda_1 \Big( \bar\sigma_0^2 + \bar\sigma_8^2 \Big) + \lambda_2 \Big( 3 \bar\sigma_0^2 + \bar\sigma_8^2 \Big) - B \sqrt{\frac{2}{3}} \bar\sigma_0 , \nonumber\\
    \Big[ m_S^2(\bar\sigma) \Big]^{88} &= \mu^2 + \lambda_1 \biggl[ \bar\sigma_0^2 - \sqrt{2} \bar\sigma_0 \bar\sigma_8 + \frac{3}{2} \bar\sigma_8^2 \biggl] + \lambda_2 \Big( \bar\sigma_0^2 + 3 \bar\sigma_8^2 \Big) + \frac{B}{\sqrt{3}} \biggl[ \frac{1}{\sqrt{2}} \bar\sigma_0 + \bar\sigma_8 \biggl], \nonumber\\
    \Big[ m_S^2(\bar\sigma) \Big]^{08} &= \Big[ m_S^2(\bar\sigma) \Big]^{80} = \lambda_1 \biggl[ 2 \bar\sigma_0 \bar\sigma_8 - \frac{1}{\sqrt{2}} \bar\sigma_8^2 \biggl] + 2 \lambda_2 \bar\sigma_0 \bar\sigma_8 + \frac{B}{\sqrt{6}} \bar\sigma_8, \nonumber\\
    \Big[ m_S^2(\bar\sigma) \Big]^{11} &= \Big[ m_S^2(\bar\sigma) \Big]^{22} = \Big[ m_S^2(\bar\sigma) \Big]^{33} \nonumber\\
    &= \mu^2 + \lambda_1 \biggl[ \bar\sigma_0^2 + \sqrt{2} \bar\sigma_0 \bar\sigma_8 + \frac{1}{2} \bar\sigma_8^2 \biggl] + \lambda_2 \Big( \bar\sigma_0^2 + \bar\sigma_8^2 \Big) + \frac{B}{\sqrt{3}} \biggl[ \frac{1}{\sqrt{2}} \bar\sigma_0 - \bar\sigma_8 \biggl], \nonumber\\
    \Big[ m_S^2(\bar\sigma) \Big]^{44} &=  \Big[ m_S^2(\bar\sigma) \Big]^{55} =  \Big[ m_S^2(\bar\sigma) \Big]^{66} =  \Big[ m_S^2(\bar\sigma) \Big]^{77} \nonumber\\
    &= \mu^2 + \lambda_1 \biggl[ \bar\sigma_0^2 - \frac{1}{\sqrt{2}} \bar\sigma_0 \bar\sigma_8 + \frac{1}{2} \bar\sigma_8^2 \biggl] + \lambda_2 \Big( \bar\sigma_0^2 + \bar\sigma_8^2 \Big) + \frac{B}{\sqrt{3}} \biggl[ \frac{1}{\sqrt{2}} \bar\sigma_0 + \frac{1}{2} \bar\sigma_8 \biggl], \nonumber\\
    {\rm (Others)} &= 0.
    \label{eq:ScalarCurvatureMasses}
\end{align}
The matrix elements $\Big[ m_S^2(\bar\sigma) \Big]^{11}$ and $\Big[ m_S^2(\bar\sigma) \Big]^{44}$ are identified as the square of the $a_0(980)$ mass $m^2_{a_0}$ and the $K^*_0(700)$ mass $m^2_{\kappa}$ respectively, and the mixing structure of $(08)$ components can be diagnolized through an orthogonal transformation
\begin{align}
    \tilde \sigma_i &= \big( O_S^{-1} \big)_{i}^{\,\, a} \sigma_a, \nonumber\\
    \Big[ \tilde m_S^2(\bar\sigma) \Big]_{(i)} \delta^{ij} &= \big( O_S^{-1} \big)^{i}_{\,\, a} \Big[ m_S^2(\bar\sigma) \Big]^{ab} \big( O_S \big)_{b}^{\,\, j},
    \label{eq:OrthogonalTransForScalarsTree}
\end{align}
where $\Big[ \tilde m_S^2(\bar\sigma) \Big]_{(i)}$ denotes the eigenvalues of the scalar mass matrix, and the transformation matrix $O_S$ reads
\begin{align}
    O_S = \pmat{ \cos{\theta_S^0} & -\sin{\theta_S^0} \\ \sin{\theta_S^0} & \cos{\theta_S^0} }.
    \label{eq:orthogonalTransScalarVacuum}
\end{align}
Then, the $f_0(500)$ and the $f_0(980)$ masses are identified through the eigenvalues
\begin{align}
    m^2[f_0(500)] = \Big[ \tilde m_S^2(\bar\sigma) \Big]_{(0)} &= \Big[ m_S^2(\bar\sigma) \Big]^{00} \cos^2 \theta_S^0 + \Big[ m_S^2(\bar\sigma) \Big]^{88} \sin^2 \theta_S^0 + 2 \Big[ m_S^2(\bar\sigma) \Big]^{08} \cos \theta_S^0 \sin \theta_S^0, \nonumber\\
    m^2[f_0(980)] = \Big[ \tilde m_S^2(\bar\sigma) \Big]_{(8)} &= \Big[ m_S^2(\bar\sigma) \Big]^{00} \sin^2 \theta_S^0 + \Big[ m_S^2(\bar\sigma) \Big]^{88} \cos^2 \theta_S^0 - 2 \Big[ m_S^2(\bar\sigma) \Big]^{08} \cos \theta_S^0 \sin \theta_S^0,
    \label{eq:TreeLevelDiagonalizationScalar}
\end{align}
and the mixing angle reads
\begin{align}
    \theta_S^0 = \frac{1}{2} \arctan \left[ \frac{2 \Big[ m_S^2(\bar\sigma) \Big]^{08}}{\Big[ m_S^2(\bar\sigma) \Big]^{00} - \Big[ m_S^2(\bar\sigma) \Big]^{88}} \right]
\end{align}
According to the previous discussion, the mixing element $\Big[ m_S^2(\bar\sigma) \Big]^{08}$ vanishes in the flavor-symmetric limit since $\bar\sigma_8 = 0$ and so as the mixing angle that $\theta_S^0 = 0$.

Similarly, the pseudo-scalar masses read
\begin{align}
     \Big[ m_P^2(\bar\sigma) \Big]^{00} &= \mu^2 + \frac{\lambda_1}{3} \Big( \bar\sigma_0^2 + \bar\sigma_8^2 \Big) + \lambda_2 \Big( \bar\sigma_0^2 + \bar\sigma_8^2 \Big) + B \sqrt{\frac{2}{3}} \bar\sigma_0, \nonumber\\
    \Big[ m_P^2(\bar\sigma) \Big]^{88} &= \mu^2 + \lambda_1 \biggl[ \frac{1}{3}\bar\sigma_0^2 - \frac{\sqrt{2}}{3} \bar\sigma_0 \bar\sigma_8 + \frac{1}{2} \bar\sigma_8^2 \biggl] + \lambda_2 \Big( \bar\sigma_0^2 + \bar\sigma_8^2 \Big) - \frac{B}{\sqrt{3}} \biggl[ \frac{1}{\sqrt{2}} \bar\sigma_0 + \bar\sigma_8 \biggl], \nonumber\\
    \Big[ m_P^2(\bar\sigma) \Big]^{08} &= \Big[ m_P^2(\bar\sigma) \Big]^{80} = \frac{\lambda_1}{3} \biggl[ 2 \bar\sigma_0 \bar\sigma_8 - \frac{1}{\sqrt{2}} \bar\sigma_8^2 \biggl] - \frac{B}{\sqrt{6}} \bar\sigma_8, \nonumber\\
    \Big[ m_P^2(\bar\sigma) \Big]^{11} &= \Big[ m_P^2(\bar\sigma) \Big]^{22} = \Big[ m_P^2(\bar\sigma) \Big]^{33} \nonumber\\
    &= \mu^2 + \frac{\lambda_1}{3} \biggl[ \bar\sigma_0^2 + \sqrt{2} \bar\sigma_0 \bar\sigma_8 + \frac{1}{2} \bar\sigma_8^2 \biggl] + \lambda_2 \Big( \bar\sigma_0^2 + \bar\sigma_8^2 \Big) - \frac{B}{\sqrt{3}} \biggl[ \frac{1}{\sqrt{2}} \bar\sigma_0 - \bar\sigma_8 \biggl], \nonumber\\
    \Big[ m_P^2(\bar\sigma) \Big]^{44} &=  \Big[ m_P^2(\bar\sigma) \Big]^{55} =  \Big[ m_P^2(\bar\sigma) \Big]^{66} =  \Big[ m_P^2(\bar\sigma) \Big]^{77} \nonumber\\
    &= \mu^2 + \frac{\lambda_1}{3} \biggl[ \bar\sigma_0^2 - \frac{1}{\sqrt{2}} \bar\sigma_0 \bar\sigma_8 + \frac{7}{2} \bar\sigma_8^2 \biggl] + \lambda_2 \Big( \bar\sigma_0^2 + \bar\sigma_8^2 \Big) - \frac{B}{\sqrt{3}} \biggl[ \frac{1}{\sqrt{2}} \bar\sigma_0 + \frac{1}{2} \bar\sigma_8 \biggl], \nonumber\\
    {\rm (Others)} &= 0.
    \label{eq:PseudoScalarCurvatureMasses}
\end{align}
The matrix elements $\Big[ m_P^2(\bar\sigma) \Big]^{11}$ and $\Big[ m_P^2(\bar\sigma) \Big]^{44}$ are identified as the square of the pion mass $m^2_\pi$ and the kaon mass $m^2_K$ respectively, and the mixing structure of $(08)$ components can be diagnolized through another orthogonal transformation
\begin{align}
    \tilde \pi_i &= \big( O_P^{-1} \big)_{i}^{\,\, a} \pi_a, \nonumber\\
    \Big[ \tilde m_P^2(\bar\sigma) \Big]_{(i)} \delta^{ij} &= \big( O_P^{-1} \big)^{i}_{\,\, a} \Big[ m_P^2(\bar\sigma) \Big]^{ab} \big( O_P \big)_{b}^{\,\, j},
    \label{eq:OrthogonalTransForPseudoScalarsTree}
\end{align}
where $\Big[ \tilde m_P^2(\bar\sigma) \Big]_{(i)}$ denotes the eigenvalues of the pseudo-scalar mass matrix, and the transformation matrix $O_P$ reads
\begin{align}
    O_S = \pmat{ \cos{\theta_P^0} & -\sin{\theta_P^0} \\ \sin{\theta_P^0} & \cos{\theta_P^0} }.
    \label{eq:orthogonalTransPseudoScalarVacuum}
\end{align}
Then, the $\eta^\prime$ and the $\eta$ masses are identified through the eigenvalues
\begin{align}
    m^2_{\eta^\prime} = \Big[ \tilde m_P^2(\bar\sigma) \Big]_{(0)} &= \Big[ m_P^2(\bar\sigma) \Big]^{00} \cos^2 \theta_P^0 + \Big[ m_P^2(\bar\sigma) \Big]^{88} \sin^2 \theta_P^0 + 2 \Big[ m_P^2(\bar\sigma) \Big]^{08} \cos \theta_P^0 \sin \theta_P^0, \nonumber\\
    m^2_{\eta} = \Big[ \tilde m_P^2(\bar\sigma) \Big]_{(8)} &= \Big[ m_P^2(\bar\sigma) \Big]^{00} \sin^2 \theta_P^0 + \Big[ m_P^2(\bar\sigma) \Big]^{88} \cos^2 \theta_P^0 - 2 \Big[ m_P^2(\bar\sigma) \Big]^{08} \cos \theta_P^0 \sin \theta_P^0,
    \label{eq:TreeLevelDiagonalizationPsuedoScalar}
\end{align}
and the mixing angle reads
\begin{align}
    \theta_P^0 = \frac{1}{2} \arctan \left[ \frac{2 \Big[ m_P^2(\bar\sigma) \Big]^{08}}{\Big[ m_P^2(\bar\sigma) \Big]^{00} - \Big[ m_P^2(\bar\sigma) \Big]^{88}} \right]
\end{align}
Again, the mixing element $\Big[ m_P^2(\bar\sigma) \Big]^{08}$ also vanishes in the flavor-symmetric limit since $\bar\sigma_8 = 0$ and so as the mixing angle that $\theta_P^0 = 0$.

%%%%%%%%%%%%%%%%%%%%%%%%%%%%%%%%%%%%%%%%%%%%%%%%%%%%%%%%%%%%%%%%%%%%%%
\section{Cornwall-Jackiw-Tomboulis formalism at finite temperature}
\label[appendix]{app:CJTFormalismAtFinitTemperature}

In this section of the appendix, we summarize the formulations in the conventional CJT formalism at finite temperature.

\subsection{Cornwall-Jackiw-Tomboulis formalism and Hartree approximation}

The CJT effective potential is given by \Cref{eq:GeneralCJTPotnetialMainText}, i.e.,
\begin{align}
    V_{\rm CJT}[\sigma, G_S, G_P] =& V(\sigma)+ \Omega_{\bar q q}(\sigma) + \frac{1}{2} \int_k \operatorname{tr} \Big[ \log G_S^{-1}(k) + \log G_P^{-1}(k) \Big] \nonumber\\
    &+ \frac{1}{2} \int_k \operatorname{tr} \Big[ \bar G_S^{-1}(k;\sigma) G_S(k) + \bar G_P^{-1}(k;\sigma) G_P(k) - 2 \Big] + V_2[\sigma, G_S, G_P],
    \label{eq:GeneralCJTPotnetial}
\end{align}
where the scalar propagator $G_S \equiv [G_S(k)]_{ab}$ and the pseudo-scalar propagator $G_P \equiv [G_P(k)]_{ab}$ are treated as independent variables, and the trace is taken in the flavor space.
In \Cref{eq:GeneralCJTPotnetial}, $V(\sigma)$ is the tree-level potential defined in \Cref{eq:LSMTreePotential}, and $V_2[\sigma, G_S, G_P]$ is the sum of all two-particle irreducible (2PI) diagrams, which are constructed by the full-meson propagators $\mathcal G_S$ and $\mathcal G_P$.
As defined in \Cref{eq:mesonicMomentumIntegral}, the shorthand notation of the momentum integral is defined as
\begin{align}
    \int_k \, f(k) = T \sum_{n = -\infty}^\infty \int \frac{\df^3 \bvec{k}}{(2\pi)^3} \, f(\omega_n, \bvec{k}),
\end{align}
where $T$ is the temperature and $\omega_n = 2 n \pi T$ is the bosonic Matsubara frequency.
The UV mesonic propagators $\bar G_S(k;\bar\sigma)$ and $\bar G_P(k;\bar\sigma)$ are defined from $S_m^{\rm eff}$ in \Cref{eq:generatingFUnctionalAfterIntegrateQuarks}, that
\begin{align}
    \Big[ \bar G_S^{-1}(k, k^\prime;\sigma) \Big]^{ab} 
    &= 
    \frac{\delta^2 S_m^{\rm eff}}{\delta \sigma_a(-k^\prime) \delta \sigma_b(k)}
    = 
    (2\pi)^4 \delta^{(4)}(k - k^\prime) \Biggl( k^2 \delta^{ab} + \Big[ m_S^2(\sigma) \Big]^{ab} + \frac{\partial^2 \Omega_{\bar q q}}{\partial \sigma_a \partial \sigma_b} \Biggl|_{\Phi = \bar\Phi} \Biggl)
    \nonumber\\
    &\equiv (2\pi)^4 \delta^{(4)}(k - k^\prime) \Big[ \bar G_S^{-1}(k;\sigma) \Big]^{ab} \, ,
    \nonumber\\
    \Big[ \bar G_P^{-1}(k, k^\prime;\sigma) \Big]^{ab} 
    &= 
    \frac{\delta^2 S_m^{\rm eff}}{\delta \pi_a(-k^\prime) \delta \pi_b(k)}
    = 
    (2\pi)^4 \delta^{(4)}(k - k^\prime) \Biggl( k^2 \delta^{ab} + \Big[ m_P^2(\sigma) \Big]^{ab} + \frac{\partial^2 \Omega_{\bar q q}}{\partial \pi_a \partial \pi_b} \Biggl|_{\Phi = \bar\Phi} \Biggl)
    \nonumber\\
    &\equiv (2\pi)^4 \delta^{(4)}(k - k^\prime) \Big[ \bar G_P^{-1}(k;\sigma) \Big]^{ab} \, ,
\end{align}
where $ m_S^2(\bar\sigma) $ and $ m_P^2(\bar\sigma) $ are the mesonic curvature masses which are given in equations~\eqref{eq:ScalarCurvatureMasses} and \eqref{eq:PseudoScalarCurvatureMasses}.

The expectation values of $\sigma$ and the propagators ${\mathcal G_S}$ and ${\mathcal G_P}$ are solved by the equations of motion with vanished source, i.e., the stationary conditions and the gap equations
\begin{align}
    \frac{\partial V_{\rm CJT}[\sigma, G_S, G_P]}{\partial \sigma_a} \Biggl|_{\sigma = \bar \sigma, G_S = \mathcal G_S, G_P = \mathcal G_P} &= 0, \nonumber\\
    \frac{\delta  V_{\rm CJT}[\sigma, G_S, G_P]}{\delta [G_S]_{ab}}\Biggl|_{\sigma = \bar \sigma, G_S = \mathcal G_S, G_P = \mathcal G_P} &= 0, \nonumber\\
    \frac{\delta  V_{\rm CJT}[\sigma, G_S, G_P]}{\delta [G_P]_{ab}}\Biggl|_{\sigma = \bar \sigma, G_S = \mathcal G_S, G_P = \mathcal G_P} &= 0. 
    \label{eq:quantumEOMsinCJT}
\end{align}
The latter two gap equations can be expressed as 
\begin{align}
        \Big[ {\mathcal G_S}^{-1}(k) \Big]^{ab} &= \Big[ \bar G_S^{-1}(k;\sigma) \Big]^{ab} + \Sigma^{ab}, \nonumber\\
        \Big[ {\mathcal G_P}^{-1}(k) \Big]^{ab} &= \Big[ \bar G_P^{-1}(k;\sigma) \Big]^{ab} + \Pi^{ab},
        \label{eq:generalGapEquations}
\end{align}
where
\begin{align}
    \Sigma^{ab} &= 2 \frac{\delta V_2[\sigma, G_S, G_P]}{\delta [G_S]_{ba}}|_{\sigma = \bar \sigma, G_S = \mathcal G_S, G_P = \mathcal G_P}, \nonumber\\
    \Pi^{ab} &= 2 \frac{\delta V_2[\sigma, G_S, G_P]}{\delta [G_P]_{ba}}|_{\sigma = \bar \sigma, G_S = \mathcal G_S, G_P = \mathcal G_P},
\end{align}
are the meson self-energies, which also depend on ${\mathcal G_S}$ and ${\mathcal G_P}$.
Notice that if we take $V_2 = 0$, the gap equations \eqref{eq:generalGapEquations} are reduced such that the propagators are replaced by the tree-level ones, thus the 2PI effective potential will be reduced to the 1PI effective action at one-loop level.

Since $V_2$ includes infinite 2PI diagrams, we have to make a truncation to close the equations of motion.
In the present work, we adopt the double-bubble diagrams, for which the truncation scheme is equivalent to the so-called Hartree approximation.
The 2PI diagrams that are taken into account are generated from the four-meson interactions in $V_0$, which read
\begin{align}
    V_2[G_S, G_P] = \mathcal{F}^{abcd} \biggl[ \int_k [G_S(k)]_{ab} \int_q [G_S(q)]_{cd} + \int_k [G_P(k)]_{ab} \int_q [G_P(q)]_{cd} \biggl] \, + \, 2 \mathcal{H}^{abcd} \int_k [G_S(k)]_{ab} \int_q [G_P(q)]_{cd},
    \label{eq:V2UnderHartree}
\end{align}
where
\begin{align}
    \mathcal{F}^{abcd} &\equiv \frac{1}{8}\frac{\partial^4 V_0}{\partial \sigma_a \partial \sigma_b \partial \sigma_c \partial \sigma_d}\Biggl|_{\Phi=0} = \frac{\lambda_1}{8} \Big( d^{abn}d^{ncd} + d^{adn}d^{nbc} + d^{acn}d^{nbd} \Big) + \frac{\lambda_2}{4} \Big( \delta^{ab}\delta^{cd} + \delta^{ad}\delta^{bc} + \delta^{ac}\delta^{bd} \Big), \nonumber\\
    \mathcal{H}^{abcd}&\equiv \frac{1}{8} \frac{\partial^4 V_0}{\partial \sigma_a \partial \sigma_b \partial \pi_c \partial \pi_d}\Biggl|_{\Phi=0} = \frac{\lambda_1}{8} \Big( d^{abn}d^{ncd} + f^{acn}f^{nbd} + f^{bcn}f^{nad} \Big) + \frac{\lambda_2}{4} \delta^{ab}\delta^{cd},
    \label{eq:definitionOfHandF}
\end{align}
and the structure constants are given by
\begin{align}
    d^{abc} = 2 \operatorname{tr} \Big[ \{ T^a, T^b \} T^c \Big], \qquad f^{abc} = -2i \operatorname{tr} \Big[ [T^a, T^b] T^c \Big].
\end{align}

\subsection{Gap equations of scalar masses}
\label[appendix]{app:CJTgapEquations}

We consider the gap equations \eqref{eq:generalGapEquations} under the two-loop truncation given in \Cref{eq:V2UnderHartree}.
Under the Hartree approximation, the loop correction does not enter the momentum-dependent part.
As in \Cref{eq:parameterizationOfFullPropagatorsMainText}, we shall parameterize the full-propagators as
\begin{align}
    \Big[ {\mathcal G_S}^{-1}(k) \Big]^{ab} &= k^2 \delta^{ab} + \Big[ M_S^2 \Big]^{ab}, \nonumber\\
    \Big[ {\mathcal G_P}^{-1}(k) \Big]^{ab} &= k^2 \delta^{ab} + \Big[ M_P^2 \Big]^{ab},
\end{align}
where $\Big[ M_S \Big]^{ab}$ and $ \Big[ M_P \Big]^{ab}$ are the dynamical meson masses.
Then, the gap equations \eqref{eq:generalGapEquations} are reduced to
\begin{align}
    \Big[ M_S^2 \Big]^{ab} &= \Big[ m_S^2(\bar\sigma) \Big]^{ab} + \frac{\partial^2 \Omega_{\bar q q}}{\partial \sigma_a \partial \sigma_b} \Biggl|_{\Phi = \bar\Phi} + 4 \mathcal{F}^{abcd} \int_k [{\mathcal G_S}(k)]_{cd} + 4 \mathcal{H}^{abcd} \int_k [{\mathcal G_P}(k)]_{cd}, \nonumber\\
    \Big[ M_P^2 \Big]^{ab} &= \Big[ m_P^2(\bar\sigma) \Big]^{ab} + \frac{\partial^2 \Omega_{\bar q q}}{\partial \pi_a \partial \pi_b} \Biggl|_{\Phi = \bar\Phi} + 4 \mathcal{F}^{abcd} \int_k [{\mathcal G_P}(k)]_{cd} + 4 \mathcal{H}^{abcd} \int_k [{\mathcal G_S}(k)]_{cd}.
    \label{eq:gapEquationsCJT}
\end{align}

At the classical level, we diagonalize the mass matrices to obtain the mass eigenstates, see equations~\eqref{eq:TreeLevelDiagonalizationScalar} and \eqref{eq:TreeLevelDiagonalizationPsuedoScalar}.
After introducing the quantum corrections to the meson propagators, we also introduce the following orthogonal transformations to diagonalize the dynamical masses, such that
\begin{align}
    \Big[ \tilde M_{S/P}^2 \Big]_{(i)} \delta^{ij} = \big( O_{S/P}^{\prime,-1} \big)^{i}_{\,\, a} \Big[ M_{S/P}^2 \Big]^{ab} \big( O_{S/P}^\prime \big)_{b}^{\,\, j}.
\end{align}
In general, the transformation matrices $O_{S/P}^\prime$ are different from those for the tree-level masses, i.e., Equations~\eqref{eq:OrthogonalTransForScalarsTree} and \eqref{eq:OrthogonalTransForPseudoScalarsTree}, due to the introduction of the quantum and thermal corrections.
Formally, the rotation angles are expressed as
\begin{align}
    \theta_{S/P} = \frac{1}{2} \arctan  \left[ \frac{2 \Big[ M_{S/P}^2 \Big]^{08}}{\Big[ M_{S/P}^2 \Big]^{00} - \Big[ M_{S/P}^2 \Big]^{88}} \right].
\end{align}
The full-propagators $\mathcal G_S$ and $\mathcal G_P$ can also be diagonalized simultaneously through $O_{S/P}^\prime$, which read
\begin{align}
    [\tilde{\mathcal G}_S(k)]_{(i)} \delta_{ij} &= \big( O_{S}^{\prime,-1} \big)_{i}^{\,\, a} [{\mathcal G_S}(k)]_{ab} \big( O_{S}^\prime \big)^{b}_{\,\, j}, \nonumber\\
    [\tilde{\mathcal G}_P(k)]_{(i)} \delta_{ij} &= \big( O_{P}^{\prime,-1} \big)_{i}^{\,\, a} [{\mathcal G_P}(k)]_{ab} \big( O_{P}^\prime \big)^{b}_{\,\, j}.
\end{align}
Then, the diagonalized meson propagators are represented fully by the mass eigenvalues, i.e.,
\begin{align}
    [\tilde{\mathcal G}_S^{-1}(k)]_{(i)} = k^2 + \Big[ \tilde M_{S}^2 \Big]_{(i)}, 
    \qquad
    [\tilde{\mathcal G}_P^{-1}(k)]_{(i)} = k^2 + \Big[ \tilde M_{P}^2 \Big]_{(i)},
\end{align}
and the loop corrections constructed by the meson propagators only include
\begin{align}
    \int_k [\tilde{\mathcal G}_S(k)]_{(i)} \quad \text{and} \quad \int_k [\tilde{\mathcal G}_P(k)]_{(i)} \, .
\end{align}

To obtain the value of the rotation angles $\theta_{S/P}$, we consider the $08$-component of the gap equations after performing the orthogonal transformation, which turns out to be the following constraints
\begin{align}
    &0 = \Big[ \tilde m_S^2(\bar\sigma) \Big]^{08} + \frac{\partial^2 \Omega_{\bar q q}}{\partial \sigma_0 \partial \sigma_8} \Biggl|_{\Phi = \bar\Phi} + \tilde \Sigma^{08}, \nonumber\\
    &0 = \Big[ \tilde m_P^2(\bar\sigma) \Big]^{08} + \frac{\partial^2 \Omega_{\bar q q}}{\partial \pi_0 \partial \pi_8} \Biggl|_{\Phi = \bar\Phi} + \tilde \Pi^{08},
    \label{eq:MixingElementOfGapEq}
\end{align}
where
\begin{align}
    \Big[ \tilde m_{S/P}^2(\bar\sigma) \Big]^{08} = \Big[ m_{S/P}^2(\bar\sigma) \Big]^{08} \cos(2\theta_{S/P}) - \biggl\{ \Big[ m_{S/P}^2(\bar\sigma) \Big]^{00} - \Big[ m_{S/P}^2(\bar\sigma) \Big]^{88} \biggl\} \cos(\theta_{S/P}) \sin(\theta_{S/P}).
\end{align}
For the lack of readability, we do not show the explicit form of the meson-loop contributions to the meson self-energies.
Part of the expressions can be found in Ref.~\cite{Guan:2025tmf}.

\subsection{Stationary conditions of VEVs}

The stationary conditions realized from the 2PI effective potential \eqref{eq:GeneralCJTPotnetial} then read
\begin{align}
    0 
    &= 
    \frac{\partial V_m(\bar\sigma)}{\partial \bar\sigma_0} 
    - 3 \mathcal A^{0bc} \biggl[ \int_k [{\mathcal G_S}(k)]_{cb} - \int_k [{\mathcal G_P}(k)]_{cb} \biggl] 
    + \frac{1}{2} \frac{\partial^3 \Omega_{\bar q q}}{\partial \sigma_0 \partial \sigma_a \partial \sigma_b} \Biggl|_{\Phi = \bar\Phi} \int_k [{\mathcal G_S}(k)]_{ba} 
    + \frac{1}{2} \frac{\partial^3 \Omega_{\bar q q}}{\partial \sigma_0 \partial \pi_a \partial \pi_b} \Biggl|_{\Phi = \bar\Phi} \int_k [{\mathcal G_P}(k)]_{ba} 
    \nonumber\\
    &\qquad + 4 \mathcal F^{0bcd} \bar\sigma_b \int_k [{\mathcal G_S}(k)]_{dc} 
    + 4 \mathcal H^{0bcd} \bar\sigma_b \int_k [{\mathcal G_P}(k)]_{dc} 
    + \frac{\partial \Omega_{\bar q q}}{\partial \sigma_0} \Biggl|_{\Phi = \bar\Phi} \, , 
    \nonumber\\
    0 
    &= 
    \frac{\partial V_m(\bar\sigma)}{\partial \bar\sigma_8} 
    - 3 \mathcal A^{8bc} \biggl[ \int_k [{\mathcal G_S}(k)]_{cb} - \int_k [{\mathcal G_P}(k)]_{cb} \biggl] 
    + \frac{1}{2} \frac{\partial^3 \Omega_{\bar q q}}{\partial \sigma_8 \partial \sigma_a \partial \sigma_b} \Biggl|_{\Phi = \bar\Phi} \int_k [{\mathcal G_S}(k)]_{ba} 
    + \frac{1}{2} \frac{\partial^3 \Omega_{\bar q q}}{\partial \sigma_8 \partial \pi_a \partial \pi_b} \Biggl|_{\Phi = \bar\Phi} \int_k [{\mathcal G_P}(k)]_{ba} 
    \nonumber\\
    &\qquad + 4 \mathcal F^{8bcd} \bar\sigma_b \int_k [{\mathcal G_S}(k)]_{dc} 
    + 4 \mathcal H^{8bcd} \bar\sigma_b \int_k [{\mathcal G_P}(k)]_{dc} 
    + \frac{\partial \Omega_{\bar q q}}{\partial \sigma_8} \Biggl|_{\Phi = \bar\Phi} \, ,
    \label{eq:stationaryCOnditionsInCJT}
\end{align}
where $V(\bar\sigma)$ is given in \Cref{eq:treePotentialAroundMeanField}, and
\begin{align}
    \mathcal A^{abc} &\equiv -\frac{1}{6} \frac{\partial^3 V_{\rm anom}}{\partial \sigma_a \partial \sigma_b \partial \sigma_c} \Biggl|_{\Phi=0} = \frac{1}{6} \frac{\partial^3 V_{\rm anom}}{\partial \sigma_a \partial \pi_b \partial \pi_c} \Biggl|_{\Phi=0} \nonumber\\
    &= \frac{B}{6} \biggl[ d^{abc} - \frac{3}{2}\Big( \delta^{a0}d^{0bc} + \delta^{b0} d^{a0c} + \delta^{c0}d^{ab0} \Big) + \frac{9}{2} d^{000}\delta^{a0}\delta^{b0}\delta^{c0} \biggl].
    \label{eq:definitionOfG}
\end{align}

To solve these equations, we have to transform the propagators into the diagonalized basis.
Then, the only contributions of the loop corrections in the mesonic sector are those from the physical meson one-loop diagrams.
As discussed in \Cref{sec:Renormalization}, we perform the vacuum subtraction such that
\begin{align}
   & \int_k [\tilde{\mathcal G}_S(k)]_{(i)} = \int \frac{\df^3 \bvec{k}}{(2\pi)^3} \frac{1}{\epsilon^{S,(i)}_{\bvec{k}}} \left( \exp \left\{ \frac{\epsilon^{S,(i)}_{\bvec{k}}}{T} \right\} -1 \right)^{-1} \nonumber\\
    &\int_k [\tilde{\mathcal G}_P(k)]_{(i)}  = \int \frac{\df^3 \bvec{k}}{(2\pi)^3} \frac{1}{\epsilon^{P,(i)}_{\bvec{k}}} \left( \exp \left\{ \frac{\epsilon^{P,(i)}_{\bvec{k}}}{T} \right\} -1 \right)^{-1},
\end{align}
where
\begin{align}
    \epsilon^{S/P,(i)}_{\bvec{k}} \equiv \sqrt{{\bvec{k}}^2 + \Big[ \tilde M_{S/P}^2 \Big]_{(i)} }.
\end{align}
In this subtraction scheme, the contributions from the vacuum loop corrections are dropped, which indicates that the vacuum information is encoded in the tree-level parameters.

\subsection{Quark loop contributions}
\label[appendix]{app:quarkloopCOntributions}

The integration of the quark degrees of freedom introduced effective vertices of the mesonic fields, which are reflected in the quark loop contribution to the grand canonical potential $\Omega_{\bar q q}$.
As mentioned in \Cref{sec:2PIFormalismAndSymmetryImprovement}, we drop the UV divergent vacuum part of $\Omega_{\bar q q}$ as the subtraction scheme, and the thermal part of $\Omega_{\bar q q}$ reads
\begin{align}
    \Omega_{\bar q q}\Big|_{T} &= - \int_p \operatorname{tr} \log \big( \Delta^q_0(p) \big)^{-1} = \nu_c T \sum_{f = u,d,s} \int_0^\infty \frac{\df^3 \bvec{p}}{(2\pi)^3} \biggl\{ \log \Big[ 1 - n_{q,f}(T,\mu_f) \Big] + \log \Big[ 1 - n_{\bar q,f}(T,\mu_f) \Big] \biggl\},
\end{align}
where $\nu_c = 2 N_c = 6$ is the internal quark degrees of freedom,
\begin{align}
    n_{q,f}(T,\mu_f) = \Big[ 1 + \exp\{ (E_{q,f} - \mu_f)/T \} \Big]^{-1}
\end{align}
is the fermionic occupation number density, and $n_{\bar q,f}(T,\mu_f) = n_{q,f}(T,-\mu_f)$ for the anti-fermions.
The single-particle dispersion relation is given by
\begin{align}
    E_{q,f} = \sqrt{|\bvec{p}|^2 + m_f^2},
\end{align}
and the flavor-dependent mass $m_f^2$ is evaluated from diagonalizing the meson field-induced mass matrix 
\begin{align}
    M^2_{ij} &= g^2 \big( T^a \big)_{ik} (\sigma_a + i \gamma_5 \pi_a)\big( T^b \big)_{kj} (\sigma_a - i \gamma_5 \pi_a) \nonumber\\
    &= g^2 \, P_R \, \big[ \Phi^\dagger \Phi \big]_{ij} + g^2 \, P_L \, \big[ \Phi \Phi^\dagger \big]_{ij} \, .
    \label{eq:quarkMassMatrixAppendix}
\end{align}
In the following formulations, we identify $\Omega_{\bar q q}$ as $\Omega_{\bar q q}\Big|_{T}$, and omit the subscription $\Big|_{T}$.

The derivatives of $\Omega_{\bar q q}$ with respect to the mesonic fields depend on the derivatives of the flavor-dependent mass $m_f^2$. We introduce the following shorthand notations for future convenience:
\begin{align}
    m_{f,a}^2 \equiv \frac{\partial m_{f}^2}{\partial \varphi_a}, \quad m_{f,ab}^2 \equiv \frac{\partial^2 m_{f}^2}{\partial \varphi_a \partial \varphi_b}, \quad m_{f,abc}^2 \equiv \frac{\partial^3 m_{f}^2}{\partial \varphi_a \partial \varphi_b \partial \varphi_c}, \qquad f \in \{u,d,s\}, \quad {\rm and} \quad \varphi_a \in \{ \sigma_a, \, \pi_a \} \, .
\end{align}
The first derivative of $\Omega_{\bar q q}$ with respect to $\varphi_a$ is
\begin{align}
    \frac{\partial \Omega_{\bar q q}}{\partial \varphi_a} = \nu_c \sum_f \int \frac{\df^3 \bvec{p}}{(2\pi)^3} \big( n_{q,f} + n_{\bar q,f} \big) \frac{m^2_{f,a}}{2 E_{q,f}}.
\end{align}
The second derivative of $\Omega_{\bar q q}$ then reads
\begin{align}
    \frac{\partial^2 \Omega_{\bar q q}}{\partial \varphi_a \partial \varphi_b} = \nu_c \sum_f \int \frac{\df^3 \bvec{p}}{(2\pi)^3} \frac{1}{2 E_{q,f}} \Biggl\{ \big(  n_{q,f} + n_{\bar q,f} \big) \biggl[ m^2_{f,ab} - \frac{m^2_{f,a} m^2_{f,b}}{2 E_{q,f}^2} \biggl] - \big(  b_{q,f} + b_{\bar q,f} \big) \frac{m^2_{f,a}m^2_{f,b}}{2 E_{q,f} T} \Biggl\},
\end{align}
where $b_{q,f}(T,\mu_f) = n_{q,f}(T,\mu_f) \big( 1 - n_{q,f}(T,\mu_f) \big)$, and $b_{\bar q,f}(T,\mu_f) = b_{q,f}(T,-\mu_f)$ respectively.
The third derivative of $\Omega_{\bar q q}$ reads
\begin{align}
    \frac{\partial^3 \Omega_{\bar q q}}{\partial \varphi_a \partial \varphi_b \partial \varphi_c} = - \nu_c \sum_f &\int \frac{\df^3 \bvec{p}}{(2\pi)^3} \frac{1}{4 E_{q,f}^2}  \nonumber\\
    &\times \Biggl\{ \frac{\big(  n_{q,f} + n_{\bar q,f} \big)}{E_{q,f}} \biggl[ m^2_{f,ab}m^2_{f,c} + m^2_{f,bc}m^2_{f,a} + m^2_{f,ca}m^2_{f,b} + \frac{ m^2_{f,a} m^2_{f,b} m^2_{f,c}}{2 E_{q,f}^2} - 2 E_{q,f}^2 m^2_{f,abc} \biggl] \nonumber\\
    &\qquad + \frac{\big(  b_{q,f} + b_{\bar q,f} \big)}{T} \biggl[ m^2_{f,ab}m^2_{f,c} + m^2_{f,bc}m^2_{f,a} + m^2_{f,ca}m^2_{f,b} - \frac{3 m^2_{f,a} m^2_{f,b} m^2_{f,c}}{2 E_{q,f}^2} \biggl] \nonumber\\
    &\qquad - \Big[ b_{q,f}\big( 1 - 2 n_{q,f} \big) + b_{\bar q,f}\big( 1 - 2 n_{\bar q,f} \big) \Big] \frac{m^2_{f,a} m^2_{f,b} m^2_{f,c}}{2 E_{q,f} T^2} \Biggl\}.
\end{align}

According to \Cref{eq:quarkMassMatrixAppendix}, since $\big[ \Phi^\dagger \Phi \big]_{ij}$ and $\big[ \Phi \Phi^\dagger \big]_{ij}$ have the same eigenvalues, the mesonic field derivatives of $m_f^2$ can be obtained after diagonalizing $\big[ \Phi^\dagger \Phi \big]_{ij}$.
We summarize the resulting derivatives of $m_f^2$ below.
For the comparison to Ref.~\cite{Schaefer:2008hk}, we adopt the same notation as
\begin{align}
    \Phi_x = 2 \bar \Phi_1 \, , \quad \Phi_y = \sqrt{2} \bar \Phi_3 \, .
\end{align}

For $m_{f,a}^2$, we define
\begin{align}
    m_{f,a}^2 \Big|_{\sigma} \equiv \frac{\partial m_f^2}{\partial \sigma_a}\, \quad m_{f,a}^2 \Big|_{\pi} \equiv \frac{\partial m_f^2}{\partial \pi_a} \, ,
\end{align}
and we have
\begin{align}
    m_{u,a}^2 \Big|_{\sigma} 
    = g^2
    \pmat{
    \frac{\Phi_x}{\sqrt{6}} \\
    -\frac{\Phi_x}{2} \textbf{1}_{3\times1} \\
    \textbf{0}_{4\times1} \\
    \frac{\Phi_x}{2 \sqrt{3}}
    } \, ,
    \qquad
    m_{d,a}^2 \Big|_{\sigma}
    = g^2
    \pmat{
    \frac{\Phi_x}{\sqrt{6}} \\
    \frac{\Phi_x}{2} \textbf{1}_{3\times1} \\
    \textbf{0}_{4\times1} \\
    \frac{\Phi_x}{2 \sqrt{3}}
    } \, ,
    \qquad
    m_{s,a}^2 \Big|_{\sigma}
    = g^2
    \pmat{
    \frac{\Phi_y}{\sqrt{3}} \\
    \textbf{0}_{3\times1} \\
    \textbf{0}_{4\times1} \\
    -\frac{\sqrt{2}}{\sqrt{3}} \Phi_y
    } \, ,
    \qquad
    m_{f,a}^2 \Big|_{\pi} 
    = 0 \, .
\end{align}

For the second derivatives, we distinguish derivatives with respect to scalar and pseudoscalar fields by writing
\begin{align}
    m_{f,ab}^2 \Big|_{\sigma} \equiv \frac{\partial^2 m_f^2}{\partial \sigma_a \partial \sigma_b}\, \quad m_{f,ab}^2 \Big|_{\pi} \equiv \frac{\partial^2 m_f^2}{\partial \pi_a \partial \pi_b} \, .
\end{align}
For the scalar mesons, the matrix form of $m_{f,a b}^2$ read
\begin{align}
    &m_{u,ab}^2 \Big|_{\sigma} 
    = g^2
    \pmat{
    \frac{1}{3} & -\frac{1}{\sqrt{6}} \textbf{1}_{1 \times 3} & \textbf{0}_{1\times4} & \frac{\sqrt{2}}{6} \\
    -\frac{1}{\sqrt{6}} \textbf{1}_{3\times1} & \frac{1}{2}\textbf{1}_{3\times3} & \textbf{0}_{3\times4} & -\frac{1}{2\sqrt{3}} \textbf{1}_{3\times1} \\
    \textbf{0}_{4\times1} & \textbf{0}_{4\times3} & \frac{\Phi_x(\Phi_x + \sqrt{2}\Phi_y)}{(\Phi_x^2 - 2 \Phi_y^2)} \textbf{1}_{4\times4} & \textbf{0}_{4\times1} \\
    \frac{\sqrt{2}}{6} & -\frac{1}{2\sqrt{3}} \textbf{1}_{1 \times 3} & \textbf{0}_{1\times4} & \frac{1}{6}
    } \, ,
    \nonumber\\ \nonumber\\
    &m_{d,ab}^2 \Big|_{\sigma} 
    = g^2
    \pmat{
    \frac{1}{3} & \frac{1}{\sqrt{6}} \textbf{1}_{1 \times 3} & \textbf{0}_{1\times4} & \frac{\sqrt{2}}{6} \\
    \frac{1}{\sqrt{6}} \textbf{1}_{3\times1} & \frac{1}{2}\textbf{1}_{3\times3} & \textbf{0}_{3\times4} & \frac{1}{2\sqrt{3}} \textbf{1}_{3\times1} \\
    \textbf{0}_{4\times1} & \textbf{0}_{4\times3} & \textbf{0}_{4\times4} & \textbf{0}_{4\times1} \\
    \frac{\sqrt{2}}{6} & \frac{1}{2\sqrt{3}} \textbf{1}_{1 \times 3} & \textbf{0}_{1\times4} & \frac{1}{6}
    } \, ,
    \nonumber\\ \nonumber\\
    &m_{s,ab}^2 \Big|_{\sigma} 
    = g^2
    \pmat{
    \frac{1}{3} & \textbf{0}_{1 \times 3} & \textbf{0}_{1\times4} & -\frac{\sqrt{2}}{3} \\
    \textbf{0}_{3\times1} & \textbf{0}_{3\times3} & \textbf{0}_{3\times4} & \textbf{0}_{3\times1} \\
    \textbf{0}_{4\times1} & \textbf{0}_{4\times3} & \frac{\Phi_y(\sqrt{2} \Phi_x + 2 \Phi_y)}{(2 \Phi_y^2 - \Phi_x^2)} \textbf{1}_{4\times4} & \textbf{0}_{4\times1} \\
    -\frac{\sqrt{2}}{3} & \textbf{0}_{1 \times 3} & \textbf{0}_{1\times4} & \frac{2}{3}
    } \, .
\end{align}
For the pseudo-scalar mesons, $m_{f,a b}^2$ read
\begin{align}
    &m_{u,ab}^2 \Big|_{\pi} 
    = g^2
    \pmat{
    \frac{1}{3} & \textbf{0}_{1 \times 3} & \textbf{0}_{1\times4} & \frac{\sqrt{2}}{6} \\
    \textbf{0}_{3\times1} & \frac{1}{2}\textbf{1}_{3\times3} & \textbf{0}_{3\times4} & \textbf{0}_{3\times1} \\
    \textbf{0}_{4\times1} & \textbf{0}_{4\times3} & \frac{\Phi_x(\Phi_x - \sqrt{2}\Phi_y)}{(\Phi_x^2 - 2 \Phi_y^2)} \textbf{1}_{4\times4} & \textbf{0}_{4\times1} \\
    \frac{\sqrt{2}}{6} & \textbf{0}_{1 \times 3} & \textbf{0}_{1\times4} & \frac{1}{6}
    } \, ,
    \nonumber\\ \nonumber\\
    &m_{d,ab}^2 \Big|_{\pi} 
    = g^2
    \pmat{
    \frac{1}{3} & \textbf{0}_{1 \times 3} & \textbf{0}_{1\times4} & \frac{\sqrt{2}}{6} \\
    \textbf{0}_{3\times1} & \frac{1}{2}\textbf{1}_{3\times3} & \textbf{0}_{3\times4} & \textbf{0}_{3\times1} \\
    \textbf{0}_{4\times1} & \textbf{0}_{4\times3} & \textbf{0}_{4\times4} & \textbf{0}_{4\times1} \\
    \frac{\sqrt{2}}{6} & \textbf{0}_{1 \times 3} & \textbf{0}_{1\times4} & \frac{1}{6}
    } \, ,
    \nonumber\\ \nonumber\\
    &m_{s,ab}^2 \Big|_{\pi} 
    = g^2
    \pmat{
    \frac{1}{3} & \textbf{0}_{1 \times 3} & \textbf{0}_{1\times4} & -\frac{\sqrt{2}}{3} \\
    \textbf{0}_{3\times1} & \textbf{0}_{3\times3} & \textbf{0}_{3\times4} & \textbf{0}_{3\times1} \\
    \textbf{0}_{4\times1} & \textbf{0}_{4\times3} & \frac{\Phi_y(\sqrt{2} \Phi_x - 2 \Phi_y)}{(\Phi_x^2 - 2 \Phi_y^2)} \textbf{1}_{4\times4} & \textbf{0}_{4\times1} \\
    -\frac{\sqrt{2}}{3} & \textbf{0}_{1 \times 3} & \textbf{0}_{1\times4} & \frac{2}{3}
    } \, .
\end{align}
For the third derivatives $m_{f, a b c}^2$, the only nonvanishing parts are those with the first index of $a = 0, 8$ for the scalar mesons and $(b,c)=(4,4),(5,5),(6,6),(7,7)$.
We thus define
\begin{align}
    m_{f,abc}^2 \Big|_{\sigma} \equiv \frac{\partial^3 m_f^2}{\partial \sigma_a \partial \sigma_b \partial \sigma_c}\, \quad m_{f,abc}^2 \Big|_{\pi} \equiv \frac{\partial^3 m_f^2}{\partial \sigma_a \partial \pi_b \partial \pi_c} \, ,
\end{align}
yielding the following components that are relevant in our computation
\begin{gather}
    m^2_{u,044}\Big|_{\sigma}
    =
    m^2_{u,055}\Big|_{\sigma}
    =
    m^2_{u,066}\Big|_{\sigma}
    =
    m^2_{u,077}\Big|_{\sigma}
    =
    \sqrt{\frac{2}{3}}
    \frac{\Phi_x+\sqrt{2}\Phi_y}{\Phi_x^2-2\Phi_y^2}, 
    \nonumber\\
    m^2_{s,044}\Big|_{\sigma}
    =
    m^2_{s,055}\Big|_{\sigma}
    =
    m^2_{s,066}\Big|_{\sigma}
    =
    m^2_{s,077}\Big|_{\sigma}
    =
    -\sqrt{\frac{2}{3}}
    \frac{\Phi_x+\sqrt{2}\Phi_y}{\Phi_x^2-2\Phi_y^2},
    \nonumber\\
    m^2_{u,044}\Big|_{\pi}
    =
    m^2_{u,055}\Big|_{\pi}
    =
    m^2_{u,066}\Big|_{\pi}
    =
    m^2_{u,077}\Big|_{\pi}
    =
    \sqrt{\frac{2}{3}}
    \frac{(\Phi_x+\sqrt{2}\Phi_y)(\Phi_x-\sqrt{2}\Phi_y)^4}
    {\left[(\Phi_x^2-2\Phi_y^2)^2\right]^{3/2}},
    \nonumber\\
    m^2_{s,044}\Big|_{\pi}
    =
    m^2_{s,055}\Big|_{\pi}
    =
    m^2_{s,066}\Big|_{\pi}
    =
    m^2_{s,077}\Big|_{\pi}
    =
    -\sqrt{\frac{2}{3}}
    \frac{(\Phi_x+\sqrt{2}\Phi_y)(\Phi_x-\sqrt{2}\Phi_y)^4}
    {\left[(\Phi_x^2-2\Phi_y^2)^2\right]^{3/2}},
    \nonumber\\
    m^2_{u,844}\Big|_{\sigma}
    =
    m^2_{u,855}\Big|_{\sigma}
    =
    m^2_{u,866}\Big|_{\sigma}
    =
    m^2_{u,877}\Big|_{\sigma}
    =
    -\sqrt{\frac{2}{3}}
    \frac{\sqrt{2}\Phi_x+\Phi_y}{(\Phi_x-\sqrt{2}\Phi_y)^2},
    \nonumber\\
    m^2_{s,844}\Big|_{\sigma}
    =
    m^2_{s,855}\Big|_{\sigma}
    =
    m^2_{s,866}\Big|_{\sigma}
    =
    m^2_{s,877}\Big|_{\sigma}
    =
    \sqrt{\frac{2}{3}}
    \frac{\sqrt{2}\Phi_x+\Phi_y}{(\Phi_x-\sqrt{2}\Phi_y)^2},
    \nonumber\\
    m^2_{u,844}\Big|_{\pi}
    =
    m^2_{u,855}\Big|_{\pi}
    =
    m^2_{u,866}\Big|_{\pi}
    =
    m^2_{u,877}\Big|_{\pi}
    =
    \sqrt{\frac{2}{3}}
    \frac{\sqrt{2}\Phi_x^3-3\Phi_x^2\Phi_y+2\Phi_y^3}
    {(\Phi_x^2-2\Phi_y^2)^2},
    \nonumber\\
    m^2_{s,844}\Big|_{\pi}
    =
    m^2_{s,855}\Big|_{\pi}
    =
    m^2_{s,866}\Big|_{\pi}
    =
    m^2_{s,877}\Big|_{\pi}
    =
    -\sqrt{\frac{2}{3}}
    \frac{\sqrt{2}\Phi_x^3-3\Phi_x^2\Phi_y+2\Phi_y^3}
    {(\Phi_x^2-2\Phi_y^2)^2},
    \nonumber\\
    \text{other}=0.
\end{gather}
%

%%%%%%%%%%%%%%%%%%%%%%%%%%%%%%%%%%%%%%%%%%%%%%%%%%%%%%%%%%%%%%%%%%%%%%
\section{Ward-Takahashi identities and Gell-Mann-Oakes-Renner relations}
\label[appendix]{app:AWTIsandGMORs}

In this appendix, we summarize the derivation of the WTIs induced by the chiral transformations in \labelcref{eq:chiralTransformation} acting on the mesonic fields \(\Phi\) and \(\Phi^\dagger\). 
The derivation follows the same strategy as in Ref.~\cite{Guan:2025tmf}. 
The main difference is that the spurion field \(\mathcal M\), namely the source of explicit chiral symmetry breaking, is coupled only linearly to the mesonic fields. 
As a result, the symmetry-breaking terms that appear in the modified WTIs take a cleaner and more transparent form.

%%%%%%%%%%%%%%%%%%%%%%%%%%%%%%%%%%%%%%%%%%%%%%%%%%%%%%%%%%%%%%%%%%%%%%%%%%%%%
\subsection{WTIs in 1PI formalism and threshold property of pseudo-NG masses}

Consider a continuous symmetry associated with an infinitesimal variation $\delta_\epsilon$ of the field $\phi$ in a linear representation of a group $G$, under which the action $S[\phi]$ is invariant:
\begin{align}
    \delta_\epsilon S[\phi]
    =
    \frac{\delta S}{\delta \phi_a}\cdot \delta_\epsilon \phi_a
    =0,
    \qquad
    \delta_\epsilon \phi_a = d_a^{\ b}\phi_b,
    \label{eq:symmetricActionUnderdeltaepsilon}
\end{align}
where $d_a^{\ b}$ is a constant matrix in field space, and
$A\cdot B \equiv \int_x A(x)B(x)$.

The generating functional associated with $S[\phi]$ in the presence of the source $J_a$ is defined by
\begin{align}
    Z[J]
    =
    \int \big[\mathcal D \phi_a\big]\,
    e^{-S[\phi]+J^a\cdot \phi_a}
    =
    e^{W[J]}.
\end{align}
Imposing the invariance of $Z[J]$ under the symmetry transformation in \Cref{eq:symmetricActionUnderdeltaepsilon}, one has
\begin{align}
    \int \big[\mathcal D \phi_a'\big]\,
    e^{-S[\phi']+J^a\cdot \phi_a'}
    \Big|_{\phi_a'=\phi_a+i\epsilon\,\delta_\epsilon\phi_a}
    =
    \int \big[\mathcal D \phi_a\big]\,
    e^{-S[\phi]+J^a\cdot \phi_a}.
\end{align}
Assuming that the path-integral measure $\big[\mathcal D\phi_a\big]$ is invariant under this transformation, i.e., that no anomaly is present, we obtain
\begin{align}
    &\int \big[\mathcal D \phi_a\big]
    \left[
    e^{-S[\phi']+J^a\cdot \phi_a'}
    -
    e^{-S[\phi]+J^a\cdot \phi_a}
    \right]_{\phi_a'=\phi_a+i\epsilon\,\delta_\epsilon\phi_a}
    \nonumber\\
    &=
    -i\epsilon
    \int \big[\mathcal D \phi_a\big]
    \left[
    \frac{\delta S}{\delta \phi_a}\cdot \delta_\epsilon\phi_a
    -
    J^a\cdot \delta_\epsilon\phi_a
    \right]
    e^{-S[\phi]+J^a\cdot \phi_a}
    +\mathcal O(\epsilon^2)
    \nonumber\\
    &=
    i\epsilon
    \Big[
    J^a\cdot \langle \delta_\epsilon \phi_a\rangle_J
    \Big]
    Z[J]
    +\mathcal O(\epsilon^2)
    \nonumber\\
    &=0,
    \label{eq:derivationOfWTIin1PI}
\end{align}
where
\begin{align}
    \langle O\rangle_J
    \equiv
    \frac{1}{Z[J,\mathcal M]}
    \int \big[\mathcal D\phi_a\big]\,
    O\,
    e^{-S[\phi,\mathcal M]+J^a\cdot \phi_a}.
\end{align}
Equation~\eqref{eq:derivationOfWTIin1PI} yields the WTI at the level of the generating functional:
\begin{align}
    J^a\cdot \langle \delta_\epsilon \phi_a\rangle_J = 0.
    \label{eq:WTIatGeneratingFunctionalLevel}
\end{align}

Defining the 1PI effective action by
\begin{align}
    \Gamma[\phi_{\rm cl}]
    =
    \sup_J \Big( J\cdot \phi_{\rm cl} - W[J] \Big),
    \label{eq:definning1PIaction}
\end{align}
we have
\begin{align}
    \frac{\delta W}{\delta J^a}
    =
    \langle \phi_a\rangle_J
    \equiv
    (\phi_{\rm cl})_a,
    \qquad
    \frac{\delta \Gamma}{\delta (\phi_{\rm cl})_a}
    =
    J^a.
\end{align}
Substituting $J_a=\delta\Gamma/\delta(\phi_{\rm cl})_a$ into \Cref{eq:WTIatGeneratingFunctionalLevel}, the WTI can be rewritten as
\begin{align}
    \frac{\delta \Gamma}{\delta (\phi_{\rm cl})_a}
    \cdot
    \delta_\epsilon (\phi_{\rm cl})_a
    =0,
    \label{eq:WTIfor1PIEA}
\end{align}
with $\delta_\epsilon (\phi_{\rm cl})_a = d_a^{\ b}(\phi_{\rm cl})_b$.

We now apply the above general WTI argument to the present LSM. First, consider the chiral limit. In this limit, the LSM action is invariant under the chiral $SU(3)_L\times SU(3)_R$ transformation
\begin{align}
    \Phi \rightarrow g_L \cdot \Phi \cdot g_R^\dagger,
    \label{eq:FieldTransformation}
\end{align}
where $g_{L/R}=\exp\!\big[i(\theta_{L/R})_\alpha T^\alpha\big]$ with $\alpha=1,\cdots,8$.
Under an $SU(3)$ axial rotation satisfying $(\theta_L)_\alpha = -(\theta_R)_\alpha = (\theta_A)_\alpha$, the fields $\sigma_a$ and $\pi_a$ transform infinitesimally as
\begin{align}
    \delta^\alpha \big(\sigma_a T^a\big)
    &=
    \{T^\alpha,\pi_a T^a\},
    \nonumber\\
    \delta^\alpha \big(\pi_a T^a\big)
    &=
    -\{T^\alpha,\sigma_a T^a\}.
\end{align}
This shows that $\sigma_a$ and $\pi_a$ transform in the adjoint representation of $SU(3)_A$, namely
\begin{align}
    \delta^\alpha \sigma_a
    =
    d^{\alpha b}_{\ \ a}\,\pi_b,
    \qquad
    \delta^\alpha \pi_a
    =
    -d^{\alpha b}_{\ \ a}\,\sigma_b,
    \label{eq:transLawOfFields}
\end{align}
where
\begin{align}
    d^{\alpha b}_{\ \ a}
    \equiv
    2\,\operatorname{tr}\!\Big[\{T^\alpha,T^b\}T_a\Big],
    \qquad
    d^{\alpha b}_{\ \ 0}
    =
    \sqrt{\frac{2}{3}}\,\delta^{\alpha b}.
\end{align}
Applying the general WTI in \Cref{eq:WTIfor1PIEA} to the 1PI effective action of the LSM, we obtain
\begin{align}
    0
    &=
    \frac{\delta \Gamma[\sigma_{\rm cl},\pi_{\rm cl}]}{\delta (\sigma_{\rm cl})_a}
    \cdot
    \delta^\alpha (\sigma_{\rm cl})_a
    +
    \frac{\delta \Gamma[\sigma_{\rm cl},\pi_{\rm cl}]}{\delta (\pi_{\rm cl})_a}
    \cdot
    \delta^\alpha (\pi_{\rm cl})_a
    \nonumber\\
    &=
    \frac{\delta \Gamma[\sigma_{\rm cl},\pi_{\rm cl}]}{\delta (\sigma_{\rm cl})_a}
    \cdot
    d^{\alpha b}_{\ \ a}(\pi_{\rm cl})_b
    -
    \frac{\delta \Gamma[\sigma_{\rm cl},\pi_{\rm cl}]}{\delta (\pi_{\rm cl})_a}
    \cdot
    d^{\alpha b}_{\ \ a}(\sigma_{\rm cl})_b.
    \label{eq:WTIofSU3ChiralLimit}
\end{align}
Performing one further functional derivative with respect to the external pseudoscalar background field $(\pi_{\rm cl}(y))_{b_1}$, we find
\begin{align}
    0 &= \frac{\delta}{\delta (\pi_{\rm cl}(y))_{b_1}}
    \biggl(
    \frac{\delta \Gamma[\sigma_{\rm cl},\pi_{\rm cl}]}{\delta (\sigma_{\rm cl})_a}
    \cdot
    d^{\alpha b}_{\ \ a}(\pi_{\rm cl})_b
    -
    \frac{\delta \Gamma[\sigma_{\rm cl},\pi_{\rm cl}]}{\delta (\pi_{\rm cl})_a}
    \cdot
    d^{\alpha b}_{\ \ a}(\sigma_{\rm cl})_b
    \biggr)
    \nonumber\\
    &=
    \frac{\delta \Gamma}{\delta (\sigma_{\rm cl})_a}
    \cdot
    \delta^{(4)}(x-y)\,d^{\alpha b_1}_{\ \ a}
    -
    \frac{\delta^2 \Gamma}{\delta (\pi_{\rm cl}(y))_{b_1}\,\delta (\pi_{\rm cl}(x))_a}
    \cdot
    d^{\alpha b}_{\ \ a}(\sigma_{\rm cl})_b.
    \label{WTI:LSM:chi}
\end{align}
Imposing the quantum equations of motion, $\delta\Gamma/\delta(\sigma_{\rm cl})_a=0$, at the background field values $\bar\Phi$, \Cref{WTI:LSM:chi} simplifies to
\begin{align}
   0 =  - \frac{\delta^2 \Gamma}{\delta (\pi_{\rm cl}(y))_{b_1}\,\delta (\pi_{\rm cl}(x))_a}
    \cdot
    d^{\alpha b}_{\ \ a}\,\bar\sigma_b
    =
    - \int_x
    \Big[\mathcal G_P^{-1}(y,x)\Big]^{b_1 a}
    d^{\alpha b}_{\ \ a}\,\bar\sigma_b,
    \label{2nd-deri}
\end{align}
where $\big(\mathcal G_P^{-1}(y,x)\big)^{b_1 a}$ denotes the full inverse propagator of the pseudoscalar mesons satisfying the gap equations in \Cref{eq:quantumEOMsinCJT}.

In the absence of external sources, the system is translationally invariant. One may therefore Fourier transform the inverse propagator in Euclidean momentum space as
\begin{align}
    \Big[\mathcal G_P^{-1}(y,x)\Big]^{b_1 a}
    =
    \int \frac{\df^4 k}{(2\pi)^4}
    \Big[\mathcal G_P^{-1}(k)\Big]^{b_1 a}
    e^{i(y-x)\cdot k}.
    \label{FT}
\end{align}
Within the Hartree approximation, as discussed in \Cref{sec:GapEquationsOfMesonPropagators} as well as in \Cref{app:CJTgapEquations}, the momentum dependence of $\mathcal P^{-1}(k)$ arises only from the canonical kinetic term:
\begin{align}
    \Big[\mathcal G_P^{-1}(k)\Big]^{b_1 a}
    =
    k^2\delta^{b_1 a}
    +
    \Big[M_P^2\Big]^{b_1 a}.
    \label{eq:parameterizationInWTIs}
\end{align}
Substituting \Cref{FT,eq:parameterizationInWTIs} into \Cref{2nd-deri}, the $x$-integration can be carried out explicitly:
\begin{align}
    - \int_x \Big[\mathcal G_P^{-1}(y,x)\Big]^{b_1 a}
    d^{\alpha b}_{\ \ a}\bar\sigma_b
    &=
    - \int \df^4 x \int \frac{\df^4 k}{(2\pi)^4}
    \Big[\mathcal G_P^{-1}(k)\Big]^{b_1 a}
    e^{i(y-x)\cdot k}
    d^{\alpha b}_{\ \ a}\bar\sigma_b
    \nonumber\\
    &=
    - \Big[\mathcal G_P^{-1}(0)\Big]^{b_1 a}
    d^{\alpha b}_{\ \ a}\bar\sigma_b
    \nonumber\\
    &=
    - \Big[M_P^2\Big]^{b_1 a}
    d^{\alpha b}_{\ \ a}\bar\sigma_b.
    \label{x-int-done}
\end{align}

Taking $b_1=1$ in \Cref{x-int-done}, the right-hand side becomes
\begin{align}
    - \Big[M_P^2\Big]^{1a}
    d^{\alpha b}_{\ \ a}
    \big(\bar\sigma_0\delta_{0b}+\bar\sigma_8\delta_{8b}\big).
\end{align}
Since the vectorial $SU(3)$ symmetry remains unbroken, the $SU(2)$ pion triplet does not mix with the kaon doublet or other channels, and hence
$\big[M_P^2\big]^{1a}=\big[M_P^2\big]^{11}\delta^{1a}$.
Using
$d^{\alpha 0}_{\ \ 1}=\sqrt{2/3}\,\delta^\alpha_1$
and
$d^{\alpha 8}_{\ \ 1}=1/\sqrt{3}\,\delta^\alpha_1$,
we obtain
\begin{align}
    - \Big[M_P^2\Big]^{11}
    d^{\alpha b}_{\ \ 1}
    \big(\bar\sigma_0\delta_{0b}+\bar\sigma_8\delta_{8b}\big)
    =
    - M_\pi^2
    \left(
    \sqrt{\frac{2}{3}}\,\bar\sigma_0
    +
    \frac{1}{\sqrt{3}}\,\bar\sigma_8
    \right)
    \delta^\alpha_1
    =
    -2M_\pi^2\bar\Phi_1\,\delta^\alpha_1.
    \label{eq:pionGMORLHS}
\end{align}

Similarly, taking $b_1=4$ in \Cref{x-int-done}, the right-hand side is
\begin{align}
    - \Big[M_P^2\Big]^{44}
    d^{\alpha b}_{\ \ 4}
    \big(\bar\sigma_0\delta_{0b}+\bar\sigma_8\delta_{8b}\big).
\end{align}
Using
$d^{\alpha 0}_{\ \ 4}=\sqrt{2/3}\,\delta^\alpha_4$
and
$d^{\alpha 8}_{\ \ 4}=-1/(2\sqrt{3})\,\delta^\alpha_4$,
we find
\begin{align}
    - \Big[M_P^2\Big]^{44}
    d^{\alpha b}_{\ \ 4}
    \big(\bar\sigma_0\delta_{0b}+\bar\sigma_8\delta_{8b}\big)
    =
    - M_K^2
    \left(
    \sqrt{\frac{2}{3}}\,\bar\sigma_0
    -
    \frac{1}{2\sqrt{3}}\,\bar\sigma_8
    \right)
    \delta^\alpha_4
    =
    - M_K^2
    \big(\bar\Phi_1+\bar\Phi_3\big)\delta^\alpha_4.
    \label{eq:kaonGMORLHS}
\end{align}
Setting $\alpha=1$ and $\alpha=4$, respectively, we therefore obtain the threshold relations for the pion and kaon masses:
\begin{align}
    M_\pi^2\bar\Phi_1 = 0,
    \qquad
    M_K^2\big(\bar\Phi_1+\bar\Phi_3\big)=0.
\end{align}
These relations reproduce the expected Nambu-Goldstone behavior when $\bar\Phi_1$ or $\bar\Phi_3$ acquires a nonzero value, i.e., when chiral symmetry is spontaneously broken.

%%%%%%%%%%%%%%%%%%%%%%%%%%%%%%%%%%%%%%%%%%%%%%%%%%%%%%%%%%%%%%%%%%%%%%%%%%%%%
\subsection{WTIs with explicitly broken $SU(3)$ chiral symmetry and GMOR relations}

In the present work, the $SU(3)$ chiral symmetry is explicitly broken by the source $\mathcal M$, and the WTI for the 1PI effective action in Eq.~\eqref{eq:WTIfor1PIEA} is accordingly modified.
Although the mesonic action $S_m$ is not invariant under the axial transformation in Eq.~\eqref{eq:FieldTransformation} with $\left( \theta_L\right)_\alpha = - \left( \theta_R \right)_\alpha$, it becomes formally invariant once the spurion field $\mathcal M$ is assigned the transformation law
\begin{align}
    \mathcal{M} \rightarrow g_L \cdot \mathcal{M} \cdot g_R^\dagger.
\end{align}
For the classical action with nonvanishing $\mathcal M$, one then has
\begin{align}
    \delta_\epsilon S[\phi]
    =
    \frac{\delta S}{\delta \phi_a} \cdot \delta_\epsilon \phi_a
    +
    \frac{\delta S}{\delta \mathcal{M}} \cdot \delta_\epsilon \mathcal{M}
    +
    \frac{\delta S}{\delta \mathcal{M}^\dagger} \cdot \delta_\epsilon \mathcal{M}^\dagger
    =
    0.
    \label{eq:symmetricActionUnderdeltaepsilonBroken}
\end{align}
At the level of the generating functional, \Cref{eq:derivationOfWTIin1PI} is modified to
\begin{align}
    &\int \left[ \mathcal{D} \phi_a \right]
    \left[
    e^{- S[\phi^\prime,\mathcal{M}] + J^a \cdot \phi_a^\prime}
    -
    e^{- S[\phi,\mathcal{M}] + J^a \cdot \phi_a}
    \right]_{\phi_a'=\phi_a+i\epsilon\,\delta_\epsilon\phi_a}
    \nonumber\\
    &=
    - i \epsilon \int \left[ \mathcal{D} \phi_a \right]
    \left[
    \frac{\delta S}{\delta \phi_a} \cdot \delta_\epsilon \phi_a
    -
    J^a \cdot  \delta_\epsilon \phi_a
    \right]
    e^{- S[\phi,\mathcal{M}] + J^a \cdot \phi_a}
    + \mathcal{O}(\epsilon^2)
    \nonumber\\
    &=
    i\epsilon \int \left[ \mathcal{D} \phi_a \right]
    \left[
    \frac{\delta S}{\delta \mathcal{M}} \cdot \delta_\epsilon \mathcal{M}
    +
    \frac{\delta S}{\delta \mathcal{M}^\dagger} \cdot \delta_\epsilon \mathcal{M}^\dagger
    +
    J^a \cdot  \delta_\epsilon \phi_a
    \right]
    e^{- S[\phi,\mathcal{M}] + J^a \cdot \phi_a}
    + \mathcal{O}(\epsilon^2)
    \nonumber\\
    &=
    i\epsilon
    \biggl[
    \biggl\langle \frac{\delta S}{\delta \mathcal{M}} \biggr\rangle_J \cdot \delta_\epsilon \mathcal{M}
    +
    \biggl\langle \frac{\delta S}{\delta \mathcal{M}^\dagger} \biggr\rangle_J \cdot \delta_\epsilon \mathcal{M}^\dagger
    +
    J^a \cdot \langle \delta_\epsilon \phi_a \rangle_J
    \biggr]
    Z[J,\mathcal{M}]
    + \mathcal{O}(\epsilon^2)
    \nonumber\\
    &= 0,
    \label{eq:derivationOfWTIAtGeneFuncLevelBroken}
\end{align}
which yields the broken WTI at the quantum level,
\begin{align}
    \biggl\langle \frac{\delta S}{\delta \mathcal{M}} \biggr\rangle_J \cdot \delta_\epsilon \mathcal{M}
    +
    \biggl\langle \frac{\delta S}{\delta \mathcal{M}^\dagger} \biggr\rangle_J \cdot \delta_\epsilon \mathcal{M}^\dagger
    +
    J^a \cdot \langle \delta_\epsilon \phi_a \rangle_J
    =
    0.
    \label{eq:WTIatQuantumLevel}
\end{align}
After introducing the 1PI effective action in Eq.~\eqref{eq:definning1PIaction}, the above identity can be rewritten as
\begin{align}
     \frac{\delta \Gamma}{\delta (\phi_{\rm cl})_a} \cdot \delta_\epsilon (\phi_{\rm cl})_a
     +
     \frac{\delta S}{\delta \mathcal{M}} \Biggl|_{\phi_a = (\phi_{\rm cl})_a + \Delta_{ab} \cdot \frac{\delta}{\delta (\phi_{\rm cl})_b}} \cdot \delta_\epsilon \mathcal{M}
     +
     \frac{\delta S}{\delta \mathcal{M}^\dagger} \Biggl|_{\phi_a = (\phi_{\rm cl})_a + \Delta_{ab} \cdot \frac{\delta}{\delta (\phi_{\rm cl})_b}} \cdot \delta_\epsilon \mathcal{M}^\dagger
     =
     0,
     \label{eq:BrokenWTIfor1PIEA}
\end{align}
where
\begin{align}
    \Delta_{ab}(x,y) = \frac{\delta^2 W}{\delta J_a(x)\,\delta J_b(y)}
\end{align}
is the full connected propagator.

With \Cref{eq:BrokenWTIfor1PIEA} at hand, we are ready to derive the GMOR relations at the 1PI level.
Since the first term in \Cref{eq:BrokenWTIfor1PIEA} has already been discussed in the previous subsection, we now focus on the explicit symmetry-breaking terms.
The infinitesimal transformation of $\mathcal M$ reads
\begin{align}
    \delta^\alpha \mathcal M_{mn} = \{ T^\alpha, \mathcal{M} \}_{mn}.
\end{align}
Decomposing the mass matrix as $\mathcal M = \mathcal M^s + i \mathcal M^p = \mathcal M^s_a T^a + i \mathcal M^p_a T^a$, the transformations of $\mathcal M^{s/p}_a$ are
\begin{align}
    \delta^\alpha \mathcal M^s_a = d^{\alpha b}_{\quad a} \mathcal M^p_b,
    \qquad
    \delta^\alpha \mathcal M^p_a = - d^{\alpha b}_{\quad a} \mathcal M^s_b.
\end{align}
Next, consider
\begin{align}
    \frac{\delta S_{qm}}{\delta \mathcal M_{mn}}
    =
    \frac{\delta S_m}{\delta \mathcal M_{mn}}
    =
    -c \big( \Phi^\dagger \big)_{nm}
    =
    -c \big( \sigma_a - i \pi_a \big) T^a_{nm}.
\end{align}
Replacing $\phi_a$ by $(\phi_{\rm cl})_a + \Delta_{ab} \cdot \frac{\delta}{\delta (\phi_{\rm cl})_b}$, we obtain
\begin{align}
    &\frac{\delta S_{qm}}{\delta \mathcal M_{mn}}
    \Biggl|_{\phi_a = (\phi_{\rm cl})_a + \Delta_{ab} \cdot \frac{\delta}{\delta (\phi_{\rm cl})_b}}
    \cdot \delta^\alpha \mathcal{M}_{mn}
    \nonumber\\
    &=
    -c \Big[ (\sigma_{\rm cl})_a - i (\pi_{\rm cl})_a \Big]
    \cdot
    \Big[  \mathcal M^p_{a_2} - i \mathcal M^s_{a_2} \Big]
    T^b_{nm} T^a_{nm} d^{\alpha a_2}_{\quad b} \, .
    \label{eq:BreakingTermOfWTIExplicit}
\end{align}
Taking one functional derivative with respect to the external pseudoscalar field $(\pi_{\rm cl}(y))_{b_1}$, this expression becomes
\begin{align}
    &\frac{\delta}{\delta (\pi_{\rm cl}(y))_{b_1}} \eqref{eq:BreakingTermOfWTIExplicit}
    \nonumber\\
    &=
    c \delta^{(4)}(x-y)\,\delta^{b_1}_a
    \cdot
    \Big[ \mathcal M^s_{a_2} + i \mathcal M^p_{a_2} \Big]
    T^b_{mn} T^a_{nm} d^{\alpha a_2}_{\quad b}
    \nonumber\\
    &=
    \frac{c}{2} \delta^{(4)}(x-y)\,\delta^{b_1}_a
    \cdot
    \Big[ \mathcal M^s_{a_2} + i \mathcal M^p_{a_2} \Big]
    \delta^{ab} d^{\alpha a_2}_{\quad b} \, .
    \label{eq:DerOfBreakingTermOfWTIExplicit}
\end{align}

Taking the background values $\bar \sigma_a T^a = \operatorname{diag}\{ \bar \Phi_1, \bar \Phi_1, \bar \Phi_3 \}$, $\mathcal M = \operatorname{diag}\{ m_l, m_l, m_s \}$, and $\bar \pi_a = 0$, we find that for $b_1 = \alpha = 1$,
\begin{align}
    \eqref{eq:DerOfBreakingTermOfWTIExplicit} = c m_l,
    \label{eq:pionGMORRHS}
\end{align}
whereas for $b_1 = \alpha = 4$,
\begin{align}
    \eqref{eq:DerOfBreakingTermOfWTIExplicit} = \frac{1}{2} \big( c m_l + c m_s \big),
    \label{eq:kaonGMORRHS}
\end{align}
after integrating over the spacetime coordinate $x$.
The Hermitian-conjugate term involving $\mathcal M^\dagger$ yields the same contribution.

Combining Eqs.~\eqref{eq:pionGMORLHS}, \eqref{eq:kaonGMORLHS}, \eqref{eq:pionGMORRHS}, and \eqref{eq:kaonGMORRHS}, we obtain the GMOR relations in the LSM:
\begin{align}
    M_\pi^2 \bar\Phi_1 &= c m_l \, , \nonumber\\
    M_K^2 \Big( \bar \Phi_1 + \bar \Phi_3 \Big) &= c m_l + c m_s \, .
    \label{eq:GMORrelations1PI}
\end{align}
In the 1PI formulation, these GMOR relations are in fact equivalent to the stationary conditions for $\bar\Phi_1$ and $\bar\Phi_3$.
As we discuss in the main text, this equivalence is no longer maintained in the 2PI formalism at vanishing local sources $J(x)$.

%%%%%%%%%%%%%%%%%%%%%%%%%%%%%%%%%%%%%%%%%%%%%%%%%%%%%%%%
\subsection{WTIs in 2PI formalism}

In this subsection of the appendix, we summarize the derivation of \Cref{eq:2PIWTIderivative}. 
We first restrict ourselves to the case without explicit symmetry breaking.

Consider the generating functional in the presence of a bilocal source $K^{ab}(x,y)$,
\begin{align}
    Z[J,K]
    =
    \int \big[ \mathcal{D} \phi_a \big]
    e^{-S[\phi] + J^a \cdot \phi_a + \frac{1}{2} \phi_a \cdot K^{ab} \cdot \phi_b}
    =
    e^{W[J,K]},
    \label{eq:bilocalGeneratingFunctional}
\end{align}
for which
\begin{align}
    \frac{\delta W[J,K]}{\delta J^a}
    =
    \langle \phi_a \rangle_{J,K}
    \equiv
    (\phi_{\rm cl})_a,
    \qquad
    \frac{\delta^2 W}{\delta J^a \delta J^b}
    =
    \langle \phi_a \phi_b \rangle_{J,K}
    -
    \langle \phi_a \rangle_{J,K}\langle \phi_b \rangle_{J,K}
    \equiv
    \Delta_{ab},
    \label{eq:FuncDerToWJK}
\end{align}
and
\begin{align}
     \frac{\delta W[J,K]}{\delta K^{ab}}
     =
     \frac{1}{2}
     \Big(
     \Delta_{ab}
     +
     (\phi_{\rm cl})_a (\phi_{\rm cl})_b
     \Big).
\end{align}
The 2PI effective action is obtained from the double Legendre transform,
\begin{align}
    \Gamma[\phi_{\rm cl}, \Delta]
    &=
    \sup_{J,K}
    \biggl[
    J^a \cdot (\phi_{\rm cl})_a
    +
    K^{ab} \cdot \frac{\delta W[J,K]}{\delta K^{ab}}
    -
    W[J,K]
    \biggr]
    \nonumber\\
    &=
    \sup_{J,K}
    \biggl[
    J^a \cdot (\phi_{\rm cl})_a
    +
    \frac{1}{2} K^{ab} \cdot
    \Big(
    \Delta_{ab}
    +
    (\phi_{\rm cl})_a (\phi_{\rm cl})_b
    \Big)
    -
    W[J,K]
    \biggr] \, .
\end{align}
Accordingly,
\begin{align}
    \frac{\delta \Gamma[\phi_{\rm cl}, \Delta]}{\delta (\phi_{\rm cl})_a}
    =
    J^a + K^{ab} \cdot (\phi_{\rm cl})_b,
    \qquad
    \frac{\delta \Gamma[\phi_{\rm cl}, \Delta]}{\delta \Delta_{ab}}
    =
    \frac{1}{2} K^{ab} \, .
    \label{eq:FuncDerToGammaJK}
\end{align}

Proceeding in the same way as in \Cref{eq:derivationOfWTIin1PI}, the symmetry variation in \Cref{eq:symmetricActionUnderdeltaepsilon}, together with the invariance of the generating functional $Z[J,K]$ in \Cref{eq:bilocalGeneratingFunctional}, leads to the WTIs at the level of $Z[J,K]$:
\begin{align}
    &\int \big[ \mathcal{D} \phi_a \big]
    \left[
    e^{-S[\phi^\prime] + J^a \cdot \phi_a^\prime + \frac{1}{2} \phi_a^\prime \cdot K^{ab} \cdot \phi_b^\prime}
    -
    e^{-S[\phi] + J^a \cdot \phi_a + \frac{1}{2} \phi_a \cdot K^{ab} \cdot \phi_b}
    \right]_{\phi_a'=\phi_a+i\epsilon\,\delta_\epsilon\phi_a}
    \nonumber\\
    &=
    -i\epsilon
    \int \big[ \mathcal{D} \phi_a \big]
    \left[
    \frac{\delta S}{\delta \phi_a} \cdot \delta_\epsilon \phi_a
    -
    J^a \cdot \delta_\epsilon \phi_a
    -
    \frac{1}{2} \phi_a \cdot K^{ab} \cdot \delta_\epsilon \phi_b
    -
    \frac{1}{2} \delta_\epsilon \phi_a \cdot K^{ab} \cdot \phi_b
    \right]
    e^{-S[\phi] + J^a \cdot \phi_a + \frac{1}{2} \phi_a \cdot K^{ab} \cdot \phi_b}
    \nonumber\\
    &\qquad
    + \mathcal{O}(\epsilon^2)
    \nonumber\\
    &=
    i \epsilon
    \biggl[
    J^a \cdot \langle \delta_\epsilon \phi_a \rangle
    +
    \frac{1}{2} K^{ab} \cdot \langle \phi_a \cdot \delta_\epsilon \phi_b \rangle
    +
    \frac{1}{2} K^{ab} \cdot \langle \delta_\epsilon \phi_a \cdot \phi_b \rangle
    \biggr]
    Z[J,K]
    + \mathcal{O}(\epsilon^2)
    \nonumber\\
    &=0.
    \label{K-WTI}
\end{align}
Using \Cref{eq:FuncDerToWJK,eq:FuncDerToGammaJK}, \Cref{K-WTI} can be rewritten as
\begin{align}
    \frac{\delta \Gamma[\phi_{\rm cl}, \Delta]}{\delta (\phi_{\rm cl})_a}
    \cdot d_a^{\,\, b} (\phi_{\rm cl})_b
    +
    \frac{\delta \Gamma[\phi_{\rm cl}, \Delta]}{\delta \Delta_{ab}}
    \cdot
    \Big(
    d_b^{\,\, c} \Delta_{ac}
    +
    d_a^{\,\, c} \Delta_{cb}
    \Big)
    =
    0.
    \label{eq:WTIat2PI}
\end{align}
This is the exact form of the WTIs in the 2PI formalism. 
Taking a further functional derivative of \Cref{eq:WTIat2PI} with respect to $(\phi_{\rm cl})_c$, we obtain
\begin{align}
    \frac{\delta^2 \Gamma[\phi_{\rm cl}, \Delta]}{\delta (\phi_{\rm cl})_c \delta (\phi_{\rm cl})_a}
    \cdot d_a^{\,\, b} (\phi_{\rm cl})_b
    +
    \frac{\delta \Gamma[\phi_{\rm cl}, \Delta]}{\delta (\phi_{\rm cl})_a}
    \cdot d_a^{\,\, c}
    +
    \frac{\delta^2 \Gamma[\phi_{\rm cl}, \Delta]}{\delta (\phi_{\rm cl})_c \delta \Delta_{ab}}
    \cdot
    \Big(
    d_b^{\,\, c} \Delta_{ac}
    +
    d_a^{\,\, c} \Delta_{cb}
    \Big)
    =
    0,
    \label{eq:2PIWTIderivativeApp}
\end{align}
which is the form quoted in \Cref{eq:2PIWTIderivative}.

In the presence of the explicit symmetry-breaking sources $\mathcal M$ and $\mathcal M^\dagger$, \Cref{K-WTI} is modified to
\begin{align}
    0
    =
    i \epsilon
    \biggl[
    \biggl\langle \frac{\delta S}{\delta \mathcal{M}} \biggr\rangle_J \cdot \delta_\epsilon \mathcal{M}
    +
    \biggl\langle \frac{\delta S}{\delta \mathcal{M}^\dagger} \biggr\rangle_J \cdot \delta_\epsilon \mathcal{M}^\dagger
    +
    J^a \cdot \langle \delta_\epsilon \phi_a \rangle
    +
    \frac{1}{2} K^{ab} \cdot \langle \phi_a \cdot \delta_\epsilon \phi_b \rangle
    +
    \frac{1}{2} K^{ab} \cdot \langle \delta_\epsilon \phi_a \cdot \phi_b \rangle
    \biggr]
    Z[J,K]
    +
    \mathcal{O}(\epsilon^2),
\end{align}
which leads to the modified 2PI WTI
\begin{align}
    \frac{\delta \Gamma[\phi, \Delta]}{\delta \phi_a} \cdot d_a^{\,\, b} \phi_b
    +
    \frac{\delta \Gamma[\phi, \Delta]}{\delta \Delta_{ab}}
    \cdot
    \Big(
    d_b^{\,\, c} \Delta_{ac}
    +
    d_a^{\,\, c} \Delta_{cb}
    \Big)
    =
    -\biggl[
    \biggl\langle \frac{\delta S}{\delta \mathcal{M}} \biggr\rangle_J \cdot \delta_\epsilon \mathcal{M}
    +
    \biggl\langle \frac{\delta S}{\delta \mathcal{M}^\dagger} \biggr\rangle_J \cdot \delta_\epsilon \mathcal{M}^\dagger
    \biggr],
    \label{eq:2PIWTIapp}
\end{align}
whose functional derivative with respect to $\phi_c$ yields \Cref{eq:2PIWTIderivative}.

%%%%%%%%%%%%%%%%%%%%%%%%%%%%%%%%%%%%%%%%%%%%%%%%%%%%%%%%%%%%%%%%%%%%%%
\section{High-temperature limit}
\label[appendix]{app:HighTemperatureLimit}

In this appendix, we summarize the formulations of the high-temperature limit of the equations of state.

We start with the discussion on the Stefan-Boltzmann (SB) limit.
To this end, we define the mass-to-temperature ratios of the fermions and the bosons as
\begin{align}
    r^\psi_T \equiv \frac{m_\psi}{T} \, , \quad r^\phi_T \equiv \frac{m_\phi}{T} \, ,
\end{align}
where $m_\psi$/$m_\phi$ denote the unified and dressed fermion/boson masses, respectively, since the mass spectrum is assumed to be degenerate in the SB limit, in the three-flavor quark-meson model.
At sufficiently high temperature, the ratios $r^\psi_T$ and $r^\phi_T$ are assumed to vanish in the SB limit
\begin{align}
    r^\psi_\infty = 0 \, , \quad r^\phi_\infty = 0 \, .
\end{align}
Up to the leading order of the high-$T$ expansion, the fermionic and bosonic one-loop integrals read
\begin{gather}
    \int_k \log \Big({\Delta}^\phi(k)\Big)^{-1}
    = 
    -2T\int \frac{\df^3 \bvec{k}}{(2 \pi)^3} \, \, \log \left[ 1 + n_B\left(\epsilon^{\phi}_{\bvec{k}}\right) \right]
    = 
    T^4 \biggl[ -\frac{\pi^2}{45} + \mathcal O (r^\phi_T) \biggl] \, , 
    \nonumber\\
    \Omega_{\bar q q}\Big|_{T} 
    = 
    \nu_c N_f T \int_0^\infty \frac{\df^3 \bvec{p}}{(2\pi)^3} \biggl\{ \log \Big[ 1 - n_{q}(T,\mu) \Big] + \log \Big[ 1 - n_{\bar q}(T,\mu) \Big] \biggl\}
    = 
    \nu_c N_f T^4 \biggl[ -\frac{7 \pi^2}{360} + \mathcal O (r^\psi_T) \biggl] \, ,
\end{gather}
where ${\Delta}^\phi(k)$ is the unified mesonic propagator.
At the one-loop level, the grand canonical potential reads
\begin{align}
    \Omega(T, \mu_B) \rightarrow \frac{N_\phi}{2} \int_k  \log \Big({\Delta}^\phi(k)\Big)^{-1} + \Omega_{\bar q q}\Big|_{T} = - T^4 \biggl[ N_\phi \frac{\pi^2}{90}+ N_c N_f \frac{7 \pi^2}{180} + \mathcal O (r^\psi_T)  + \mathcal O (r^\phi_T)\biggl] \, ,
\end{align}
where $N_\phi$ is the total number of the scalar and the pseudo-scalar mesons.
In the three-flavor quark-meson model, we take $N_c = 3$, $N_f = 3$, and $N_\phi = 2 N_f^2 = 18$.
Thus, the normalized pressure in the one-loop SB limit with fixed baryon chemical potential reads
\begin{align}
    \lim_{T\rightarrow\infty}\frac{P_{\rm SB}^{\rm 1-loop}}{T^4} = \frac{\pi^2}{5}+ \frac{7 \pi^2}{20} \simeq 5.4282 \, .
\end{align}

In the CJT formalism employed in the present formulation, the two-loop contribution is taken into account.
In the unified limit where the meson spectrum is degenerate, the mesonic Gaussian contraction term, as well as the 2PI contributions, read
\begin{align}
    &\frac{1}{2} \int_k \operatorname{tr} \Big[ \bar G_S^{-1}(k;\sigma) G_S(k) + \bar G_P^{-1}(k;\sigma) G_P(k) - 2 \cdot \textbf{1}_{\rm adj} \Big] + V_2
    \nonumber\\
    &\qquad \rightarrow \frac{N_\phi}{2} (m^2 - M^2) \int_k {\Delta}^\phi(k) + \frac{9\lambda_{\rm tot}}{2} \biggl[ \int_k {\Delta}^\phi(k) \biggl]^2
    \nonumber\\
    &\qquad = - \frac{9\lambda_{\rm tot}}{2} \biggl[ \int_k {\Delta}^\phi(k) \biggl]^2 \, ,
    \label{eq:2loopUnifiedLimit}
\end{align}
where $\lambda_{\rm tot} \equiv 12 \lambda_1 + 20 \lambda_2$, $M$ is the unified meson dynamical mass, and $m$ includes the tree-level mass with quark one-loop contribution.
In the last line of \Cref{eq:2loopUnifiedLimit}, we have imposed the gap equation.
Performing the high-$T$ expansion, the mesonic tadpole one-loop integral reads
\begin{align}
    \int_k {\Delta}^\phi(k) = \int \frac{\df^3 \bvec{k}}{(2 \pi)^3} \frac{n_B\left(\epsilon^{\phi}_{\bvec{k}}\right)}{\epsilon^{\phi}_{\bvec{k}}} = T^2 \biggl[ \frac{1}{12} + \mathcal O (r_T^\phi) \biggl] \, .
\end{align}
Thus, the normalized pressure in the two-loop SB limit with fixed baryon chemical potential reads
\begin{align}
    \lim_{T\rightarrow\infty}\frac{P_{\rm SB}^{\rm 2-loop}}{T^4} = \frac{\pi^2}{5}+ \frac{7 \pi^2}{20} + \frac{9}{2} \lambda_{\rm tot} \left( \frac{1}{12} \right)^2 \simeq 31.2895 \, .
\end{align}

We note here, however, that the SB limit for the mesons does not work well because the mass-temperature ratio remains sufficiently large in the high-temperature limit
\begin{align}
    r^\phi_\infty > 1,
\end{align}
such that the high-$T$ expansion exceeds the convergence radius.
Thus, in the current analysis for the CJT formalism, we perform the high-$T$ estimation at finite $r^\phi_\infty$.
For the latter convenience, we define
\begin{gather}
    I_\phi(r_T^\phi) 
    \equiv 
    \frac{1}{T^2}\int_k {\Delta}^\phi(k) 
    = 
    \frac{1}{2 \pi^2}\int \df x \, \frac{x^2}{\sqrt{x^2 +  \big( r_T^\phi \big)^2}} \frac{1}{\exp \Big\{ \sqrt{x^2 +  \big( r_T^\phi \big)^2} \Big\} - 1 } \, ,
    \nonumber\\
    I_\psi(r_T^\psi) 
    \equiv  
    \int \frac{\df^3 \bvec{p}}{(2 \pi)^3} \frac{n_{q}(T,0)}{E_q}
    = 
    \frac{1}{2 \pi^2}\int \df y \, \frac{y^2}{\sqrt{y^2 +  \big( r_T^\psi \big)^2}} \frac{1}{\exp \Big\{ \sqrt{y^2 +  \big( r_T^\psi \big)^2} \Big\} + 1 } \, ,
\end{gather}
where $x \equiv |\bvec{k}|/T$ and $y \equiv |\bvec{p}|/T$, respectively.

Consider the unified gap equation
\begin{align}
    M^2 = m^2 + \lambda_{\rm tot} \int_k {\Delta}^\phi(k) \, ,
    \label{eq:unifiedGapEquation}
\end{align}
where $m^2 = m_{\rm tree}^2 + m_\psi^2$, and 
\begin{align}
    m_\psi^2 = \frac{\partial^2 \Omega_{\bar q q}}{\partial \phi \partial \phi} \Biggl|_{\Phi = \bar\Phi} = \nu_c N_f \frac{g^2}{3}  \int \frac{\df^3 \bvec{p}}{(2 \pi)^3} \frac{1}{2 E_q}\Big[ n_{q}(T,\mu) + n_{\bar q}(T,\mu) \Big] + \mathcal O (\bar \Phi^2)\, .
\end{align}
Dividing $T^2$ on both sides of \Cref{eq:unifiedGapEquation} and taking the $T \to \infty$ limit, we obtain
\begin{align}
    \big( r_\infty^\phi \big)^2 = m_\psi^2 \Big|_{r_\infty^\psi \to 0} + \lambda_{\rm tot} I_\phi(r_\infty^\phi) = \nu_c N_f \frac{g^2}{3}I_\psi(0) + \lambda_{\rm tot} I_\phi(r_\infty^\phi) \, ,
    \label{eq:HighTempGapEquation}
\end{align}
where in the last equality we have used the fact that
\begin{align}
    \lim_{T \to \infty}\frac{\bar \Phi}{T} = \lim_{T \to \infty}\frac{\mu}{T} = 0 \, .
\end{align}
In \Cref{eq:HighTempGapEquation}, $I_\psi(0)$ is the fermionic tadpole one-loop integral at vanishing quark mass and $T \to \infty$ limit, and has the value $I_\psi(0) = 1/24$.
For the efficiency of solving $r_\infty^\phi$ numerically, we employ the identity
\begin{align}
    I_\phi(r_T^\phi) &= \frac{1}{2 \pi^2}\int \df x \, \frac{x^2}{\sqrt{x^2 +  \big( r_T^\phi \big)^2}} \frac{\exp \Big\{ -\sqrt{x^2 +  \big( r_T^\phi \big)^2} \Big\}}{1 - \exp \Big\{ - \sqrt{x^2 +  \big( r_T^\phi \big)^2} \Big\}  }
    \nonumber\\
    &= \frac{1}{2 \pi^2}\int \df x \, \frac{x^2}{\sqrt{x^2 +  \big( r_T^\phi \big)^2}} \sum_{\ell = 1}^{\infty} \exp \Big\{ - \ell \sqrt{x^2 +  \big( r_T^\phi \big)^2} \Big\}
    \nonumber\\
    &= \frac{1}{2 \pi^2} \sum_{\ell = 1}^{\infty} \frac{r_T^\phi}{\ell} K_1(\ell \, r_T^\phi) \, ,
\end{align}
where $K_\alpha(x)$ is the modified Bessel function of the second kind.
We checked that this series converges sufficiently up to $\ell = 20$.
Insert this form into \Cref{eq:HighTempGapEquation}, we arrive at
\begin{align}
    \big( r_\infty^\phi \big)^2 = \frac{N_c N_f g^2}{36} + \frac{\lambda_{\rm tot}}{2 \pi^2} \sum_{\ell = 1}^{\infty} \frac{r_\infty^\phi}{\ell} K_1(\ell \, r_\infty^\phi) \, .
\end{align}
For the parameter setup of $m[f_0(500)] = 800$ MeV, the numerical solution yields
\begin{align}
    r_\infty^\phi \simeq 3.6674 \, ,
\end{align}
for $N_c = 3$ and $N_f = 3$.
We note here that this numerical value is model-dependent, but the strategy is rather universal for the self-energy resummed prescriptions.

Once the $r_\infty^\phi$ is fixed, the normalized pressure is straightforwardly evaluated.
The mesonic one-loop yields
\begin{align}
    \lim_{T \to \infty}\frac{1}{T^4}\int_k \log \Big({\Delta}^\phi(k)\Big)^{-1} = \frac{1}{\pi^2} \int\df x \, \log\Big[ 1 - \exp\Big\{ x^2 + \big( r_\infty^\phi \big)^2 \Big\} \Big] \, ,
\end{align}
and thus the normalized pressure reads
\begin{align}
    \lim_{T \to \infty}\frac{P_{\rm High-T}}{T^4} = \frac{7 \pi^2}{20} - \frac{N_\phi}{2\pi^2} \int\df x \, \log\Big[ 1 - \exp\Big\{ x^2 + \big( r_\infty^\phi \big)^2 \Big\} \Big] + \frac{9}{2}\lambda_{\rm tot} \Big[ I_\phi(r_\infty^\phi) \Big]^2 \simeq 3.8207\, .
\end{align}

For the symmetry-improved prescriptions, the external sources are suppressed as the temperature increases. 
Thus, all pressure prescriptions in the SICJT formalism converge to the same high-$T$ asymptotics as in the CJT formalism.

Similarly, the High-$T$ limit for the normalized entropy density, the baryon chemical potential, and the energy density reads
\begin{gather}
    \lim_{T\to \infty}\frac{n_{B,{\rm High-T}}}{T^3} = 0 \, ,
    \nonumber\\
    \lim_{T\to \infty}\frac{s_{\rm High-T}}{T^3} = 4 \lim_{T \to \infty}\frac{P_{\rm High-T}}{T^4} \simeq 15.2829 \, ,
    \nonumber\\
    \lim_{T\to \infty}\frac{\varepsilon_{\rm High-T}}{T^4} = 3 \lim_{T \to \infty}\frac{P_{\rm High-T}}{T^4} \simeq 11.4622 \, .
\end{gather}
%

%%%%%%%%%%%%%%%%%%%%%%%%%%%%%%%%%%%%%%%%%%%%%%%%%%%%%%%%%%%%%%%%%%%%%%
\section{Results on thermodynamic quantities}
\label[appendix]{app:ResultsOnEquationsOfState}

In this appendix, we collect additional results for the thermodynamic quantities associated with the different pressure prescriptions introduced in \Cref{sec:PressureAndThermodynamicAmbiguity}. 
The baryon number density, entropy density, and energy density are shown in normalized form, with their temperature power. 
The derivatives are evaluated using the fixed-source prescription discussed in \Cref{sec:ThermodynamicObservablesOnTheSIManifold}. 
These plots are included for completeness and provide consistency checks for the equations of state used in the main text.

%%%%%%%%%%%%%%
%%%%%%%%%%%%%%
\begin{figure*}
    \centering
    \subfloat[]{
        \centering
        \includegraphics[width=0.32\linewidth]{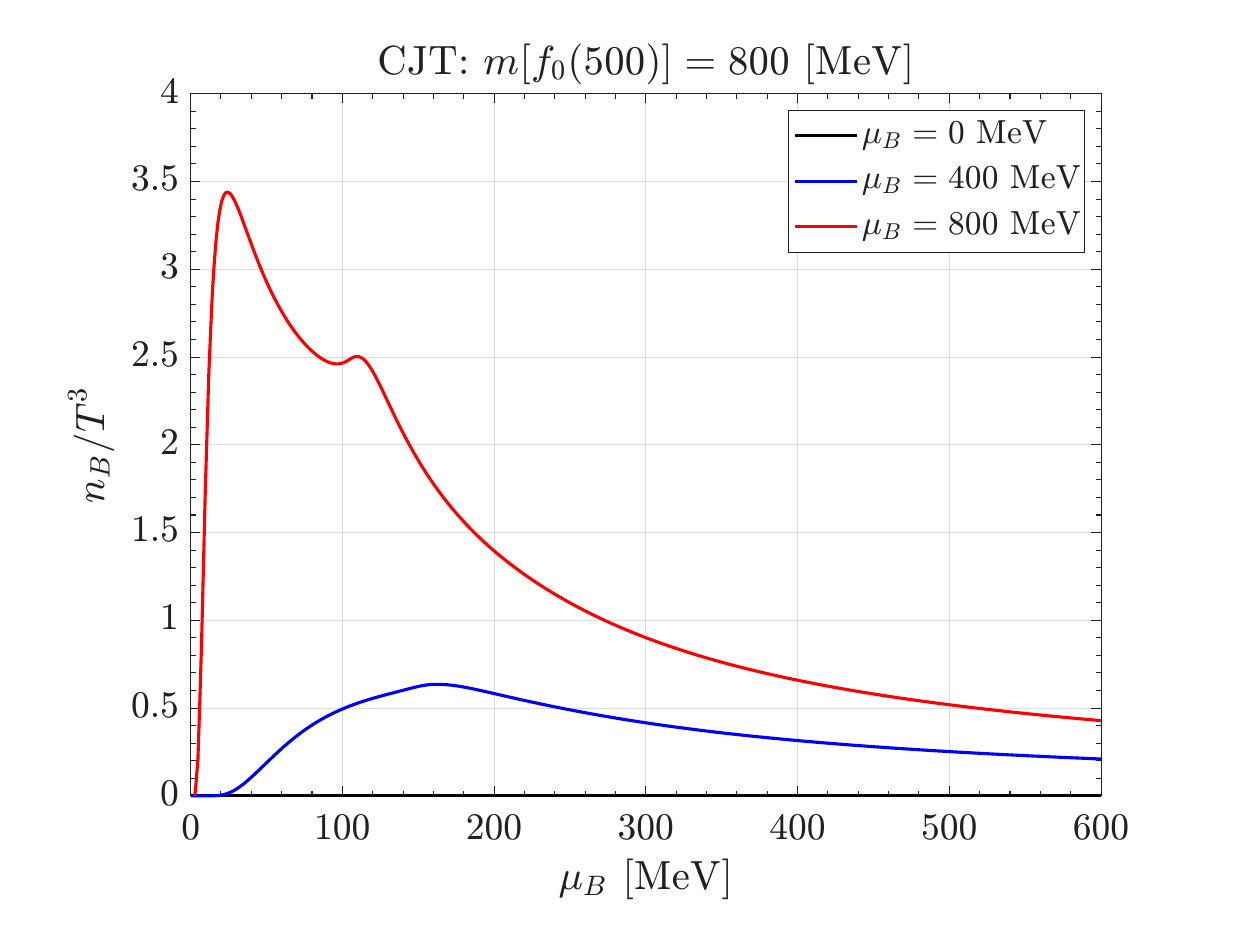}
        \includegraphics[width=0.32\linewidth]{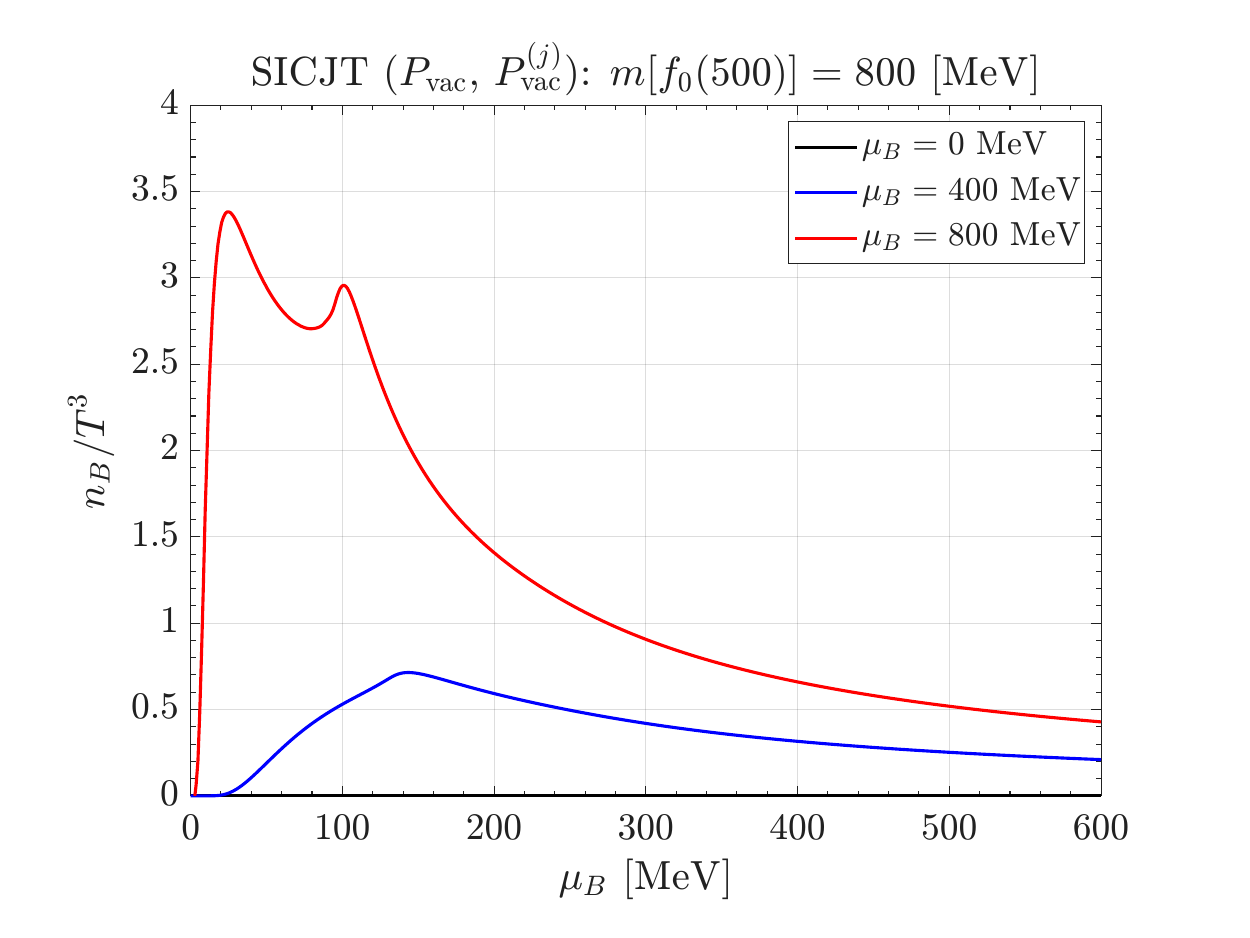}
        \includegraphics[width=0.32\linewidth]{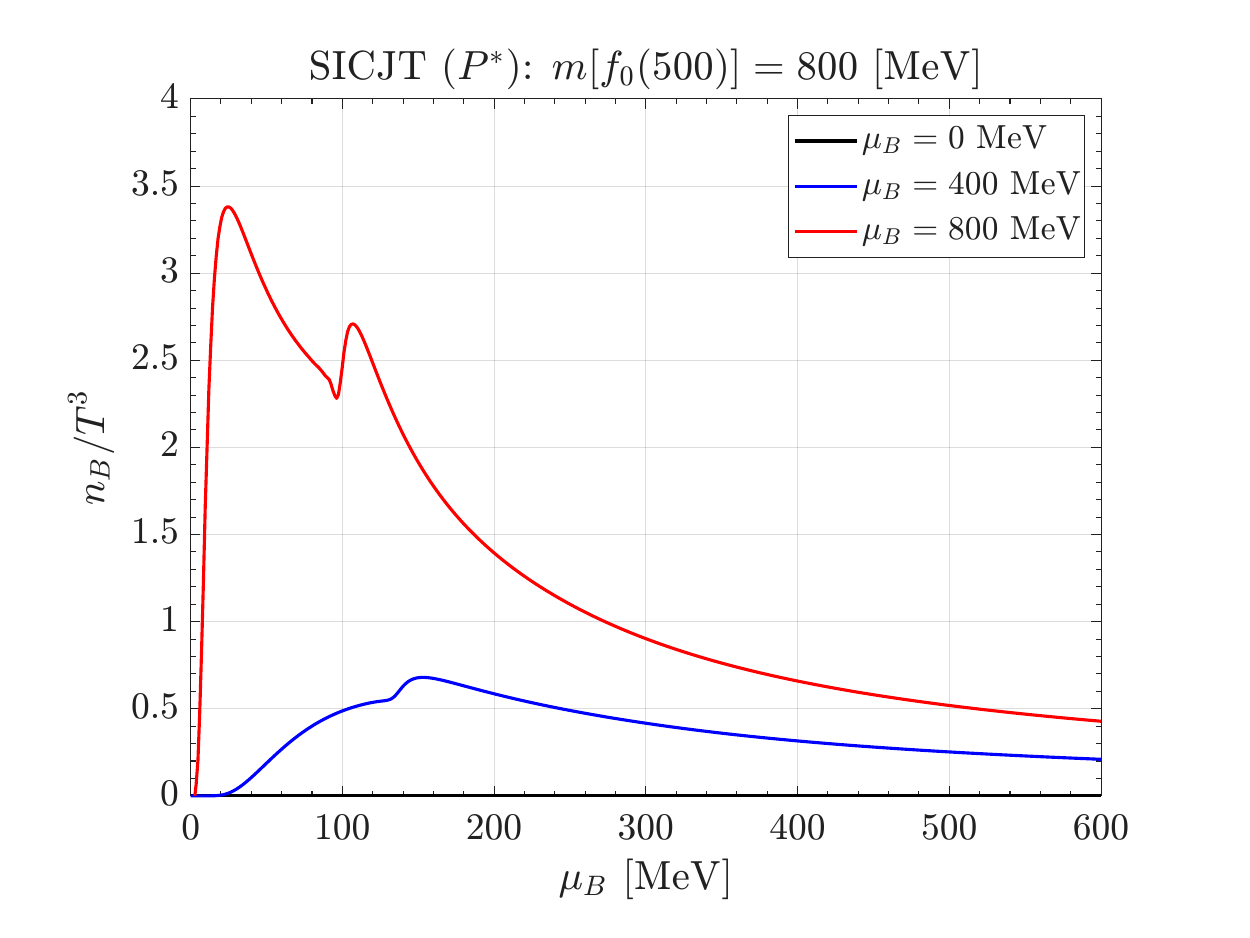}
        \label{fig:BaryonDensity}
    }
    \\
    \subfloat[]{
        \centering
        \includegraphics[width=0.32\linewidth]{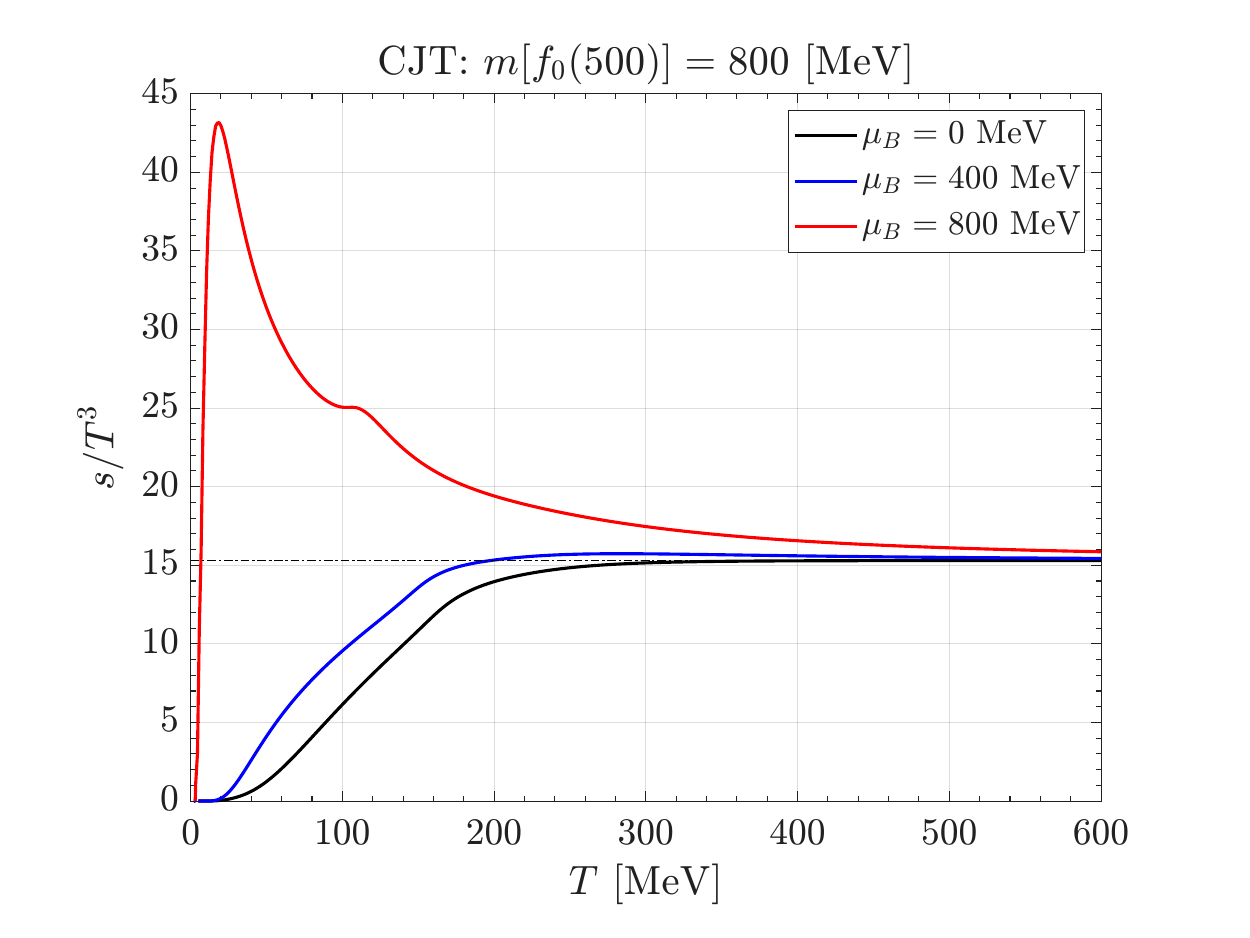}
        \includegraphics[width=0.32\linewidth]{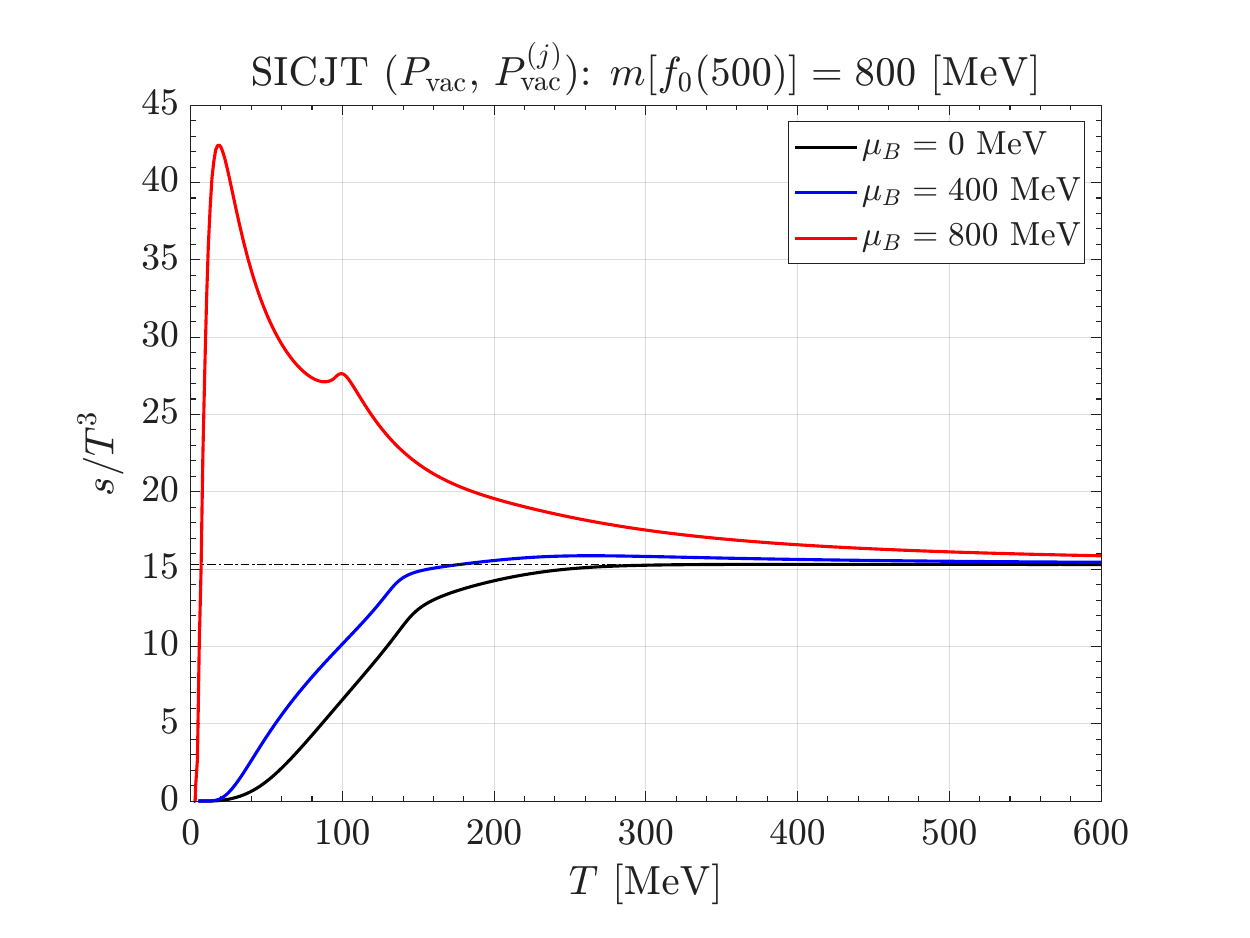}
        \includegraphics[width=0.32\linewidth]{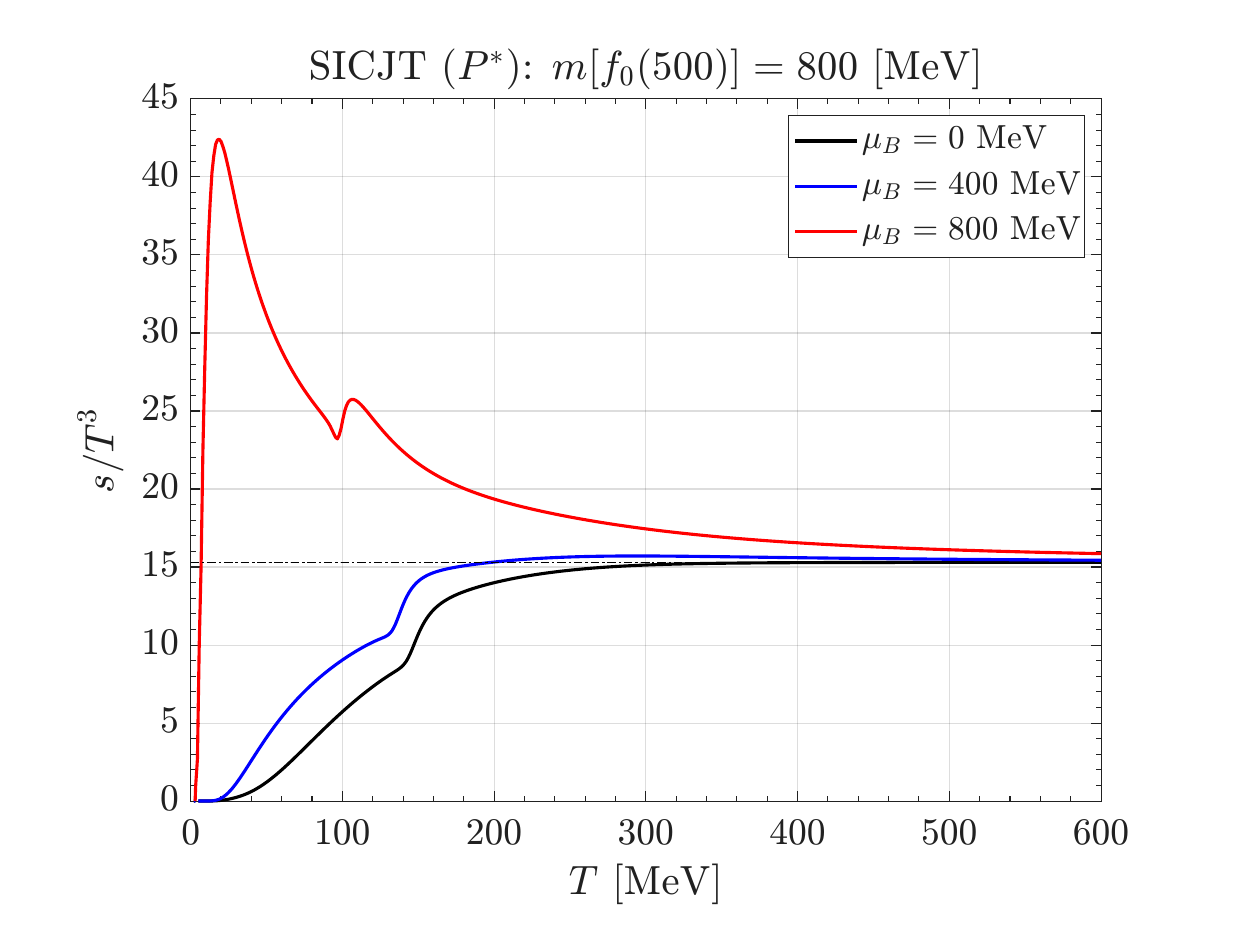}
        \label{fig:EntropyDensity}
    }
    \\
    \subfloat[]{
        \begin{minipage}{1\textwidth}
            \centering
            \includegraphics[width=0.4\linewidth]{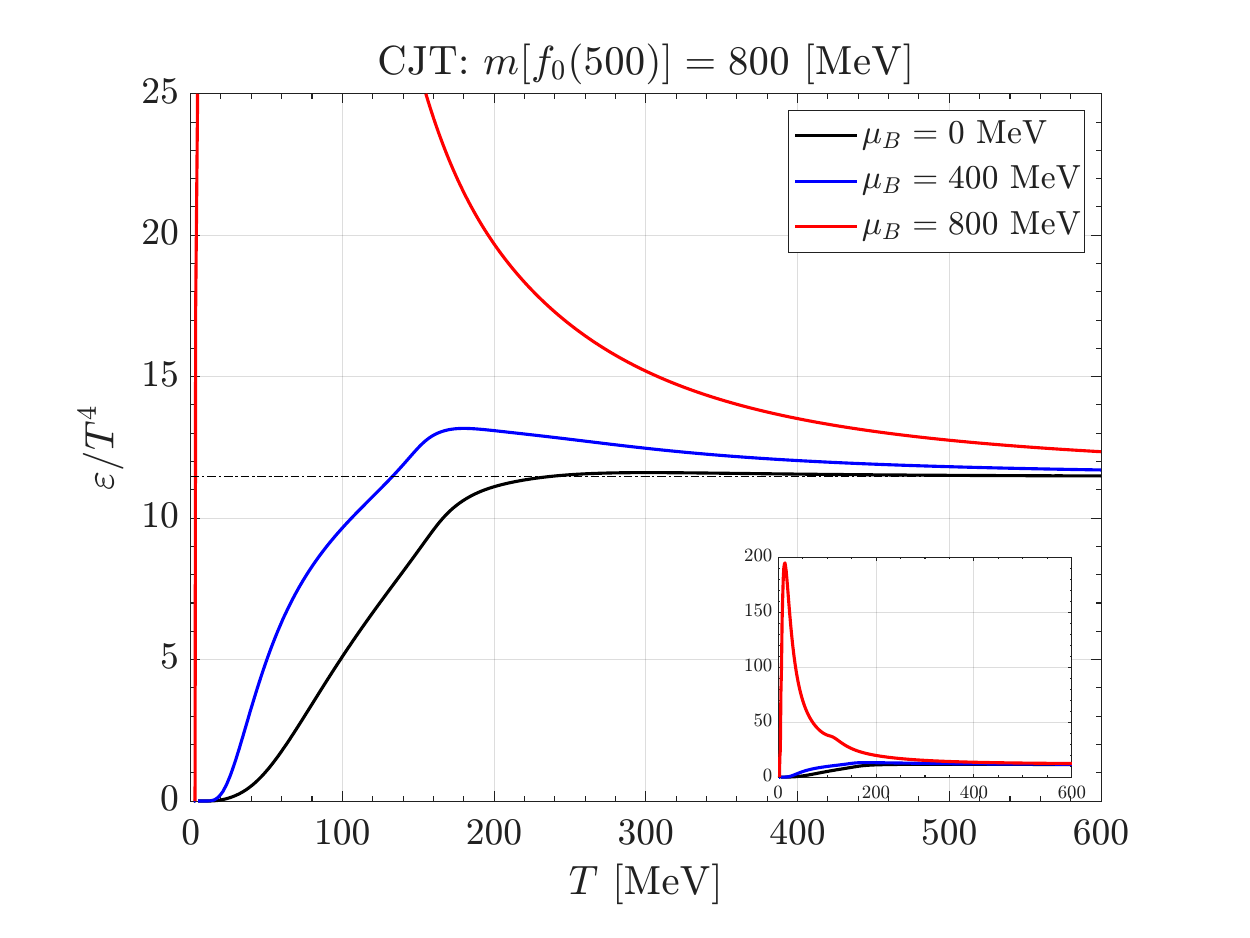}
            \includegraphics[width=0.4\linewidth]{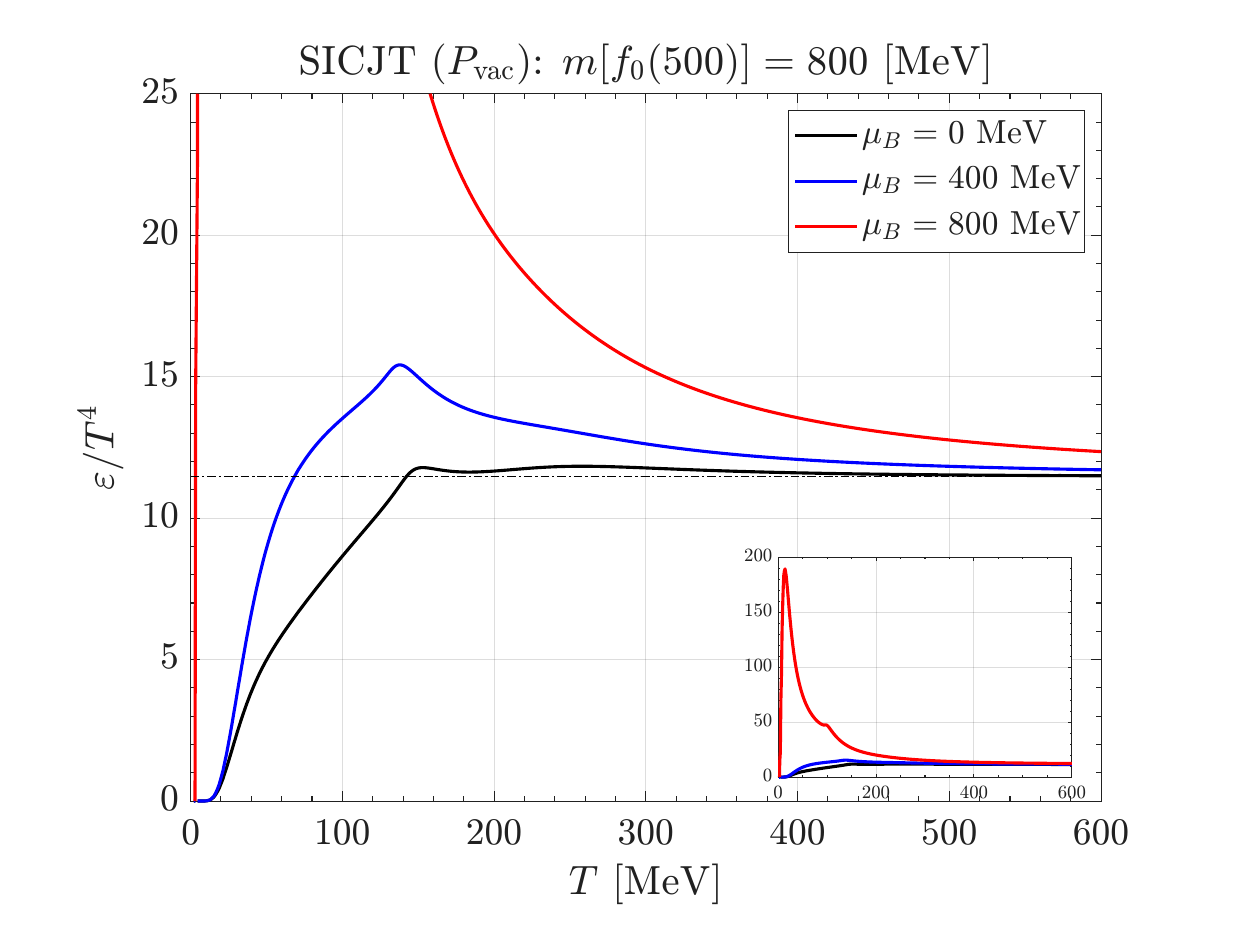}
            \includegraphics[width=0.4\linewidth]{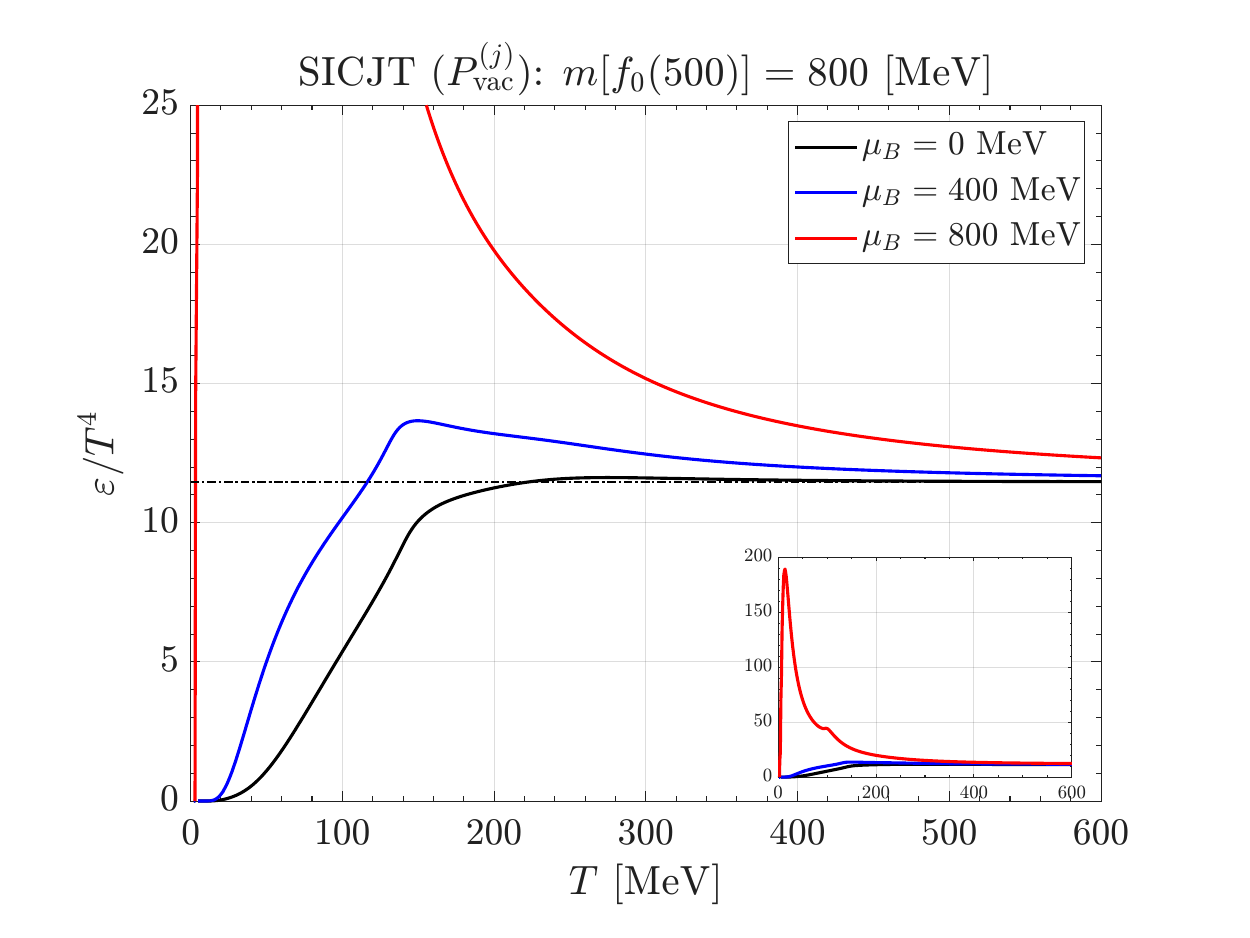}
            \includegraphics[width=0.4\linewidth]{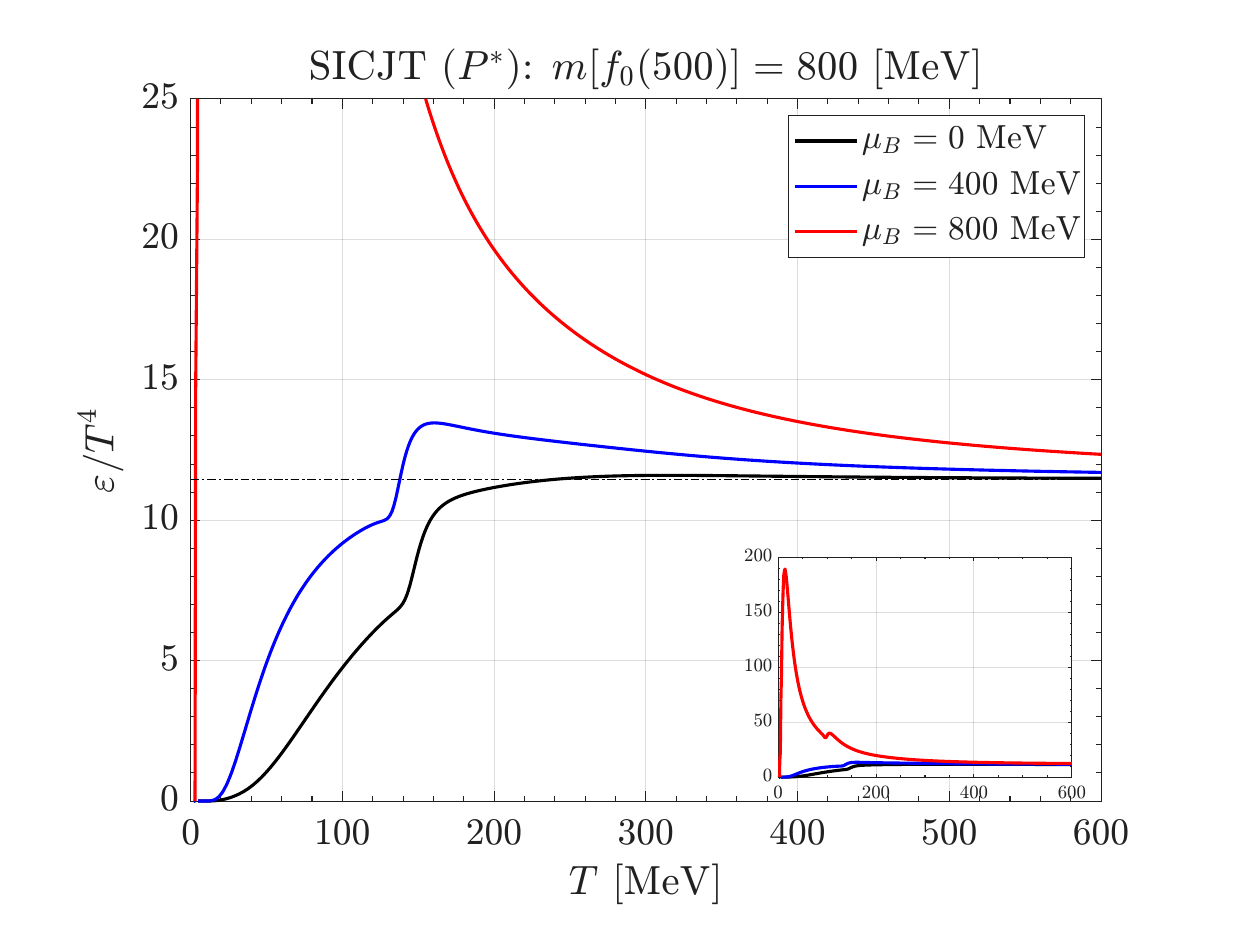}
        \end{minipage}
        \label{fig:EnergyDensity}
    }
    \caption{
    Normalized thermodynamic quantities obtained from the CJT formalism and from the different pressure prescriptions of the SICJT formalism: 
    (a) the baryon number density $n_B/T^3$; 
    (b) the entropy density $s/T^3$; 
    (c) the energy density $\epsilon/T^4$.
    }
\end{figure*}
%%%%%%%%%%%%%%
%%%%%%%%%%%%%%

\newpage 

%%%%%%%%%%%%%%%%%%%%%%%%%%%%%%%%%%%%%%%%%%%%%%%%%%%%%%%%%%%%%%%%%%%%%%
\section{Comparisons to other quantitative approaches}
\label[appendix]{app:ComparisonsToOtherQuantitativeApproaches}

In this appendix, we place the numerical results of the CJT and SICJT formalisms in the context of representative mean-field, functional
renormalization-group, and lattice QCD calculations. 
They provide an external benchmark for the location and shape of the chiral crossover, and help separate the intrinsic thermodynamic ambiguity associated with the symmetry-improving sources from differences caused by the microscopic degrees of freedom and the approximation scheme. 
Since the absolute transition scales and high-temperature limits are model dependent, we use dimensionless normalizations whenever appropriate.
% In particular, temperatures are divided by the pseudocritical temperature of the corresponding approach, while the
% thermodynamic observables discussed in \Cref{app:EOSCompare} are also presented with their respective model-specific asymptotic normalizations.

For all CJT and SICJT curves shown in this appendix, we employ the
$m_\sigma=800\,{\rm MeV}$ parameter set used in the main numerical analysis.
The external mean-field, fRG, and lattice curves retain the parameter sets and scale-setting prescriptions of their respective source calculations.
Accordingly, the comparison is not intended as a common-parameter benchmark;
normalizing by each approach's own $T_{\rm pc}$ isolates the crossover shape from the sizable differences in the absolute transition scale.

%%%%%%%%%%%%%%%%%%%%%%%%%%%%%%%%%%%%%%%%%%%%%%%%%%%%%%%%%%%%%%%%%%%%%%%%%%%%%
\subsection{Chiral condensates and pseudo-critical temperature at vanishing baryon chemical potential}
\label[appendix]{app:chiralCondensatesCompare}

%%%%%%%%%%%%%%%%%%%
%%%%%%%%%%%%%%%%%%%
\begin{table*}[t]
    \centering
    \caption{
    Comparison of pseudocritical temperatures at vanishing baryon
    chemical potential.
    Entries belonging to the same broad method but employing different
    numerical setups are listed separately.
    The pseudocritical temperatures identify characteristic inflection
    points or susceptibility peaks of chiral observables.
    The QM-MF value $\simeq139.5\,{\rm MeV}$ is obtained from the
    inflection point of the standard-MFA condensate curve digitized from
    Ref.~\cite{Mitter:2013fxa}; it is not quoted explicitly in that
    reference.
    }
    \label{tab:chiralTpcComparison}
    \begin{tabular}{lcc}
        \hline
        \hline
        Approach and setup
        & $T_{\rm pc}\,[{\rm MeV}]$
        & Criterion
        \\
        \hline
        CJT (this work)
        & 164.4
        & inflection of $\Delta_{l,s}$
        \\
        SICJT (this work)
        & 143.3
        & inflection of $\Delta_{l,s}$
        \\
        $N_f=2+1$ QM-MF~\cite{Mitter:2013fxa}
        & $\simeq139.5$
        & inflection of $\sigma_x$
        \\
        $N_f=2+1$ QM-MF
        ($m_\pi=220\,{\rm MeV}$)~\cite{Schaefer:2009ui}
        & 164
        & inflection of $\sigma_x$
        \\
        $N_f=2+1$ PQM-MF
        (log; $m_\pi=220\,{\rm MeV}$)~\cite{Schaefer:2009ui}
        & 215
        & inflection of $\sigma_x$
        \\
        $N_f=2+1$ PQM-FRG~\cite{Herbst:2013ail}
        & 172
        & $\max[-\partial_T\Delta_{l,s}]$
        \\
        $N_f=2+1$ PQM-FRG~\cite{Fu:2018qsk}
        & 176.5
        & inflection of $\Delta_{LS}$
        \\
        Lattice QCD (WB)~\cite{Borsanyi:2010bp}
        & $157(3)(3)$
        & inflection of $\Delta_{l,s}$
        \\
        Lattice QCD
        (HotQCD continuum)~\cite{HotQCD:2018pds}
        & $156.5(1.5)$
        & average over five chiral criteria
        \\
        \hline
        \hline
    \end{tabular}
\end{table*}
%%%%%%%%%%%%%%%%%%%
%%%%%%%%%%%%%%%%%%%

To compare the chiral order parameters between the effective models and
lattice QCD, we employ the reduced light--strange condensate
\begin{align}
    \Delta_{l,s}(T)=
    \frac{\langle\bar l l\rangle_T-(m_l/m_s)\langle\bar s s\rangle_T}
    {\langle\bar l l\rangle_0-(m_l/m_s)\langle\bar s s\rangle_0}.
    \label{eq:DeltaLSComparison}
\end{align}
For the present three-flavor linear sigma model, the corresponding
combination is constructed as
\begin{align}
    \Delta_{l,s}^{\rm CJT/SICJT}(T)=
    \frac{\bar\Phi_1(T)-(c m_l/c m_s)\bar\Phi_3(T)}
    {\bar\Phi_1(0)-(c m_l/c m_s)\bar\Phi_3(0)}.
    \label{eq:DeltaLSCJT}
\end{align}
This definition removes the overall normalization of the condensates and suppresses the additive mass-dependent contribution, making it a suitable quantity for comparison with the PQM and lattice results \cite{Schaefer:2009ui, Herbst:2013ail, Borsanyi:2010bp}.

For the comparison in this appendix, the CJT and SICJT pseudocritical
temperatures are determined using the same reduced-condensate criterion as the Wuppertal-Budapest (WB) lattice result,
\begin{align}
    T_{\rm pc}^{(\Delta)}
    =
    \mathop{\rm arg\,max}_T
    \left[-\frac{\partial\Delta_{l,s}(T)}{\partial T}\right],
    \label{eq:TpcDeltaCriterion}
\end{align}
which is equivalent to the inflection point of $\Delta_{l,s}(T)$. This gives
\begin{align}
    T_{\rm pc,CJT}^{(\Delta)}=164.365\,{\rm MeV},
    \qquad
    T_{\rm pc,SICJT}^{(\Delta)}=143.309\,{\rm MeV}.
    \label{eq:TpcDeltaCjtSicjt}
\end{align}
For comparison, defining the crossover through the inflection point of the light condensate $\bar\Phi_1(T)$ gives $T_{\rm pc,CJT}^{(\bar\Phi_1)}=164.394\,{\rm MeV}$ and $T_{\rm pc,SICJT}^{(\bar\Phi_1)}=143.298\,{\rm MeV}$. 
Changing the definition therefore shifts the CJT value by $-0.030\,{\rm MeV}$ and the SICJT value by $+0.011\,{\rm MeV}$. These differences are numerically negligible at the precision used in \Cref{tab:chiralTpcComparison}, so the rounded entries remain $164.4\,{\rm MeV}$ and $143.3\,{\rm MeV}$, respectively. 
The small shifts arise because the strange condensate varies more slowly across the light crossover and enters \eqref{eq:DeltaLSCJT} with the small coefficient $c m_l/c m_s=0.03268315$.

The mean-field calculations determine their pseudocritical temperatures from the inflection point of a nonstrange chiral order parameter. 
The PQM-fRG result uses the maximum of $-\partial_T\Delta_{l,s}$, while the lattice value associated with the points shown below is obtained from the inflection point of $\Delta_{l,s}$. Hence, the present CJT/SICJT, PQM-fRG, and WB entries now use the same reduced-condensate criterion, whereas the mean-field entries use the closely related nonstrange-condensate criterion.

%%%%%%%%%%%%%%%%%%%
%%%%%%%%%%%%%%%%%%%
\begin{figure}[t]
    \centering
    \includegraphics[width=0.72\textwidth]{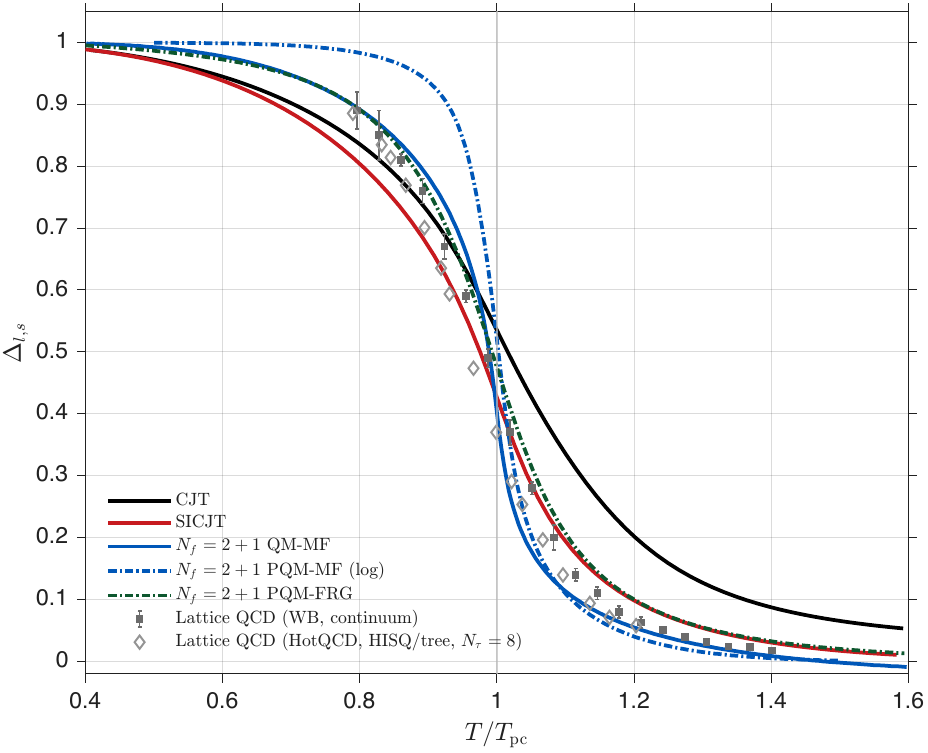}
    \caption{
    Reduced light-strange chiral condensate $\Delta_{l,s}(T)$, normalized to its zero-temperature value, as a function of $T/T_{\rm pc}$, where $T_{\rm pc}$ is determined separately within each approach. 
    The black and red solid curves denote, respectively, the conventional CJT and SICJT calculations of the present work.
    The blue solid curve shows the $N_f=2+1$ quark-meson model in the mean-field approximation (QM-MF)~\cite{Mitter:2013fxa}, while the blue dash-dotted curve denotes the $N_f=2+1$ Polyakov-quark-meson model in the mean-field approximation with the logarithmic Polyakov-loop potential (PQM-MF (log))~\cite{Schaefer:2009ui}. 
    The dark-green dash-dotted curve corresponds to the $N_f=2+1$ Polyakov-quark-meson model treated with the functional
    renormalization group (PQM-FRG)~\cite{Herbst:2013ail}. 
    The gray points with error bars show the lattice QCD result of the
    WB Collaboration~\cite{Borsanyi:2010bp}.
    The gray diamonds represent the HotQCD HISQ/tree results at $N_\tau=8$ lattices~\cite{Bazavov:2011nk}.
    }
    \label{fig:chiralCondensateComparison}
\end{figure}
%%%%%%%%%%%%%%%%%%%
%%%%%%%%%%%%%%%%%%%

Figure~\ref{fig:chiralCondensateComparison} compares the reduced condensates as functions of $T/T_{\rm pc}$, with each curve normalized by its own pseudocritical temperature. 
After rescaling by the respective crossover temperatures, all approaches display the same qualitative interpolation between the chirally broken and approximately restored regimes. 
The remaining differences quantify the width and asymmetry of the crossover. 
In particular, the mean-field curves tend to be steeper, whereas mesonic fluctuations and the lattice result produce a smoother crossover. 
The CJT and SICJT curves should therefore be compared primarily in terms of their crossover width and shape, rather than through an
unqualified point-by-point identification with full QCD.

The source-specific pseudocritical temperatures are collected in \Cref{tab:chiralTpcComparison}. 
We list the condensate and equation-of-state benchmark setups separately whenever they differ. 
This distinction is necessary for the QM-MF and PQM-fRG comparisons: the references providing the condensate curves are not the same numerical setups as those used for the thermodynamic benchmark curves. 
The spread among the effective-model entries reflects differences in particle content, vacuum parameterization, and fluctuation treatment. 
The WB value $157(3)(3)\,{\rm MeV}$ is retained for the lattice points in \Cref{fig:chiralCondensateComparison}, while the more recent continuum HotQCD analysis gives $T_{\rm pc}=156.5\pm1.5\,{\rm MeV}$ from several consistent chiral criteria~\cite{HotQCD:2018pds}.

%%%%%%%%%%%%%%%%%%%%%%%%%%%%%%%%%%%%%%%%%%%%%%%%%%%%%%%%%%%%%%%%%%%%%%%%%%%%%
\subsection{Thermodynamic quantities at vanishing baryon chemical potential}
\label[appendix]{app:EOSCompare}

\begin{figure*}[t]
    \centering
    \subfloat[]{
        \includegraphics[width=0.48\linewidth]
        {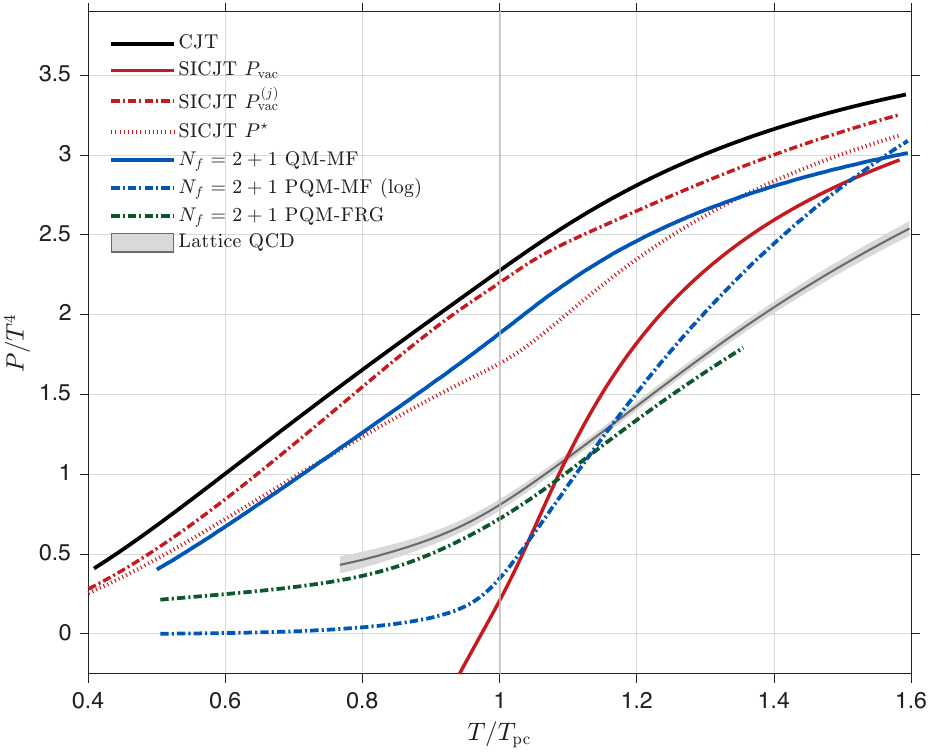}
        \label{fig:EOSComparisonPressure}
    }
    \hfill
    \subfloat[]{
        \includegraphics[width=0.48\linewidth]
        {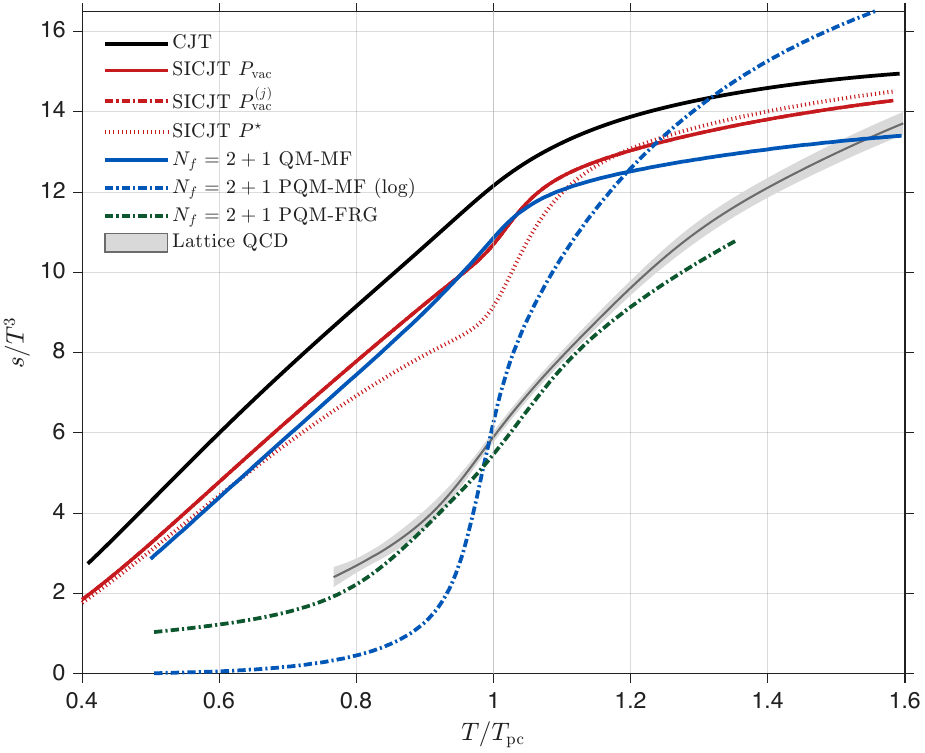}
        \label{fig:EOSComparisonEntropy}
    }
    \\
    \subfloat[]{
        \includegraphics[width=0.48\linewidth]
        {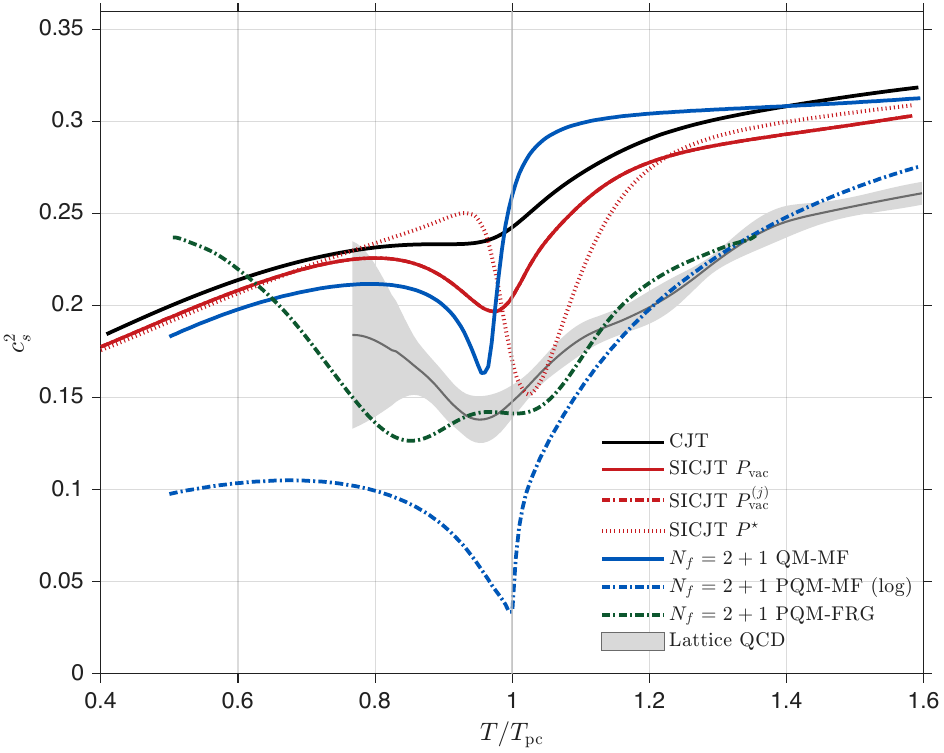}
        \label{fig:EOSComparisonSoundVelocity}
    }
    \hfill
    \subfloat[]{
        \includegraphics[width=0.48\linewidth]
        {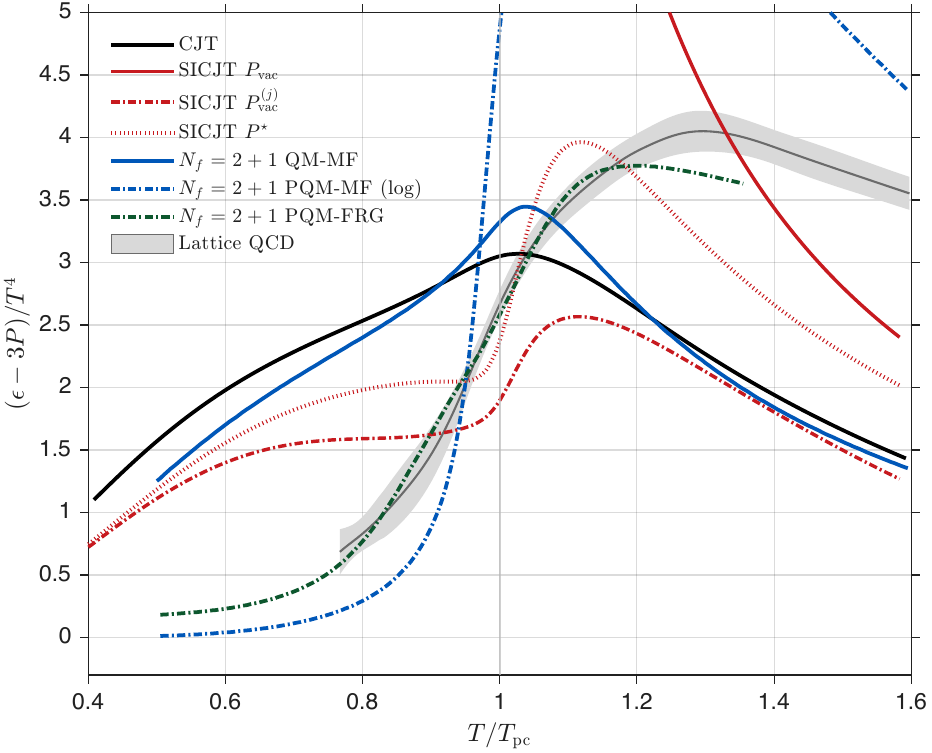}
        \label{fig:EOSComparisonTraceAnomaly}
    }
    \caption{
    Thermodynamic observables at $\mu_B=0$ as functions of
    $T/T_{\rm pc}$:
    (a) pressure $P/T^4$,
    (b) entropy density $s/T^3$,
    (c) squared sound velocity $c_s^2$, and
    (d) trace anomaly $(\varepsilon-3P)/T^4$.
    The pseudocritical temperature is determined separately within each
    approach. The effective-model and lattice benchmarks are taken from
    Refs.~\cite{Schaefer:2009ui,Fu:2018qsk,Borsanyi:2013bia,HotQCD:2014kol}.
    }
    \label{fig:EOSComparisonDimensionless}
\end{figure*}

\begin{figure}[t]
    \centering
    \subfloat[]{
        \includegraphics[width=0.9\linewidth]
        {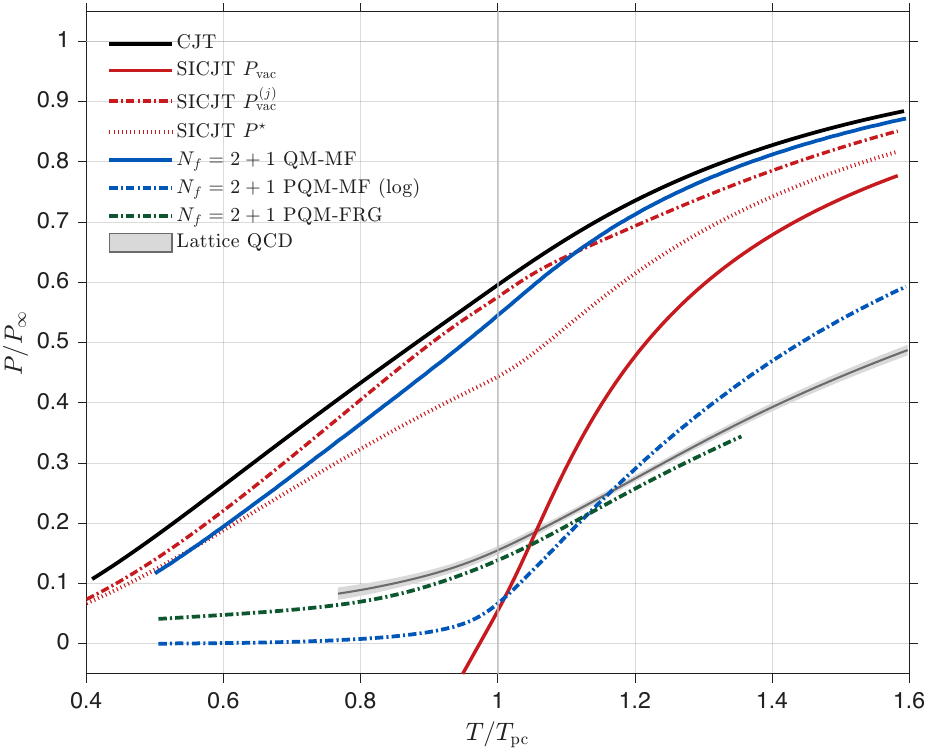}
        \label{fig:EOSComparisonPressureNormalized}
    }
    \hfill
    \subfloat[]{
        \includegraphics[width=0.9\linewidth]
        {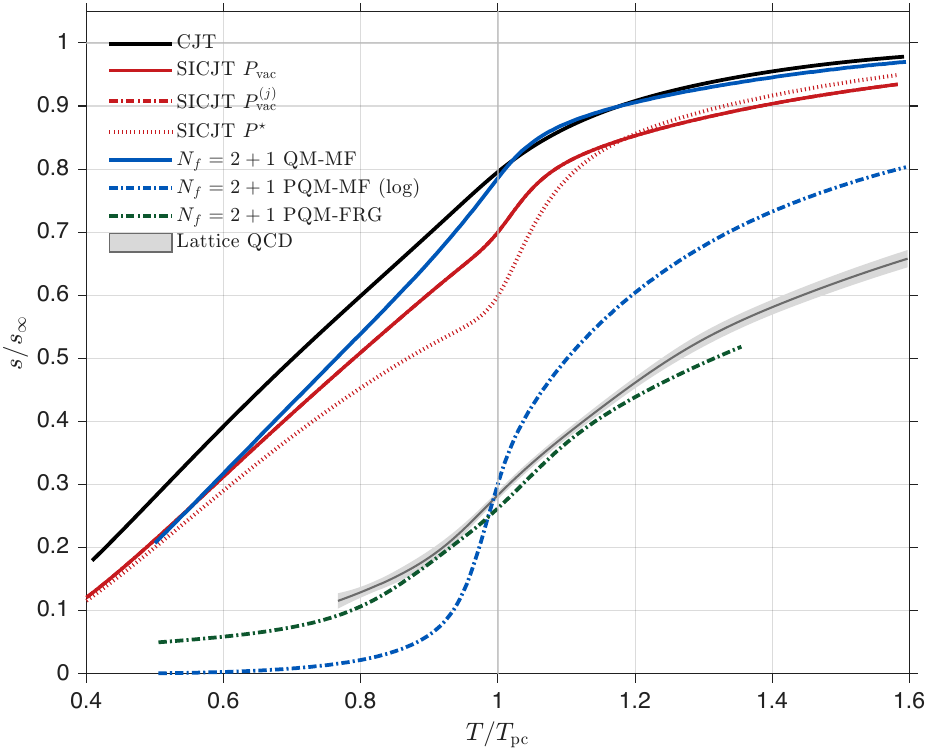}
        \label{fig:EOSComparisonEntropyNormalized}
    }
    \hfill
    \subfloat[]{
        \includegraphics[width=0.9\linewidth]
        {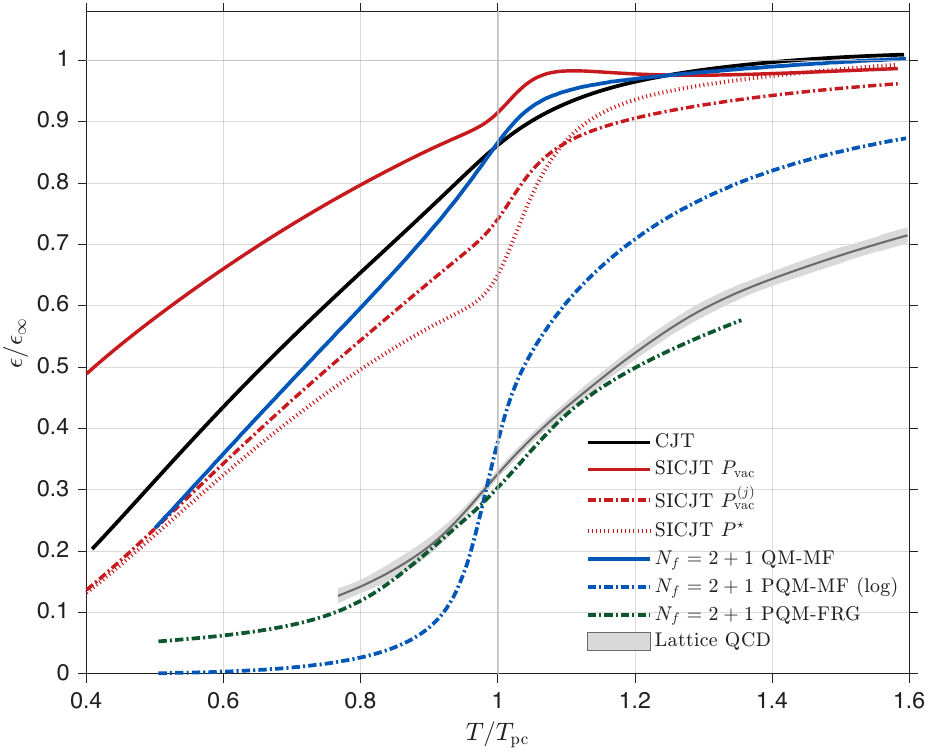}
        \label{fig:EOSComparisonEnergyNormalized}
    }
    \caption{
    Thermodynamic observables at $\mu_B=0$, normalized by the corresponding
    model-specific high-temperature limits:
    (a) pressure,
    (b) entropy density, and
    (c) energy density.
    Each curve is shown as a function of its own $T/T_{\rm pc}$.
    }
    \label{fig:EOSComparisonNormalized}
\end{figure}

We compare the CJT and SICJT equations of state at $\mu_B=0$ with representative three-flavor QM-MF and PQM-MF~\cite{Schaefer:2009ui}, PQM-FRG~\cite{Fu:2018qsk}, and lattice-QCD\cite{Borsanyi:2013bia,HotQCD:2014kol} results in \Cref{fig:EOSComparisonDimensionless,fig:EOSComparisonNormalized}.
The temperature is normalized by the pseudocritical temperature determined within each approach. The external results retain their original particle content, parameterization, and scale setting; the comparison therefore probes the crossover shape rather than a common-parameter equation of state.

%%%%%%%%%%%%%%%%%%%
%%%%%%%%%%%%%%%%%%%

The absolute normalizations exhibit a sizable model dependence above the crossover, reflecting the different microscopic degrees of freedom and high-temperature limits. 
Nevertheless, all approaches show the characteristic softening of $c_s^2$ and enhancement of the trace anomaly in the transition region. 
Their depths, widths, and peak positions remain approximation
dependent.

Within the SICJT calculation, $P_{\rm vac}^{(j)}$ and $P^\star$ remain close to the CJT baseline over most of the temperature range. The conventional vacuum-subtracted pressure $P_{\rm vac}$ shows a substantially larger deviation, most prominently in the pressure and trace anomaly. 
This behavior is expected because an additive source-dependent pressure shift enters $\varepsilon-3P$ directly, whereas the source-matched and pulled-back prescriptions remove most of this contribution.

Normalizing each observable by its own asymptotic value substantially reduces the spread caused by the different high-temperature normalizations. 
The CJT, $P_{\rm vac}^{(j)}$, and $P^\star$ curves then display a similar crossover evolution, while the remaining differences relative to the functional and lattice results mainly characterize the width and asymmetry of the transition.
The distinct behavior of $P_{\rm vac}$ confirms that its conventional vacuum subtraction is not well adapted to normalization-sensitive observables in the source-based SICJT formulation.

%%%%%%%%%%%%%%%%%%%%%%%%%%%%%%%%%%%%%%%%%%%%%%%%%%%%%%%%%%%%%%%%%%%%%%%%%%%%%
\subsection{Isentropic trajectories}
\label[appendix]{app:IsenCompare}

\begin{figure}[t]
    \centering
    \subfloat[]{
        \includegraphics[width=0.9\linewidth]
        {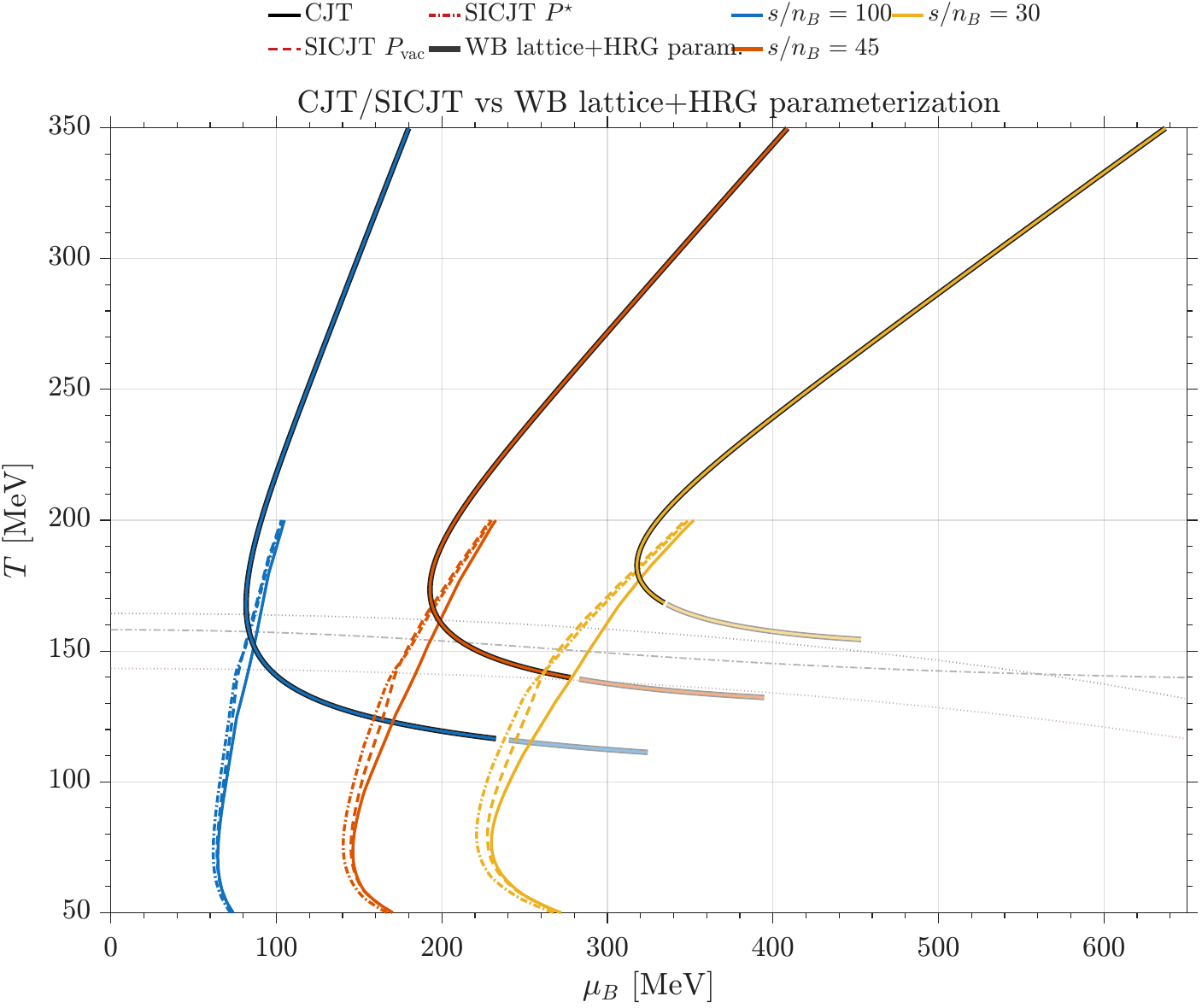}
        \label{fig:IsentropicComparisonLattice}
    }
    \hfill
    \subfloat[]{
        \includegraphics[width=0.9\linewidth]
        {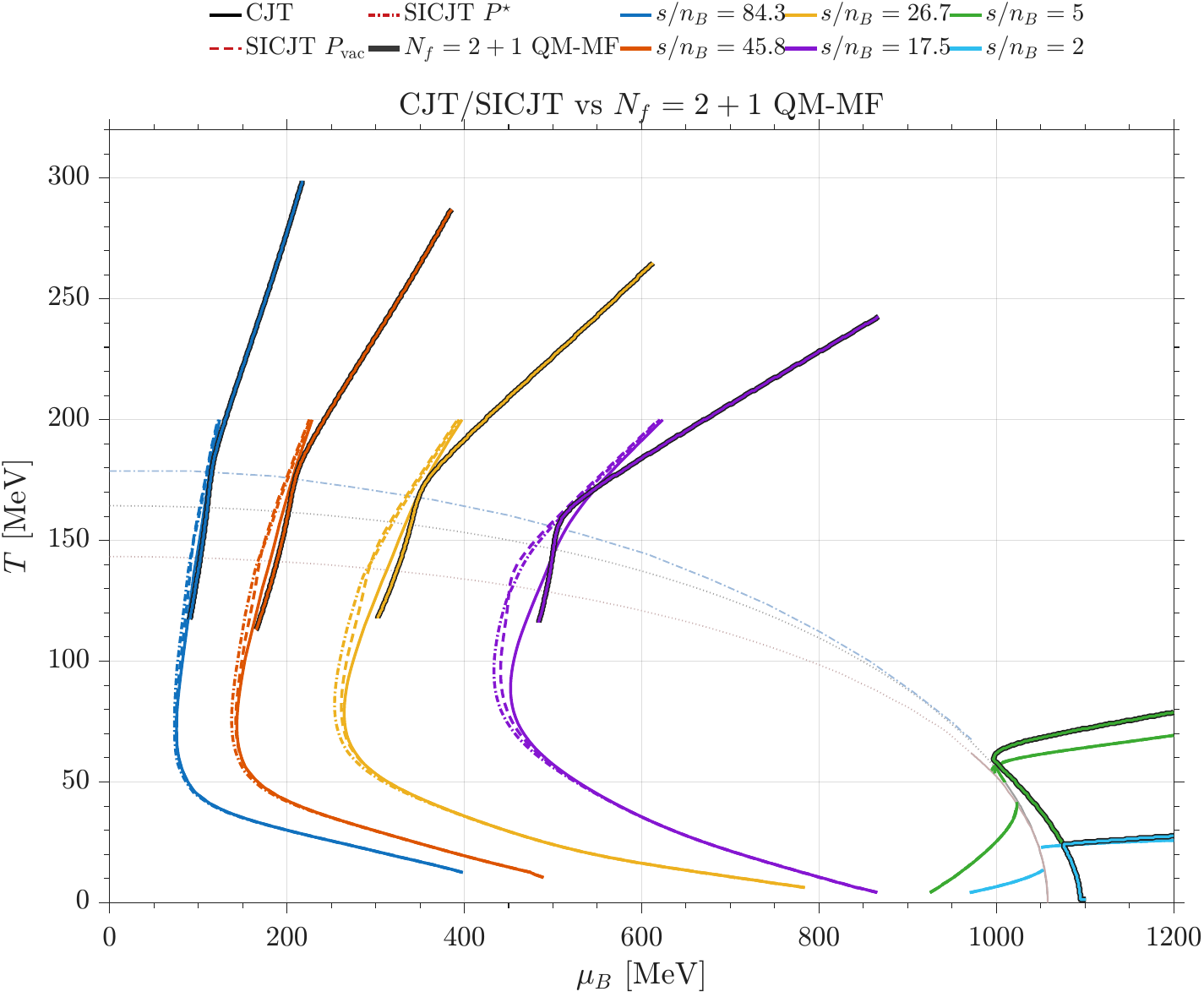}
        \label{fig:IsentropicComparisonQM}
    }
    \hfill
    \subfloat[]{
        \includegraphics[width=0.9\linewidth]
        {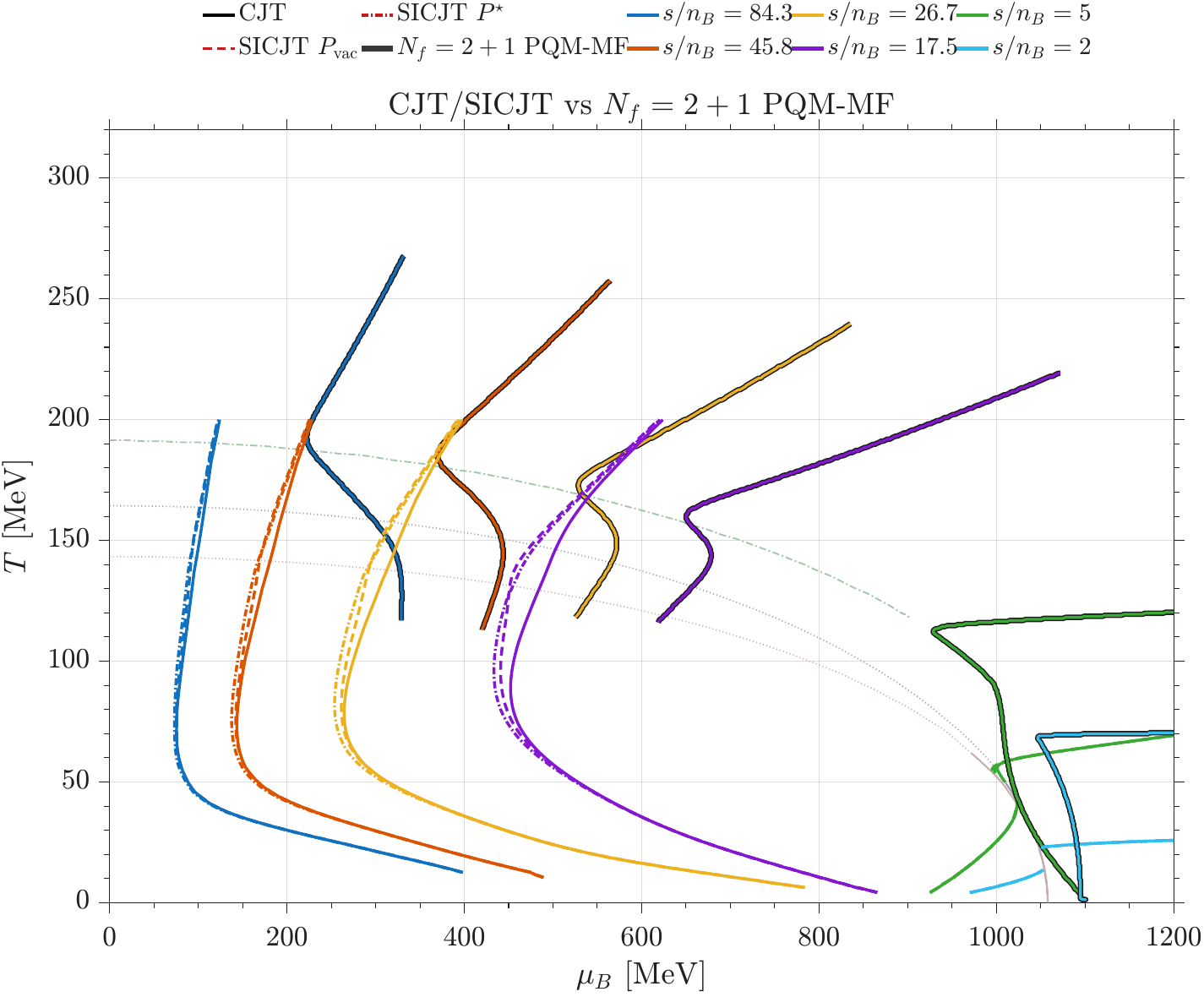}
        \label{fig:IsentropicComparisonPQM}
    }
    \caption{
    Isentropic trajectories in the physical $(\mu_B,T)$ plane, in comparison with
    (a)the full-baryonic lattice+HRG parameterization~\cite{Borsanyi:2012cr};
    (b)the $N_f=2+1$ QM-MF~\cite{Motta:2020cbr} trajectories;
    (c)the $N_f=2+1$ PQM-MF~\cite{Motta:2020cbr}.
    Colors identify $s/n_B$, while line styles distinguish the different formalisms and pressure prescriptions.
    The individual phase boundaries are presented as thin-dotted or -dashed lines.
    }
    \label{fig:IsentropicExternalComparison}
\end{figure}

Figure~\ref{fig:IsentropicExternalComparison} compares the CJT and SICJT isentropic trajectories with lattice+HRG~\cite{Borsanyi:2012cr}, QM-MF as well as the PQM-MF~\cite{Motta:2020cbr} results directly in the physical $(\mu_B,T)$ plane. No model-dependent normalization of either axis is applied. 
For each external calculation, the CJT and SICJT contours are evaluated at the same values of $s/n_B$. 
The $P_{\rm vac}^{(j)}$ trajectories are not displayed separately because their thermodynamic derivatives, and therefore their isentropes, coincide with those obtained from $P_{\rm vac}$.

The CJT and SICJT trajectories remain close over the full range displayed, showing that the thermodynamic ambiguity associated with the pressure prescription produces a smaller effect than the change of the underlying model.
The QM-MF trajectories exhibit a similar ordering and curvature,
particularly for the moderate and large values of $s/n_B$. 
At smaller $s/n_B$, all trajectories enter the high-density region and become more sensitive to the location and treatment of the first-order boundary.

The PQM-MF trajectories show a larger displacement toward high $\mu_B$ and a stronger bending near the transition region. 
This difference indicates that the Polyakov-loop sector and the associated modification of the thermal degrees of freedom can affect the isentropic structure more strongly than the choice among the CJT and SICJT pressure prescriptions. 
The lattice+HRG parameterization passes through a comparable crossover domain, but its global curvature differs from the effective-model trajectories. 
Since it is based on a finite-density Taylor parameterization, this comparison should be regarded as qualitative rather than as a precision test at large $\mu_B/T$.

\twocolumngrid

%%%%%%%%%%%%%%%%%%%
%%%%%%%%%%%%%%%%%%%

%%%%%%%%%%%%%%%%%%%
%%%%%%%%%%%%%%%%%%%

\clearpage

\bibliographystyle{JHEP}
\bibliography{refs}

\providecommand{\href}[2]{#2}\begingroup\raggedright\begin{thebibliography}{10}

\bibitem{Boyanovsky:2006bf}
D.~Boyanovsky, H.~J. de~Vega and D.~J. Schwarz, \emph{{Phase transitions in the early and the present universe}}, \href{https://doi.org/10.1146/annurev.nucl.56.080805.140539}{\emph{Ann. Rev. Nucl. Part. Sci.} {\bfseries 56} (2006) 441} [\href{https://arxiv.org/abs/hep-ph/0602002}{{\ttfamily hep-ph/0602002}}].

\bibitem{Addazi:2022whi}
A.~Addazi, T.~Lundberg, A.~Marcian{\`o}, R.~Pasechnik and M.~{\v{S}}umbera, \emph{{Cosmology from Strong Interactions}}, \href{https://doi.org/10.3390/universe8090451}{\emph{Universe} {\bfseries 8} (2022) 451} [\href{https://arxiv.org/abs/2204.02950}{{\ttfamily 2204.02950}}].

\bibitem{Fraga:2015xha}
E.~S. Fraga, A.~Kurkela and A.~Vuorinen, \emph{{Neutron star structure from QCD}}, \href{https://doi.org/10.1140/epja/i2016-16049-6}{\emph{Eur. Phys. J. A} {\bfseries 52} (2016) 49} [\href{https://arxiv.org/abs/1508.05019}{{\ttfamily 1508.05019}}].

\bibitem{Fukushima:2025ujk}
K.~Fukushima, \emph{{QCD phase diagram and astrophysical implications}}, \href{https://doi.org/10.1016/j.jspc.2025.100066}{\emph{J. Subatomic Part. Cosmol.} {\bfseries 3} (2025) 100066} [\href{https://arxiv.org/abs/2501.01907}{{\ttfamily 2501.01907}}].

\bibitem{Foka:2016vta}
P.~Foka and M.~A. Janik, \emph{{An overview of experimental results from ultra-relativistic heavy-ion collisions at the CERN LHC: Bulk properties and dynamical evolution}}, \href{https://doi.org/10.1016/j.revip.2016.11.002}{\emph{Rev. Phys.} {\bfseries 1} (2016) 154} [\href{https://arxiv.org/abs/1702.07233}{{\ttfamily 1702.07233}}].

\bibitem{Fukushima:2020yzx}
K.~Fukushima, B.~Mohanty and N.~Xu, \emph{{Little-Bang and Femto-Nova in Nucleus-Nucleus Collisions}}, \href{https://doi.org/10.1007/s43673-021-00002-7}{\emph{AAPPS Bull.} {\bfseries 31} (2021) 1} [\href{https://arxiv.org/abs/2009.03006}{{\ttfamily 2009.03006}}].

\bibitem{Stephanov:1998dy}
M.~A. Stephanov, K.~Rajagopal and E.~V. Shuryak, \emph{{Signatures of the tricritical point in QCD}}, \href{https://doi.org/10.1103/PhysRevLett.81.4816}{\emph{Phys. Rev. Lett.} {\bfseries 81} (1998) 4816} [\href{https://arxiv.org/abs/hep-ph/9806219}{{\ttfamily hep-ph/9806219}}].

\bibitem{Stephanov:2008qz}
M.~A. Stephanov, \emph{{Non-Gaussian fluctuations near the QCD critical point}}, \href{https://doi.org/10.1103/PhysRevLett.102.032301}{\emph{Phys. Rev. Lett.} {\bfseries 102} (2009) 032301} [\href{https://arxiv.org/abs/0809.3450}{{\ttfamily 0809.3450}}].

\bibitem{Stephanov:2024xkn}
M.~Stephanov, \emph{{QCD critical point: Recent developments}}, \href{https://doi.org/10.1051/epjconf/202431400042}{\emph{EPJ Web Conf.} {\bfseries 314} (2024) 00042} [\href{https://arxiv.org/abs/2410.02861}{{\ttfamily 2410.02861}}].

\bibitem{Borsanyi:2012cr}
S.~Borsanyi, G.~Endrodi, Z.~Fodor, S.~D. Katz, S.~Krieg, C.~Ratti et~al., \emph{{QCD equation of state at nonzero chemical potential: continuum results with physical quark masses at order $mu^2$}}, \href{https://doi.org/10.1007/JHEP08(2012)053}{\emph{JHEP} {\bfseries 08} (2012) 053} [\href{https://arxiv.org/abs/1204.6710}{{\ttfamily 1204.6710}}].

\bibitem{Borsanyi:2013bia}
S.~Borsanyi, Z.~Fodor, C.~Hoelbling, S.~D. Katz, S.~Krieg and K.~K. Szabo, \emph{{Full result for the QCD equation of state with 2+1 flavors}}, \href{https://doi.org/10.1016/j.physletb.2014.01.007}{\emph{Phys. Lett. B} {\bfseries 730} (2014) 99} [\href{https://arxiv.org/abs/1309.5258}{{\ttfamily 1309.5258}}].

\bibitem{HotQCD:2014kol}
{\scshape HotQCD} collaboration, A.~Bazavov et~al., \emph{{Equation of state in ( 2+1 )-flavor QCD}}, \href{https://doi.org/10.1103/PhysRevD.90.094503}{\emph{Phys. Rev. D} {\bfseries 90} (2014) 094503} [\href{https://arxiv.org/abs/1407.6387}{{\ttfamily 1407.6387}}].

\bibitem{HotQCD:2018pds}
{\scshape HotQCD} collaboration, A.~Bazavov et~al., \emph{{Chiral crossover in QCD at zero and non-zero chemical potentials}}, \href{https://doi.org/10.1016/j.physletb.2019.05.013}{\emph{Phys. Lett. B} {\bfseries 795} (2019) 15} [\href{https://arxiv.org/abs/1812.08235}{{\ttfamily 1812.08235}}].

\bibitem{Nagata:2021ugx}
K.~Nagata, \emph{{Finite-density lattice QCD and sign problem: Current status and open problems}}, \href{https://doi.org/10.1016/j.ppnp.2022.103991}{\emph{Prog. Part. Nucl. Phys.} {\bfseries 127} (2022) 103991} [\href{https://arxiv.org/abs/2108.12423}{{\ttfamily 2108.12423}}].

\bibitem{Fukushima:2010bq}
K.~Fukushima and T.~Hatsuda, \emph{{The phase diagram of dense QCD}}, \href{https://doi.org/10.1088/0034-4885/74/1/014001}{\emph{Rept. Prog. Phys.} {\bfseries 74} (2011) 014001} [\href{https://arxiv.org/abs/1005.4814}{{\ttfamily 1005.4814}}].

\bibitem{Braun:2018svj}
J.~Braun, M.~Leonhardt and J.~M. Pawlowski, \emph{{Renormalization group consistency and low-energy effective theories}}, \href{https://doi.org/10.21468/SciPostPhys.6.5.056}{\emph{SciPost Phys.} {\bfseries 6} (2019) 056} [\href{https://arxiv.org/abs/1806.04432}{{\ttfamily 1806.04432}}].

\bibitem{Zhang:2017icm}
H.~Zhang, D.~Hou, T.~Kojo and B.~Qin, \emph{{Functional renormalization group study of the quark-meson model with $\omega$ meson}}, \href{https://doi.org/10.1103/PhysRevD.96.114029}{\emph{Phys. Rev. D} {\bfseries 96} (2017) 114029} [\href{https://arxiv.org/abs/1709.05654}{{\ttfamily 1709.05654}}].

\bibitem{Osman:2024xkm}
M.~Osman, D.~Hou, W.~Wang and H.~Zhang, \emph{{Functional renormalization group study of the quark-meson model with $\omega $ and $\rho $ vector mesons}}, \href{https://doi.org/10.1140/epjc/s10052-025-14493-3}{\emph{Eur. Phys. J. C} {\bfseries 85} (2025) 804} [\href{https://arxiv.org/abs/2402.15474}{{\ttfamily 2402.15474}}].

\bibitem{Drews:2013hha}
M.~Drews, T.~Hell, B.~Klein and W.~Weise, \emph{{Thermodynamic phases and mesonic fluctuations in a chiral nucleon-meson model}}, \href{https://doi.org/10.1103/PhysRevD.88.096011}{\emph{Phys. Rev. D} {\bfseries 88} (2013) 096011} [\href{https://arxiv.org/abs/1308.5596}{{\ttfamily 1308.5596}}].

\bibitem{Fu:2018qsk}
W.-j. Fu, J.~M. Pawlowski and F.~Rennecke, \emph{{Strangeness Neutrality and QCD Thermodynamics}}, \href{https://doi.org/10.21468/SciPostPhysCore.2.1.002}{\emph{SciPost Phys. Core} {\bfseries 2} (2020) 002} [\href{https://arxiv.org/abs/1808.00410}{{\ttfamily 1808.00410}}].

\bibitem{Hansen:2019lnf}
H.~Hansen, R.~Stiele and P.~Costa, \emph{{Quark and Polyakov-loop correlations in effective models at zero and nonvanishing density}}, \href{https://doi.org/10.1103/PhysRevD.101.094001}{\emph{Phys. Rev. D} {\bfseries 101} (2020) 094001} [\href{https://arxiv.org/abs/1904.08965}{{\ttfamily 1904.08965}}].

\bibitem{Andersen:2004fp}
J.~O. Andersen and M.~Strickland, \emph{{Resummation in hot field theories}}, \href{https://doi.org/10.1016/j.aop.2004.09.017}{\emph{Annals Phys.} {\bfseries 317} (2005) 281} [\href{https://arxiv.org/abs/hep-ph/0404164}{{\ttfamily hep-ph/0404164}}].

\bibitem{Curtin:2016urg}
D.~Curtin, P.~Meade and H.~Ramani, \emph{{Thermal Resummation and Phase Transitions}}, \href{https://doi.org/10.1140/epjc/s10052-018-6268-0}{\emph{Eur. Phys. J. C} {\bfseries 78} (2018) 787} [\href{https://arxiv.org/abs/1612.00466}{{\ttfamily 1612.00466}}].

\bibitem{Niemi:2021qvp}
L.~Niemi, P.~Schicho and T.~V.~I. Tenkanen, \emph{{Singlet-assisted electroweak phase transition at two loops}}, \href{https://doi.org/10.1103/PhysRevD.103.115035}{\emph{Phys. Rev. D} {\bfseries 103} (2021) 115035} [\href{https://arxiv.org/abs/2103.07467}{{\ttfamily 2103.07467}}].

\bibitem{Cornwall:1974vz}
J.~M. Cornwall, R.~Jackiw and E.~Tomboulis, \emph{{Effective Action for Composite Operators}}, \href{https://doi.org/10.1103/PhysRevD.10.2428}{\emph{Phys. Rev. D} {\bfseries 10} (1974) 2428}.

\bibitem{Baym:1962sx}
G.~Baym, \emph{{Selfconsistent approximation in many body systems}}, \href{https://doi.org/10.1103/PhysRev.127.1391}{\emph{Phys. Rev.} {\bfseries 127} (1962) 1391}.

\bibitem{Petropoulos:1998gt}
N.~Petropoulos, \emph{{Linear sigma model and chiral symmetry at finite temperature}}, \href{https://doi.org/10.1088/0954-3899/25/11/305}{\emph{J. Phys. G} {\bfseries 25} (1999) 2225} [\href{https://arxiv.org/abs/hep-ph/9807331}{{\ttfamily hep-ph/9807331}}].

\bibitem{Lenaghan:1999si}
J.~T. Lenaghan and D.~H. Rischke, \emph{{The O(N) model at finite temperature: Renormalization of the gap equations in Hartree and large N approximation}}, \href{https://doi.org/10.1088/0954-3899/26/4/309}{\emph{J. Phys. G} {\bfseries 26} (2000) 431} [\href{https://arxiv.org/abs/nucl-th/9901049}{{\ttfamily nucl-th/9901049}}].

\bibitem{Lenaghan:2000ey}
J.~T. Lenaghan, D.~H. Rischke and J.~Schaffner-Bielich, \emph{{Chiral symmetry restoration at nonzero temperature in the SU(3)(r) x SU(3)(l) linear sigma model}}, \href{https://doi.org/10.1103/PhysRevD.62.085008}{\emph{Phys. Rev. D} {\bfseries 62} (2000) 085008} [\href{https://arxiv.org/abs/nucl-th/0004006}{{\ttfamily nucl-th/0004006}}].

\bibitem{Roder:2003uz}
D.~Roder, J.~Ruppert and D.~H. Rischke, \emph{{Chiral symmetry restoration in linear sigma models with different numbers of quark flavors}}, \href{https://doi.org/10.1103/PhysRevD.68.016003}{\emph{Phys. Rev. D} {\bfseries 68} (2003) 016003} [\href{https://arxiv.org/abs/nucl-th/0301085}{{\ttfamily nucl-th/0301085}}].

\bibitem{Pilaftsis:2013xna}
A.~Pilaftsis and D.~Teresi, \emph{{Symmetry Improved CJT Effective Action}}, \href{https://doi.org/10.1016/j.nuclphysb.2013.06.004}{\emph{Nucl. Phys. B} {\bfseries 874} (2013) 594} [\href{https://arxiv.org/abs/1305.3221}{{\ttfamily 1305.3221}}].

\bibitem{Mao:2013gva}
H.~Mao, \emph{{On the symmetry improved CJT formalism in the $O(4)$ linear sigma model}}, \href{https://doi.org/10.1016/j.nuclphysa.2014.02.011}{\emph{Nucl. Phys. A} {\bfseries 925} (2014) 185} [\href{https://arxiv.org/abs/1305.4329}{{\ttfamily 1305.4329}}].

\bibitem{Pilaftsis:2015bbs}
A.~Pilaftsis and D.~Teresi, \emph{{Symmetry-Improved 2PI Approach to the Goldstone-Boson IR Problem of the SM Effective Potential}}, \href{https://doi.org/10.1016/j.nuclphysb.2016.03.018}{\emph{Nucl. Phys. B} {\bfseries 906} (2016) 381} [\href{https://arxiv.org/abs/1511.05347}{{\ttfamily 1511.05347}}].

\bibitem{Garbrecht:2015cla}
B.~Garbrecht and P.~Millington, \emph{{Constraining the effective action by a method of external sources}}, \href{https://doi.org/10.1016/j.nuclphysb.2016.02.022}{\emph{Nucl. Phys. B} {\bfseries 906} (2016) 105} [\href{https://arxiv.org/abs/1509.07847}{{\ttfamily 1509.07847}}].

\bibitem{Kobayashi:1970ji}
M.~Kobayashi and T.~Maskawa, \emph{{Chiral symmetry and eta-x mixing}}, \href{https://doi.org/10.1143/PTP.44.1422}{\emph{Prog. Theor. Phys.} {\bfseries 44} (1970) 1422}.

\bibitem{Kobayashi:1971qz}
M.~Kobayashi, H.~Kondo and T.~Maskawa, \emph{{Symmetry breaking of the chiral u(3) x u(3) and the quark model}}, \href{https://doi.org/10.1143/PTP.45.1955}{\emph{Prog. Theor. Phys.} {\bfseries 45} (1971) 1955}.

\bibitem{tHooft:1976rip}
G.~'t~Hooft, \emph{{Symmetry Breaking Through Bell-Jackiw Anomalies}}, \href{https://doi.org/10.1103/PhysRevLett.37.8}{\emph{Phys. Rev. Lett.} {\bfseries 37} (1976) 8}.

\bibitem{tHooft:1976snw}
G.~'t~Hooft, \emph{{Computation of the Quantum Effects Due to a Four-Dimensional Pseudoparticle}}, \href{https://doi.org/10.1103/PhysRevD.14.3432}{\emph{Phys. Rev. D} {\bfseries 14} (1976) 3432}.

\bibitem{Osipov:2002wj}
A.~A. Osipov and B.~Hiller, \emph{{Path integral bosonization of the 't Hooft determinant: fluctuations and multiple vacua}}, \href{https://doi.org/10.1016/S0370-2693(02)02041-5}{\emph{Phys. Lett. B} {\bfseries 539} (2002) 76} [\href{https://arxiv.org/abs/hep-ph/0204182}{{\ttfamily hep-ph/0204182}}].

\bibitem{Osipov:2003xu}
A.~A. Osipov and B.~Hiller, \emph{{Path integral bosonization of the 't Hooft determinant: Quasiclassical corrections}}, \href{https://doi.org/10.1140/epjc/s2004-01779-3}{\emph{Eur. Phys. J. C} {\bfseries 35} (2004) 223} [\href{https://arxiv.org/abs/hep-th/0307035}{{\ttfamily hep-th/0307035}}].

\bibitem{Osipov:2006xa}
A.~A. Osipov, B.~Hiller, J.~Moreira and A.~H. Blin, \emph{{Stationary phase corrections in the process of bosonization of multi-quark interactions}}, \href{https://doi.org/10.1140/epjc/s2006-02488-7}{\emph{Eur. Phys. J. C} {\bfseries 46} (2006) 225} [\href{https://arxiv.org/abs/hep-ph/0601074}{{\ttfamily hep-ph/0601074}}].

\bibitem{Blaizot:2021ikl}
J.-P. Blaizot, J.~M. Pawlowski and U.~Reinosa, \emph{{Functional renormalization group and 2PI effective action formalism}}, \href{https://doi.org/10.1016/j.aop.2021.168549}{\emph{Annals Phys.} {\bfseries 431} (2021) 168549} [\href{https://arxiv.org/abs/2102.13628}{{\ttfamily 2102.13628}}].

\bibitem{Kapusta:2006pm}
J.~I. Kapusta and C.~Gale, \emph{{Finite-temperature field theory: Principles and applications}}, Cambridge Monographs on Mathematical Physics. Cambridge University Press, 2011, \href{https://doi.org/10.1017/CBO9780511535130}{10.1017/CBO9780511535130}.

\bibitem{Schaefer:2008hk}
B.-J. Schaefer and M.~Wagner, \emph{{The Three-flavor chiral phase structure in hot and dense QCD matter}}, \href{https://doi.org/10.1103/PhysRevD.79.014018}{\emph{Phys. Rev. D} {\bfseries 79} (2009) 014018} [\href{https://arxiv.org/abs/0808.1491}{{\ttfamily 0808.1491}}].

\bibitem{Wetterich:1992yh}
C.~Wetterich, \emph{{Exact evolution equation for the effective potential}}, \href{https://doi.org/10.1016/0370-2693(93)90726-X}{\emph{Phys. Lett. B} {\bfseries 301} (1993) 90} [\href{https://arxiv.org/abs/1710.05815}{{\ttfamily 1710.05815}}].

\bibitem{Guan:2025tmf}
Y.~Guan, M.~Kawaguchi, S.~Matsuzaki and A.~Tomiya, \emph{{Columbia plot based on symmetry-improved CJT formalism in linear sigma model}},  \href{https://arxiv.org/abs/2508.02257}{{\ttfamily 2508.02257}}.

\bibitem{Scavenius:2000qd}
O.~Scavenius, A.~Mocsy, I.~N. Mishustin and D.~H. Rischke, \emph{{Chiral phase transition within effective models with constituent quarks}}, \href{https://doi.org/10.1103/PhysRevC.64.045202}{\emph{Phys. Rev. C} {\bfseries 64} (2001) 045202} [\href{https://arxiv.org/abs/nucl-th/0007030}{{\ttfamily nucl-th/0007030}}].

\bibitem{Kawaguchi:2020qvg}
M.~Kawaguchi, S.~Matsuzaki and A.~Tomiya, \emph{{Analysis of nonperturbative flavor violation at chiral crossover criticality in QCD}}, \href{https://doi.org/10.1103/PhysRevD.103.054034}{\emph{Phys. Rev. D} {\bfseries 103} (2021) 054034} [\href{https://arxiv.org/abs/2005.07003}{{\ttfamily 2005.07003}}].

\bibitem{Cohen:2003kd}
T.~D.~. Cohen, \emph{{Functional integrals for QCD at nonzero chemical potential and zero density}}, \href{https://doi.org/10.1103/PhysRevLett.91.222001}{\emph{Phys. Rev. Lett.} {\bfseries 91} (2003) 222001} [\href{https://arxiv.org/abs/hep-ph/0307089}{{\ttfamily hep-ph/0307089}}].

\bibitem{Fu:2016tey}
W.-j. Fu, J.~M. Pawlowski, F.~Rennecke and B.-J. Schaefer, \emph{{Baryon number fluctuations at finite temperature and density}}, \href{https://doi.org/10.1103/PhysRevD.94.116020}{\emph{Phys. Rev. D} {\bfseries 94} (2016) 116020} [\href{https://arxiv.org/abs/1608.04302}{{\ttfamily 1608.04302}}].

\bibitem{Fu:2019hdw}
W.-j. Fu, J.~M. Pawlowski and F.~Rennecke, \emph{{QCD phase structure at finite temperature and density}}, \href{https://doi.org/10.1103/PhysRevD.101.054032}{\emph{Phys. Rev. D} {\bfseries 101} (2020) 054032} [\href{https://arxiv.org/abs/1909.02991}{{\ttfamily 1909.02991}}].

\bibitem{Blaschke:2017boi}
D.~Blaschke and D.~Ebert, \emph{{Variational path-integral approach to back-reactions of composite mesons in the Nambu{\textendash}Jona-Lasinio model}}, \href{https://doi.org/10.1016/j.nuclphysb.2017.06.013}{\emph{Nucl. Phys. B} {\bfseries 921} (2017) 753} [\href{https://arxiv.org/abs/1703.08964}{{\ttfamily 1703.08964}}].

\bibitem{Rennecke:2021ovl}
F.~Rennecke and R.~D. Pisarski, \emph{{Moat Regimes in QCD and their Signatures in Heavy-Ion Collisions}}, \href{https://doi.org/10.22323/1.400.0016}{\emph{PoS} {\bfseries CPOD2021} (2022) 016} [\href{https://arxiv.org/abs/2110.02625}{{\ttfamily 2110.02625}}].

\bibitem{Fu:2024rto}
W.-j. Fu, J.~M. Pawlowski, R.~D. Pisarski, F.~Rennecke, R.~Wen and S.~Yin, \emph{{QCD moat regime and its real-time properties}}, \href{https://doi.org/10.1103/PhysRevD.111.094026}{\emph{Phys. Rev. D} {\bfseries 111} (2025) 094026} [\href{https://arxiv.org/abs/2412.15949}{{\ttfamily 2412.15949}}].

\bibitem{Mitter:2013fxa}
M.~Mitter and B.-J. Schaefer, \emph{{Fluctuations and the axial anomaly with three quark flavors}}, \href{https://doi.org/10.1103/PhysRevD.89.054027}{\emph{Phys. Rev. D} {\bfseries 89} (2014) 054027} [\href{https://arxiv.org/abs/1308.3176}{{\ttfamily 1308.3176}}].

\bibitem{Schaefer:2009ui}
B.-J. Schaefer, M.~Wagner and J.~Wambach, \emph{{Thermodynamics of (2+1)-flavor QCD: Confronting Models with Lattice Studies}}, \href{https://doi.org/10.1103/PhysRevD.81.074013}{\emph{Phys. Rev. D} {\bfseries 81} (2010) 074013} [\href{https://arxiv.org/abs/0910.5628}{{\ttfamily 0910.5628}}].

\bibitem{Herbst:2013ail}
T.~K. Herbst, J.~M. Pawlowski and B.-J. Schaefer, \emph{{Phase structure and thermodynamics of QCD}}, \href{https://doi.org/10.1103/PhysRevD.88.014007}{\emph{Phys. Rev. D} {\bfseries 88} (2013) 014007} [\href{https://arxiv.org/abs/1302.1426}{{\ttfamily 1302.1426}}].

\bibitem{Borsanyi:2010bp}
{\scshape Wuppertal-Budapest} collaboration, S.~Borsanyi, Z.~Fodor, C.~Hoelbling, S.~D. Katz, S.~Krieg, C.~Ratti et~al., \emph{{Is there still any $T_c$ mystery in lattice QCD? Results with physical masses in the continuum limit III}}, \href{https://doi.org/10.1007/JHEP09(2010)073}{\emph{JHEP} {\bfseries 09} (2010) 073} [\href{https://arxiv.org/abs/1005.3508}{{\ttfamily 1005.3508}}].

\bibitem{Bazavov:2011nk}
A.~Bazavov et~al., \emph{{The chiral and deconfinement aspects of the QCD transition}}, \href{https://doi.org/10.1103/PhysRevD.85.054503}{\emph{Phys. Rev. D} {\bfseries 85} (2012) 054503} [\href{https://arxiv.org/abs/1111.1710}{{\ttfamily 1111.1710}}].

\bibitem{Motta:2020cbr}
M.~Motta, R.~Stiele, W.~M. Alberico and A.~Beraudo, \emph{{Isentropic evolution of the matter in heavy-ion collisions and the search for the critical endpoint}}, \href{https://doi.org/10.1140/epjc/s10052-020-8218-x}{\emph{Eur. Phys. J. C} {\bfseries 80} (2020) 770} [\href{https://arxiv.org/abs/2003.04734}{{\ttfamily 2003.04734}}].

\end{thebibliography}\endgroup

\end{document}